\documentclass[a4,useAMS,usenatbib,usegraphicx]{mn2e}
\usepackage{subfig}
\usepackage{lscape}
\usepackage{natbib}
\usepackage{amssymb,amsmath}
\usepackage{fixltx2e}
\usepackage{longtable}
\usepackage[T1]{fontenc}
\usepackage{hyperref}
\hypersetup{
    colorlinks=true,
    linkcolor=blue,
    filecolor=magenta,      
    urlcolor=blue,
}

\title[Detached H\textsc{i} clouds]{Attack of the Flying Snakes : Formation of Isolated H\textsc{i} Clouds By Fragmentation of Long Streams}

\author[R. Taylor, J. I. Davies, P. J\'{a}chym, O. Keenan, R. F. Minchin, J. Palou\v{s}, R. Smith, R. W\"{u}nsch]{R. Taylor$^{1}$\thanks{Email: rhyst@naic.edu}, J. I. Davies$^2$, P. J\'{a}chym$^1$, O. Keenan$^2$, R. F. Minchin$^3$, J. Palou\v{s}$^1$, \newauthor R. Smith$^4$, R. W\"{u}nsch$^1$\\
$^1$Astronomical Institute of the Czech Academy of Sciences, Bocni II 1401/1a, 141 31 Praha 4\\
$^2$School of Physics \& Astronomy, Cardiff University, Queens Buildings, The Parade, Cardiff CF24 3AA, U.K.\\
$^3$Arecibo Observatory, HC03 Box 53995, Arecibo, Puerto Rico 00612\\
$^4$Yonsei University, Graduate School of Earth System Sciences-Astronomy-Atmospheric Sciences, Yonsei-ro 50, Seoul 120-749, Korea;\\}

\begin{document}

\newcommand{\HI}{H\textsc{i}}
\newcommand{\Msolar}{M$_{\odot}$}
\newcommand{\kms}{km\,s$^{-1}$}

\date{2016}

\pagerange{\pageref{firstpage}--\pageref{lastpage}} \pubyear{2015}

\maketitle

\label{firstpage}

\begin{abstract}
The existence of long ($>$ 100 kpc) \HI{} streams and small ($<$ 20 kpc) free-floating \HI{} clouds is well-known. While the formation of the streams has been investigated extensively, and the isolated clouds are often purported to be interaction debris, little research has been done on the formation of optically dark \HI{} clouds that are not part of a larger stream. One possibility is that such features result from the fragmentation of more extended streams, while another idea is that they are primordial, optically dark galaxies. We test the validity of the fragmentation scenario (via harassment) using numerical simulations. In order to compare our numerical models with observations, we present catalogues of both the known long \HI{} streams (42 objects) and free-floating \HI{} clouds suggested as dark galaxy candidates (51 objects). In particular, we investigate whether it is possible to form compact features with high velocity widths ($>$ 100 \kms{}), similar to observed clouds which are otherwise  intriguing dark galaxy candidates. We find that producing such features is possible but extremely unlikely, occurring no more than 0.2\% of the time in our simulations. In contrast, we find that genuine dark galaxies could be extremely stable to harassment and remain detectable even after 5 Gyr in the cluster environment (with the important caveat that our simulations only explore harassment and do not yet include the intracluster medium, heating and cooling, or star formation). We also discuss the possibility that such objects could be the progenitors of recently discovered ultra diffuse galaxies.

\end{abstract}

\begin{keywords}
galaxies: evolution - surveys: galaxies.
\end{keywords}

\section{Introduction}
\label{sec:intro}
The ``missing satellite'' problem is a discrepancy between the number of galaxies detected in the Local Group and the number predicted in cold dark matter (CDM) simulations. In particular, \cite{moore} found that such simulations easily reproduced the correct number of low-mass galaxies on the mass scale of a cluster, but failed on the scale of individual galaxies (with the simulations predicting about a factor 10 more dwarf galaxies than were actually observed). There are two broad solutions to this problem. Either the standard CDM model is fundamentally flawed (perhaps simply because the baryonic physics is not accounted for correctly), a possibility which we do not consider here, or the small dark matter halos do exist but are not detected observationally. In the latter case, one idea is that some of the smaller dark matter halos never accumulate sufficient gas for star formation, but may have sufficient gas to be detectable in neutral atomic hydrogen (\HI{}) surveys (\citealt{d06}).

Over the years, a number of objects have been detected which have been proposed as (optically) `dark galaxy' candidates; for a summary of these we refer the reader to section \ref{sec:obs2}. None of the candidates have been widely regarded as ideal. It is well-known that large amounts of gas can be removed to large distances from their parent galaxies (see section \ref{sec:streams}), a fact that makes it very difficult to decide if a candidate is a primordial gas cloud embedded in a dark matter halo, or simply unusual tidal debris, possibly mimicking the effects of rotation (\citealt{bekki}, \citealt{duc}). None of the candidates are thought to be sufficiently isolated that an interaction origin can be entirely ruled out.

Although optically bright tidal dwarf galaxies have been investigated in considerable detail (e.g. \citealt{ductdgs} and references therein), their abundance of stars and low dynamical to baryonic mass ratios (no higher than 2-3 - \citealt{bou07}) makes them distinctly different objects to the clouds we describe here. Very little work has been done to numerically simulate the formation of dark galaxy candidates via interactions, and is largely limited to \cite{bekki} and \cite{duc}. \cite{duc} describe the formation of the well-known VIRGOHI21, a 14 kpc long \HI{} overdensity with a high velocity gradient embedded in a 200 kpc stream. However in their model the velocity gradient of the feature is much shallower, not actually reaching the observed total velocity width of 200 \kms{} even over the $\gtrsim$100 kpc of the simulated feature. \cite{bekki} describe the formation of objects which have double-horn velocity profiles, normally taken to be a feature resulting from a flat rotation curve, but in their simulated clouds resulting from streaming motions along the line of sight. However their objects are, as in \citealt{duc}, extremely extended ($\gtrsim$ 100 kpc) and are far larger than most galaxies. Thus, owing to the observational evidence available at the time, both existing studies have examined the formation of objects with very extended components and shallow velocity gradients. 

What appears to be lacking is a mechanism by which much smaller ($\sim$20 kpc) features may be formed without an associated extended stream but still showing the high velocity widths which may be mistaken for (or truly indicative of) rotation. Small ($<$ 20 kpc diameter), low mass ($\sim$10$^{7}$ \Msolar{}) \HI{} clouds with wide velocity widths ($\sim$150 \kms{}) were discovered in the Virgo cluster by \cite{me12} - hereafter AGES V - as part of the Arecibo Galaxy Environment Survey (see \citealt{auld}). While the cluster environment means they can hardly be described as `isolated', no associated extended streams are detected that could indicate their possible origin. Indeed, no such extended features were detected in the AGES 20 deg$^{2}$ region at all, despite the fact that such streams have been detected (and resolved by Arecibo) in other parts of the cluster (see section \ref{sec:streams}) and many galaxies there are highly \HI{} deficient - they have much less gas than similar galaxies in the field (see section \ref{sec:gasfate}).

The combination of the lack of extended \HI{} features and high deficiency of many galaxies looks especially strange given the results of numerical simulations. Many previous works have found that the gas deficiency can be explained by ram pressure stripping, but this typically results in the formation of very extended ($>$\,100 kpc) streams that should be detectable to contemporary \HI{} surveys. For example, \cite{roe08} report that typical column densities are approximately 2$\times$10$^{18}$ cm$^{-2}$, (4$\sigma$ for the ALFALFA survey - \citealt{m33}; 13$\sigma$ for AGES - \citealt{olivia}) which can persit over timescales $\gtrsim$ 500-1000 Myr at distances $>$ 100 kpc from the stripped galaxy. \cite{ton10} predict even higher column densities ($>$ 10$^{19}$ cm$^{-2}$) at similar distances and timescales. Thus if such extended features were common, they should have been detected.

There are many possible reasons why long \HI{} streams appear to be rare, despite the high fraction of galaxies with significant \HI{} deficiency (76\% with \HI{}$_{def}$ $>$ 0.3, \citealt{me12}). The streams may expand and become undetectable simply due to their own velocity dispersion (though cooling of the gas makes the tails narrower - hence \citealt{ton10}, who use cooling, predict higher column densities than \citealt{roe08}, who do not), some of the gas may cool and form molecular gas (\citealt{pavel}) or stars (e.g. \citealt{kap09}), or it might be heated and ionized by the hot intracluster medium (ICM). \cite{ton10} mention this latter possibility, but argue against significant heat conduction from the ICM as their simulations, which neglect it, successfully reproduce the very short ($<$ 30 kpc) streams described in \cite{chung}. However the observed rarity of the long streams predicted by their simulations may suggest that conduction is important, at least for the low column density gas (but see section \ref{sec:gasfate}). Alternatively the observed streams could simply be an early phase of gas stripping, but this does not readily explain the extreme rarity of the streams (we discuss this further in sections \ref{sec:gasfate} and \ref{sec:virgoclouds}).

We here explore another environmental effect that has thus far largely been neglected in studying the evolution of the stripped gas : harassment - repeated rapid tidal encounters with cluster galaxies (\citealt{mooreharas}). While no extended \HI{} features were detected in the AGES areas, a total of eight clouds (as described above) were found. If these small clouds are not primordial objects, then there are two possibilities~: 1) They were detached from their parent galaxies without the formation of a more extended \HI{} component (i.e. not via ram pressure stripping, which predicts much larger features\footnote{The only possible exception would be if the ram pressure occurred only for a very short timescale, perhaps due to local overdensities in the ICM.}); 2) The \HI{} stream has since dispersed and these are the last surviving relics of what were initially much larger structures, i.e. those predicted by ram-pressure stripping simulations but not observed in reality.

It is the latter possibility we consider here. Our primary goal is to establish if, in principle, a long \HI{} stream can fragment to produce detached clouds that match the observed properties of dark galaxy candidates, particularly those described in AGES V. This could then explain why these relatively compact clouds are detected without the much larger features seen in previous simulations, and potentially reconcile the absence of streams with the observed \HI{} deficiencies of many galaxies.

Here we study only the fragmentation process of the stream in a cluster environment. We leave the more ambitious prospect of modelling the formation process of a stream to a future study, though we do base the simulated streams on those known to exist. We also seek to establish whether the resulting clouds can survive for long enough that we would expect a reasonable chance of detecting them. We do not simulate other possible mechanisms for the formation of the clouds in this study.

We regard this current work as the first in a series, in which we will examine the different physical processes both separately and acting in combination. Ram pressure stripping has been the subject of many previous studies (e.g. \citealt{vol01}, \citealt{roe08}, \citealt{kap09}, \citealt{ton10}), but far less work has been done on the effects of harassment - particularly regarding the stripped gas. Our main interest is in producing the high velocity widths of the observed clouds, which the results of \cite{bekki} and \cite{duc} suggest is something harassment may be able to accomplish (albeit with shallower velocity gradients in the previous works). Therefore we take the unusual approach of examining harassment without ram pressure stripping. While it may seem strange to ignore such an important process, it is - as we shall show - probably equally wrong to consider ram pressure stripping without harassment (as most previous investigations have done), at least over certain timescales and locations within a cluster.

The rest of this paper is organised as follows. In section \ref{sec:obs1}, we present a catalogue of  the largest observed streams and comment on their possible origins given the observational evidence available. We present a similar analysis for the `dark galaxy' candidates in section \ref{sec:obs2}. We use these observational results to help establish plausible initial and final conditions for the simulations we describe in section \ref{sec:sims}, and we also discuss possible explanations for the objects based solely on the observations. Our conclusions are presented and discussed in section \ref{sec:conc}.

\section{Observations I : A catalogue of extended H\textsc{i} structures}
\label{sec:obs1}
Determining the nature of the streams we wish to simulate, as well as the isolated clouds we have postulated may result from their fragmentation, is crucial. We want to establish whether there is \textit{any} plausible stream (i.e. with parameters comparable to those observed) which can fragment to produce the observed clouds. Observations will therefore form the starting point for the properties of the initial stream; if it is necessary to alter these parameters, then it is essential to consider how far we must deviate them from those of actual streams before the observed clouds may form.

\label{sec:streams}
We have conducted literature searches for both long \HI{} streams (our initial conditions) and isolated clouds (our hypothesised end state). For the `streams', our criteria to include them in our catalogue is very simple : the \HI{} must span at least 100 kpc in projection. This is a somewhat arbitrary number, but it is designed to ensure that the resulting clouds are likely to be sufficiently far from their parent galaxy that they will be clearly separated from it observationally (see section \ref{sec:virgoclouds}). It has the additional advantage of excluding almost all galaxies - thus reducing the features to a manageable number - without having to define any more complex parameter such as axial ratio. The disadvantage is that it is distance-dependent, a quantity on which there is often a large margin of error.

Our resulting catalogue is shown in table \ref{tab:stab}. Some additional details are given in appendix \ref{sec:ap1}. The most important point is that there are caveats to almost every parameter - nonetheless, it is certainly better than having no catalogue at all. A major limitation is that the raw data is only rarely available, so we must rely on the author's measurements rather than performing them in a consistent way ourselves. Worse, the column density sensitivities of the different observations vary by many orders of magnitude (10$^{17}$ - 10$^{20}$ cm$^{-2}$). Precise measurements at the same column density levels are impossible without the raw data; as it is, we often have to resort to using author's figures when their measurement procedure is not clearly defined (i.e. what does the length of the stream refer to - a particular feature outside the parent galaxy, or the whole \HI{} envelope ?)\footnote{While far from ideal, the errors in such an approach should be smaller than the intrinsic variation in the nature of the streams and the unknown projection effects.}. Our procedure in this case is to consider the whole envelope at whatever sensitivity the data has. We convert the angular size into diameter using the median redshift-independent distance determination given in the NASA Extragalactic Database (NED), or if this is not available, we assume Hubble flow with H$_{o}$ = 71 \kms{}Mpc$^{-1}$.

\begin{table*}
\centering
\tiny
\caption[stab]{Major properties, where available, of all \HI{} features spanning $>$ 100 kpc in projected extent. The columns are as follows : (1) Name of object or parent galaxy in major catalogue; (2) Description of environment; (3) Code describing the nature of the object : 0 - undisturbed giant galaxy; 1 - stream from one galaxy, or ring (R) around one galaxy - for rings, length refers to the circumference of the ring and diameter to its thickness; 2 - bridge between two galaxies; 3 - filament or envelope containing three or more galaxies, complex, or ring with multiple (but discrete) optical counterparts; 4 - collection of discrete clouds; 5 - unclear; 6 - stream of both \HI{} and stars); (4) Angular length of the \HI{} in arcminutes; (5) Typical diameter in arcminutes, or range if highly variable; (6) Total velocity width of the feature in \kms{}; (7) Flux of the object where available; (8) Assumed distance in Mpc; (9) Total \HI{} mass in solar masses; (10) Projected physical length in kpc; (11) Projected physical diameter or range of diameters in kpc; (12) Major reference, omitting `et al' for the sake of space. For explanations of the various flags in each column, see appendix \ref{sec:ap1}.}
\label{tab:stab}
\begin{tabular}{c c c c c c c c c c c c c c c c c c c c c c c c c c c c c c c c c c c c }
  \hline
  \multicolumn{1}{c}{(1)} &
  \multicolumn{1}{c}{(2)} &
  \multicolumn{1}{c}{(3)} &
  \multicolumn{1}{c}{(4)} &
  \multicolumn{1}{c}{(5)} &
  \multicolumn{1}{c}{(6)} &
  \multicolumn{1}{c}{(7)} &
  \multicolumn{1}{c}{(8)} &
  \multicolumn{1}{c}{(9)} &
  \multicolumn{1}{c}{(10)} &
  \multicolumn{1}{c}{(11)} &
  \multicolumn{1}{c}{(12)} \\
  \multicolumn{1}{c}{Name} &
  \multicolumn{1}{c}{Environment} &
  \multicolumn{1}{c}{Code} &
  \multicolumn{1}{c}{Length} &
  \multicolumn{1}{c}{Diameter} &
  \multicolumn{1}{c}{Velocity width} &
  \multicolumn{1}{c}{Flux} &
  \multicolumn{1}{c}{Distance} &
  \multicolumn{1}{c}{M\HI{}} &
  \multicolumn{1}{c}{Length} &
  \multicolumn{1}{c}{Diameter} &
  \multicolumn{1}{c}{Reference} \\
  \multicolumn{1}{c}{} &
  \multicolumn{1}{c}{} &
  \multicolumn{1}{c}{} &
  \multicolumn{1}{c}{(arcmin)} &
  \multicolumn{1}{c}{(arcmin)} &
  \multicolumn{1}{c}{(\kms{})} &
  \multicolumn{1}{c}{Jy \kms{}} &
  \multicolumn{1}{c}{Mpc} &
  \multicolumn{1}{c}{\Msolar{}} &
  \multicolumn{1}{c}{(kpc)} &
  \multicolumn{1}{c}{(kpc)} &
  \multicolumn{1}{c}{} \\  
  
\hline
  VIRGOHI21 & Virgo & 1 & 50 & 2.4 & 463 & 2.7 & 17.0 & 1.8E8 & 250 & 11 & Minchin 2007\\
  HI1225+01 & Near Virgo & 1 & 37 & 1.3-8.8 & 60 & 42.37T & 20.0 & 4.0E9 & 215 & 8-51 & Giovanelli 1989\\
  Koopmann & Virgo & 1B & 100 & $<$3.5 & 290 & 5.9 & 17.0 & 4.0E8 & 500 & $<$17 & Koopmann 2008\\
  NGC3193Stream & Group & 1B & 41 & 1.95 & 300 & 3.61 & 25.0 & 5.3E8 & 300 & 14 & Serra 2013\\
  LeoRing & Group & 3RB & 236 & $<$3.5-9 & 338G & 70.9 & 11.1 & 2.0E9 & 762 & $<$11-29 & Stierwalt 2009\\
  KentComplex & Virgo & 4 & 35 & 2.7V & 127 & 7.83 & 17.0 & 5.3E8 & 173 & 13 & Kent 2009\\
  M33/M31 & Local Group & 2B & 1200 & 300 & 121 & ? & 0.75 & >2E6 & 260 & 65 & Braun 2004\\
  NGC877 & Group & 3 & 20U & 16U & $\sim$500G & 40.8E & 50.0 & 2.4E10 & 291 & 232 & Lee-Waddel 2014\\
  NGC7448 & Group & 3 & 47 & $<$3.5 & 342 & 144.0E & 28.6 & 2.8E10 & 391 & 29 & Taylor 2014\\
  Malin1 & Field & 0 &  & 1.75 & 455 & 3.5 & 348.0 & 1E11 &  & 178 & Bothun 1987\\
  NGC262 & Field & 0,1 &  & 12 & 60 & 17.78 & 63.0 & 1.7E10 &  & 220 & Morris 1980\\
  NGC4388Plume & Virgo & 1 & 25 & 2.5-7.5 & 550 & 5.63D & 17.0 & 3.8E8 & 124 & 12-37 & Oosterloo 2005\\
  MagallenStream & Local Group & 5 & 12000 & 600-1800 & 700 & 6.9E5D & 0.055 & 4.9E8 & 192 & 10-30 & For 2014\\
  Vela & Group & 4 & 15 & 6 & 167 & 10.34 & 38.0 & 3.5E9 & 166 & 66 & English 2010\\
  HPJ0731-69 & Group & 1 & 40U & $<$15 & 270 & 18.0 & 15.5 & 1.0E9 & 180 & $<$68 & Ryder 2001\\
  VGS\_31 & Void & 3 & 6 & $<$0.5 & 200E &  & 87.0 &  & 152 & $<$13 & Beygu 2013\\
  ESO381-47 & Group ETG & 3R & 14 & 5 & 105 & 7.6 & 73.4 & 9.7E9 & 299 & 106 & Donovan 2009\\
  NGC895 & Field & 1 & 16U & 2 & 231G & 45.0D & 28.8 & 2.0E9 & 134 & 17 & Pisano 2002\\
  NGC691 & Group & 3 & 30U & $<$1-4.6 & 118 & 43.0TD & 36.0 & 1.3E10 & 314 & 10-50 & van Moorsel 1988\\
  NGC5395 & Pair & 1,2 & 9U & 1.25 & $\sim$600E & 7.7D & 57.6 & 6.0E9 & 151 & 21 & Kaufman 1999\\
  NGC4038 & Merger & 6 & 20 & $<$0.6-2.5 & $\sim$200E & 37.3 & 21.5 & 4.1E9 & 125 & 4-13 & Hibbard 2001\\
  NGC4676 & Pair & 6 & 4 & 0.5 & 420G & 3.7D & 93.0 & 7.5E9 & 108 & 13 & Hibbard 1996\\
  NGC7252 & Merger & 6 & 20 & 1.7 & 280G & 3.8D & 58.6 & 3.1E9 & 340 & 29 & Hibbard 1996\\
  NGC3690 & Triple & 6 & 10 & 1.7 & 190 & 6.1 & 44.0 & 2.8E9 & 128 & 22 & Hibbard 1999\\
  NGC3424 & Pair & 2 & 19U & $<$1.4-5.2 & 356G & 62.6G & 29.6 & 1.3E10 & 164 & 43 & Nordgren 1997\\
  NGC7125 & Pair & 2 & 9 & 7 &  &  & 55.0 &  & 143 & 111 & Nordgren 1997B\\
  VV784 & Group & 5 & 4 & $<$0.3-1.0 & 80 &  & 127.0 &  & 147 & 12-37 & Higdon 1996\\
  NGC3561 & Triple & 1 & 3 & 1.2 & $\sim$260 &  & 124.0 &  & 108 & 43 & Duc 1997\\
  IC2006 & Field & 6R & 22 & 1 & 270E & 3.0 & 19.7 & 2.7E8 & 126 & 6 & Franx 1994\\
  NGC5903 & Pair & 1 & 12 & $<$0.8-1.6 & 126E & 7.7D & 33.6 & 2.1E9 & 117 & 8-14 & Appleton 1990\\
  NGC5291 & Triple & 3R? & 15U & 1.5 & 464G & 22.7D & 59.7 & 1.9E10 & 260 & 26 & Malphrus 1997\\
  Arp295 & Group & 6 & 9 & $<$0.5 & 520G & 1.2 & 108.5 & 2.7E10 & 256 & $<$14 & Hibbard 1996\\
  EA1 & Group & 1,2 & 2 & $<$0.4 & 150E & 0.3T & 315.0 & 6.7E9 & 183 & $<$37 & Chang 2001\\
  NGC1241 & Pair & 2 & 10U & 2V & 436G & 10.71U & 50.4 & 6.4E9 & 146 & 29 & Nordgren 1997\\
  NGC5218 & Pair & 6 & 9 & $<$0.5 & 377 & 4.9 & 53.3 & 3.3E9 & 140 & $<$8 & Cullen 2007\\
  Arp314 & Triple & 6 & 7 & $<$0.5 & 104G & 2.8 & 51.9 & 1.8E9 & 106 & $<$8 & Nordgren 1997\\
  NGC3995 & Triple & 3 & 12 & 5V & 300E &  & 45.8 &  & 160 & 67 & Wilcots 2004\\
  NGC6872 & Pair & 5,6 & 8 & 1.2 & 950 & 17.5 & 64.2 & 1.7E10 & 149 & 22 & Horellou 2007\\
  NGC3227 & Pair & 5 & $>$25 & $<$1 & 520 & 13.0D & 20.1 & 1.2E9 & 146 & 6 & Mundell 1995\\
  AF7448\_035X & Pair & 2 & 14.8 & $<$3.5 & 217G & 20.8 & 34.4 & 5.8E9 & 134 & $<$35 & Taylor 2014\\
  AF7448\_059X & Pair & 2 & 10.6 & $<$3.5 & 33G & 2.5 & 102.4 & 6.2E9 & 298 & $<$104 & Taylor 2014\\
  AF7448\_208X & Field & 5 & 3.5 & $<$3.5 &  & 1.0 & 161.2 & 6.1E9 & 150 & $<$164 & Taylor 2014\\
  AF7448\_245X & Triple & 5 & 17.1 & $<$3.5 & 9G & 3.5 & 170.5 2.4E10 & 848 & $<$173 & Taylor 2014\\
\hline
\end{tabular}
\end{table*}

Another important caveat is that the nature of the objects differs widely. We have attempted to define broad categories to describe the objects as simply as possible, for the sake of clarity, but it is important to remember that there is a great deal of variation even in these small sub-samples. For example, some features are complete rings around individual galaxies (e.g. ESO381-47), while others are incomplete ring-like structures extending through galaxy groups (such as the Leo Ring) - whether such features share a common origin, or even have similar 3D geometry, is beyond the scope of this work. Some objects we have classified as streams are (partly or entirely) associated with stellar streams, but note that stellar streams which are devoid of gas also exist (e.g. \citealt{mihos05}, \citealt{me13} - hereafter AGES VI).

Since it is difficult to judge precisely where a galaxy ends and an extended stream begins, we have avoided making this choice for including objects in the catalogue. This means we also include a few giant galaxies, where the \HI{} appears to be entirely in a disc-like distribution - the only unusual feature being that it is much larger than in more typical galaxies. Readers wishing to use the table for their own purposes are thus advised not to assume the table is representative of only one class of object. We hope that our labels will provide a useful starting point for anyone wishing to select a particular type of object, but we strongly encourage readers to examine the source references for themselves - we make no claim that the table is anything more than a rough guide. Owing to the highly diverse nature of both the objects catalogued and the observations used, we have not attempted to quantify the errors on any of the parameters.

Finally, the distance error makes the completeness of the catalogue very difficult to assess - even a small measurement error would mean that some of the objects listed are actually smaller than our 100 kpc inclusion criteria. Conversely, there are almost certainly some objects not included as their assumed distance makes them (apparently) slightly smaller than 100 kpc, but they may in reality exceed this size. A different, but related, issue concerns how people report their discoveries. We have used the NRAO's ``\HI{} Rogue's Gallery'' (\citealt{rogue}) as one of our starting points; we would hope that many features would be noteworthy enough by their sheer size to be included here. We restrict the catalogue to only objects described in refereed journals - from the Rogue's Gallery alone, we found 15 objects which were never formally published. We have limited the inclusion of features described in \cite{me14} (AGES VII) to those where we can clearly measure the extent of the \HI{} in intergalactic space.

\subsection{The nature of the streams}
\label{sec:streamnat}

\begin{figure*}
\centering 
  \subfloat[]{\includegraphics[height=45mm]{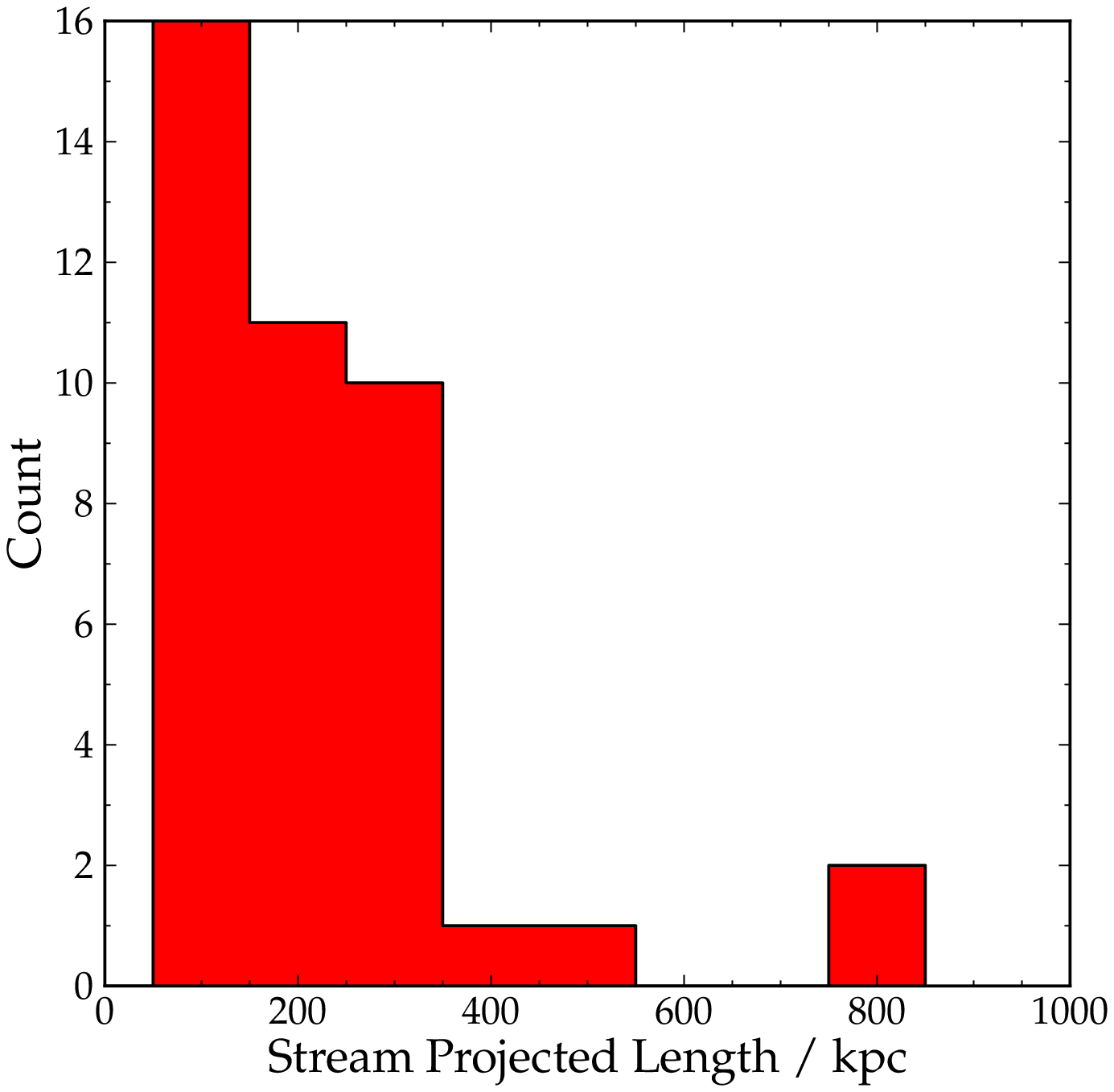}}
  \subfloat[]{\includegraphics[height=45mm]{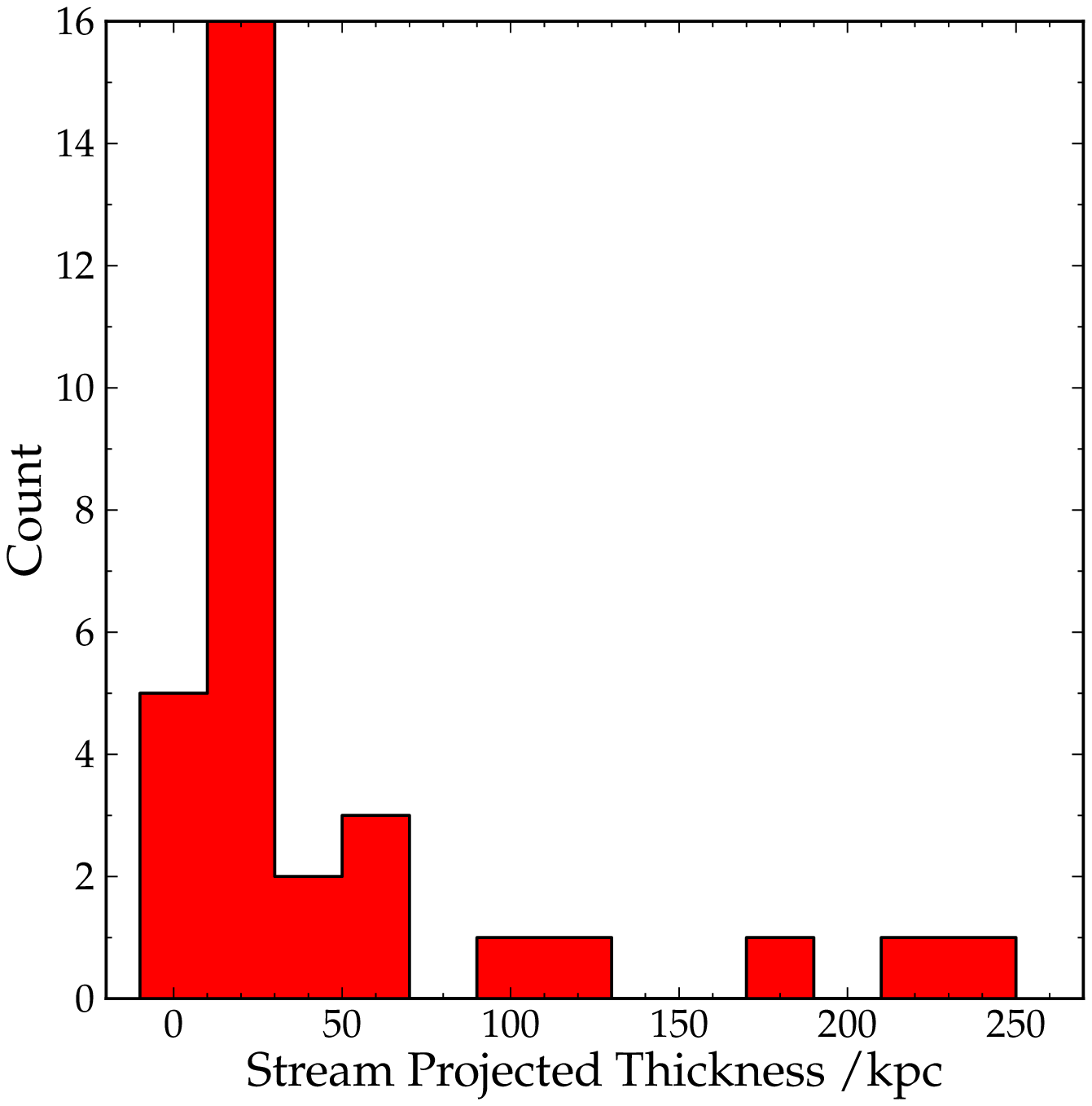}} 
  \subfloat[]{\includegraphics[height=45mm]{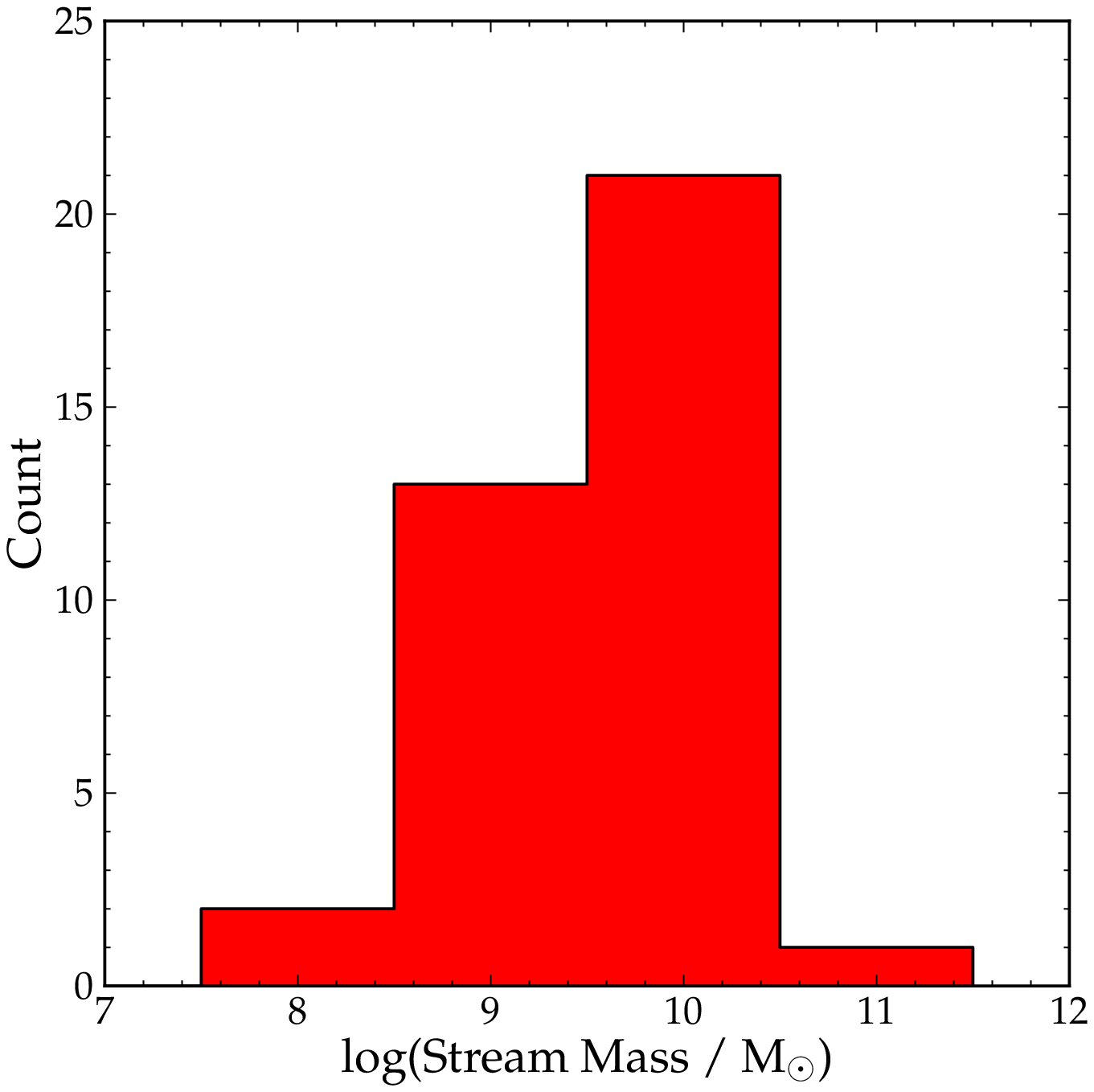}}
  \subfloat[]{\includegraphics[height=45mm]{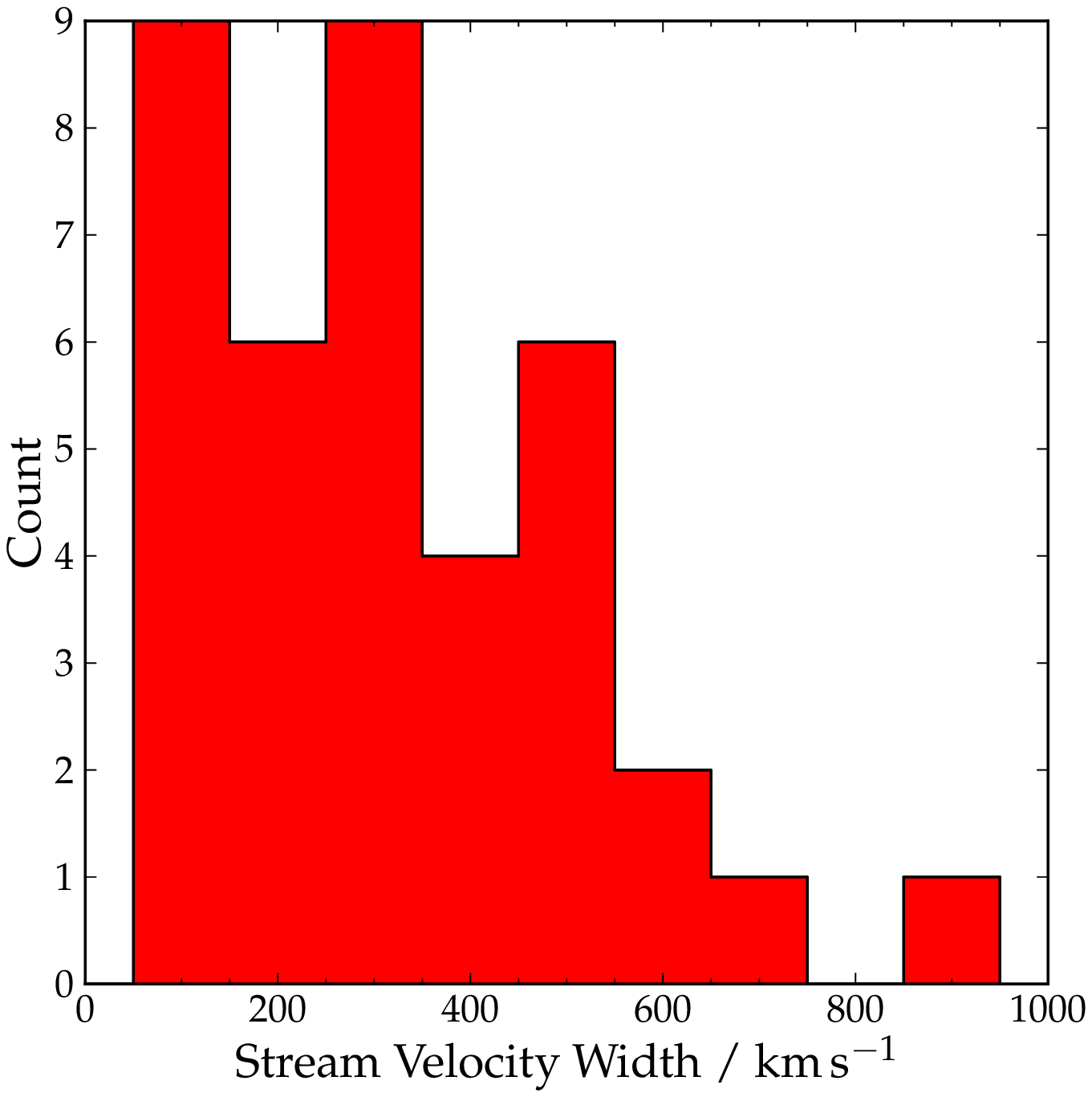}}  
\caption[Streams]{Distribution of the major properties of the extended \HI{} features.}
\label{fig:sprop}
\end{figure*}

Leaving aside these many difficulties, a catalogue, however flawed, is a necessary first step in our investigation. We can at least place some broad constraints on the properties of the streams, and while many properties vary significantly (see table \ref{tab:sprop}), the distributions are non-uniform - there is a clear bias towards certain values (see figure \ref{fig:sprop}).

\begin{table}
\centering
\caption[sprop]{Properties of all of the \HI{} features more than 100 kpc in extent.}
\label{tab:sprop}
\begin{tabular}{l l l l l}\\
\hline
  \multicolumn{1}{c}{Parameter} &
  \multicolumn{1}{c}{Min.} &
  \multicolumn{1}{c}{Med.} &
  \multicolumn{1}{c}{Max.} &
  \multicolumn{1}{c}{$\sigma$} \\  
\hline
  Length / kpc & 106 & 165 & 762 & 127 \\
  Min. diameter / kpc (J2000) & 4 & 17 & 232 & 55 \\
  log(M\HI{} / \Msolar{}) & 6.4 & 9.4 & 11.0 & 0.8 \\
  Velocity width / \kms{} & 60 & 290 & 950 & 194 \\
\hline
\end{tabular}
\end{table}

We may therefore define a ``typical'' stream from which we may begin our investigation, bearing in mind the large variations. HI1225+01 is a reasonable example of such a feature : about 200 kpc in extent, 10 kpc in diameter, with a mass of 4$\times$10$^{9}$ \Msolar{}. Only its velocity width is atypical, being just 60 \kms{} compared to the median of 290 \kms. The three-dimensional nature of this structure is much harder to quantify, as we discuss in section \ref{sec:sims}. An early hope when compiling the table was the minimum diameter might indicate some real minimum thickness to the features - unfortunately, the multifarious nature of the objects and varying resolution of the observations has rendered this prospect untenable. The observed length:diameter ratio ranges from 1 to 70, with a median of 10, but the different resolutions of the data sets makes it difficult to comment on the significance of this.

A review of the formation mechanisms for the long streams is beyond the scope of this paper. Broadly there are three main mechanisms : removal by ram-pressure stripping (e.g. \citealt{vol01}), removal by tidal encounters (e.g. \citealt{rory}), or accretion (\citealt{sancisi}). We do not consider the latter possibility here. Our focus is on the cluster environment (for reasons that will become apparent in section \ref{sec:virgoclouds}) where gas removal is expected to dominate over accretion except at large clustercentric radii (\citealt{ton07}). Our interest here is less on how the streams form and more on what happens to them afterwards. The most important parameter from the formation perspective is the gas removal timescale - the longer it takes to produce a stream, the longer any \HI{} overdensities must persist to be detected after the stream disperses. Conversely if the streams can be produced rapidly, then the resulting debris need not be stable over long periods - or even stable at all, as we shall see in section \ref{sec:unbound}.

Timescale estimates depend on the galaxy-galaxy relative velocities for streams formed tidally, and the galaxy velocity through the intracluster medium for ram pressure stripping. However, there is a consensus that long streams may persist for several hundreds of Myr even in clusters : \cite{oo05} describe a 100 kpc plume in Virgo that has persisted for $>$ 100 Myr; \cite{kent09} suggest the formation of an extended cloud complex in Virgo that occurred at least 500 Myr ago; the formation model of VIRGOHI21 of \cite{duc} has an encounter 750 Myr ago. The formation of the Leo Ring has been suggested to have begun 1.2 Gyr ago in the model of \cite{dansac}. Clearly, large \HI{} structures can be very long-lived. An important caveat is that there may be a selection effect at work since we do not know how many long streams have already been destroyed. This is a possibility we will examine further throughout this study. We leave the prospect of modelling the formation of the streams to future studies and here only examine the evolution of individual streams after the gas removal has occurred. 

\subsection{The fate of the stripped gas}
\label{sec:gasfate}

It is important to recognise that long \HI{} streams are by no means common. In the whole of the Virgo cluster - one of the nearest rich clusters (854 spectroscopically confirmed members in GOLDMine, see \citealt{gm}) and therefore one of the most studied regions in the nearby Universe - despite the high deficiency of many of its 355 late-type galaxies, the total number of long \HI{} streams known to date is a mere four (five including HI1225+01, which is in the cluster outskirts). As noted by \cite{oo05}, given the long survival time of some \HI{} plumes, more should exist if the deficiency is a result of gas stripping (though there is as yet no quantitative estimate as to exactly how many we might expect to observe). The temperature of \HI{} is equivalent to velocity dispersion of a few \kms{},
insufficient to disperse a typical 4$\times$10$^{9}$ \Msolar{} stream (the escape velocity would be $\sim$20 \kms{}, but see also section \ref{sec:simsetup}).  

We have already described that many ram-pressure stripping simulations predict the existence of long \HI{} streams, but none were detected in the 3.0$\times$0.6 Mpc VC1 region (AGES V) despite many galaxies being \HI{} deficient. \HI{} deficiency is defined (\citealt{haynes84}) as :
\begin{equation}
\HI{}_{def} = log(M\HI{}_{expected}) - log(M\HI{}_{observed})
\label{eqt:hidef}
\end{equation} 
Where the expected \HI{} mass is calculated based on field galaxies of the same morphology and optical diameter :
\begin{equation}
M\HI{}_{expected} = a + b\,log(d)
\label{eqt:hiexp}
\end{equation}
Where $a$ and $b$ depend weakly on the galaxy's morphology (\citealt{solanes}) and $d$ is its optical diameter in kpc.

We further add that about 25\% of galaxies in that region are non-deficient. This combination of deficient (some with deficiencies as high as 2.0 - gas fraction compared to a field galaxy is given by $10^{-\HI{}_{def}}$, so a deficiency of 2.0 is equivalent to possessing just 1\% of the \HI{} of a field galaxy) and non-deficient galaxies makes it almost certain that some galaxies are currently in the process of losing gas, making the lack of detected streams even stranger. 

It seems that the gas removal process therefore cannot always result in such long-lived streams, even when the amount of gas removed is very large (e.g. 21 galaxies with deficiencies equivalent to $>$10$^{9}$ \Msolar{} of missing \HI{} in AGES V). Gas which falls into clusters can be heated up to 10$^{8}$ K by the release of gravitational potential energy (e.g. \citealt{takah}), and the interaction of this hot gas with cold gas in galaxies can cause heating via conduction and turbulent shocks (\citealt{ton10}). Furthermore, \cite{borth15} and \cite{borth10} note that there is a sharp cutoff in the observed \HI{} column densities in galactic discs at 2$\times$10$^{19}$ cm$^{-2}$. They interpret this to mean that gas below this density is vulnerable to ionization by the cosmic UV background (see also \citealt{maloney}). 

Yet even in the Virgo cluster, both the stream described in \cite{k08} and HI1225+01 (\citealt{chen}) are detected at lower column densities than this threshold ($\sim$10$^{18}$ cm$^{-2}$). In the Local Group the disc of M33 extends to even lower column densities, as does the nearby newly-discovered ring-shaped cloud of comparable angular size to M33 (\citealt{olivia}), while the M33-M31 bridge and Magellanic Streams are detected at the level of $\sim$10$^{17}$ cm$^{-2}$ (\citealt{web}). It is possible that these column densities are lower than the true values due to the large beam size (see section \ref{sec:simresultsmassive}), though this is not very likely for M33 and its nearby clouds where the diffuse \HI{} appears to be well-resolved. Thus it is not at all obvious whether ionization can explain the lack of detected streams (see also \citealt{yoshida}).

Conversely, \cite{oo05} find that the \HI{} in the plume associated with NGC 4388 can only account for about 10\% of its missing \HI{} (see equations \ref{eqt:hidef} and \ref{eqt:hiexp}) - they speculate that the rest may be in the form of (much colder) molecular hydrogen.

We note that the major alternative to the formation of streams by gas removal is cold accretion from the cosmic web (see \citealt{web}, \citealt{wolfe}). While this is not a possibility we consider here, we note that there appears to be no correlation between the \HI{} deficiency of a galaxy and the presence of a stream - indeed, some galaxies are non-deficient but, perhaps paradoxically, have clearly associated massive streams (see AGES VII, also \citealt{n2442} and \citealt{don}). It is also interesting to note that the case of VIRGOHI21 demonstrates the detectability of low-mass streams in the Virgo environment and its likely parent galaxy (NGC 4254) actually has a \textit{negative} deficiency. Yet those galaxies which \textit{are} strongly deficient, sometimes having lost $>$ 10$^{9}$ \Msolar{} of \HI{}, very rarely show streams (NGC 4388 being an exception). Examples where a long stream has sufficient \HI{} mass to account for the high deficiency of its parent galaxy appear to be extremely rare. It seems almost certain that the gas removal process is far more complex than simply removing the neutral gas.

This problem has also previously been studied by \cite{vol07}. They propose a scenario where the detected \HI{} in partially stripped cluster galaxies corresponds to the cold, dense \HI{} that was in the galactic disc prior to stripping, with only the warm, low-density outer gas being removed by ram-pressure stripping. At the time, NGC 4388 was the only galaxy known in the Virgo cluster with a long \HI{} tail. In their scenario, there is a narrow time window of $\sim$ 200 Myr during which the stripped outer \HI{} can be detected before it is ionized by the hot ICM (thus NGC 4388 could be the single exception which happens to exist in that narrow period in which the \HI{} is detectable). 

The \cite{vol07} scenario now looks difficult to maintain. As described in section \ref{sec:intro}, simulations have predicted that stripped \HI{} should be long-lived. We showed in AGES VI that the most strongly \HI{}-deficient galaxies have narrower velocity widths than are predicted by the Tully-Fisher relation, which can be explained as stripping the inner part of the disc (where the rotation curve is rising). Yet even these highly deficient galaxies in which, according to that scenario, even the colder dense gas has been stripped do not show evidence of extended \HI{} streams. Moreover, while the AGES and ALFALFA surveys have greatly increased the amount and depth of \HI{} observations (covering most of the cluster), the number of streams has increased only to four. Two of these (VIRGOHI21 and the stream in \citealt{k08}) are believed to be the result of tidal encounters, while a third (\citealt{kent09}) is completely detached from its parent galaxy (which remains unknown). Thus, arguably, the new observations have not uncovered any additional \textit{long} ram-pressure \HI{} streams at all.

It is true that more (shorter) features have been discovered which likely result from ram pressure (\citealt{chung}), but as discussed in section \ref{sec:intro}, ram-pressure stripping simulations predict that low column density gas should be detectable at much greater distances (though the simulations explain the high density gas tails very well). While these simulations have not included harassment, they have included other physics such as star formation, heating and cooling, and ionization (e.g. \citealt{kap09}, \citealt{ton10}).

In short the fate of the stripped gas is still not well understood, even when the mechanisms for its removal are readily apparent. Gas loss in clusters is common, but \HI{} streams appear to be surprisingly rare (but, importantly, they are not non-existent) given the known physics of ram-pressure stripping and ionization. The processes of ionization and/or cooling to molecular gas might be able to explain the lack of streams, but it is unclear which (if either) of these mechanisms dominates. We note that of the long \HI{} streams, 88\% occur in low density, low velocity dispersion environments (from voids to groups), with the remainder found in the Virgo cluster (see also section \ref{sec:env}). This strongly suggests that whatever process suppresses stream formation or promotes their destruction is unique to the cluster environment.

\section{Observations II : Dark galaxy candidates}
\label{sec:obs2}
\begin{figure*}
\centering 
  \subfloat[]{\includegraphics[height=45mm]{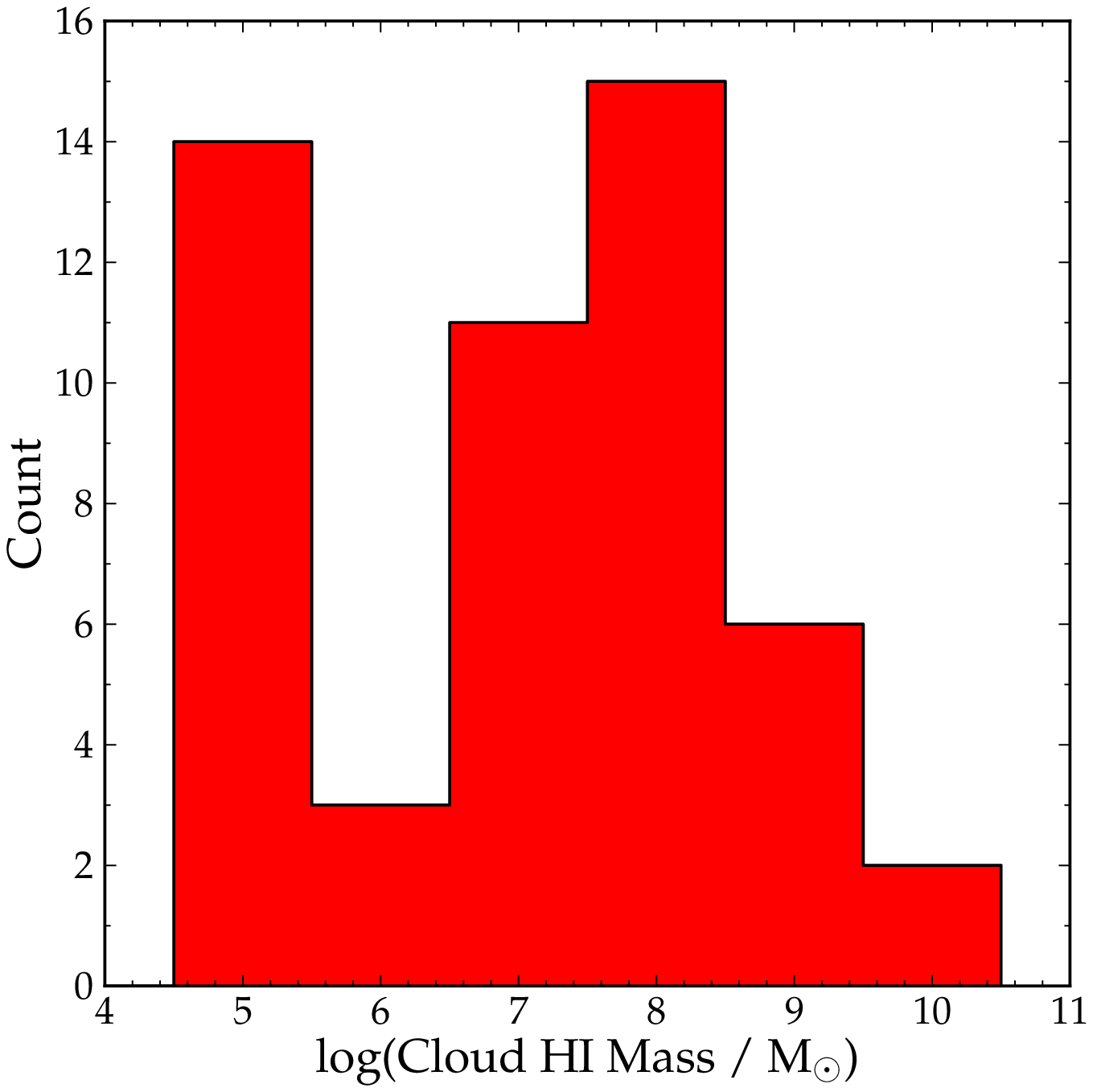}}
  \subfloat[]{\includegraphics[height=45mm]{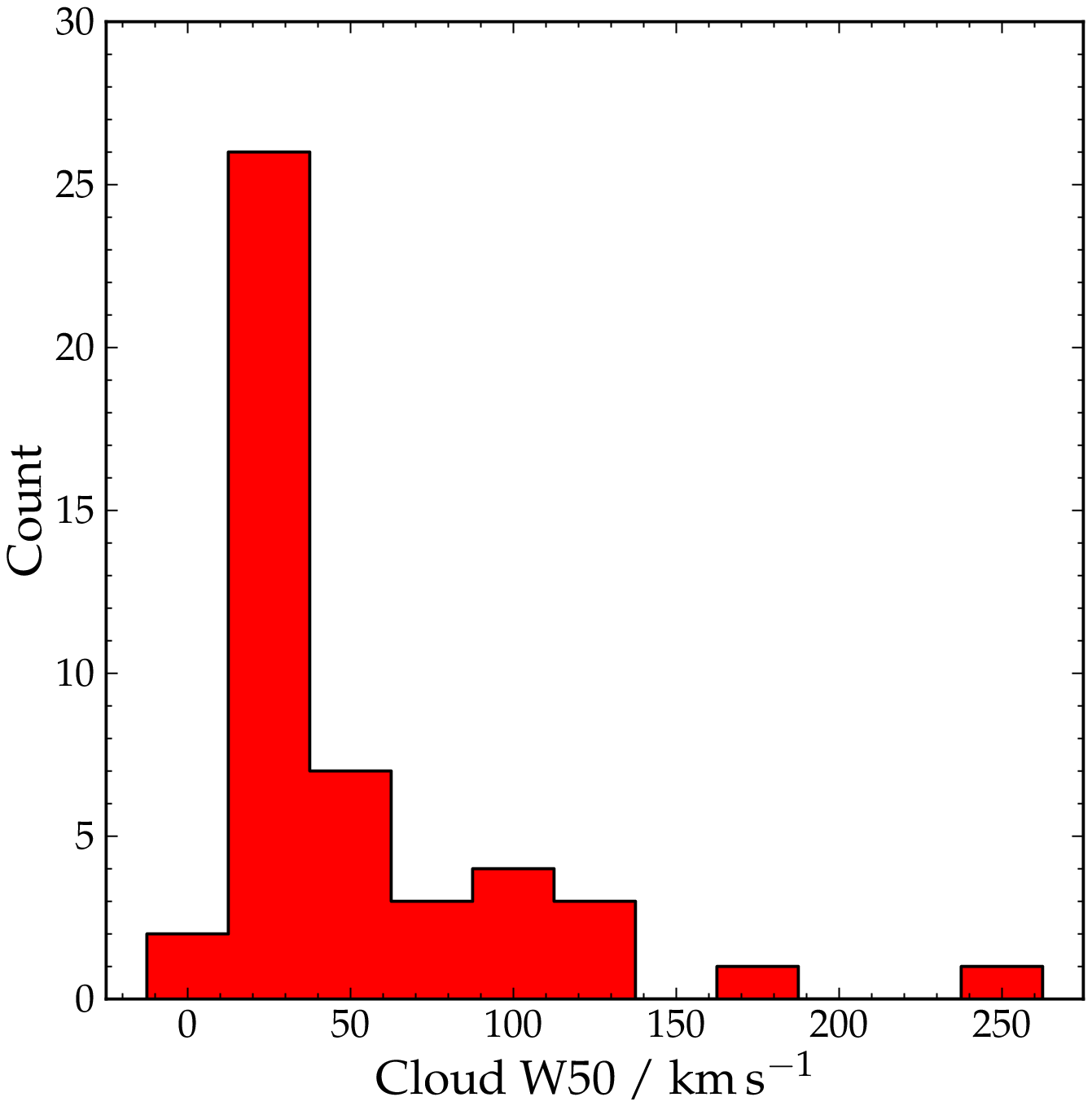}} 
  \subfloat[]{\includegraphics[height=45mm]{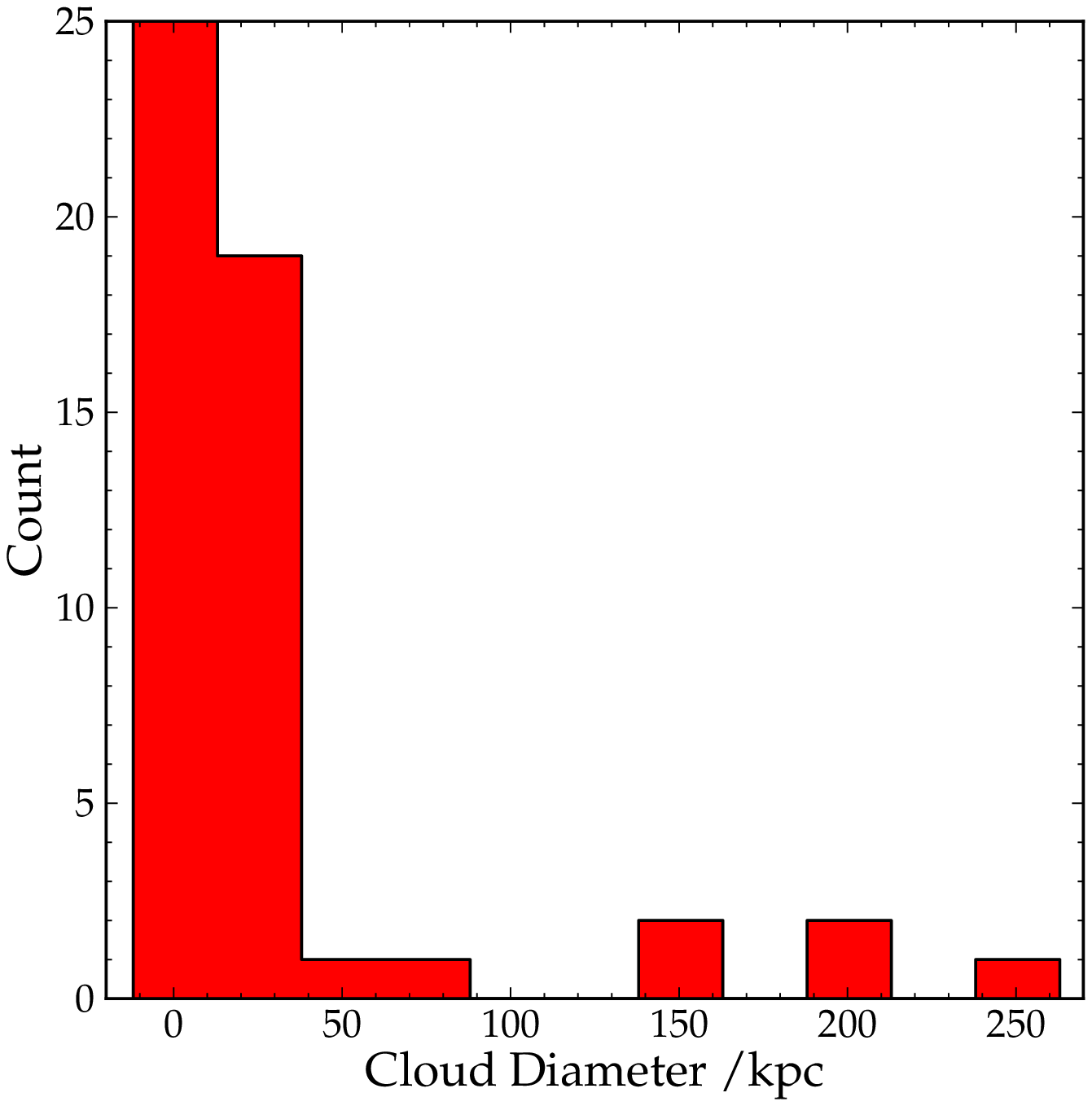}}  
  \subfloat[]{\includegraphics[height=45mm]{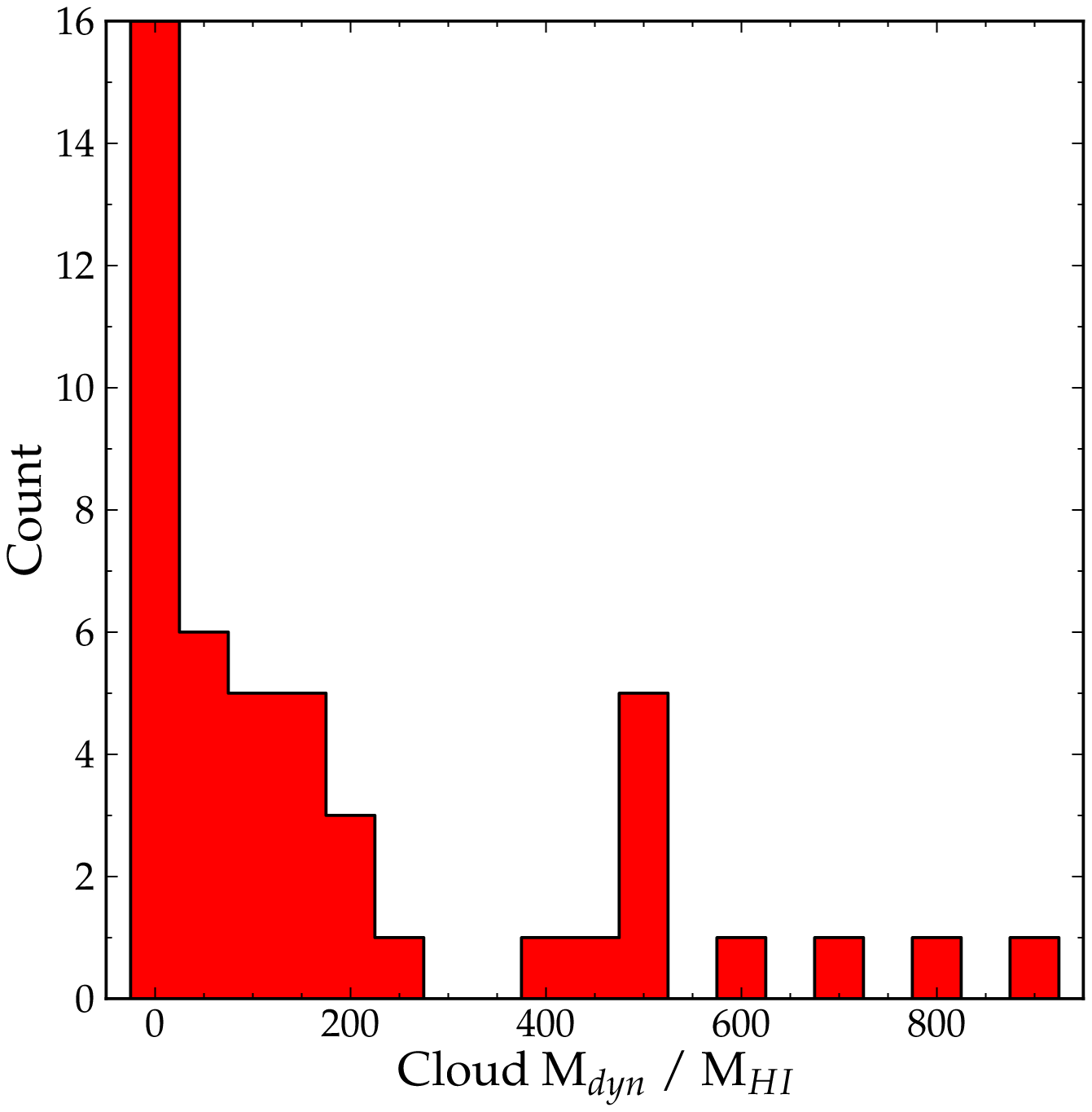}}  
\caption[Streams]{Distribution of the major properties of the proposed dark galaxy candidates.}
\label{fig:cloudd}
\end{figure*}

With a standard stream defined as a starting point, we should also consider the end result we are attempting to reproduce via numerical simulations. We wish to know if it is possible to create clouds that may be mistaken for optically dark galaxies. There are many proposed candidates for such objects, shown in table \ref{tab:dgal}. Our criteria for inclusion is far less well-defined than our stream catalogue - broadly, we include any object proposed as a dark galaxy, regardless of the reason it is (or was) thought to be a suitable candidate. These reasons vary (see appendix \ref{sec:ap2}), the only common feature being that they all lack optical counterparts (it should be emphasised that neither we nor the discoverers necessarily support dark galaxies as the most likely interpretation of the objects). The median parameters of the objects are shown in table \ref{tab:medd} and the distributions shown in figure \ref{fig:cloudd}.

\begin{table*}
\centering
\tiny
\caption[dgal]{Major properties, where available, of extragalactic objects proposed as dark galaxy candidates. The columns are as follows : (1) Name in a major catalogue; (2),(3) Spatial coordinates J2000; (4) Heliocentric systemic velocity in \kms{}; (5), (6) Velocity width in \kms{}; (7) Total flux in Jy \kms{}; (8) Distance in Mpc; (9) \HI{} mass in solar masses; (10) Diameter in arcmin; (11) Diameter in kpc; (12) Dynamical mass in solar masses as calculated with equation \ref{eqt:mdyn}; (13) Dynamical mass to \HI{} mass ratio; (14) Type of object, as follows : 1 - isolated cloud, 2 - part of a complex of presumably related clouds, 3 - overdensity in a stream, 4 - stream, 5 - HVC with unusual properties. For references see appendix \ref{sec:ap2}.}
\label{tab:dgal}
\begin{tabular}{c c c c c c c c c c c c c c c c c c c c c c c c c c c c c c c c c c c c c c c c c}
\hline
  \multicolumn{1}{c}{(1)} &
  \multicolumn{1}{c}{(2)} &
  \multicolumn{1}{c}{(3)} &
  \multicolumn{1}{c}{(4)} &
  \multicolumn{1}{c}{(5)} &
  \multicolumn{1}{c}{(6)} &
  \multicolumn{1}{c}{(7)} &
  \multicolumn{1}{c}{(8)} &
  \multicolumn{1}{c}{(9)} &
  \multicolumn{1}{c}{(10)} &
  \multicolumn{1}{c}{(11)} &
  \multicolumn{1}{c}{(12)} &
  \multicolumn{1}{c}{(13)} &
  \multicolumn{1}{c}{(14)} \\
  \multicolumn{1}{c}{Name} &
  \multicolumn{1}{c}{RA} &
  \multicolumn{1}{c}{Dec} &
  \multicolumn{1}{c}{Vel} &
  \multicolumn{1}{c}{W50} &
  \multicolumn{1}{c}{W20} &
  \multicolumn{1}{c}{Flux} &
  \multicolumn{1}{c}{Dist} &
  \multicolumn{1}{c}{M\HI{}} &
  \multicolumn{1}{c}{D} &
  \multicolumn{1}{c}{D} &
  \multicolumn{1}{c}{M$_{dyn}$} &
  \multicolumn{1}{c}{M$_{dyn}$/M\HI{}} &
  \multicolumn{1}{c}{Code} \\
  \multicolumn{1}{c}{} &
  \multicolumn{1}{c}{} &
  \multicolumn{1}{c}{} &
  \multicolumn{1}{c}{\kms{}} &
  \multicolumn{1}{c}{\kms{}} &
  \multicolumn{1}{c}{\kms{}} &
  \multicolumn{1}{c}{Jy \kms{}} &
  \multicolumn{1}{c}{Mpc} &
  \multicolumn{1}{c}{\Msolar{}} &
  \multicolumn{1}{c}{arcmin} &
  \multicolumn{1}{c}{kpc} &
  \multicolumn{1}{c}{\Msolar{}} &
  \multicolumn{1}{c}{} &
  \multicolumn{1}{c}{} \\
\hline
  AGESVC1\_231 & 12:18:17.9 & 07:21:40 & 1911 & 36 & 152 & 0.173 & 32.0 & 4.2E7 & <3.5 & <32.6 & <2.2E10 & <521.2 & 1 \\
  AGESVC1\_247 & 12:24:59.2 & 08:22:38 & 1087 & 22 & 33 & 0.183 & 23.0 & 2.3E7 & <3.5 & <23.4 & <7.4E8 & <32.2 & 1 \\
  AGESVC1\_257 & 12:36:55.1 & 07:25:48 & 1580 & 131 & 157 & 0.199 & 17.0 & 1.4E7 & <3.5 & <17.3 & <1.2E10 & <885.2 & 1 \\
  AGESVC1\_258 & 12:38:07.2 & 07:30:45 & 1786 & 32 & 120 & 0.2 & 17.0 & 1.4E7 & <3.5 & <17.3 & <7.2E9 & <517.1 & 1 \\
  AGESVC1\_262 & 12:32:27.2 & 07:51:52 & 1322 & 104 & 146 & 0.16 & 23.0 & 2.0E7 & <3.5 & <23.4 & <1.4E10 & <724.8 & 1 \\
  AGESVC1\_266 & 12:36:06.5 & 08:00:07 & 1691 & 77 & 173 & 0.245 & 17.0 & 3.2E7 & <3.5 & <17.3 & <1.5E10 & <470.2 & 1 \\
  AGESVC1\_274 & 12:30:25.6 & 08:38:05 & 1297 & 22 & 35 & 0.107 & 17.0 & 7.3E6.0 & <3.6 & <17.3 & <6.2E8 & <84.4 & 1 \\
  AGESVC1\_282 & 12:25:24.1 & 08:16:54 & 943 & 69 & 164 & 0.351 & 23.0 & 4.4E7 & <3.5 & <23.4 & <1.8E10 & <408.6 & 1 \\
  AGES628\_011 & 01:39:56.2 & 15:31:35 & 17343 & 105 & 214 & 0.42 & 244.3 & 5.9E9 & <3.5 & <248.7 & <3.3E11 & <56.1 & 1 \\
  AGES1376\_004 & 11:48:09.1 & 19:21:09 & 11252 & 171 & 179 & 0.34 & 158.5 & 2.0E9 & <3.5 & <161.4 & <1.5E11 & <75.14 & 1 \\
  VIRGOHI21 & 12:17:52.9 & 14:47:19 & 2005 & 87 & 200 & 0.6 & 17.0 & 4.1E7 & 3.0 & 14.8 & 2.0E10 & 487.8 & 3 \\
  HI1225+01 & 12:27:46.3 & 01:36:01 & 1292 & - & 60 & 42.37 & 20.0 & 4.0E9 & 35.0 & 203.6 & 2.1E10 & 5.5 & 4 \\
  GBT1355+5439 & 13:54:50.6 & 54:37:50 & 210 & - & 41 & 1.1 & 6.9 & 1.2E7 & 5.0 & 10.0 & 4.9E8 & 40.7 & 1 \\
  AGC749170 & 02:17:50.3 & 14:24:40 & 3905 & - & 56 & 2.4 & 50.0 & 1.4E9 & 0.26 & 3.7 & 3.2E9 & 2.3 & 2 \\
  SmithCloud & 19:49:32.2 & -01:05:59 & 99 & - & 16 &  & 0.0124 & 1.0E6 & 1018.0 & 3.0 & 2.2E7 & 22.3 & 5 \\
  AAK1\_C1 & 12:30:25.8 & 09:28:01 & 488 & 62 & - & 2.48 & 17.0 & 1.7E8 & 2.7 & 13.4 & 2.3E9 & 13.5 & 2 \\
  AAK1\_C2 & 12:31:19.0 & 09:27:49 & 607 & 56 & - & 0.72 & 17.0 & 4.9E7 & 1.5 & 7.4 & 3.0E8 & 6.2 & 2 \\
  AAK1\_C3 & 12:29:42.8 & 09:41:54 & 524 & 116 & - & 1.16 & 17.0 & 7.9E7 & <3.5 & <17.3 & <6.8E9 & <85.6 & 2 \\
  AAK1\_C4 & 12:30:19.4 & 09:35:18 & 603 & 252 & - & 2.56 & 17.0 & 1.7E8 & <3.5 & <17.3 & <3.2E10 & <187.8 & 2 \\
  AAK1\_C5 & 12:31:26.7 & 09:18:52 & 480 & 53 & - & 0.91 & 17.0 & 6.2E7 & <3.5 & <17.3 & <1.4E9 & <22.8 & 2 \\
  AAK2\_C1N & 12:08:47.6 & 11:55:57 & 1234 & 22 & - & 0.29 & 17.0 & 2.0E7 & 1.0 & 5.0 & 6.2E7 & 3.1 & 1 \\
  AAK2\_C1S & 12:08:47.4 & 11:54:48 & 1225 & 20 & - & 0.39 & 17.0 & 2.7E7 & 1.4 & 6.9 & 4.1E7 & 1.6 & 1 \\
  AAK2\_C2N & 12:13:42.5 & 12:54:50 & 2237 & 13 & - & 0.14 & 32.0 & 3.4E7 & 2.5 & 23.2 & 3.7E7 & 1.1 & 1 \\
  AAK2\_C2W & 12:13:33.1 & 12:52:44 & 2205 & 41 & - & 0.25 & 32.0 & 6.0E7 & 2.4 & 22.3 & 1.1E9 & 18.0 & 1 \\
  AAK2\_C2S & 12:13:41.9 & 12:51:16 & 2234 & 6 & - & 0.05 & 32.0 & 1.2E7 & 0.8 & 7.5 & 9.2E6 & 0.7 & 1 \\
  Engima & 07:49:49.6 & 04:30:20 & 48 & 3 & - & 2.966 &  &  & 6.4 &  &  &  & 5 \\
  AAM33\_01 & 01:34:36.9 & 30:59:35 & -83 & 22 & - & 0.63 & 0.84 & 9.5E4 & 6.5 & 1.6 & 1.4E7 & 148.1 & 2 \\
  AAM33\_08 & 01:36:15.0 & 29:58:10 & -158 & 15 & - & 0.35 & 0.84 & 5.5E4 & 6.5 & 1.6 & 6.5E6 & 130.8 & 2 \\
  AAM33\_09 & 01:33:27.8 & 29:14:49 & -185 & 22 & - & 0.38 & 0.84 & 6.4E4 & 5.0 & 1.2 & 1.4E7 & 219.8 & 2 \\
  AAM33\_16 & 01:35:28.9 & 30:43:17 & -292 & 15 & - & 0.21 & 0.84 & 3.6E4 & 4.5 & 1.1 & 6.5E6 & 181.6 & 2 \\
  AAM33\_17 & 01:29:52.0 & 31:04:23 & -298 & 14 & - & 0.24 & 0.84 & 3.8E4 & 4.0 & 1.0 & 5.7E6 & 149.9 & 2 \\
  AAM33\_18 & 01:34:39.9 & 30:09:20 & -326 & 23 & - & 1.21 & 0.84 & 1.1E5 & 10.6 & 2.6 & 1.5E7 & 139.8 & 2 \\
  AAM33\_19 & 01:33:26.6 & 29:29:42 & -327 & 24 & - & 1.56 & 0.84 & 1.3E5 & 13.0 & 3.2 & 1.7E7 & 128.8 & 2 \\
  AAM33\_20 & 01:31:12.5 & 30:24:04 & -341 & 25 & - & 0.61 & 0.84 & 8.0E4 & 8.5 & 2.1 & 1.8E7 & 227.0 & 2 \\
  VelaA & 10:28:35.0 & -44:04:00 & 2842 & 41 & - & 0.47 & 37.6 & 1.6E8 & 3.7 & 40.0 & 3.9E9 & 24.4 & 2 \\
  VelaB & 10:28:30.0 & -44:09:00 & 2885 & 97 & - & 4.2 & 37.6 & 1.4E9 & 6.1 & 67.0 & 3.7E10 & 26.4 & 2 \\
  VelaC & 10:28:20.0 & -44:17:00 & 2813 & 20 & - & 4.9 & 37.6 & 1.6E9 & 12.7 & 138.3 & 3.2E9 & 2.0 & 2 \\
  HGC44\_CS & 10:18:42.0 & 21:50:00 & 1400 & <100 & - & 0.82 & 25.0 & 1.2E8 & 0.9 & 6.4 &  &  & 2 \\
  HJ1021+6842 & 10:21:00.0 & 68:42:00 & 46 & 120 & - & 39.72 & 4.0 & 1.5E8 & 25.8 & 30.0 & 1.3E10 & 83.7 & 2 \\
  HPJ0731-69 & 07:31:39.4 & -69:01:36 & 1481 & - & 270 & 18.0 & 15.5 & 1.0E9 & 44.4 & 200.0 & 4.2E10 & 42.4 & 4\\
  ComplexH & 02:03:00.0 & 62:31:54 & -175 & 20 & - & 114000.0 & 0.027 & 2.0E7 & 1200.0 & 9.4 & 1.1E8 & 5.5 & 5 \\
  GEMS\_N3783\_2 & 11:31:27.0 & -36:18:37 & 2731 & 40 & 50 & 1.242 & 47.8 & 6.7E8 & <1.5 & <20.9 & <1.5E9 & <2.3 & 1\\
  HIJASS1219+46 & 12:19:52.8 & 46:35:27 & 392 & 34 & 47 & 2.36 & 6.6 & 2.4E7 & <13.1 & <25.2 & <1.6E9 & <67.4 & 3 \\
  AGC198606 & 09:30:02.5 & 16:38:08 & 51 & 25 & - & 14.8 & 0.42 & 6.2E5 & 23 & 3.2 & 5.8E7 & 93.7 & 5 \\
  AGC208602 & 10:54:37.4 & 17:38:06 & 1093 & 50 & - & 1.45 & 15.3 & 8.1E7 & 5.3 & 23.6 & 1.7E9 & 21.2 & 1 \\
  M31Wolfe\_01 & 01:24:41.6 & 37:24:00 & -298 & 22 & - & 0.84 & 0.8 & 1.3E5 & 18.9 & 4.4 & 6.2E7 & 487.3  & 3 \\
  M31Wolfe\_02 & 01:23:21.7 & 37:18:45 & -223 & 28 & - & 0.30 & 0.8 & 4.5E4 & 18.9 & 4.4 & 9.1E7 & 2025.4 & 3 \\
  M31Wolfe\_04 & 01:19:15.8 & 37:29:15 & -228 & 37 & - & 0.57 & 0.8 & 8.6E4 & 15.9 & 3.7 & 1.5E8 & 1711.8 & 3 \\
  M31Wolfe\_05 & 01:16:53.7 & 36:49:00 & -309 & 22 & - & 0.52 & 0.8 & 7.8E4 & 30.1 & 7.0 & 9.8E7 & 1262.4 & 3 \\
  M31Wolfe\_06 & 01:08:29.6 & 37:45:00 & -279 & 34 & - & 2.60 & 0.8 & 3.9E5 & 30.9 & 7.2 & 2.4E8 & 617.1  & 3 \\
  M31Wolfe\_07 & 01:05:00.8 & 36:21:00 & -210 & 27 & - & 0.83 & 0.8 & 1.3E5 & 21.1 & 4.9 & 1.0E8 & 824.0  & 3 \\
  M31Wolfe\_09 & 01:01:24.6 & 36:12:15 & -341 & 21 & - & 0.58 & 0.8 & 8.7E4 & 15.0 & 3.5 & 4.5E7 & 515.6  & 3 \\
\hline
\end{tabular}
\end{table*}

\begin{table}
\centering
\caption[medd]{Median properties of the dark galaxy candidates. Upper limits have been excluded from the diameter and dynamical mass measurements.}
\label{tab:medd}
\begin{tabular}{l l l l l}\\
\hline
  \multicolumn{1}{c}{Parameter} &
  \multicolumn{1}{c}{Min.} &
  \multicolumn{1}{c}{Med.} &
  \multicolumn{1}{c}{Max.} &
  \multicolumn{1}{c}{$\sigma$} \\  
\hline
  log(M\HI{}) / \Msolar{} & 4.56 & 7.38 & 9.77 & 1.54 \\
  W50 / \kms{} & 3 & 32 & 252 & 48 \\
  Diameter / kpc & 1.0 & 13.4 & 248.7 & 55.2 \\
  log(M$_{dyn})$ / \Msolar{} & 6.76 & 8.00 & 10.62 & 1.16 \\
  M$_{dyn}$/M\HI{} & 0.7 & 84.4 & 2025.4 & 428.13\\
\hline
\end{tabular}
\end{table}

\subsection{Major candidates}
\label{sec:cloudoverview}

While we give a little more detail of the nature of the individual objects in appendix \ref{sec:ap2}, it is useful to give a brief review of some of the more well-known objects. The Smith Cloud is a high velocity cloud that may have survived a passage through the Milky Way disc - \cite{nich} find through numerical simulations that this is very difficult to do unless it has a dark matter halo. Recently \cite{Fox} have found the object's metallicity makes a primordial origin unlikely and instead favours an origin from the outer galactic disc. HI1225+01 is a stream-like feature near the Virgo cluster. When discovered (\citealt{ghcloud}) no optical counterpart was known, though subsequently a dwarf galaxy has been found at the northern end of the feature (\citealt{ghcloud2}). However this dwarf galaxy is an order of magnitude too small to be the source of all of the gas, and the southern part of the stream is still regarded as a dark galaxy candidate (\citealt{ghcloud3}). VIRGOHI21 is an overdensity in a stream showing signs of ordered rotation; it was proposed as being a dark galaxy which disturbed the gas in the nearby spiral NGC 4254 (\citealt{m07}). Subsequently it was shown that the object is actually in the middle, not at the end, of a much larger stream (\citealt{haynesV21}) and could be explained as the result of an interaction by an ordinary galaxy (\citealt{duc}).

The idea that some Galactic high velocity clouds might be dark galaxies is not new (e.g. \citealt{blitz}), but has been gaining traction in recent years thanks to the ALFALFA survey. \cite{uchvcs} and \cite{adams} have uncovered a population of `ultra-compact high velocity clouds' which may satisfy the criteria to solve the missing satellites problem in the Local Group. Recently, \cite{bellaz} discovered an optical counterpart to one of these UCHVCs with a matching redshift to the \HI{} detection. Their interpretation is that this object is not a Local Group minihalo at all, but an extremely faint galaxy in the outskirts of the Virgo cluster. Given the uncertainties inherent to these objects, we choose to exclude this (rather large - 62 objects) population of UCHVCs from our table. We do, however, include one particularly noteworthy example (AGC 198606) that has been studied in greater detail (\citealt{adams15}). Similarly, we include the well-studied Smith Cloud, as described above.

\subsection{Unusual \HI{} clouds in the Virgo cluster}
\label{sec:virgoclouds}

The clouds described in AGES V and in more detail in AGES VI in the Virgo cluster are of particular interest to us. Free-floating \HI{} clouds are sometimes described as being tidal debris (e.g. AGC 208602 in \citealt{cannon}; \citealt{kent09}), yet without a stream connecting the purported debris to its proposed parent, this remains a supposition (especially in cases where the nearby galaxy does not seem to be involved in any interaction, e.g.  \citealt{oo13}). 

The AGES Virgo clouds are especially interesting thanks to a combination of properties : lack of any streams indicating an obvious tidal origin, their high velocity widths ($\sim$150 \kms{}), low \HI{} masses ($\sim$3$\times$10$^{7}$ \Msolar{}) \kms{} and unresolved nature (implying diameters $<$ 17 kpc). They are also relatively isolated, typically 100-150 kpc from the nearest \HI{}-detected galaxy. Two of these clouds have much lower velocity widths ($\sim$ 30 \kms{}) than the others ($\gtrsim$100 \kms{}); we hereafter refer to the high-width clouds as `type 1' and the low-width clouds as `type 2'. Since the velocity dispersion of \HI{} is typically around 10 \kms{}, it is not immediately obvious how such large velocity widths may arise in such compact debris\footnote{As apparent in figure \ref{fig:cloudd}, few other clouds have similar velocity widths. Clouds C3 and C4 in \cite{kent09} have similar widths, but they are clearly part of a larger, extended feature which the authors state is likely to have a tidal origin. The clouds in \cite{kent10} have smaller velocity widths but are even more isolated and it is not easy to explain them as tidal debris.}.

\begin{figure*}
\centering
\includegraphics[width=170mm]{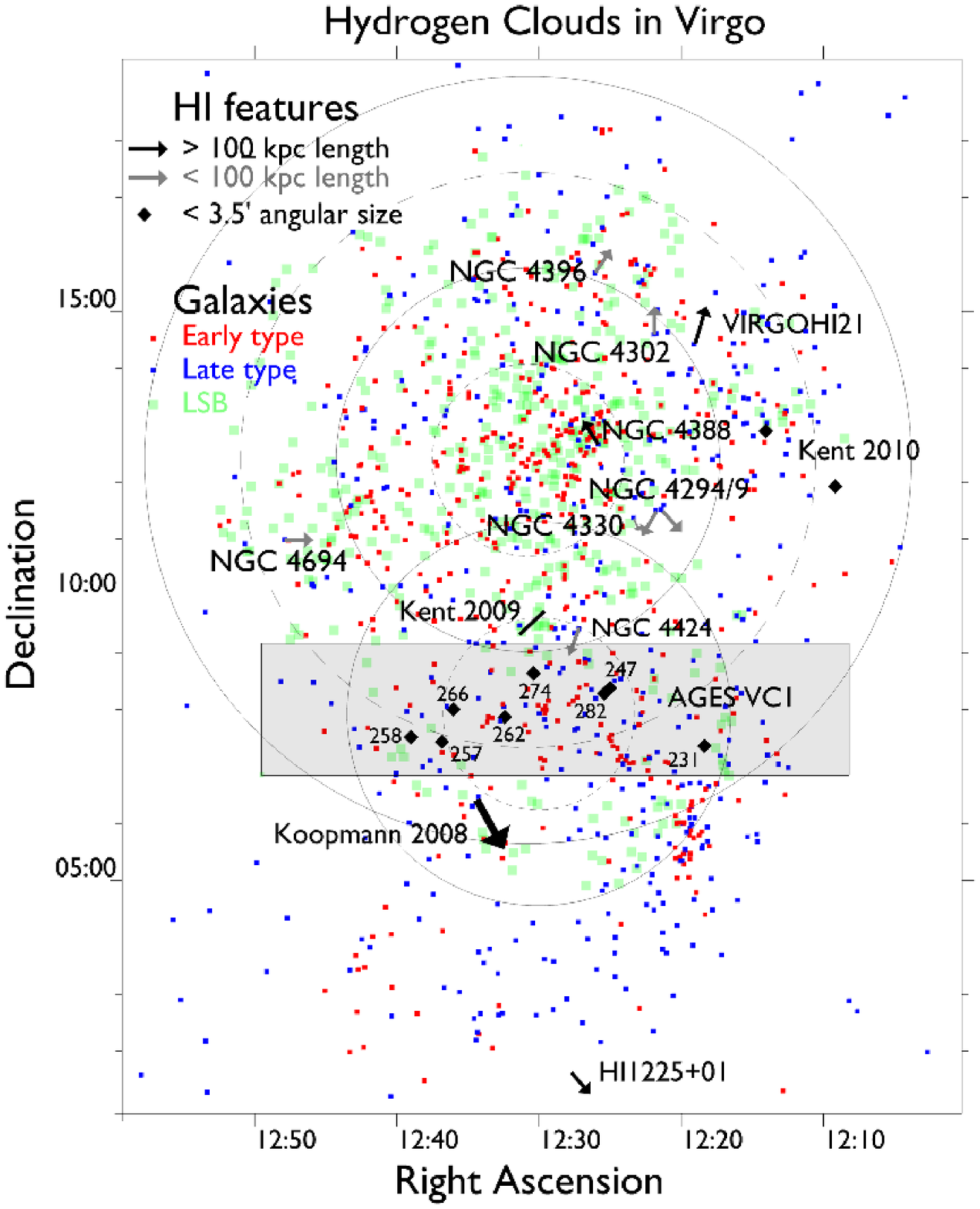}
\caption[vmap]{Map of the optical and \HI{} content of the Virgo cluster. Red and blue squares indicate spectroscopically confirmed cluster members; red for early and blue for late type galaxies (data from GOLDMine, see \citealt{gm}). Faint green squares indicate low surface brightness galaxies identified in \cite{d15} as likely cluster members. Arrows indicate the approximate direction of a stream away from its progenitor galaxy. For features larger than 100 kpc (black) the size of the arrow is to the same scale as the cluster, while for features smaller than this (grey) the size is not meaningful. Data for the latter objects are from \cite{chung} except NGC 4694 which is from \cite{duc07}. Circles show the distance of 0.5, 1.0, 1.5 and 2.0 Mpc away from M87 (north) and M49 (south). The light grey area is the AGES VC1 survey region. Black diamonds indicate dark galaxy candidates, with those detected by AGES labelled with their catalogue number.}
\label{fig:VirgoMap}
\end{figure*}

In figure \ref{fig:VirgoMap} we show the galaxies (from the Virgo Cluster Catalogue of \citealt{bing}), \HI{} streams and clouds which have been discovered thus far in the Virgo cluster (the LSB galaxies are described in section \ref{sec:env}). Since it is the clouds in this region which are our main interest, and as the area is well-studied and relatively small, we also show the known shorter streams (from \citealt{chung} and the bridge between VCC 2062 and VCC 2066 as described in \citealt{duc07}). Despite the historical importance of the cluster, the census of extended \HI{} features remains incomplete. The most sensitive large-area survey, ALFALFA, has released a point-source catalogue from +3:00 to +16:00 degrees declination (\citealt{haynes11}), but more detailed catalogues only as far south as declination +08:00 (\citealt{kent08}). Additionally, extended emission is very difficult to measure if it is shorter than about two beam widths. For the 3.5$'$ beam of Arecibo (17.3 kpc at the distance of Virgo) this makes it difficult to resolve features less than 35 kpc across. The VIVA survey was able to detect features $\lesssim$ 30 kpc (\citealt{chung}) thanks to the higher resolution of the VLA (15''), but only targeted 53 galaxies. Thus this map represents only our current state of the art; there are almost certainly more \HI{} streams (and clouds) which remain to be discovered, particularly ones which are comparable to or shorter in extent than the Arecibo beam.

In AGES V we demonstrated that the AGES clouds lack obvious optical counterparts in the Sloan Digital Sky Survey (SDSS). They appear to be categorically different from other objects in the sample. In AGES VI, we showed that \HI{} detections which had what we regarded as obvious, plausible optical counterparts obeyed the standard optical or baryonic Tully-Fisher relation (TFR) - even though some of those objects are optically extremely faint (M$_{g}$ = -10.0) and are not listed in the VCC. In contrast, although optical counterpart candidates can be found for the AGES clouds, they are all too faint to obey the standard TFR\footnote{We also considered the possibility that the clouds are not in the Virgo cluster, but this would require them to be foreground objects which is not plausible given their redshifts - see AGES VI for details.}. Although the \HI{} masses of the clouds are similar to other objects in the sample, their offset from the TFR may indicate that their nature is very different to that of ordinary, faint galaxies.

\begin{figure}
\centering  
  \subfloat[AGESVC1 247]{\includegraphics[height=40mm]{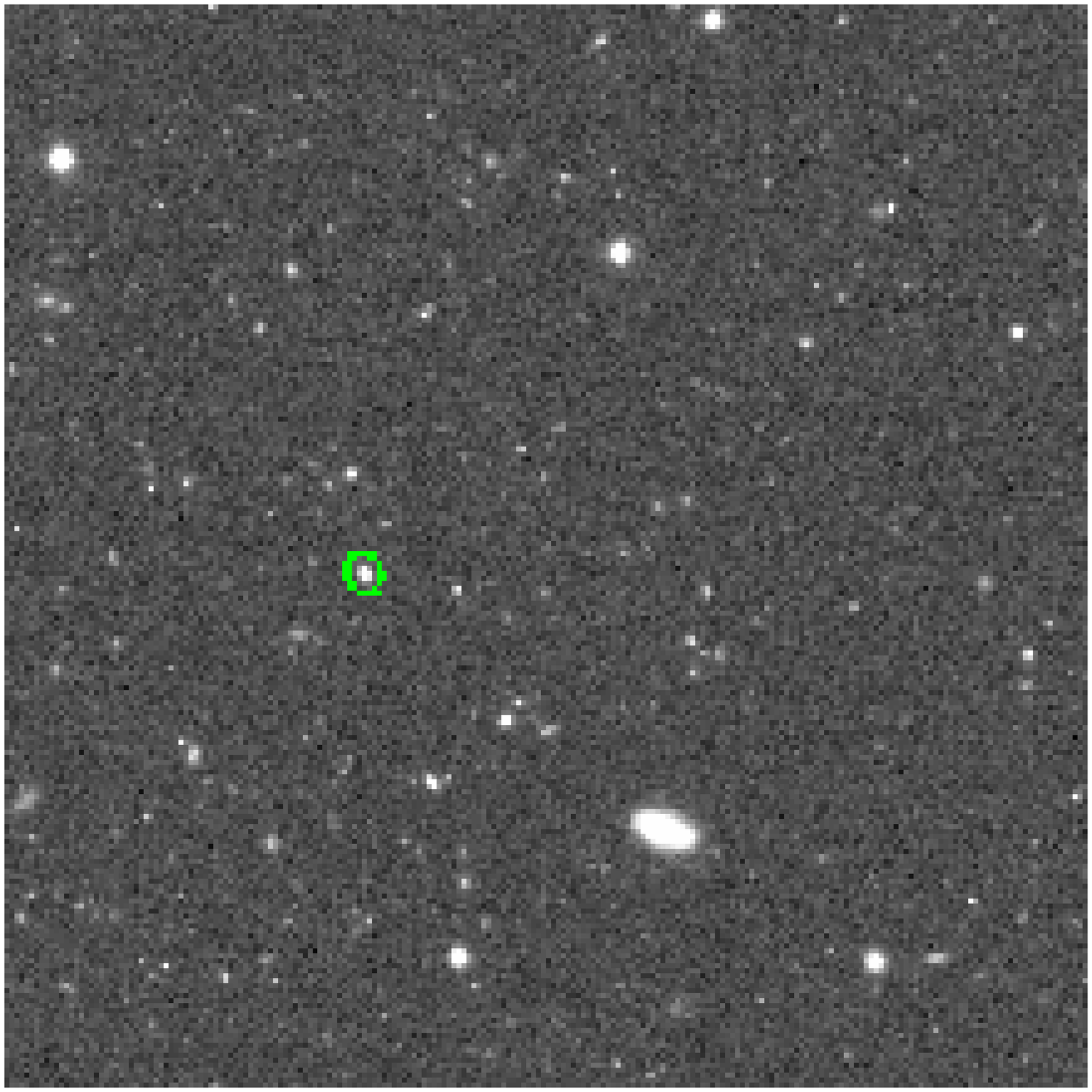}}
  \subfloat[AGESVC1 257]{\includegraphics[height=40mm]{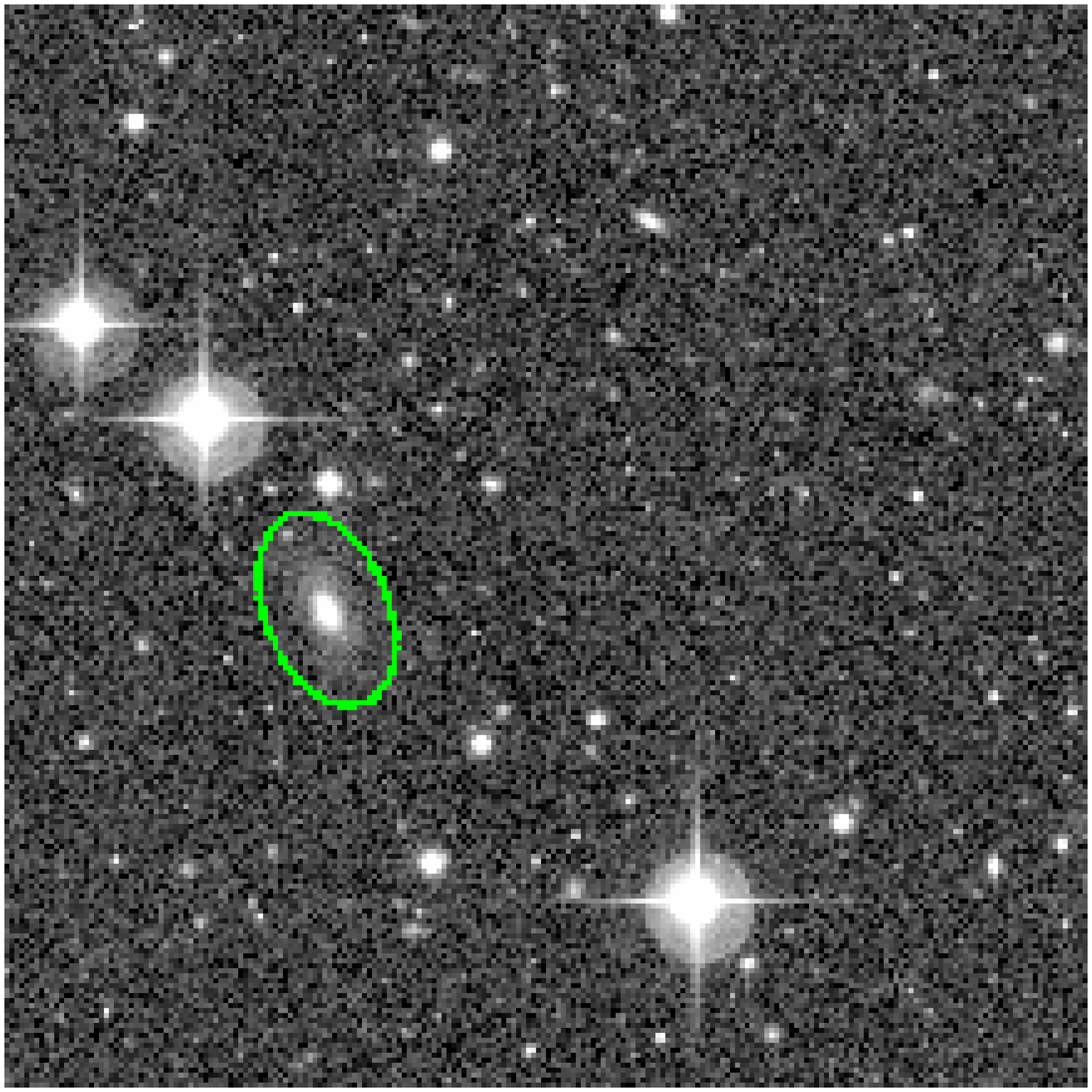}}\\ 
  \subfloat[AGESVC1 262]{\includegraphics[height=40mm]{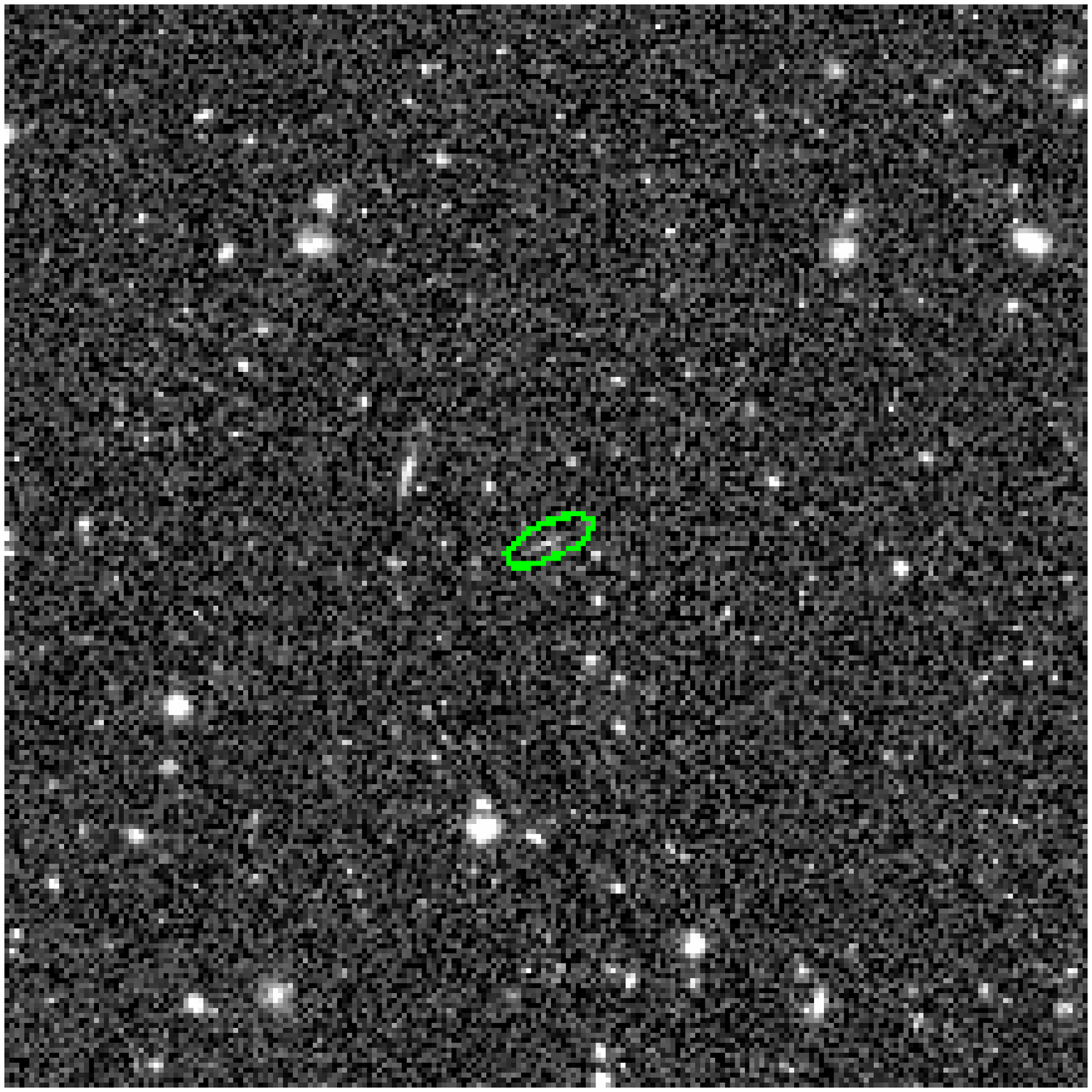}}
  \subfloat[AGESVC1 266]{\includegraphics[height=40mm]{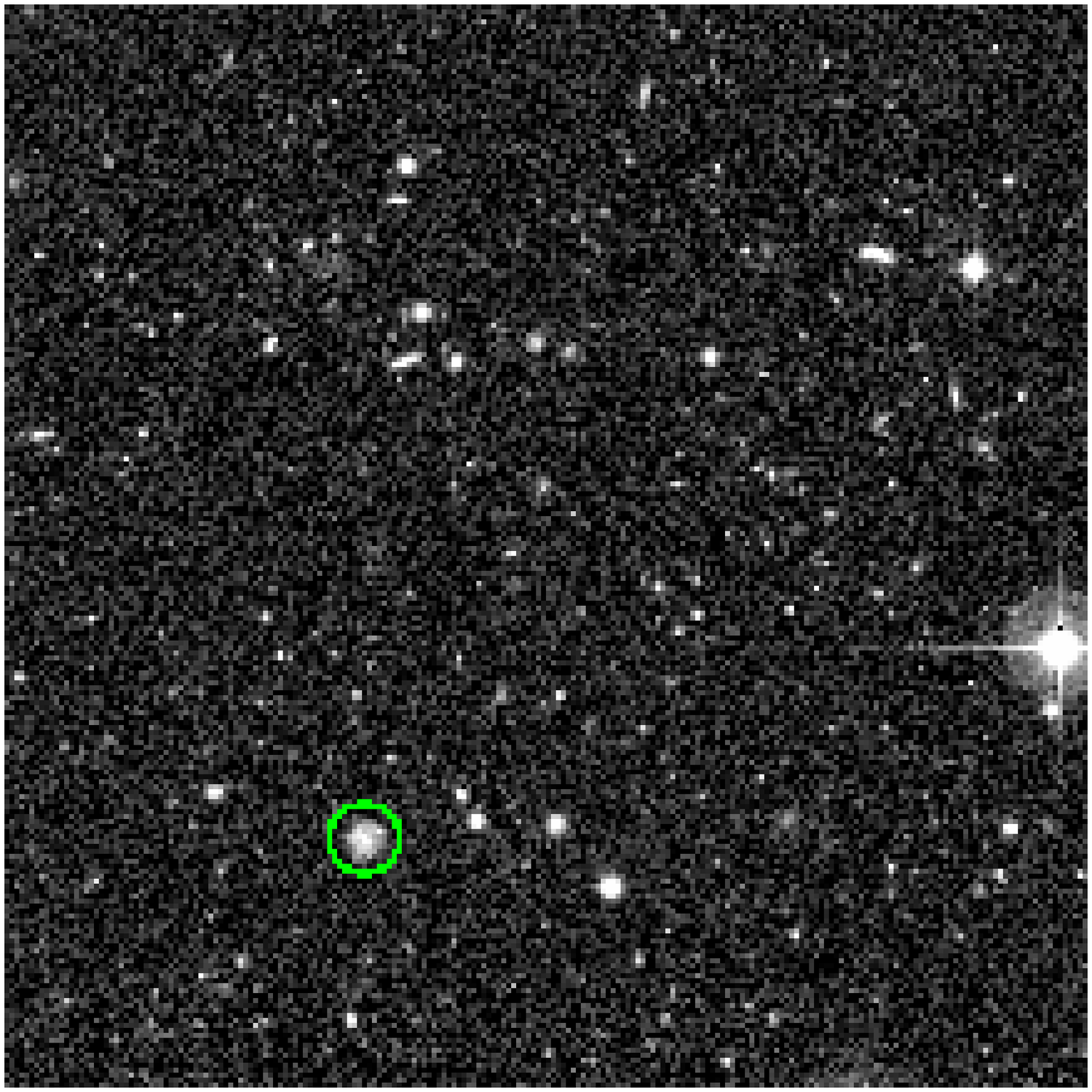}}\\ 
  \subfloat[AGESVC1 274]{\includegraphics[height=40mm]{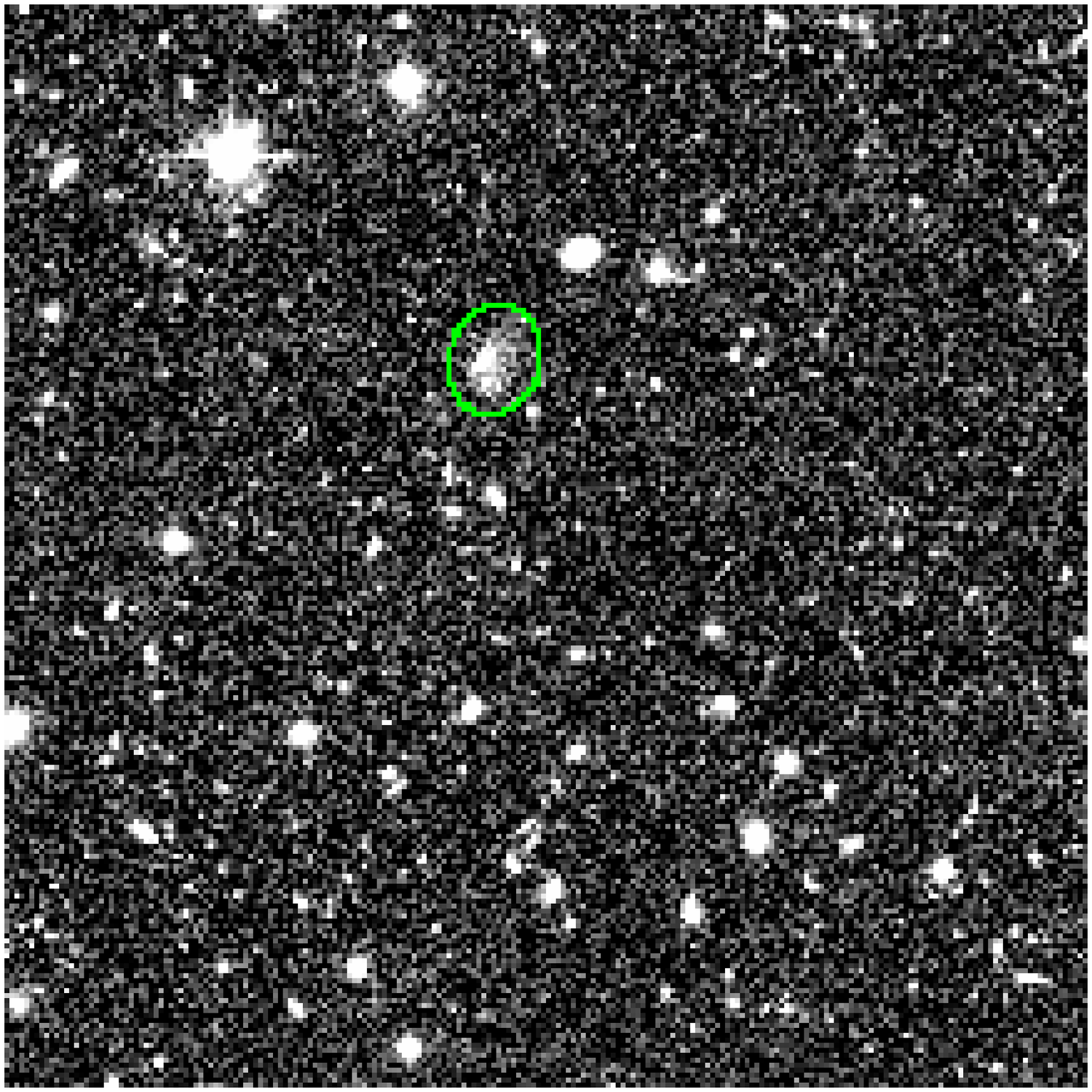}}
  \subfloat[AGESVC1 282]{\includegraphics[height=40mm]{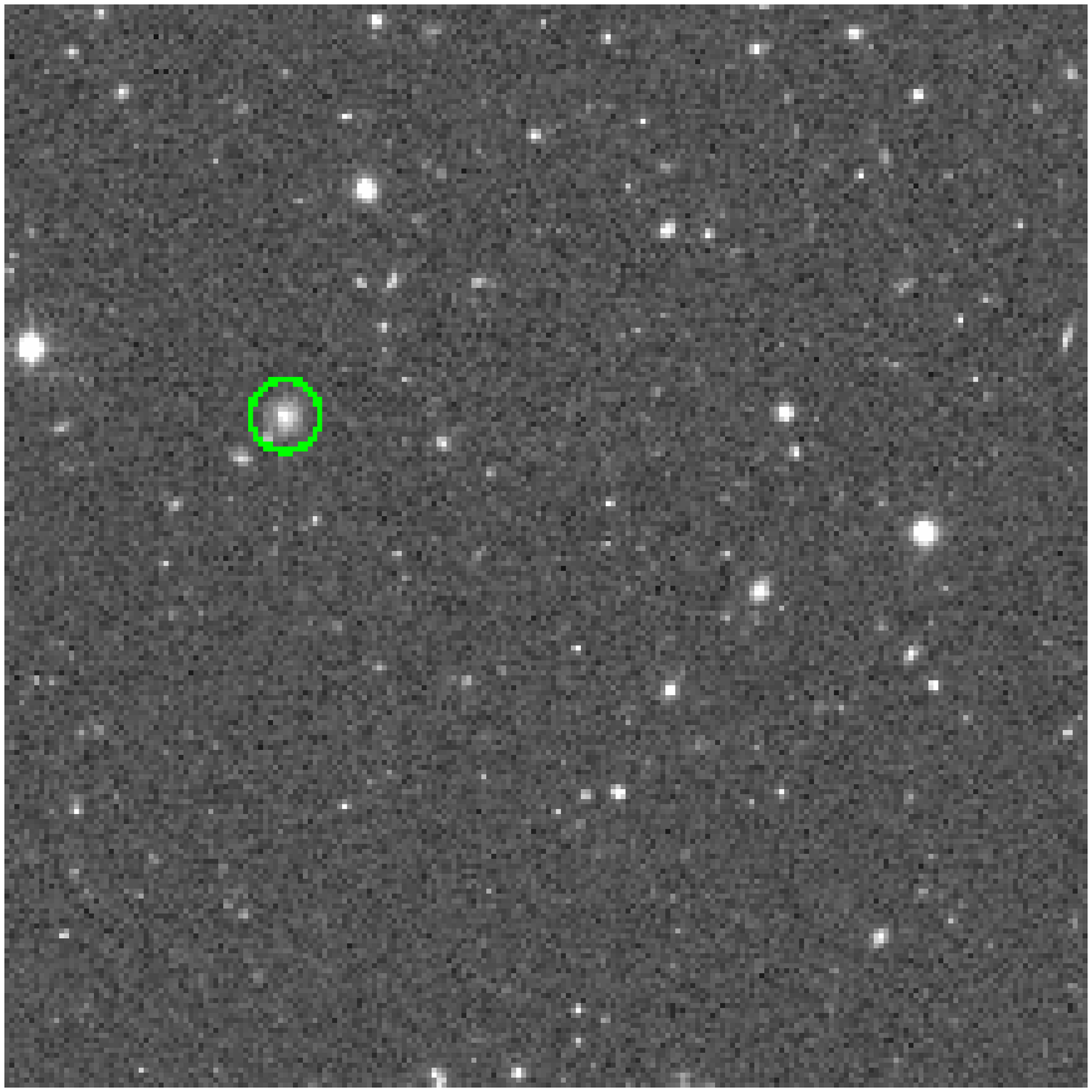}}   
\caption[Streams]{NGVS \textit{g}-band images for \HI{} clouds reported as optically dark based on SDSS data. Each image is 3.5$'$ across (the size of the Arecibo beam - the clouds are unresolved to Arecibo); north is up and east is to the left. The images are centred on the \HI{} coordinates. The green apertures show our identified optical counterpart candidate.}
\label{fig:dgNGVS}
\end{figure}

We here confirm this using data from the much deeper Next Generation Virgo Survey (NGVS; \citealt{ngvs}). Two of the clouds (AGESVC1 231 and AGESVC1 258) are outside the NGVS footprint. We show images for the remaining six in figure \ref{fig:dgNGVS}. All of the visible objects are too faint by 3-4 magnitudes to lie on the TFR - and this is without correcting the velocity widths for the inclination, which could only make the deviation stronger. For AGESVC1 257 and 274 extended optical objects are visible that may be within the Virgo cluster (based on their size and morphologies). Without higher resolution \HI{} observations and optical spectroscopy, we cannot say for certain if these objects are really associated with the \HI{} or not.

We are applying for VLA observation time for these objects, which will have sufficient resolution to determine if the \HI{} is really currently associated with the purported optical counterparts. We refrain from commenting further on the optical properties of these objects until the observations are obtained, except to note that if they are associated, these objects would be galaxies that do not obey the baryonic TFR. Recently \cite{lelli} have described six tidal dwarf galaxies which also deviate from the TFR but in the opposite sense to the Virgo clouds : their objects have higher baryonic masses than the TFR predicts, whereas the Virgo clouds have lower baryonic masses (see also \citealt{jan15}).

\subsubsection{\HI{} clouds in the Virgo cluster - self-bound, pure \HI{}}
\label{sec:selfbound}

Several possible interpretations for the AGES clouds are outlined in table \ref{tab:agesclouds}. First, in principle the clouds could be gravitationally self-bound and stable without dark matter if they are sufficiently small. The radius for this is given in column 6, assuming that we are viewing edge-on rotating discs. The radii for the type 1 clouds correspond to column densities about a thousand times greater than that typically found in dwarf galaxies (6 \Msolar{}/pc$^{2}$ from \citealt{leroy}), so high that we may safely dismiss this scenario. In the type 2 clouds the radii correspond to much lower column densities, around 10 times greater than in dwarf galaxies - values which are compatible with the observed densities of star-forming molecular gas in spiral galaxies ($>$ 10 \Msolar{}/pc$^{2}$, \citealt{leroy}). 

\begin{table*}
\centering
\caption[agescloud]{Derived properties of the dark \HI{} clouds detected in the Virgo Cluster by AGES. The columns are as follows : (1) Name in AGES V; (2) Assumed distance as described in AGES V; (3) \HI{} mass in solar units; (4) Velocity width in \kms{}; (5) Peak S/N; (6) Radius at which the \HI{} would be in stable, self-bound rotation; (7) Radius (in kpc) at which the \HI{} would have a column density of 9.4$\times$10$^{20}$ cm$^{-2}$ - typical for dwarf galaxies as described in \cite{leroy}; (8) Maximum radius of the \HI{} (since they are unresolved by the Arecibo beam) in kpc; (9) Time in Myr to expand to the Arecibo beam size assuming their velocity width corresponds to the expansion velocity; (10) Distance (kpc) and (11) angular distance (arcmin) travelled across the cluster assuming a velocity of 590 \kms{}.}
\scriptsize
\label{tab:agesclouds}
\begin{tabular}{c c c c c c c c c c c}
\hline
  \multicolumn{1}{c}{(1)} &
  \multicolumn{1}{c}{(2)} &
  \multicolumn{1}{c}{(3)} &
  \multicolumn{1}{c}{(4)} &
  \multicolumn{1}{c}{(5)} &
  \multicolumn{1}{c}{(6)} &
  \multicolumn{1}{c}{(7)} &
  \multicolumn{1}{c}{(8)} &
  \multicolumn{1}{c}{(9)} &
  \multicolumn{1}{c}{(10)} &
  \multicolumn{1}{c}{(11)} \\
  \multicolumn{1}{c}{Name} &
  \multicolumn{1}{c}{Distance} &
  \multicolumn{1}{c}{M\HI{}} &
  \multicolumn{1}{c}{W20} &
  \multicolumn{1}{c}{SN} &
  \multicolumn{1}{c}{Size to be} &
  \multicolumn{1}{c}{Size at dwarf} &
  \multicolumn{1}{c}{Arecibo beam} &
  \multicolumn{1}{c}{Lifespan} &
  \multicolumn{1}{c}{Travelled} &
  \multicolumn{1}{c}{Travelled} \\
  \multicolumn{1}{c}{} &
  \multicolumn{1}{c}{Mpc} &
  \multicolumn{1}{c}{\Msolar{}} &
  \multicolumn{1}{c}{\kms{}} &
  \multicolumn{1}{c}{} &
  \multicolumn{1}{c}{self bound; kpc} &
  \multicolumn{1}{c}{galaxy N\HI; kpc} &
  \multicolumn{1}{c}{size; kpc} &
  \multicolumn{1}{c}{Myr} &
  \multicolumn{1}{c}{kpc} &
  \multicolumn{1}{c}{arcmin} \\
\hline
  AGESVC1\_231 & 32 & 4.2E7 & 152 & 7.8 & 0.031 & 1.3 & 16.3 & 192 & 116 & 12\\
  AGESVC1\_247 & 23 & 2.3E7 & 33 & 15.9 & 0.363 & 1.0 & 11.7 & 635 & 383 & 57\\
  AGESVC1\_257 & 17 & 1.4E7 & 157 & 5.4 & 0.010 & 0.8 & 8.7 & 98 & 59 & 11\\
  AGESVC1\_258 & 17 & 1.4E7 & 120 & 7.6 & 0.017 & 0.8 & 8.7 & 128 & 78 & 16\\
  AGESVC1\_262 & 23 & 2.0E7 & 146 & 7.2 & 0.016 & 0.9 & 11.7 & 144 & 87 & 13\\
  AGESVC1\_266 & 17 & 3.2E7 & 173 & 6.4 & 0.018 & 1.2 & 8.7 & 85 & 51 & 10\\
  AGESVC1\_274 & 17 & 7.3E6 & 35 & 14.2 & 0.103 & 0.6 & 8.7 & 452 & 273 & 55\\
  AGESVC1\_282 & 23 & 4.4E7 & 164 & 11.4 & 0.028 & 1.4 & 11.7 & 123 & 74 & 11\\
\hline								   
\end{tabular}
\end{table*}

Alternatively the clouds could be gravitationally bound but collapsing. Since the clouds are optically dark, it is worth considering their free fall times as collapse would eventually mean they reach density high enough to trigger star formation. Ignoring their line widths for just a moment, the free fall time for the uniform density spherical clouds is given by :
\begin{equation}
t_{ff} = \sqrt{\frac{\pi^{2}r^{3}}{8 G M}} 
\label{eqt:freefall2}
\end{equation}
For a radius of 1 kpc the free fall time is about 100 Myr, though that rises dramatically to over 1 Gyr for 5 kpc. However this will change depending on what their line width really represents. We will consider the possibility of expansion and rotation in sections \ref{sec:unbound} and \ref{sec:darkgals} respectively (see also section \ref{sec:dgsurvival}). In the collapsing scenario, the typical velocity width of 150 \kms{} corresponds to a collapse time of $<$100 Myr assuming the source fills the Arecibo beam (hence the upper limit). Thus the clouds are very unlikely to be self-bound by their \HI{} alone : if they were stable this would require fantastically high column densities; if they were collapsing they would quickly form stars. Hence this scenario is unlikely since there is no evidence for very young galaxies in the Virgo cluster. Stability without star formation implies extra mass, which we discuss in section \ref{sec:darkgals}.

\subsubsection{\HI{} clouds in the Virgo cluster - unbound debris}
\label{sec:unbound}
If the clouds have column densities typically observed in galaxies, they cannot be self-bound by the mass of their \HI{} alone. In this case it is interesting to estimate the dispersal timescale for the clouds. We do not know the true 3D geometry of the clouds, making it difficult to estimate how long the clouds would remain detectable if we are seeing streaming motions along the line of sight (as in \citealt{bekki}). However it is straightforward to consider the case of an unbound disc viewed edge-on : in this case our measured velocity widths correspond to the expansion velocity (see \citealt{borth10} for a similar analysis). 

We assume that the cloud's initial size corresponds to the size in column 7 of table \ref{tab:agesclouds} (corresponding to the typical column density in dwarf galaxies); since this is much smaller than the Arecibo beam it makes very little difference if we allow even higher initial column densities. Since there is no sign of any extended component the objects cannot have expanded to a size larger than the Arecibo beam (column 8). This gives an upper limit of the time in column 9. With a median `lifespan' of 125 Myr for type 1 clouds, this simple calculation shows that unbound debris can, in principle, survive for extended periods even with the highest velocity widths observed. We have neglected self-gravity in this calculation since the escape velocity would be 11 \kms{} at a size of 1 kpc, about seven times less than the expansion velocity.

Since we do not know the proper motions of the objects, our estimates of their lifespans cannot constrain how far they may have travelled from their parent galaxies. We assume, for the sake of a guide, a proper motion of 590 \kms{} - the velocity dispersion of subcluster A according to \cite{mei}. This seems a reasonable assumption since \cite{duc} have shown that encounters at relative velocities $>$ 1,000 \kms{} can remove similar (or even greater) masses of hydrogen (\citealt{oo05} perform a similar exercise). Combined with the lifespans of the objects, this gives us the estimates of the distances travelled in columns 10 and 11. Searching NED within these radii, we find that there are giant spiral galaxies present in most cases that could potentially be the source of the \HI{} (there are two exceptions but both have galaxies present a few arcminutes outside the search radius). So indeed unbound debris can, in principle, travel sufficiently far from its parent galaxies that it can appear isolated before it disperses and becomes an extended source (or even undetectable, which we discuss below), though this neglects the ICM - see section \ref{sec:hirps}.

We assumed above that the clouds would expand to no more than an Arecibo beam size. It is interesting to note that the type 1 clouds, given their S/N values, could expand at least a factor of two before they become undetectable (possibly more since larger clouds will be easier to detect even at low S/N levels). If they are unbound debris, then, we have to wonder why we only detect small clouds despite that larger clouds are potentially detectable for twice as long as our lifespan estimates. Additionally the type 2 clouds apparently have much lower expansion velocities so should be detectable for longer, yet we detect six type 1 clouds and only two type 2 clouds. We would also expect type 2 clouds to be more common since their lower velocity widths are much closer to the typical 10 \kms{} velocity dispersion of \HI{}). Small number statistics could explain this.

Alternatively, if we are seeing inclined (rather than edge-on) discs, or spheres, then a more modest increase in diameter would render them undetectable (for the extreme case, a disc seen face-on - or a sphere - which fills the beam would only have to expand by a factor $\sqrt{2}$ for its S/N to decrease by a factor of two). If they are inclined discs, then discs seen face-on should have higher S/N since we will only measure the velocity dispersion of the gas, not its rotation (i.e. we detect the same total flux but in fewer channels). The type 2 clouds indeed have the highest S/N measurements, though we again caution the danger of small number statistics. Nevertheless, the idea of the clouds being rotating, expanding discs is consistent with the observations.

The lack of detected streams may be more problematic for this scenario. It is unclear how such a \HI{} cloud could be removed from a galaxy without disrupting a larger part of its disc into a stream, nor is it obvious that the stream would become undetectable more quickly than any overdensities within it. We note that the dispersal timescale estimates are of the same order as the formation timescales for the streams (see section \ref{sec:streams}), so the lack of associated streams is surprising.

Despite these caveats, the most important point from this exercise is that unbound debris can persist, even if expanding at velocities $\sim$100 \kms{}, for timescales $\sim$100 Myr. Thus it is not necessary for the clouds to be self-bound. We will examine whether fragmentation and dispersal of the stream can be responsible for the formation of such features, as well as the survival of such features once they are formed, in section \ref{sec:sims}.

Another variation of the interpretation of the clouds as unbound debris would be that each one is actually composed of several different clouds. These sub-clouds could actually be overdensities in a stream or entirely unrelated clouds that are merely aligned by chance. In either case, the individual clouds could have low velocity widths (which are known to exist in systems in which the tidal origin is certain, e.g. \citealt{lelli}), and only the superposition makes them appear as a single cloud of a high velocity width. However these scenarios scarcely offer any advantages : if the clouds within one beam are all separate parts of the same stream, some mechanism must still be found to separate the clouds in velocity whilst keeping them within the 17 kpc Arecibo beam area. Multiple clouds with independent origins could explain this. However given the extremely low number of clouds known in total in the cluster, the chance of several clouds with unrelated origins happening to align so closely both in space and velocity is negligible.

\subsubsection{\HI{} clouds in the Virgo cluster - the ICM and ram-pressure stripping}
\label{sec:hirps}
It is very difficult to quantify the exact effect of the intracluster medium without a numerical simulation, which we plan for a future study. However, using some simple approximations, we can at least quantify whether or not we expect ram pressure to be significant given the properties of the blobs.

Following \cite{vol01}, we approximate the ICM density using the following :
\begin{equation}
\rho = \rho_{C}\Big(1 + \frac{r^{2}}{r_{c}^{2}}\Big)^{-\frac{3}{2}\beta}
\label{eqt:icmrho}
\end{equation}
Where $\rho$ is the ICM density at a given distance $r$ from the cluster centre, $\rho_{C}$ is the central density (4$\times$10$^{-2}$cm$^{-3}$ from \citealt{vol01}), $r_{c}$ is the radius of the cluster core (13.4 kpc from \citealt{vol01}), and $\beta$ is the slope parameter (0.5 from \citealt{vol01}). We also use \cite{vol01}'s approximation that the ICM temperature $T$ is a uniform 10$^{7}$ K (see also \citealt{shib}). Thermal pressure $P$ in the ICM is given by :
\begin{equation}
P_{therm} = \frac{\rho\,k_{B}\,T}{m_{p}}
\label{eqt:picm}
\end{equation}
Where $k_{B}$ is the Boltzman constant and $m_{p}$ is the average particle mass which we take to be one atomic mass unit. We find that for a temperature of 10,000 K for the clouds (the maximum temperature of \HI{}, above which most of the gas is expected to be ionized), and a radius of 1 kpc (see table \ref{tab:agesclouds}) assuming them to be uniform-density spheres, the ICM pressure is comparable to that in the clouds ($\sim$1750 k$_{B}$\,K\,cm$^{-3}$) at a distance of 500 kpc from the cluster centre; at 1 Mpc the ICM pressure is only slightly lower (about 80\% of the value at 500 kpc). The ram pressure is given by :
\begin{equation}
P_{ram} = \rho\,v^{2} 
\label{eqt:picm}
\end{equation}
Which we can re-arrange to calcuate the velocity at which the ram-pressure equals the thermal pressure : 290 \kms{}. Therefore we expect the ICM to be highly significant for the evolution of the clouds, both from its static thermal pressure and ram pressure at all but the lowest velocities.

One method to estimate the effects of ram-pressure stripping is to use Newton's impact formula to estimate the penetration depth (see also \citealt{pavel07}) :
\begin{equation}
D \approx L_{cloud} \frac{\rho_{cloud}}{\rho_{ICM}} 
\label{eqt:newton}
\end{equation}
Where $D$ is the penetration depth, L is the diameter of the cloud, and $\rho_{cloud}$ and  $\rho_{ICM}$ are the volume densities of the cloud and the ICM respectively. We can use equation \ref{eqt:icmrho} to find the approximate ICM density at 500 kpc from the cluster centre. Assuming spherical clouds of mass 3$\times$10$^{7}$ \Msolar{} and radii 1, 5 and 10 kpc, we calculate penetration depths of 3.3 Mpc, 133 kpc, and 33 kpc respectively. Unless the clouds are so dense that they should be forming stars, they should be very close to the site where they were initially deposited into the intracluster medium (with the caveat that this is strongly dependent on the exact size of the clouds). Coupled with the long estimates for the survival timescales of ram-pressure stripped streams described in section \ref{sec:intro}, the lack of any streams detected from the nearest galaxies is indeed surprising. However we also caution that the dark-matter dominated galaxies could penetrate much further into the ICM than any deposited material, which could explain the large distances of the clouds from their parent galaxies. 

In the above calculations we assumed a density for the clouds equal to or lower than that in dwarf galaxies, which is comparable to the observed thresholds for star formation (\citealt{schaye}). Smaller radii imply that star formation should occur and the clouds would not be optically dark (but see also section \ref{sec:darkgals}). Yet even at this density the clouds are vulnerable to ram-pressure stripping. They seem almost paradoxical : they should either be forming stars or rapidly destroyed by ram-pressure stripping, but they have somehow avoided doing both.

The major unknown is whether the ram-pressure would rapidly break the clouds apart or if pressure confinement would keep them stable over long periods. Recently, \cite{burk} have pointed out that the clouds would be in pressure equilibrium with the ICM given their velocity widths, the local ICM density, and assuming radii of 1-10 kpc. These radii are certainly compatible with the Arecibo observations so higher resolution studies are necessary to determine the clouds' true size and morphology. In the \cite{burk} model, the clouds are free of dark matter and remain stationary within the cluster, so ram pressure stripping is not important. This is also consistent with the preferential detection of such clouds in clusters (see section \ref{sec:env}), where the external pressure is expected to be greater than in less dense environments. Additionally, \cite{vilnar} have run galaxy cluster formation SPH simulations and find that there are of order a few (maximum ten) starless \HI{} clouds (free of dark matter) produced in each simulated cluster. While this number compares well with the total number found by AGES in Virgo, AGES covered just 10\% of the cluster. \cite{vilnar} also note that they are uncertain how many of the clouds may be numerical artifacts, e.g. caused by low resolution.

Little further quantitative analysis is possible without numerical simulations to examine the role of the ICM in detail. We note, though, that if the clouds are indeed pure pressure-confined gas we would not expect them to be long-lived. Their internal pressure can only arise from thermal pressure or dynamic pressure (i.e. turbulence). Thermal pressure would mean temperatures $>$10$^{5}$ K to produce the observed line widths ($\sim$150 \kms{}), which is too hot for \HI{}. There is no obvious mechanism that could maintain the bulk internal motions required for dynamic pressure, while the ICM would continuously act to reduce the velocities (so reducing the observed line widths). Even if there was some driving source for the turbulence, it is unclear if this would cause the gas to collapse and form stars (\citealt{burk} propose that the clouds are on the cusp of forming molecular gas) or simply disperse more rapidly. We are preparing a set of high resolution hydrodynamic simulations to quantify the detectable timescales of such turbulent clouds (Taylor \& W\"{u}nsch in preparation). This solution is arguably more plausible for some of the non-AGES clouds (which have lower line widths that can be readily explained from thermal motions), though without examining the details of each one, we note that the tidal debris scenario appears to be a valid alternative (see section \ref{sec:sims}; \citealt{wolfe} discuss the pressure confinement scenario for the clouds between M33 and M31).

\subsubsection{\HI{} clouds in the Virgo cluster - ionization}
\label{sec:ions}
Based on \cite{cowie}, \cite{borth10} attempt to constrain how long \HI{} clouds can remain neutral given the presence of a hot intracluster medium. We here repeat this analysis for the AGES Virgo clouds. The ratio of classical to saturated heat flux, which determines if classical or saturated evaporation applies, is given by :
\begin{equation}
\sigma_{0}\simeq \frac{(T_{ICM} / 1.5\times10^{7})^{2}}{\rho_{ICM}\,R_{cloud}}
\label{eqt:heatratio}
\end{equation}
Where $T_{ICM}$ is the temperature of the ICM in Kelvin, $\rho_{ICM}$ is the density of the ICM in cm$^{-3}$, and $R_{cloud}$ is the radius of the \HI{} cloud in parsecs (assuming a spherical cloud). For the Virgo cluster, $T_{ICM}$$\approx$1$\times10^{7}$ K and $\rho_{ICM}$ is 1.75$\times10^{-4}$ cm$^{-3}$ at 500 kpc from the cluster centre (\citealt{vol01}). We only know for certain that $2R_{cloud}$ $\leq$ 17.3 kpc. Assuming the true size is actually that given in column 7 of table \ref{tab:agesclouds}, about 1 kpc, then $\sigma_{0}$ = 2.5. $\sigma_{0}$ $>$ 1.0 indicates that the evaporation timescale can be estimated based on the saturated approximation :
\begin{equation}
t_{evap} \sim 10^{6}\:\rho_{cloud}\:\rho_{ICM}^{-1}\:R_{cloud}\:T_{ICM}^{-1/2}
\label{eqt:satrate}
\end{equation}
Where $t_{evap}$ is in years if $R_{cloud}$ is in parsecs. For a typical cloud mass of 3$\times$10$^{7}$ \Msolar{} with  $R_{cloud}$ = 1.0 kpc, $\rho_{cloud}$ = 0.29 cm$^{-3}$, giving an evaporation time of 530 Myr - rather longer than the typical `lifespan' estimated from expansion and detectability.

However, this will be significantly affected by the cloud radius. At the typical minimum column density observed in galaxies of 2$\times$10$^{19}$ cm$^{-2}$, which \cite{borth10} say indicate the threshold for ionization by the cosmic UV background, the size of the cloud will be 7.3 kpc. Equation \ref{eqt:heatratio} then predicts the evaporation will occur in classical mode since $\sigma_{0}$ $<$ 1.0, and the evaporation timescale will then be given by :
\begin{equation}
t_{evap} = 3.3\times10^{20}\:\rho_{cloud}\:R_{cloud}^{2}\:T_{ICM}^{-5/2}\:ln(\Lambda/30)
\label{eqt:classrate}
\end{equation}
Where $ln(\Lambda) = 29.7 + ln(\rho_{ICM}^{-1/2})\times T_{ICM}/10^{6}$ if $T_{ICM}$ in in Kelvin. At 7.3 kpc radius the evaporation time would be 2.9 Gyr, while at the maximum 8.7 kpc radius (17.3 kpc diameter - the beam width) this drops slightly to 2.4 Gyr. These estimates are considerably more than the $\sim$280 Myr detectable lifetime based on expansion - evaporation is essentially negligible except at the smallest sizes, where their density should be high enough for star formation. A caveat is that if the clouds are moving through the ICM, the ICM density will be continuously varying, so the evaporation timsescales are subject to large uncertainties. If the clouds are themselves expanding as in \ref{sec:unbound}, their density will be continuously decreasing, making their true survival time even more difficult to estimate.

\subsubsection{\HI{} clouds in the Virgo cluster - dark galaxies}
\label{sec:darkgals}
As discussed above, it appears unlikely that the clouds are self-bound by their \HI{} alone, and the idea that they are unstable debris also has difficulties. The alternative is that these are stable objects in dark matter halos. The arguments in section \ref{sec:unbound} that the clouds are rotating discs (there are fewer type 2 than type 1 clouds, and type 2 clouds have higher S/N levels - consistent with their being inclined discs) still apply, with the advantage that they could also be long-lived. It would also explain why fewer type 2 clouds are detected than type 1, which is surprising in the case of the clouds being unbound since those expanding more slowly should be detectable for longer.

The model of \cite{d06} predicted the fraction of \HI{} detections expected to be optically dark galaxies for various \HI{} surveys - for AGES, this was estimated at 23\%. This was shown in AGES V not to be the case - the fraction is, at most 7\% of detections in the Virgo cluster, 1\% of the whole survey, and very plausibly 0\%. However, as well as the ALFALFA observational evidence described in section \ref{sec:cloudoverview}, the recent EAGLE simulation (Evolution and Assembly of Galaxies and their Local Environments : \citealt{eagle}) has revived the idea that such objects are an important part of cosmological models (\citealt{saw}), albeit at the scale of dwarf galaxies rather than giants. If we can show that some \HI{} detections cannot be explained as tidal debris (see section \ref{sec:sims}), it could imply that many other objects purported to be tidal debris are actually primordial objects. As these are generally far more massive than the missing halos required to satisfy $\Lambda$CDM simulations of the Local Group, the existence of even a few such dark galaxies would have important consequences for cosmological models.

In AGES VI, we estimated the dynamical masses for the clouds assuming them to be rotating discs viewed edge-on, with a radius determined by the typical density of \HI{} in dwarf galaxies (given here in column 7 of table \ref{tab:agesclouds}). We calculate dynamical masses assuming the objects are discs in stable circular rotation, thus : 
\begin{equation}
M_{dyn} = \frac{r\,v_{c}^{2}}{G}
\label{eqt:mdyn}
\end{equation} 
Where $v_{c}$ is the circular velocity and assumed to be equal to half the line width W20.

Dynamical mass to \HI{} (M$_{dyn}$/M\HI{}) ratios range from 20-60 for type 1 clouds, but only around 3 for the type 2 clouds (but note also the values in table \ref{tab:dgal}, which give the upper limits using the beam size for the diameter of the clouds and as such are ten times higher). Type 2 clouds were also shown to obey the standard Tully-Fisher relation for field galaxies, whereas type 1 cloud's velocity widths are too large given their baryonic masses. Intriguingly, \cite{adams15} find that AGC 198606 shows a similar deviation and also shows ordered rotation - though its velocity width is comparable to the type 2 clouds.

It is interesting to consider whether we should expect these clouds to be stable against star formation. Without higher resolution observations we can only roughly estimate the Jeans length of the objects since we do not know their true geometry/densities or velocity dispersion (i.e. temperature). The Jeans length is given by the following equation~:
\begin{equation}
L_{J} = c_{s}\sqrt{\frac{15}{4\pi\,G\,\rho}} 
\label{eqt:jeanlength1}
\end{equation}
Where $c_{s}$ is the sound speed and is given by $\sqrt{k_{B}T / \overline{m_{p}}}$, $k_{B}$ is the Boltzman constant, T the temperature and $\overline{m_{p}}$ the average particle mass, so for a spherical cloud :
\begin{equation}
L_{J} = \sqrt{\frac{5 k_{B} T R_{cloud}^3}{G M m_{H}}}
\label{eqt:jeanlength1}
\end{equation}
Where, assuming pure hydrogen, $m_{H}$ is the mass of the hydrogen atom, $M$ the total mass of the cloud and $R_{cloud}$ is its radius. Since $L_{J}$ scales as $R_{cloud}^{1.5}$, if a cloud of a particular mass is Jeans stable at a particular size it will also be stable at larger sizes. For spherical clouds of the smallest plausible size (1 kpc, see table \ref{tab:agesclouds} and section \ref{sec:unbound}), at 10,000~K (approximately equal to the 10 \kms{} velocity dispersion observed in most galaxies, and above which most of the \HI{} becomes ionized), the Jeans length is 1.6 kpc. Hence 10,000 K clouds are Jeans stable at any plausible size. At 1,500 K , the Jeans length for a 1 kpc cloud is 0.3 kpc. Thus, for most plausible sizes and temperatures of the clouds, we expect them to be stable against Jeans fragmentation and star formation.

The clouds would also be less susceptible to stripping if they were embedded in dark matter halos, simply due to the greater restoring force from the extra mass (see \citealt{pavel}). Hence they could travel much further through the ICM than in the scenario of section \ref{sec:hirps}, which would readily explain their locations far from the nearest galaxies. In this scenario the lack of detected streams from \HI{}-deficient galaxies is an unrelated problem.

One difficulty for the dark galaxies hypothesis is that while this would allow the objects to be dynamically stable, it would not prevent them being ionized as discussed in section \ref{sec:ions}. The evaporation timescales will be altered since the discs will have higher volume densities than spherical clouds. Without resolved observations to determine the disc thickness we cannot say how much of a difference this will make to the evaporation timescales.
  
\subsection{Environment of the clouds and streams}
\label{sec:env}
\begin{figure}
\centering
\includegraphics[width=84mm]{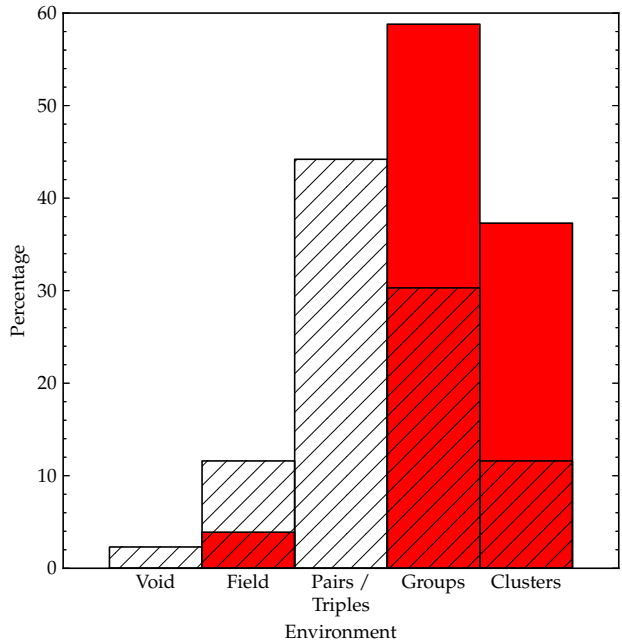}
\caption[hienv]{Distribution of the environments of the catalogued \HI{} clouds (red) and streams (hatching).}
\label{fig:hienv}
\end{figure}

We show the distribution of the environments of the clouds and streams in figure \ref{fig:hienv}. There is some evidence that the clouds favour the denser cluster environment while the streams are found preferentially in lower-density regions. However, because of the small samples sizes, there are major caveats to this. All of the clouds found in clusters are actually found exclusively in the Virgo cluster. Features of similar mass would be very difficult to detect in more distant clusters : for instance at the $\sim$100 Mpc distance of the Coma cluster, the Arecibo beam would be $\sim$100 kpc across, and a survey as sensitive of AGES would only have a \HI{} mass sensitivity of approximately 10$^{8}$ \Msolar{}. Despite this, it is clear that streams are preferentially found in low-density environments and are lacking in the Virgo cluster, as discussed previously.

We noted in section \ref{sec:virgoclouds} that the census of \HI{} clouds and streams in the Virgo cluster is certainly not complete. As discussed in section \ref{sec:darkgals} it is now very unlikely that there is a large population of massive, \HI{}-rich galaxies that have escaped detection by optical surveys. However there is now a wealth of evidence for a population of galaxies which are optically faint (but not totally dark) and \HI{}-poor. \cite{vandok} reported the discovery of 47 galaxies in the Coma cluster with approximately the same physical extent as the Milky Way but about a thousand times fainter. \cite{koda} extended this analysis and found nearly a thousand LSB (low surface brightness) galaxies in Coma, about 300 of which are as large as the Milky Way. Both groups speculate that to survive in the cluster environment (especially near the cluster core) the galaxies must be extremely dark matter dominated. Additionally, \cite{van16} find evidence that these ultra-diffuse galaxies (UDGs) are a common feature of clusters.

A large population of UDGs has also been found in the Virgo cluster. \cite{d15} describe 303 galaxies discovered using deep optical data that had not been listed in the well-known Virgo Cluster Catalogue of \cite{bing}. They did not detect galaxies as large as those discovered in Coma, though this may relate to how the data was reduced. A small number (three) of such galaxies have been detected by \cite{mihos15} in Virgo, and since that data only covered 15 square degrees, it is likely that many more such galaxies await discovery in Virgo.

It is interesting to note that the VC1 area has a very low density of these UDGs compared to the rest of the cluster, and it is in this region that most of the cluster's optically dark clouds are found (see figure \ref{fig:VirgoMap}). Since these clouds may have the high amounts of dark matter that other authors have suggested would be necessary for the survival of UDGs in clusters, and two of them have possible optical counterparts, it is very tempting to speculate that these clouds are the progenitors of the small LSB galaxies. We are applying for further Arecibo time to survey an additional large area of the Virgo cluster to the same sensitivity as AGES, which may allow us to study the possible relationship between the dark clouds and LSB galaxies in more detail. At present the limited extent of the AGES observations makes it impossible to say whether the clouds are unique to this LSB-poor area or are also found alongside the LSB galaxies.

Kinematic information on these ultra-diffuse galaxies is very limited, but recently \cite{bee} have measured the kinematics of one such object, VCC 1287. They found a ratio of dark to luminous matter of 3,000 - considerably higher even than the maximum of $\sim$900 possible for the Virgo \HI{} clouds. This object is extremely intriguing. Its outermost globular cluster has a galactocentric radius of 8.1 kpc, making it the same size as the maximum possible size of the optically dark Virgo \HI{} clouds. Its stellar mass of 2.8$\times$10$^{7}$ \Msolar{} is also strikingly similar to the baryonic mass ($\sim$3.0$\times$10$^{7}$ \Msolar{}) of the \HI{} clouds. The measured dynamical mass of VCC 1287 (2.6$\times$10$^{9}$ \Msolar{}) is rather lower than the maximum of the Virgo clouds ($\sim$1.3$\times$10$^{10}$ \Msolar{}), however \cite{bee} note that its true virial mass may be as high as 7.3$\times$10$^{10}$ \Msolar{}. Perhaps most interestingly of all, figure 5 of \cite{bee} shows that VCC 1287 has an unusually high dynamical mass given its baryon content : exactly the same situation as our Virgo \HI{} clouds (see AGES VI figure 7 and also this work figure \ref{fig:tfrclouds}).

It is important to note that not all of the optically dark \HI{} clouds are found in clusters or in systems with clear signatures of interactions. \cite{jan15} describe a system of clouds around a field spiral galaxy (but perhaps not associated with it owing to their different velocities), and \cite{oo13} describe a cloud 150 kpc from the nearest spiral galaxy with no detectable extended \HI{} stream (see also appendix \ref{sec:ap2}). Thus while extremely rare, there do exist \HI{} clouds for which no conventional explanation (tidal debris or gas-rich LSB galaxies) is currently sufficient. This supports the possibility that clouds in denser environments might also be primordial objects that have been misidentified as tidal debris.

Yet the tidal debris explanation is not without its own advantages. Since, as described in section \ref{sec:gasfate}, many galaxies in clusters are strongly \HI{} deficient, the lack of streams in clusters cannot be because the gas is more difficult to remove in a cluster. It suggests instead that the gas is rendered undetectable more rapidly in a cluster than in other environments. We therefore propose a connection between the streams and clouds, with the clouds being remnants of much larger features that are now undetectable as \HI{} (either because the gas disperses and becomes too faint to detect, or it is ionized, cools and becomes molecular, or a combination of factors). In particular, the stream described in \cite{k08} (to our knowledge uniquely) shows several features of M\HI{} $\sim$10$^{7}$\Msolar{} with velocity widths exceeding 100 \kms{}, demonstrating that parts of the streams can attain high velocity widths. The tidal debris idea could simultaneously explain the optically dark \HI{} clouds and the lack of streams - if instead the clouds are primordial dark galaxies, the lack of streams requires a different explanation.

This tidal debris hypothesis is, in principal, valid for the cluster environment (we present a preliminary examination of this in the next section), but much harder to support for clouds detected elsewhere. Our catalogues have allowed us to estimate the frequency at which known streams and clouds are detected in different environments, but this is not the same as the frequency at which different environments contain streams and clouds (i.e. 30\% of all long streams are in groups, but this does not mean that 30\% of all groups contain long streams).

\section{Numerical simulations}
\label{sec:sims}
We have discussed some of the physical processes that may be affecting the streams and clouds throughout this work~: ram-pressure stripping, harassment, heating and cooling, star formation, and ionization. Clearly a wide range of different processes may be important here. In order to understand them, our approach is to simplify the problem as much as possible. As already described, there have been a number of different studies exploring the effects of ram-pressure stripping in detail, but relatively few that consider harassment. Our interest is on the clouds that are known not to form stars, so we may safely neglect star formation. For the sake of simplicity, we will ignore heating, cooling, and ionization for the present study.

We have suggested the hypothesis that the long \HI{} streams are rare in clusters because they fragment and leave behind clouds (sometimes of high velocity widths) as their observable remnants. We seek to test this in the highly limited situation of a galaxy cluster. We examine the effects of harassment (tidal encounters between galaxies and the stripped gas) but neglect the intracluster medium. Our intention is that this first (very simplified) study will be the beginning of a larger project and we will gradually add in additional physics in an incremental approach, in order to understand the importance of each process.

\subsection{Simulation setup}
\label{sec:simsetup}
To study the effects of harassment, we use the gravitational field of a Virgo-mass cluster simulated and described in \cite{warnick} and \cite{rory15}. We refer the reader to those papers for full details. In brief, the simulation used up to 512$^{3}$ dark matter particles (no baryonic physics is employed) in a 64 $h^{-1}$Mpc box in cosmological expansion, allowed to evolve for 6.5 Gyr. From this, 400 subhalos were identified using the halo finding algorithm described in \cite{gill}. For our simulations, we use the positions of these subhalos and approximate them using the Navarro-Frenk-White potential (\citealt{nfw}) based on their total mass and concentration, which vary with time based on the original particle data.

We then set up a cylinder of gas particles at different locations in the cluster, using the smooth particle hydrodynamics code `gf' described in \cite{gf}. The cylinder is used as a toy model to approximate the stripped wake from a galaxy. We do not include the progenitor galaxy (we discuss the significance of this in section \ref{sec:simresults}). In order to minimize computation time we only use 10,000 gas particles (however in limited tests with 100,000 particles we found no significant differences from the smaller runs). There are no dark matter or star particles - the latter seems a reasonable assumption given the observations already discussed, the former is justified by previous simulations (e.g. \citealt{roe08}, \citealt{ductdgs}). The cylinder's length is set to be 200 kpc based on the catalogue described earlier. Unfortunately well-resolved observations that can constrain the radial profile are rarer; we have used the \cite{chen} observations of HI1225+01 to determine a scale height (radius) of 4.2 kpc. We use a gravitational softening length of 100 pc, so the cylinder is very well resolved gravitationally. Each SPH kernel contains 50 particles. In all the simulations which follow we assume the gas is isothermal.

For the radial density profile of the cylinder we follow \cite{inut} :
\begin{equation}
\rho(r) = \rho_{c}\Big(1 +\Big(\frac{r}{h_{0}}\Big)^{2}\Big)^{-2}  
\label{eqt:densityprofile}
\end{equation}
For an isothermal cylinder in radial equilibrium :
\begin{equation}
h_{0} = \sqrt{\frac{2c_{s}^{2}}{\pi G \rho_{c}}}
\label{eqt:hosize}
\end{equation}
Where $\rho$ is the density, $\rho_{c}$ is the central (peak) density, $h_{0}$ is the scale height, and $c_{s}$ is the sound speed. For an isothermal gas :
\begin{equation}
c_{s} = \sqrt{\frac{\gamma k_{B}T}{m_{p}}}
\label{eqt:soundspeed}
\end{equation}
Where $\gamma$ is the ratio of specific heats, $k_{B}$ is the Boltzman constant, T is the temperature, and m$_{p}$ is the average mass of a particle (which we take to be equal to one atomic mass unit). We cannot know the central density $\rho_{c}$ from observations directly, however \cite{inut} state that the condition for radial equilibrium (with no external pressure) is that :
\begin{equation}
M_{line} \leq \frac{2c_{s}^{2}}{G}
\label{eqt:linemass}
\end{equation}
Where $M_{line}$ is simply the mass per unit length of the stream, which is easily obtained via observations. Thus the line mass gives us the temperature (sound speed) and since $h_{0}$ is set independently from observations equation \ref{eqt:hosize} can be re-arranged to give $\rho_{c}$. Of course, there is no particular reason to think that the stream should be in equilibrium (indeed their very absence makes this unlikely) which is something we will explore in the simulations. We keep the initial length and scale height of the stream the same in all simulations, but we have varied the temperature and total mass of the stream (which we will describe in section \ref{sec:simresults}). The stream initially has a uniform density along its length, which is obviously idealised. 

Unfortunately \textit{gf} does not allow periodic gravity, so edge effects prevent us from testing if our idealised stream would be truly stable in isolation. We ran simulations of two isolated streams, one of mass 4$\times$10$^{9}$ \Msolar{} in equilibrium (5,100 K), and the other of mass 4$\times$10$^{8}$ \Msolar{} with the same radial density profile but a temperature of 1,500 K (see section \ref{sec:lowmassstreams}). The massive stream slowly collapses lengthwise (maintaining the same radius), approximately halving in length in 5 Gyr. The low-mass stream slowly disperses, becoming undetectable to an ALFALFA-class survey (see section \ref{sec:simanalysis}) in about 3.5 Gyr.

\begin{figure}
\centering
\includegraphics[width=84mm]{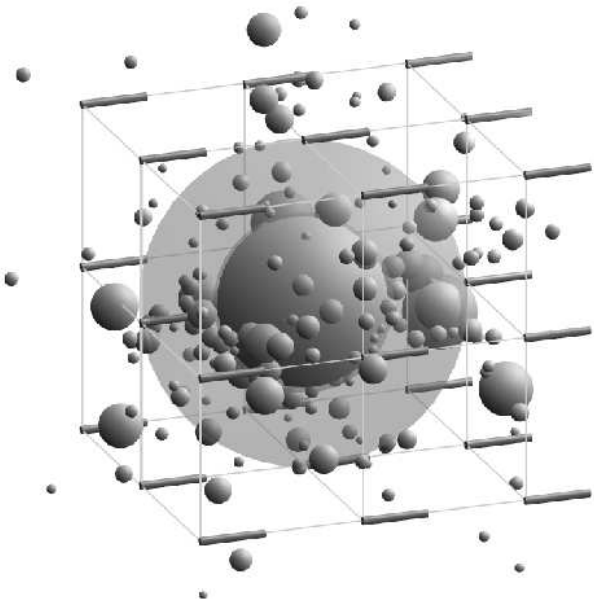}
\caption[hienv]{Illustration of all possible initial positions of each stream in the simulated cluster. Galaxies are represented by the spheres with their size proportional to their viral radius; the semi-transparent sphere represents the whole cluster. Streams are shown as cylinders. The grey lines serve to guide the eye. The various components displayed are not to scale.}
\label{fig:clustersetup}
\end{figure}

Since the harassment experienced by the stream may depend strongly on its path through and orientation with respect to the cluster, we run multiple simulations for streams of each given mass and temperature. Specifically, each stream is allowed to start from 26 initial positions determined by a simple uniform grid, as shown in figure \ref{fig:clustersetup}. We also vary the size of the grid such that the initial position of the stream is either $\approx$500 kpc or $\approx$1 Mpc (though we did not investigate the latter for every case since harassment is much weaker at this distance). Although the AGES clouds we discussed earlier are further from the cluster centre (M87) than this, the real Virgo cluster is more complex. As we shall see, harassment's effects are weaker at greater distances, so if we cannot form such clouds at these lower distances, we will be able to confidently rule out harassment as their formation mechanism.

The orbits of the individual streams vary strongly, and it is difficult to describe a `typical' orbit through the cluster. Most streams reach a pericentre at least once, some twice, a very few do so three times. The first pass near the centre does not happen until at least 1 Gyr. Thereafter streams pass through the centre in a more-or-less random, uniform distribution of times. All this makes it very difficult to quantify where and when harassment typically becomes important (but see section \ref{sec:simresults} and particularly appendix \ref{sec:ap3}). In fact since the streams are not given any angular momentum relative to the cluster centre they are doomed to fall through the central region, and thus will experience the maximum possible amount of harassment. Thus these simulations should give a lower limit on the timescale by which harassment can render the streams undetectable.

Although we earlier quantified the typical velocity gradients of the observed streams, we don't know what this line of sight velocity gradient represents in terms of true 3D velocity. We therefore begin with all the particles in the stream moving at 0 \kms{} with respect to the cluster centre, so that we may investigate if the observed gradient arises naturally as the stream falls into the cluster. We investigate the possibility of different initial velocity gradients in section \ref{sec:needmorephysics}.

\subsection{Simulation analysis}
\label{sec:simanalysis}
Since our hypothesis that dark galaxy candidates are actually the fragments of larger streams is observationally motivated, it is essential to compare the results of the simulations with observations. We do this by gridding the data to create virtual position-velocity data cubes of various spatial and velocity resolutions and sensitivity levels. To save memory, each cube is of a fixed size (1 Mpc) with the centre tracking the mean particle position. We check the position of each particle and determine which cell it should occupy by the following :
\begin{equation}
c_{x,y} = int\Big(\frac{p - p_{min}}{S_{p}}\Big)
\label{eqt:cellpos}
\end{equation}
Where $c_{x,y}$ is the cell number along the x or y axis, $p$ is the particle position, $p_{min}$ is the minimum position of the grid (i.e. 0.5 Mpc from the centre) and $S_{p}$ is the size of the pixels in physical units and is determined simply by :
\begin{equation}
S_{p} = \Big(\frac{S_{res}/60.0}{360.0}\Big)2.0\pi D
\label{eqt:sp}
\end{equation}
Where $S_{res}$ is the spatial resolution of the data cube in arc minutes and $D$ is the assumed distance to the cluster. 

The procedure for determining the velocity cell number is essentially the same. We arbitrarily select one velocity axis and extract the velocity of each particle along that axis directly from the simulation, choosing this to correspond to the observer's line of sight (the size of the velocity cell, equivalent to $S_{p}$, is fixed since this depends only upon the instrument). In principle we could choose any axis for this measurement. However, since the cluster is not strongly asymmetric, and the initial orientation of the streams is fixed with respect to the cluster, it is very unlikely that this would dramatically alter our results.

Once we have determined which cell a particle occupies, we add it to an array containing the number of particles in each cell. This gives us the total mass in each cell. The total \HI{} mass is determined observationally by the standard equation :
\begin{equation}
M_{HI} = 2.36\times10^{5}D^{2}F_{T}
\label{eqt:himass}
\end{equation}
Where $M_{HI}$ is the \HI{} mass in solar masses, $D$ is the distance in Mpc, and $F_{T}$ is the total integrated flux in Jy \kms{}. For a tophat function, $F_{T} = SNR\:rms\:\Delta V$, where $SNR$ is the signal to noise ratio, $rms$ is the noise in Jy, and $\Delta V$ is the velocity width of the detection in \kms{}. In each cell $\Delta V$ is simply the velocity resolution of the survey. Ideally, we might inject noise into the data cubes and run source-finding algorithms to see what would be measured. This is impractical as the number of simulations is necessarily large and we would have to process every timestep of each simulation. Instead, we re-arrange equation \ref{eqt:himass} given that the detections in each cell correspond to a tophat function to estimate the signal to noise ratio :
\begin{equation}
SNR_{cell} = \frac{n_{p}\:M_{p}}{2.36\times10^{5}D^{2}\;rms\;v_{res}}
\label{eqt:cellsnr}
\end{equation}
Where $SNR_{cell}$ is the signal to noise ratio in a particular cell, $n_{p}$ is the number of particles in that cell, $M_{p}$ is the mass of each particle (all particles are given equal mass hence this depends only on the total mass of the stream and the number of particles), and $v_{res}$ is the velocity resolution. We assume that the $rms$ noise level is constant and uniform. $v_{res}$ and $rms$ are survey-dependent parameters and we have calculated the results for surveys equivalent in capabilities to ALFALFA and AGES (as well as the beam size, which is 3.5$'$ for both surveys). From experience in using automatic source extraction algorithms and visual methods, we choose to define a pixel as detected if it has $SNR$ $\geq$ 4.0.

Once we have an array of all the detectable pixels, the final step is to select those pixels which are most isolated. As we are particularly concerned about the AGES dark clouds, which are both unresolved and isolated, our condition for isolation is very strict : we require there to be no other detectable pixels within a search box of $\pm$100 kpc (corresponding to the isolation of the observed clouds described in section \ref{sec:virgoclouds}). We do not impose any isolation criteria along the velocity axis, though in practice this turns out to be unimportant. Pixels which satisfy the condition for isolation are added to another array so that the properties of the isolated clouds which form {(see section \ref{sec:simresults}) can be monitored throughout the simulation. We measure their properties independently in each timestep - no attempt is made to examine how individual clouds evolve over time. Our underlying goal is to answer the question, `given these simulated conditions, would an observer measure anything that resembles reality, and if so for how long~?'. Thus we do not care if an individual cloud briefly resembles one of the observed clouds - we want to know for how much time the simulations produce any AGES-like clouds at all.

We do not define isolated clouds by their velocity widths, though that property is measured. This allows us to compare the number of dark galaxy candidates (which we here define to be clouds with $W50$ $\geq$ 50 \kms{}, generously low compared to the AGES dark clouds) with the number of isolated clouds produced overall.

With arrays containing the detectable pixels and isolated clouds, it is straightforward to automatically measure many of the key properties : total detected mass (both overall and in the clouds), peak SNR, and velocity width. Velocity width is measured by examining the adjacent channels  to the identified peak and measuring their SNR - if it exceeds the threshold (normally half the peak flux) the velocity width is increased by the width of one channel, and this is repeated until a pixel is found which is below the threshold. In addition to the standard $W50$ parameter, for streams we also measure the width at a fixed SNR level of 4.0, which we call $W4$. For massive streams this can give a very different estimate of the velocity width to the standard $W50$. A SNR of 4.0 corresponds to 40 particles for the 10,000 particle low-mass streams at 10 km/s velocity resolution.

Given our concern with reproducing the high velocity widths of the AGES Virgo clouds, we restrict our analysis of the isolated clouds to those with the highest velocity width in each simulation (at each timestep - we are not following the evolution of individual clouds). This cloud is found automatically and its $W50$ and detectable mass are recorded.

Although the underlying physics is relatively simple, we have explored a large, complex data set and thus we necessarily rely mainly on the automatic measurement techniques described above. Note that we measure all parameters assuming surveys of equal capabilities to both ALFALFA and AGES. However we also construct movies of each simulation so we have used at least a basic visual inspection of the data as well. Examples are shown in section \ref{sec:simresults} (the serpentine morphology of some of the streams being responsible for the title of this work). We examine not only the raw particle positions, but also the peak SNR along our chosen line of sight, as well as following the positions of the identified isolated clouds. We have also inspected the raw particle data in a number of simulations where we wished to measure very specific parameters of particular features in a more customized way (e.g. measuring the column density using different beam sizes).

We discuss the results of the simulations in the next section. Note that there is a strong scatter in most trends due to the very different harassment experienced by individual streams, so it is necessary to talk in generalisations.

\subsection{Simulation results}
\label{sec:simresults}
\subsubsection{Massive streams}
\label{sec:simresultsmassive}
We began by using a stream as described in section \ref{sec:streamnat} based on the full set of observed streams : 200 kpc long, 4.2 kpc scale height, of mass 4$\times$10$^{9}$\Msolar{}. This is similar to HI1225+01, on the outskirts of the Virgo cluster. The line mass gave an equilibrium sound speed (equation \ref{eqt:linemass}) of 6.5 \kms{} (temperature 5,100 K), close to the 10 \kms{} dispersion typically observed in galaxies.

The results from dropping the massive stream in from 500 kpc and 1 Mpc, measured using either an ALFALFA or AGES level of sensitivity, are broadly very similar. The detectable mass in the stream drops by at most a factor of three, so a massive `stream' (or at least a structure of some sort) is still easily detectable (SNR $>$ 100) after 5 Gyr. Numerous instabilities are produced in the streams which become compact, self-gravitating blobs. Although some of these features do have high velocity widths (see figure \ref{fig:massivestream0.5Mpc}(b)), they are not isolated clouds since so much of the stream remains detectable.

A very few truly isolated clouds are produced, but none at all with $W50$ $>$ 50.0 \kms{} to a survey of the sensitivity level of AGES. From the view of ALFALFA, one simulation did manage to produce three isolated clouds of the necessary velocity width (two with $W50$ $\geq$ 100 \kms{}), but only for three (well-separated) output timesteps (each output timestep is 25 Myr; the simulation timestep is 2.5 Myr). Given the suite of 26 simulations each lasting 5 Gyr, the fraction of time ALFALFA would have recorded dark galaxy candidates is 0.09\% and zero for AGES (since it can detect fainter gas than ALFALFA, clouds which look isolated/unresolved to ALFALFA do not look isolated/unresolved to AGES). We show the evolution of the parameters in appendix \ref{sec:ap3}.

It is interesting to note that the low angular resolution of Arecibo becomes important here. We measured the properties of ten individual blobs in several random simulations. The mass of the blobs is typically $\sim$5$\times$10$^{8}$\Msolar{} ($\sim$1,200 particles), which with the 17.3 kpc beam of Arecibo (at the Virgo distance) would imply a column density of 2.7$\times$10$^{20}$ cm$^{-2}$ - not an extraordinarily high value. This is highly misleading as in fact, the blobs are much smaller than the Arecibo beam. In fact their true column density is very much higher - on average 10$^{24}$ cm$^{-2}$, not varying by more than a factor of a few. Such extreme densities mean that in reality, this stream would become a star-forming wake rather than the optically dark \HI{} clouds we are trying to explain.

\subsubsection{Low-mass streams}
\label{sec:lowmassstreams}
Since harassment appears wholly unable to explain either the disappearance of the massive \HI{} streams or their fragmentation into dark galaxy candidates, we next examined the case of lower mass streams. Disruption of lower-mass streams should more readily render them undetectable and any resulting clouds should also have lower mass. The long streams observed in Virgo, with the exception of HI1225+01 which is on the cluster outskirts, all have masses $\sim$10$^{8}$\Msolar{}. Additionally, if we assume that the AGES dark clouds fill the Arecibo beam, their line mass would be equivalent to a 200 kpc stream with this total mass. We therefore used a 200 kpc long stream with a scale hight of 4.2 kpc and a total mass of 4$\times$10$^{8}$\Msolar{}.

These constraints mean that the stream cannot consist of neutral atomic hydrogen in equilibrium. The equilibrium conditions (equations \ref{eqt:hosize}-\ref{eqt:linemass}) give a temperature of approximately 500 K, at which the gas is expected to be molecular, not atomic. For comparison with the massive streams, we used initial temperatures of both 5,100 K and 1,500 K (the lowest temperature at which the gas is expected to remain atomic\footnote{However, the transition between molecular and \HI{} is also expected to be strongly affected by the pressure and ambient radiation (\citealt{elme}), and \cite{heiles} note that \HI{} can indeed be found at hotter and colder temperatures than those we have assumed. Additionally, \cite{ton10} note that the additional processes of small-scale turbulence, cosmic rays and magnetic fields may provide additional pressure at low temperatures. Our choice of the lower temperature is somewhat arbitrary, and we leave the non-trivial task of accurately assessing the expected temperature of the stripped \HI{} gas to a future study.}). These streams are thus strongly out of equilibrium. We again place them initially at 500 kpc and 1 Mpc from the cluster centre (104 simulations in total). 

\paragraph*{\textnormal{\textit{5,100 K streams}}}\mbox{}\\

\noindent Again, we find that the production of isolated high velocity width clouds is negligible. Since the streams are not in radial equilibrium but tend to expand in isolation, the extremely dense blobs seen in the massive streams no longer form. Instead, the isolated clouds which do form are merely slight overdensities in the stream and their velocity widths are due to streaming motions along the line of sight. They exist for even less of the total time than in the high mass stream case, just 0.08\% of the time. Velocity widths of the clouds are also on the low side - although they exceed our 50 \kms{} criteria to be included as dark galaxy candidates, they never exceed 100 \kms{}.

For an ALFALFA sensitivity level, a much greater fraction of the low-mass streams becomes undetectable compared to the high-mass case, though the amount depends strongly on the initial temperature and clustercentric distance. The most extreme case of this is when the stream is at a temperature of 5,100 K and starts at 1 Mpc from the cluster centre. Here the stream evolves in near-isolation, and since it is strongly out of equilibrium, it simply expands radially. In about 1 Gyr, the entire thing becomes undetectable (this is faster than the isolated case since the stream's length increases due to tidal stretching by the cluster potential). This means that when it eventually reaches the cluster centre, harassment is occurring on a very low-density stream. Consequently any structures produced by harassment are rarely sufficiently dense to become detectable.

Importantly, the situation is very different for this same stream when observed with AGES depth. Although the detectable mass does drop significantly, in most simulations there is still gas (5$\times$10$^{7}$\Msolar{} in the median case) detectable after 5 Gyr. Thus although long streams might become undetectable to an ALFALFA-class survey, they should not escape detection with AGES.

\paragraph*{\textnormal{\textit{1,500 K streams}}}\mbox{}\\

\noindent The cooler streams are still out of equilibrium, but don't expand as rapidly as the hotter ones. Even starting 1 Mpc from the cluster centre, streams rarely become entirely undetectable to ALFALFA after 5 Gyr, though the fraction of mass remaining can be very low ($\sim$10\%). With AGES the detectable mass decreases rather less, with over 10$^{8}$\Msolar{}$\;$still detectable after 5 Gyr.

As with the hotter streams, the dark galaxy candidates which are produced are merely slight overdensities in the streams rather than kinematically distinct features, so again their velocity widths are due to streaming motions. The cloud masses are typically $\sim$10$^{7}$\Msolar{}, much lower than in the massive streams and comparable to those which are actually observed. Although slightly more common than in the previous cases (0.23\% for starting at 500 kpc from the cluster centre), the velocity widths of the clouds are too low ($<$ 100 \kms{}) to explain the AGES Virgo clouds. Interestingly about three times as many clouds are produced which are visible to ALFALFA as to AGES - the greater sensitivity of AGES means that fewer features appear isolated or unresolved. The column density of the clouds is typically $\sim$3$\times$10$^{19}$cm$^{-2}$, and does not vary by more than a factor of a few.

\subsubsection{Other features of the streams}
It is interesting to note that some of the initially linear streams become highly distorted due to harassment, to the point where they do not resemble anything seen in reality (see appendix \ref{sec:ap3}). This occurs for both the high and low mass streams. Although the variety of observed stream morphologies (in reality) is very large in general, in Virgo they are all comparatively simple, linear features. The wide variety of features in the simulated streams may be due to the effect of missing physics. However, it does indicate (along with the decrease in detectable mass already described) that harassment has a significant effect on the streams - simulations of ram-pressure stripping which ignore it are missing a very important process\footnote{It is difficult to say which effect will dominate. While the streams remain largely unaffected by harassment until $\sim$1 Gyr after the start of the simulation, once harassment begins to take effect changes can occur very rapidly (e.g. some of the isolated clouds persist for $<$50 Myr)}. 

Due to the diverse nature of the features produced we have not made any attempt to rigorously quantify the lengths of the streams - indeed for some features `stream' becomes an inappropriate label. Visual inspection of the movies (see appendix \ref{sec:ap3}) reveals that the detectable features rarely exceed 1 Mpc in length even for high mass streams, though they can approach this size. Low mass streams (their detectable emission) remain shorter, typically $<$ 200 kpc, usually only exceeding this for short periods, though we have only examined the morphologies using an ALFALFA sensitivity level. In reality only one feature is known in the Virgo cluster which exceeds 200 kpc in length.

We have already described that very few isolated, unresolved high velocity width clouds are produced in any of the simulations. However, if we ignore the condition that the clouds must be unresolved, then the situation improves somewhat. As shown in appendix \ref{sec:ap3}, in some simulations there are many features produced with $W50$ $\geq$ 50 \kms{} - in a few cases they can appear for a significant fraction ($\geq$10\%) of the simulation time. However these features would not make for particularly good dark galaxy candidates. They are rarely sudden increases in the velocity width. Rather they are manifestations of the slowly-varying velocity width along the stream : a pixel in which there is a high width is usually surrounded by other pixels of similar width. Features such as VIRGOHI21, which is a very sharp and distinctive velocity `kink' in a stream, are only rarely seen in our simulations (though we leave a quantitative measurement of this to a future work).

Just as in \cite{bekki} and \cite{duc}, \textit{large} features with high velocity widths are readily produced. It is the combination of small size and isolation (see section \ref{sec:needmorephysics}) that is difficult to achieve, not necessarily the high velocity width. Simulations typically produce no more than 10 such clouds of \textit{any} velocity width at any timestep, and generally far fewer ($\lesssim$1). Thus our simulations are consistent with the previous studies, but emphasise that this mechanism cannot reproduce the observed dark galaxy candidates. It is not a minor detail that the previous works produce features otherwise similar to the dark clouds but larger - producing smaller features with these high velocity widths is extremely difficult.

\subsubsection{Deviations from the Tully-Fisher relation}
One of the distinctive features of the type 1 clouds described earlier is that they are offset from the baryonic Tully Fisher relation, regardless of whether one accepts the possibility that they have very faint optical counterparts or not. Our simulated clouds do not reproduce this, as shown in figure \ref{fig:tfrclouds} - they do not deviate from the TFR as much as the observed dark galaxy candidates. However, they do at least have higher velocity widths than the standard TFR given their mass - hence the idea of `tidal debris' is certainly not obviously wrong. It is the quantitative match to the real objects that is difficult to reproduce, not the qualitative shift from the TFR. It should also be again emphasised that these fragmented clouds are extremely rare and transient features in our simulations.

\begin{figure}
\centering
\includegraphics[width=84mm]{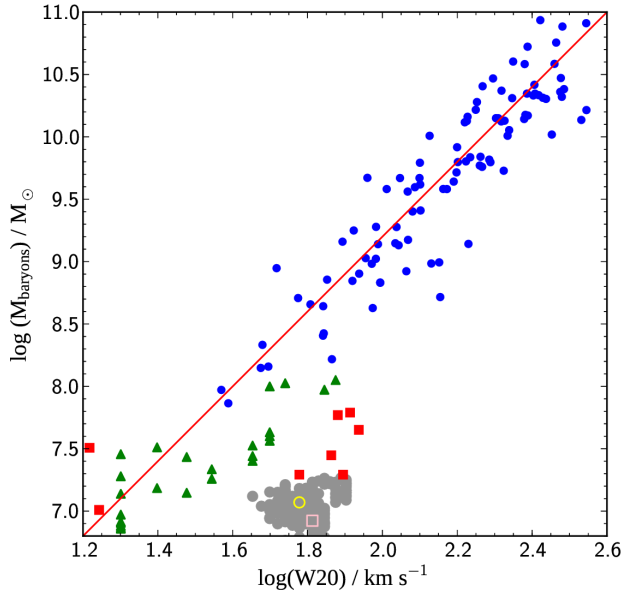}
\caption[hienv]{The baryonic Tully-Fisher relation for ordinary galaxies (taken from AGES V and VI) shown as blue circles, optically dark \HI{} clouds (also from AGES V and VI) shown as red squares, and our simulated clouds (using the 4$\times$10$^{8}$\Msolar{}, 1,500 K stream) shown as green triangles. The red line shows the relation from \cite{MG00}. We assume the circular velocity for the dark clouds is half their observed line width. We have chosen not to include the (negligible) mass of their possible optical counterparts in the baryonic mass calculation, though we include the correction for helium as 1.4$\times$M\HI. Grey circles show the median properties of our simulated dark galaxies (see section \ref{sec:dgsurvival}) throughout their 5 Gyr of evolution in the cluster, with the open yellow circle indicating the initial disc and the pink square showing the final timestep.}
\label{fig:tfrclouds}
\end{figure}

\subsubsection{Impact of the limited physics and analysis procedures}
\label{sec:needmorephysics}
\paragraph*{\textnormal{\textit{Lack of intracluster medium}}}\mbox{}\\
As discussed earlier it is impossible to quantify the exact effect of ram-pressure stripping without a numerical simulation. We are limited here to noting that the ICM thermal pressure is at least an order of magnitude greater than the pressure in the streams, so we expect them to be strongly affected by ram-pressure stripping (assuming that they are moving supersonically through the ICM). We have already noted in \ref{sec:hirps} that unless the observed blobs are so dense that they should be forming stars, they are vulnerable to ram-pressure stripping. On the other hand, resistance from the ICM could mean the \HI{} is essentially deposited into the ICM at the site of stripping without continuing to move through it. Pressure confinement could mean that the streams would not expand and thus would remain detectable for longer (eventually shrinking and perhaps fragmenting), but heating from the ICM could also ionize the stripped gas.

The major limitation of our study - as in previous harassment studies - is that the streams are free to orbit through the cluster with no resistance. It is likely that this has exaggerated the velocity widths of the streams and clouds, especially at pericentre passage, since the velocity widths in our simulations arise purely from streaming motions. The presence of the ICM would exert resistance to any expanding feature, so the velocity widths produced by pure harassment may be upper limits. An important caveat is that we cannot quantify the effects of turbulence in the ICM.

\paragraph*{\textnormal{\textit{Lack of a progenitor galaxy}}}\mbox{}\\
The presence of the galaxy from which the gas is lost might be expected to have an important role in the evolution of the stripped gas owing to its large mass and close proximity. In fact this is unlikely. The free fall time to a Milky Way-mass (10$^{12}$\Msolar) galaxy at 200 kpc is 1.5 Gyr, much less than the $\sim$25 Myr timescales of the persistence of the detectable isolated clouds produced by the fragmenting streams. Additionally since the gas is being removed from the galaxy this 1.5 Gyr will be a lower limit. The parent galaxy is therefore unlikely to significantly affect any of our conclusions. 

\paragraph*{\textnormal{\textit{Lack of velocity gradient across the stream}}}\mbox{}\\
The streams we have modelled initially have velocity dispersions of a few \kms{} due to their temperature, but do not include the large-scale, non-thermal velocity gradient often observed (see table \ref{tab:stab}) in real streams. This gradient can occur simply due to the gas moving through the deep gravitational potential well of the galaxy. We do not expect this to be significant for the varying overall properties of the stream (i.e. detectability). With a velocity gradient of 200 \kms{} it would take an isolated 200 kpc stream approximately 1 Gyr to double in size, comparable to the timescales of the effects of harassment. 

However, the velocity gradient may be important for the formation of small-scale high velocity width features : on the scale of 17 kpc (the Arecibo beam size) a velocity width of 17 \kms{} would arise from the large-scale gradient. This is small compared to the $\geq$100 \kms{} of the AGES Virgo clouds, but with larger overall widths (up to 950 \kms{} has been observed - see table \ref{tab:sprop} - corresponding to a width of 76 \kms{} across 17 kpc) it may be significant. Perturbations of streams with pre-existing velocity gradients thus may give different results to those without velocity gradients.

The counter-argument to this is that large-scale velocity gradients naturally develop in the simulations anyway. Although we have not measured the total velocity width of the streams, a visual inspection shows that total widths $\gtrsim$200 \kms{} typically develop within 500-1,000 Myr - well before pericentre passage.

We have attempted to account for the effects of an initial velocity gradient in the streams prior to their infall into the cluster through another batch of simulations. Since we do not know the true 3D velocity structure of the streams we restrict ourselves to three cases : 1) a line of sight  (perpendicular to the major axis of the stream) velocity gradient of 300 \kms{} (the median observed gradient); 2) a velocity gradient of 300 \kms{} along the major axis of the stream (i.e. the stream is expanding along its length); 3) both line of sight and major axis velocity gradients of 300 \kms{}. We used the 1,500 K, 4$\times$10$^{8}$\Msolar{} stream starting at 0.5 Mpc from the cluster centre.

We found that adding a velocity gradient changes the properties of the detected streams, but makes little difference to the dark galaxy candidates. The streams become less detectable on shorter timescales (by a factor of a few) and though a greater number of isolated clouds are produced, there is essentially no difference in the number of isolated clouds with W50 $>$ 50 \kms{}.

\paragraph*{\textnormal{\textit{Lack of star formation, heating and cooling}}}\mbox{}\\
The effects of these are much more difficult to quantify. Star formation is expected to be negligible in the low-mass streams, which always remain at a low density, but may be dominant in the high-mass streams. The effect of cooling on ram-pressure stripped streams has been previously investigated by \cite{ton10}, who found that cooling led to narrower tails with a wider range of densities and temperatures. They found that during the stripping phase this gave a better match to the observed long tails, but it is very difficult to predict how this would be affected with harassment. 

\paragraph*{\textnormal{\textit{Relaxing the isolation criteria}}}\mbox{}\\
Thus far our isolation criteria have been relatively strict in order to match the observed clouds, as described in section \ref{sec:virgoclouds}. In fact the situation is slightly more complicated as some of the observed AGES clouds, though far from galaxies, are closer to each other than our 100 kpc exclusion region. Many more clouds are produced than our strictly isolated clouds, so in principle there could be larger numbers of high velocity width clouds among them. Therefore, using the 1,500 K, 4$\times$10$^{8}$\Msolar{} stream (starting at 0.5 Mpc from the cluster centre) simulation, we re-analysed the results using a $\pm$34 kpc (two Arecibo beams) exclusion zone.

The number of dark galaxy candidates changes significantly with this more relaxed isolation criteria, though they are still a rarity. Clouds with W50 $\geq$ 50 \kms{} exist for a total of 4$\%$ of the time, though for clouds $\geq$ 100 \kms{} that falls to just 0.3$\%$. Although ideally we require an estimate of how frequently streams are produced, it is difficult to imagine how harassment of the streams could be responsible for the six type 1 clouds observed (four of which have W20 $\geq$ 150 \kms{}).

\subsection{Survival of Dark Galaxies}
\label{sec:dgsurvival}
We have hitherto considered the formation of dark galaxy candidates that are actually simply unbound \HI{} clouds. As discussed analytically in section \ref{sec:unbound}, such features should rapidly disperse. This is borne out in our above simulations, but we also performed an idealised test in which we begin with a rotating (but unbound) \HI{} disc which is then subjected to harassment. The \HI{} indeed becomes undetectable in $\sim$100 Myr, as predicted. By the time it experiences an interaction with one of the sub-halos, its density is extremely low and it never becomes detectable again during the simulation. In short, harassment makes no difference to this scenario.

It is more interesting to examine the case of \textit{actual} dark galaxies, where the \HI{} disc is a stable bound disc embedded in a dark matter halo (as opposed to the various sorts of `fake' dark galaxies discussed previously). If either the fake or genuine dark galaxies are more stable than the other, this could indicate the likely nature of the observed clouds. Unfortunately our observational parameters are limited to the \HI{} mass, velocity width and an upper size limit, though we can also place a lower limit on the size based on typical \HI{} column densities. This means we have a wide range of dynamical mass estimates which vary by an order of magnitude. Fortunately, since the velocity width is observationally fixed, the dynamical mass depends only on the size of the \HI{} disc.

We choose parameters based on the AGES clouds detected in Virgo. We consider two possible disc radii : 4.33 kpc and 8.65 kpc (the latter is the maximum based on the Arecibo beam size). We use the median \HI{} mass of 3$\times$10$^{7}$\Msolar{} and a maximum circular velocity of 75 \kms{}. This gives a dynamical mass of 5.2$\times$10$^{9}$\Msolar{} and 1.0$\times$10$^{10}$\Msolar{} for the radii of 4.33 and 8.65 kpc respectively. We assume a temperature of 1,500 K (as before the gas is assumed to be isothermal) and set the disc thickness arbitrarily to be one-tenth of the radius. 

The dark matter halo consists of 10,000 particles with a spherical logarithmic density profile, truncated at the radius of the gas disc (this means the discs are `maximally harassable' - more extended dark matter halos would only make it more difficult to remove the disc material). The gas consists of 10,000 particles distributed in a uniform density disc. The rotation curve of the gas disc is set by considering the radial accelerations and assuming a circular velocity, i.e. $v_{circ}^{2}$ = $r\,a_{r}$ where $r$ is the radial distance from the centre and $a_{r}$ is the acceleration along the radial vector. We allow each dark galaxy to first evolve in isolation for 5 Gyr to test for stability, which typically results in no more than $\sim$20\% of particles migrating beyond the edge of the initial disc. Thus the discs are stable on long timescales so any gas lost during the cluster infall can be attributed to harassment.

Figure \ref{fig:enddiscs} shows the discs after 5 Gyr in the cluster. We placed the discs at the same initial locations as for the streams, with a clustercentric distance of 500 kpc. The largest discs (8.65 kpc radius) survive virtually unscathed, with typically $\gtrsim$60\% of the particles remaining within the disc. Synthetic observations at the ALFALFA or AGES levels would show the disc as nothing more than a single detectable pixel that slowly decreases in mass\footnote{Although some extremely long ($>$500 kpc) tidal tails are produced, as well as occasional detached clouds, the surface density and total mass of these features is extremely low - they would be well below the AGES sensitivity level.}.

We have not included star formation in these simulations, but we can estimate its likely importance by measuring the density of the gas over time. At 17 kpc diameter these discs have surface densities over 60 times less than in typical star-forming dwarf galaxies. The median density varies significantly throughout the simulations (see figure \ref{fig:discrho}), but we found that typically $<$10\% of the particles in these simulations even slightly exceed the threshold for star formation. Therefore the choice to neglect star formation appears to be justified.

Using smaller 4.33 kpc discs, we found that even fewer particles were removed from the discs by harassment - with typically $\gtrsim$90\% of the particles remaining within the initial radius. While smaller discs would therefore be even more robust against dispersal due to harassment, they are conversely more susceptible to star formation because of their higher gas density. Even in isolation approximately 25\% of the particles exceed the density threshold for star formation, and this can rise to over 50\% during cluster infall. We thus expect even smaller discs to be dominated by star formation and therefore not remain optically dark. There are two important caveats to this. Firstly, feedback from star formation might keep the density low and prevent a detectable number of stars from forming (this would be better examined using another code such as \textsc{gadget}, which has a more sophisticated treatment of star formation than \textit{gf}). Secondly, smaller dark galaxies could have thicker gas discs or even be significantly supported by pressure as well as rotation, so they would not necessarily be denser. We leave exploration of this large parameter space to a future work.

\begin{figure*}
\centering
\includegraphics[width=180mm]{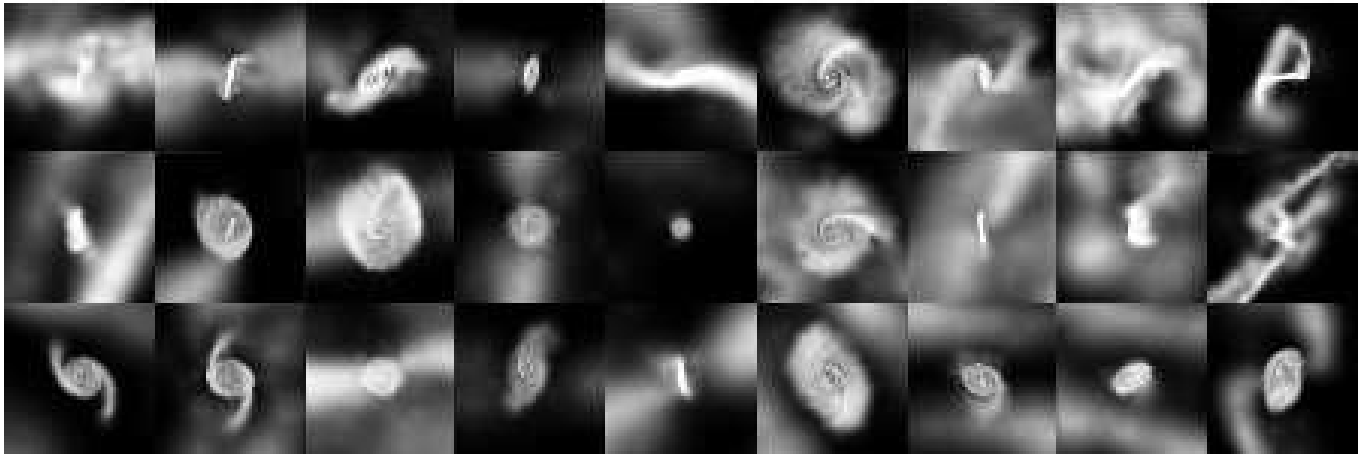}
\caption[discend]{The result of 5 Gyr in the cluster on 17 kpc diameter discs, with M\HI{} = 3$\times$10$^{7}$\Msolar{}, v$_{circ}$ = 75 \kms and M$_{dark}$ = 1.3$\times$10$^{10}$\Msolar{}. The field of view is 50 kpc in all cases. Movies for all of the harassed dark galaxies can be seen at \href{http://tinyurl.com/jq8llqs}{this url} : http://tinyurl.com/jq8llqs.}
\label{fig:enddiscs}
\end{figure*}

\begin{figure}
\centering
\includegraphics[width=84mm]{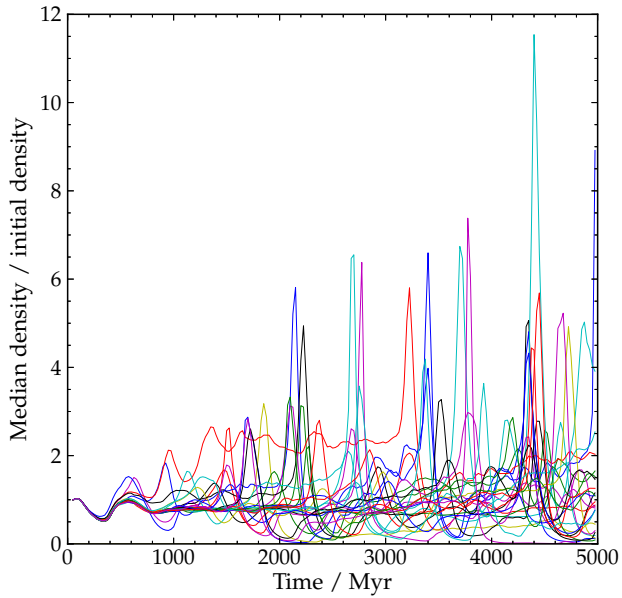}
\caption[discrho]{Median density evolution of the 17 kpc discs. We extract the density measured for every particle and find the median at each timestep, for each simulation.}
\label{fig:discrho}
\end{figure}

Since we chose parameters for the initial discs which are based on the observed clouds, they deviate from the Tully-Fisher relation in roughly the same way as the observed clouds do. Due to the evolution of the simulated discs in isolation, the offset from the TFR is not quite the same as that of the real clouds - the isolated discs expand slightly (thus some gas reaches a column density too low to detect) and some particles reach higher velocities than their initial circular velocity (thus reducing S/N by spreading the flux over more channels). The net result is that when injected into the cluster, the simulated discs already have a factor of a few less detectable mass than the real clouds. However, as shown in figure \ref{fig:tfrclouds}, the discs remain detectable and maintain their deviation from the TFR throughout the entire 5 Gyr of the simulation - harassment does not significantly affect them. One should also bear in mind that we do not know how long the real clouds have been in the cluster, so their initial conditions might also be different.

Harassment can cause the discs to deviate in both directions away from the standard TFR, but generally they maintain significantly greater velocities than the standard TFR predicts. In a very few cases (not shown in figure \ref{fig:tfrclouds} since we have plotted median values), the measured W20 of the simulated discs decreases so much that they would appear to lie exactly on the TFR, but these cases are rare exceptions - additionally, even these rare cases take approximately 2.5 Gyr to reach their minimum velocity width.

In contrast, although the clouds produced by fragmenting streams do also deviate towards higher velocities, they never do so sufficiently as to lie as far from the TFR as the observed clouds, though a few do come close. Importantly though, the fragmented stream clouds rarely remain detectable for $>$ 50 Myr, so if this really was the explanation for the observed clouds, we would be witnessing six clouds simultaneously in a very peculiar stage of their evolution. The simulated dark galaxies, on the other hand, not only remain detectable but also almost always deviate from the TFR in roughly the same way as the real clouds. Thus dark galaxies appear to be a significantly better (though not perfect) explanation for the observed clouds.

In conclusion, genuine dark galaxies can be stable to harassment both in terms of disruption (very little material is lost so they remain detectable) and star formation (little density increase occurs so they could remain optically dark or very faint) - even over timescales of 5 Gyr. Therefore dark galaxies are consistent with the observed \HI{} clouds in the Virgo cluster.

\section{Summary and Discussion}
\label{sec:conc}
We have shown that long \HI{} streams are rare in clusters, contrary to predictions from numerical simulations. However, we know from deficiency measurements that gas is being lost in cluster galaxies, and streams are readily detected in less dense environments. The mix of strongly deficient and non-deficient galaxies in clusters implies that some of them should be currently losing gas, yet in the VC1 region not a single galaxy showed evidence of a long \HI{} stream even at a column density sensitivity level $\sim$10$^{17}$ cm$^{-2}$.

Since \HI{} streams appear to be more common in galaxy groups, we argue that there must be some process acting which is unique to clusters that destroys streams soon after their formation (noting that long \HI{} streams are rare \textit{but not absent} in Virgo). A phase change of the \HI{} to either H\textsc{ii} or H$_{2}$ could render most of the gas undetectable, however as different authors have suggested both heating and cooling to change the phase of the gas it is not clear which might be dominant. Still, it is likely that there is more to gas stripping than simply gas removal, since there appears to be no correlation with the \HI{} deficiency of a galaxy and the existence of a long stream.

We have also discussed the idea of optically dark galaxies as a solution to the missing satellite problem, an idea which has been resurrected in recent years in part due to the discoveries of ultra-compact high velocity clouds. We have catalogued known dark galaxy candidates beyond the Local Group, and find that the clouds known in the Virgo cluster have some of the highest velocity widths. They are also relatively isolated, compact, and there is no sign of any extended \HI{} stream in their vicinity. Observationally it is not easy to explain them away as `tidal debris'. Similarly there are now many known optically faint UDGs which also appear to be primordial, dark matter-dominated objects, and the \HI{} clouds are compatible with being their progenitors.

The dark galaxy hypothesis is not without its own problems. Far too few candidates have been found to explain the missing satellite problem, which is the main reason they were proposed in the first place, and no truly isolated candidates have been found. Given the lack of expected streams in the Virgo cluster but the presence of \HI{} clouds, we have postulated that there might be a connection between the two. Streams which are subject to galaxy harassment, a process which is specific to clusters, might be torn apart and some of the fragments might gain the high velocity widths that could be mistaken for rotation.

We have conducted a suite of numerical simulations in an attempt to determine whether harassment can cause the high velocity widths of some of the observed clouds. Taking an incremental approach, we have simplified the situation as much as possible and considered only the effects of harassment on the stripped gas, ignoring star formation, heating, cooling, and the intracluster medium. Whereas previous simulations (\citealt{bekki}, \citealt{duc}) have demonstrated  that harassment alone \textit{can} produce features similar to those observed, we have attempted to quantify (for the first time) how \textit{likely} this is to actually occur.

We find that harassment by itself is not able to explain either the absence of the streams or the existence of optically dark clouds, though it is clearly a very important effect that has previously been neglected. Harassment can cause far more dramatic changes in stream morphology than what appears in simulations of pure ram-pressure stripping. It can also cause significant decreases in how much gas in the stream is detectable - not enough by itself to explain the absence of streams, but in combination with the effects of the intracluster medium it may be sufficient.

From these simulations, the already problematic explanation of the clouds as tidal debris looks even more dubious. Although high velocity width features are produced, they are usually part of much larger extended structures (which is also the case in previous simulations), and in isolated clouds they are extremely rare - far too rare to be a plausible explanation for the eight clouds detected in the VC1 region. Our simulations also failed to reproduce the strong deviations from the baryonic Tully Fisher relation of the observed clouds.

In short, our simulations reproduce (albeit briefly) all of the features of the streams and clouds we observe in the Virgo cluster, but not at the same time or even in the same structures. Long streams are indeed sometimes fragmented by harassment, but not quickly enough to explain the absence of observed streams in Virgo, nor are isolated clouds of sufficient velocity width produced with a frequency that would make this a plausible explanation of the observed clouds. In contrast, our simulations showed that optically dark galaxies are extremely robust to both dispersal and star formation due to harassment and could remain detectable for many gigayears. Thus from these results the dark galaxies hypothesis appears to be a more viable explanation for the observed clouds than tidal debris, though we caution that including more physics could change the results.

We have also discussed from a purely observational perspective the prospect that the clouds might instead be optically dark galaxies instead of tidal debris. There are arguments for and against this idea, which we summarise below.
\mbox{}\\
\noindent
1) Observational reasons why AGES dark galaxy candidates might be tidal debris rather than dark galaxies :
\begin{itemize}
\item Their \HI{} masses are well below the mass typically found in long streams, so they are consistent with the idea of a fragmented stream.
\item Streams are known to exist which are long enough that fragmentation could make the debris appear isolated.
\item At least one long stream is known containing high velocity width clumps that is believed to have formed from a tidal encounter (\citealt{k08}).
\item Gas removal does not necessarily result in a long-lived stream, suggesting rapid fragmentation is possible.
\item Even if the debris is unbound, it will still persist for $\sim$100 Myr. Given the typical velocity through the cluster, this is sufficient for it to appear isolated from its parent galaxies.
\item No platonic ideal dark galaxy candidate has ever been found, i.e. a double-horn profile (or resolved disc with ordered motions) too isolated to explain as debris. Almost all dark galaxy candidates are detected in relatively dense environments, where a tidal origin is at least possible.
\item Velocity widths of some candidates are incompatible with the baryonic TFR; we might expect all primordial objects to obey this relation.
\item Too few clouds have been detected to explain the missing satellite problem, which was the reason they were originally proposed to exist.
\end{itemize}

\noindent
2) Observational reasons why AGES dark galaxy candidates might \textit{not} be tidal debris but are instead dark galaxies :
\begin{itemize}
\item Known \HI{} streams are believed to be long-lived; if the objects are transient debris they should not survive longer than the streams. Therefore the lack of detected associated streams is surprising.
\item They cannot be self-bound by their \HI{} alone as this would imply extraordinarily high column densities and star formation rates.
\item If they are unbound, AGES should detect clouds up to twice the Arecibo beam size but none are detected larger than a single beam.
\item If they are unbound, high velocity width clouds should disperse more rapidly, yet we detect more of these than we do of low velocity width clouds.
\item Relatively isolated candidates do exist which are very difficult to explain as tidal debris.
\item The objects SNR and velocity widths are consistent with their being rotating discs. A dark halo would make the objects much more stable and so increase the chance of their being detected; they could also be the progenitors of recently-uncovered small UDGs in the cluster.
\item While massive candidates are relatively rare, new surveys have uncovered larger populations of small objects that may yet explain the missing satellite problem.
\end{itemize}

Clearly there are many unanswered questions still to address. Higher resolution observations may be able to answer many of these regarding the AGES Virgo clouds - we still don't know what they really are. As we improve the numerical simulations we will be able to examine the dark galaxies or tidal debris question more broadly. Solving the origin of these mysterious objects is not trivial, and they should not always be dismissed as ``tidal debris''. On the contrary, our results indicate that this is not a sensible explanation within clusters, at least for the six high velocity width clouds. Optically dark galaxies would be robust to the effects of harassment - our simulations show that they would suffer neither significant tidal disruption nor significant levels of star formation - and could potentially explain the newly-discovered ultra-diffuse galaxies in several clusters.

Finally, we note that objects like the AGES Virgo clouds, the various populations of LSB galaxies recently discovered, and the system discovered by \cite{jan15} (which has objects which are apparently tidal debris but deviate from the TFR in the opposite sense to the Virgo clouds, and no obvious source of the \HI), demonstrate that there is still much to be learned about the baryon cycle of galaxies. An AGES-depth survey of the whole Virgo cluster could greatly improve our understanding both of the unusual objects and the environmental processes influencing the typical galaxies present.

\section*{Acknowledgments}

We thank Joachim Koeppen, Blakesley Burkhart and Abraham Loeb for their insightful and constructive discussions. We also thank the anonymous referee whose comments improved the quality of this manuscript.

This work was supported by the Tycho Brahe LG14013 project, the Czech Science Foundation projects P209/12/1795 and RVO 67985815. R.S. acknowledges support from Brain Korea 21 Plus Program (21A20131500002) and the Doyak Grant (2014003730).

This work is based on observations collected at Arecibo Observatory. The Arecibo Observatory is operated by SRI International under a cooperative agreement with the National Science Foundation (AST-1100968), and in alliance with Ana G. M\'{e}ndez-Universidad Metropolitana, and the Universities Space Research Association. 
This research has made use of the NASA/IPAC Extragalactic Database (NED) which is operated by the Jet Propulsion Laboratory, California Institute of Technology, under contract with the National Aeronautics and Space Administration. 

This work has made use of the SDSS. Funding for the SDSS and SDSS-II has been provided by the Alfred P. Sloan Foundation, the Participating Institutions, the National Science Foundation, the U.S. Department of Energy, the National Aeronautics and Space Administration, the Japanese Monbukagakusho, the Max Planck Society, and the Higher Education Funding Council for England. The SDSS Web Site is http://www.sdss.org/.

The SDSS is managed by the Astrophysical Research Consortium for the Participating Institutions. The Participating Institutions are the American Museum of Natural History, Astrophysical Institute Potsdam, University of Basel, University of Cambridge, Case Western Reserve University, University of Chicago, Drexel University, Fermilab, the Institute for Advanced Study, the Japan Participation Group, Johns Hopkins University, the Joint Institute for Nuclear Astrophysics, the Kavli Institute for Particle Astrophysics and Cosmology, the Korean Scientist Group, the Chinese Academy of Sciences (LAMOST), Los Alamos National Laboratory, the Max-Planck-Institute for Astronomy (MPIA), the Max-Planck-Institute for Astrophysics (MPA), New Mexico State University, Ohio State University, University of Pittsburgh, University of Portsmouth, Princeton University, the United States Naval Observatory, and the University of Washington.

{}

\appendix
\section{Notes on \HI{} streams}
\label{sec:ap1}
Given the problems in compiling this catalogue, there are many flags to the parameters of table \ref{tab:stab}, which we describe here.

\textbf{Code} : Some of the objects show multiple extended feature, we separate the flags with commas. Thus NGC 5395 is given the code, `1,2', meaning that two galaxies are joined by a bridge of \HI{} with a stream from one of them extending into intergalactic space. NGC 262 is given the seemingly self-contradictory code `0,1' - giant undisturbed galaxy with a stream - as the disc in this case appears to be largely undisturbed and itself larger than 100 kpc in extent, but there is also an even larger stream present.`B' indicates the stream is broken into several sections and not detected along its entire length, though this may only be due to sensitivity limits.

\textbf{Length} : `U' indicates that it is particularly difficult to determine where the galaxy ends and the stream begins. A lower limit is given for NGC 3227 as the extended \HI{} reaches the edge of the area observed.

\textbf{Diameter} : Upper limits are given when the diameter is close to the resolution of the observations. `V' indicates that the diameter is highly variable along the stream - sometimes a range is given when the stream is continuous, but this is not possible when the stream is broken or has a very complicated structure (e.g. the Vela cloud in \citealt{vela}). `U' indicates the diameter is unclear - for NGC 877 the difference between single-dish and VLA flux shows that there is a more extended component present than detected in the high resolution observations, but the lower resolution observations are not sufficient to estimate the diameter.

\textbf{DeltaV} : `G' indicates that the authors give measurements for the velocity width of the galaxies but do not specify the width of the stream, and no data or images are available to measure the stream separately. `U' indicates it is not clear when the author's measurements refer to. `E' indicates that we have made an estimate of the stream's velocity width using the raw data or published figures.

\textbf{Flux} : `T' is given when the authors only present the total flux in the galaxies and the extended \HI{} component, and the raw data is not available. `U' indicates that it is unclear if the authors refer to the galaxies, the stream, or both. `D' indicates that the authors only give the \HI{} mass; we have derived the flux using their distance estimate. `E' indicates that we used the raw data for our own estimate of the flux in the stream following the procedures in \cite{me14}.

\section{Notes on optically dark \HI{} clouds}
\label{sec:ap2}
There is some overlap between the dark galaxy candidates of table \ref{tab:dgal} and the extended features in table \ref{tab:stab}. A few very extended features were originally proposed as dark galaxy candidates in their own right (notably HI1225+01), others are long streams in which the dark galaxy candidates are embedded. In this section we give brief descriptions of the objects, and the reasons why they were proposed as potential dark galaxy candidates (excepting the AGESVC1 clouds, HI1225+01 and the Smith Cloud, as these are described in the main text).

\textbf{HI1232+20} : Discovered in the ALFALFA survey and re-observed with the WSRT and with deep optical imaging using the WIYN, 3.5 telescope, described in \cite{jan15}. A complex of three clouds (AGC 229383, AGC 229385, and AGC 229384) in close proximity (50-100 kpc projected distance) to a spiral galaxy AGC 222741. However the spiral's velocity differs by 500 \kms{} to the clouds so its association is unclear. The clouds have \HI{} masses from 6$\times$10$^{7}$-7$\times$10$^{8}$\Msolar{} at the assumed 25 Mpc distance. The most massive cloud has an optical counterpart also detected in the UV with GALEX. Two of the clouds have velocity widths lower than expected based on the baryonic TFR. The spiral galaxy appears undisturbed so the source of the \HI{} is not obvious. The authors conclude that ``this system defies conventional explanations.''

\textbf{AGES628\_011} : Described in \cite{auld}, AGESJ013956+153135 is a potentially massive \HI{} cloud with no obvious optical counterparts in the SDSS data, though it is only a 5$\sigma$ detection.

\textbf{AGES1376\_004} : AGESJ114809+192109 in \cite{c08}. A weak 4$\sigma$ detection, but confirmed with L-wide follow-up observations. There are several faint galaxies within 1$'$, making this an unlikely dark galaxy candidate.

\textbf{VIROGHI21} : Discovered in relatively shallow, low resolution HIJASS data (\citealt{d04}), subsequent Westerbork observations (\citealt{m07}) revealed an overdensity at the end of a 125 kpc stream with indications of ordered rotation. It was postulated to be a massive dark galaxy that had disturbed the gas in the one-armed spiral NGC 4254. More sensitive, but lower resolution, ALFALFA data (\citealt{haynesV21}) revealed the overdensity is actually in the middle of a 250 kpc stream, making the dark galaxy model much less likely. Modelling by \cite{duc} describes how the whole stream, complete with overdensity, may be formed by a high velocity encounter between two normal, optically bright galaxies. However the sharp velocity gradient observed is not reproduced in the simulations.

\textbf{GBT1355+5439} : Discovered in a GBT (Green Bank Telescope) survey (\citealt{mihos}), it was subsequently re-observed at higher resolution with Westerbork (\citealt{oo13}). It has a projected distance of 150 kpc from the nearest galaxy (M101). There is a velocity gradient across the object along an axis perpendicular to the direction of M101, contrary to what would be expected if it were tidal debris. It systemic velocity differs from known Galactic HVCs by at least 150 \kms{}. No optical counterpart is detected down to $\mu$ $<$ 29 mag arcsec$^{-2}$.

\textbf{AGC749170} : The NGC 877 galaxy group was observed by ALFALFA, which discovered a large \HI{} envelope indicating that many of the group members are interacting (\citealt{n877}). GMRT observations reveal AGC749170 as an overdensity within this envelope, but do not recover all of the single-dish flux. There is a small velocity gradient ($<$ 40 \kms{}) across the object. A very faint (m$_{g}$ = 22.6) optical counterpart is detected; though no optical redshift has been determined it is an unlikely dark galaxy candidate.

\textbf{AAK} : The AAK1 clouds are described in \cite{kent09} and the AAK2 clouds in \cite{kent10}. Both sets of clouds were originally detected by ALFALFA and re-observed with the VLA. The AAK1 clouds are part of a complex, spanning about 170 kpc at the 17.0 Mpc distance of the Virgo cluster, which shows no ordered motions. The authors find that a ram-pressure stripping event $>$10$^{8}$ Myr ago is a plausible explanation and they are therefore unlikely dark galaxy candidates. 
The AAK2 clouds are in the outskirts of the Virgo cluster. We have listed their north and south components separately but not all of the flux is recovered by the VLA observations. AAK2\_C1 is isolated, with the nearest late-type galaxy being one degree away. While the authors do not rule out these clouds as stripped debris, their being dark galaxies seems at least a plausible alternative.

\textbf{Enigma} : Detected in an Arecibo survey this HVC was re-observed with the VLA (\citealt{enigma}). Although it shows ordered motions that might indicate rotation, the velocity gradient is $<$ 1 \kms{}. Since the systemic velocity of the source is just 47 \kms{} its true distance cannot be ascertained. Thus its \HI{} and dynamical mass are unknown, hence it is difficult to comment further on this object.

\textbf{AAM33} : This collection of clouds around M33 was detected as part of the ALFALFA survey and with deeper Arecibo observations (\citealt{m33}). The main reason to postulate these as potential dark galaxy candidates was that their numbers are consistent with the predicted number of satellite galaxies within the same distance from M33. However, deeper observations over a larger area (\citealt{olivia}) have shown that the situation is more complex - the clouds do not possess ordered motions, may be part of a larger extended feature, and have a strongly asymmetrical distribution around M33.

\textbf{Vela} : The very complicated morphological structure, as seen with ATCA and described in \cite{vela}, of this cloud makes it particularly difficult to parametrise. It is composed of several different sub-structures which run perpendicular to, and are offset from, the disc of NGC 3263. Given that the nearby spiral galaxies show clear signs of disturbances it is much more likely that this is tidal debris than a dark galaxy.

\textbf{HCG44} : This is reported as a giant 300 kpc structure in the NGC 3193 group but it actually consists of four discrete clouds, observed with Westerbork (\citealt{hcg44}). There is some evidence of an ordered velocity gradient across the C$_{S}$ cloud but the authors do not discuss the possibility of a dark galaxy, citing the model of \cite{bekki} as a sufficient explanation.

\textbf{HJ1021+6842} : Discovered in HIJASS observations (\citealt{boyce}) this was re-observed with the VLA (\citealt{walt}) which resolved seven distinct sub-structures. Tidal debris cannot be ruled out, but the cloud complex shows a velocity gradient which may indicate that each cloud is part of a single bound structure 30 kpc in extent. A dark galaxy seems at least a valid alternative in this case.

\textbf{HPJ0731-69} : This massive cloud was discovered in HIPASS data (\citealt{n2442}) and the low resolution of Parkes makes it difficult to accurately determine the size of the cloud. Given the optically disturbed nature of the nearby spiral NGC 2442, a tidal origin seems likely.

\textbf{ComplexH} : A massive starless cloud in the Local Group, this is probably at a distance just beyond the disc of the Milky Way. \cite{complexH} note that it fulfils many of the criteria to be one of the missing satellites predicted by CDM simulations, except that its velocity width is too narrow.

\textbf{GEMS\_N3783\_2} : This dwarf galaxy-sized \HI{} cloud in the NGC 3783 group is described in \cite{kil} using observations from Parkes and ATCA. It is approximately 500 kpc from the nearest spiral galaxy ESO 378G 003, but a tidal origin cannot be ruled out. The authors conclude that it is unlikely to be a dark galaxy, however this conclusions rests on the fact that similar detections are rare.

\textbf{HIJASS1219+46} : Discovered in a HIPASS study of Ursa Major (\citealt{wolfinger}), this `cloud' cannot be resolved from the nearby spiral NGC 4288 due to the large size of the Jodrell beam. NGC 4288 is known to have other associated \HI{} clouds so a tidal origin is difficult to rule out, especially given the low resolution of the observations.

\textbf{AGC198606} : This HVC shows a clear, ordered velocity structure with a gradient of 25 \kms{}  across its apparent disc. It has no optical counterpart to M$_{i}$ $\simeq$ -6.6. Detected in the ALFALFA survey, it was re-observed with Westerbork and described in \cite{adams15}, who note that it is quite similar to the clouds described in \cite{adams}. The authors favour a dark minihalo over a galactic \HI{} cloud as its properties do not match those of other known HVCs.

\textbf{AGC208602} : The ALFALFA survey has discovered 200 \HI{} sources without optical counterparts (out of a total of 15,855). \cite{cannon} present observations for five of these from a VLA follow-up survey of 50 objects (the remainder are described as being `likely tidal'). Of the five presented so far, four have optical counterparts so we exclude them here. AGC 208602 is likely tidal : four other galaxies are known within 36$'$ (400 kpc) and its morphology and velocity structure are irregular.

\clearpage 
\section{Evolution of the properties of streams and clouds}
\label{sec:ap3}

\begin{figure*}
\nopagebreak[4]
\centering
  \subfloat[]{\includegraphics[height=55mm]{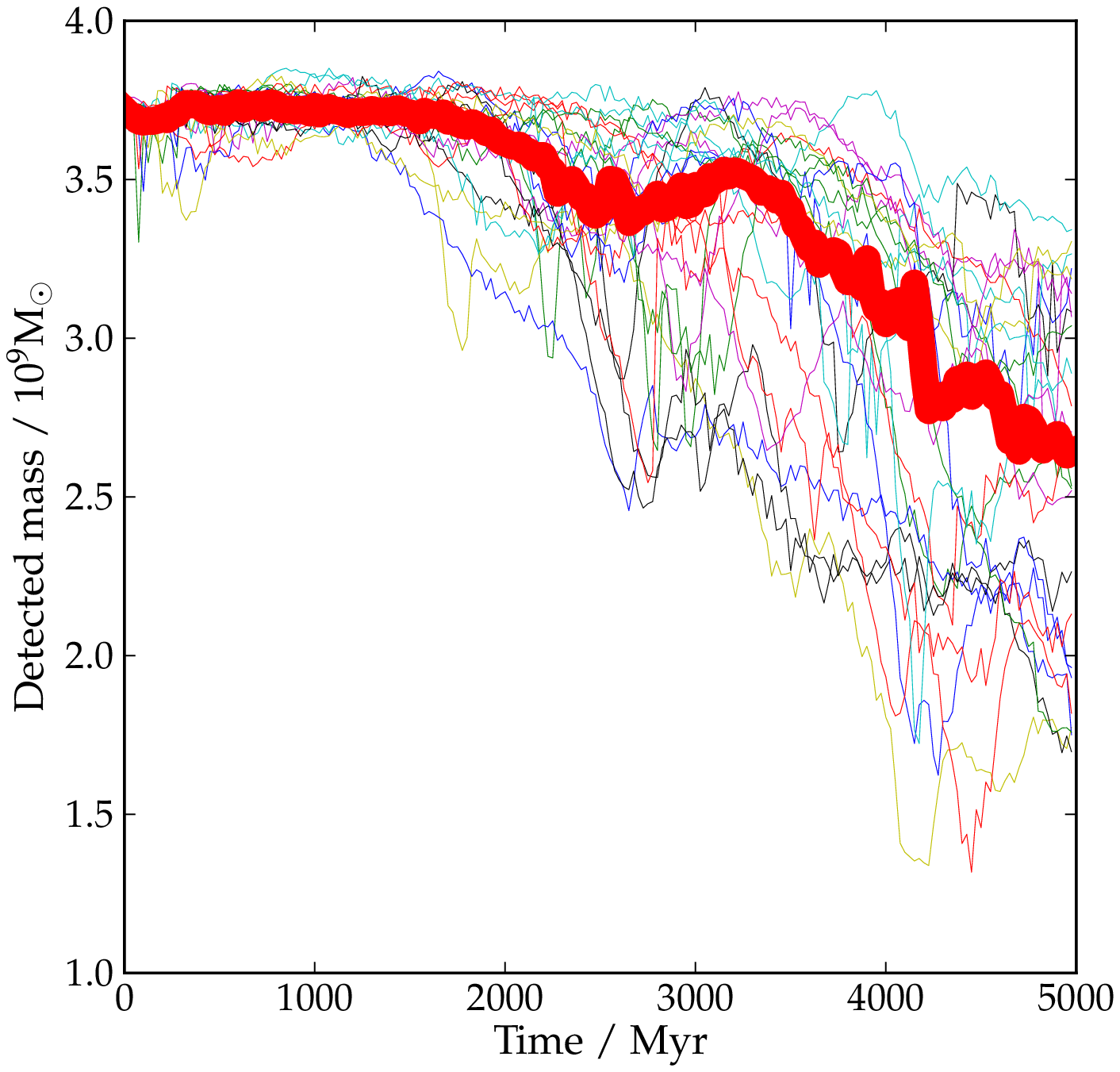}}
  \subfloat[]{\includegraphics[height=55mm]{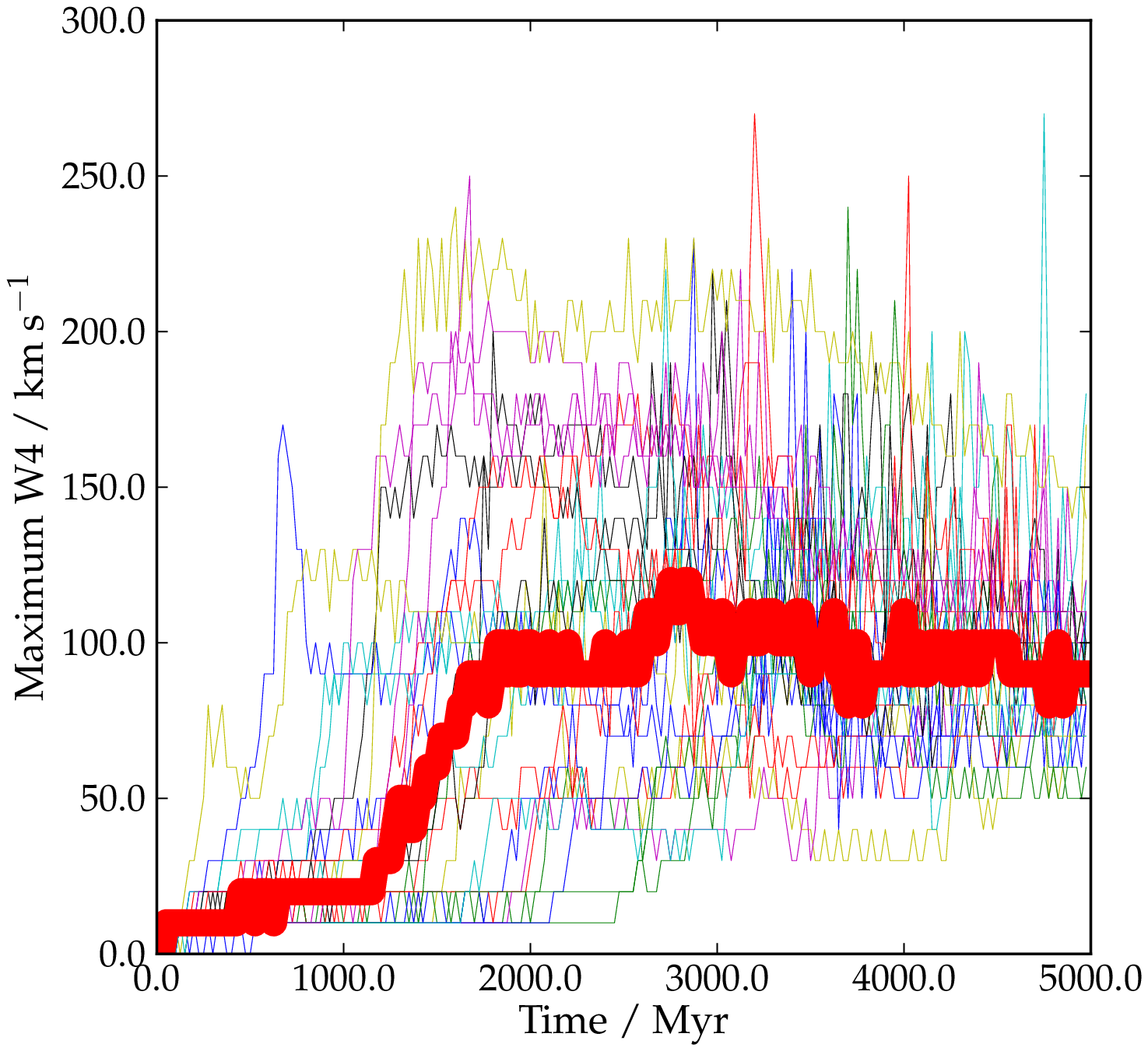}} 
  \subfloat[]{\includegraphics[height=55mm]{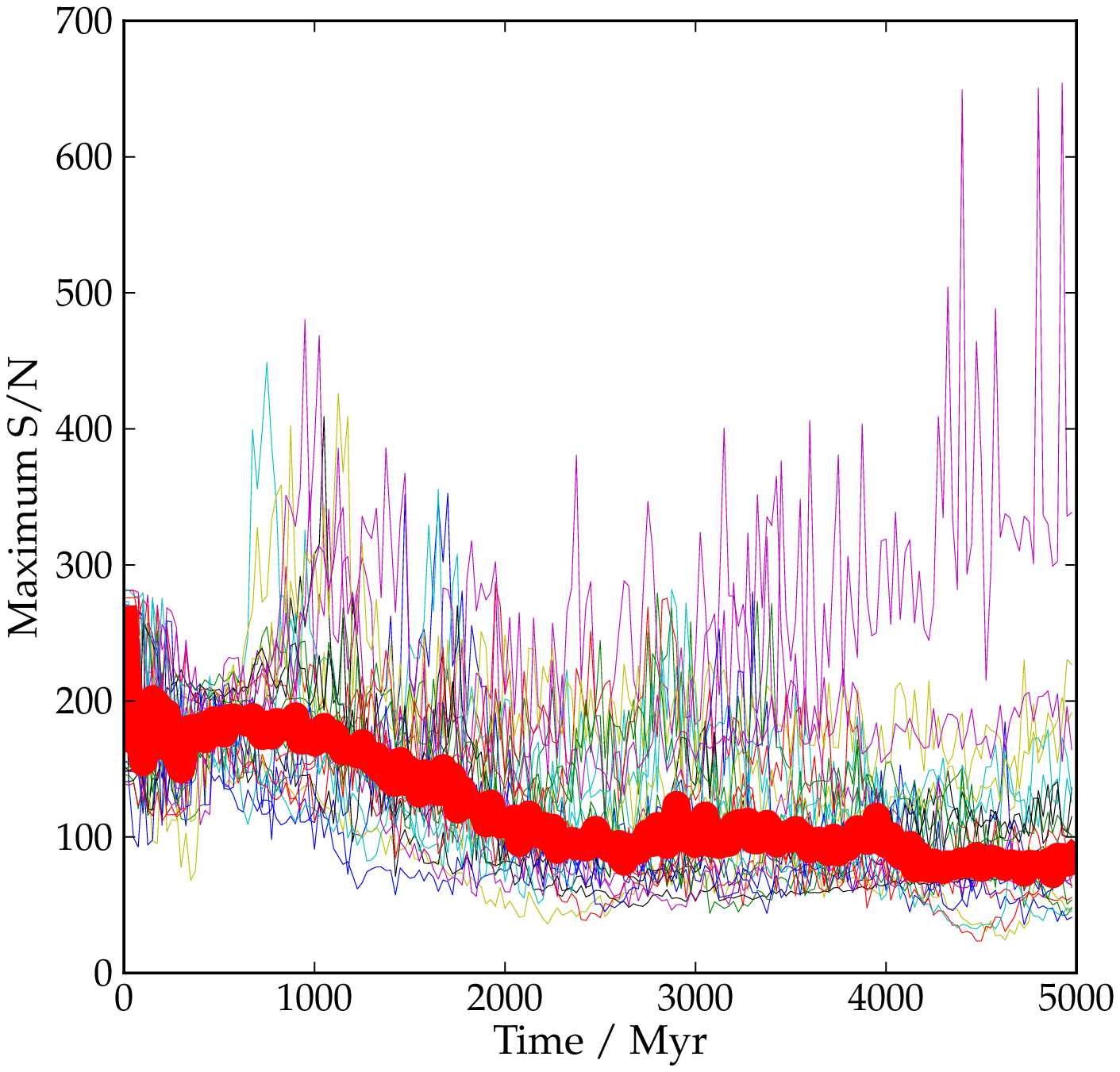}}\\
  \subfloat[]{\includegraphics[height=55mm]{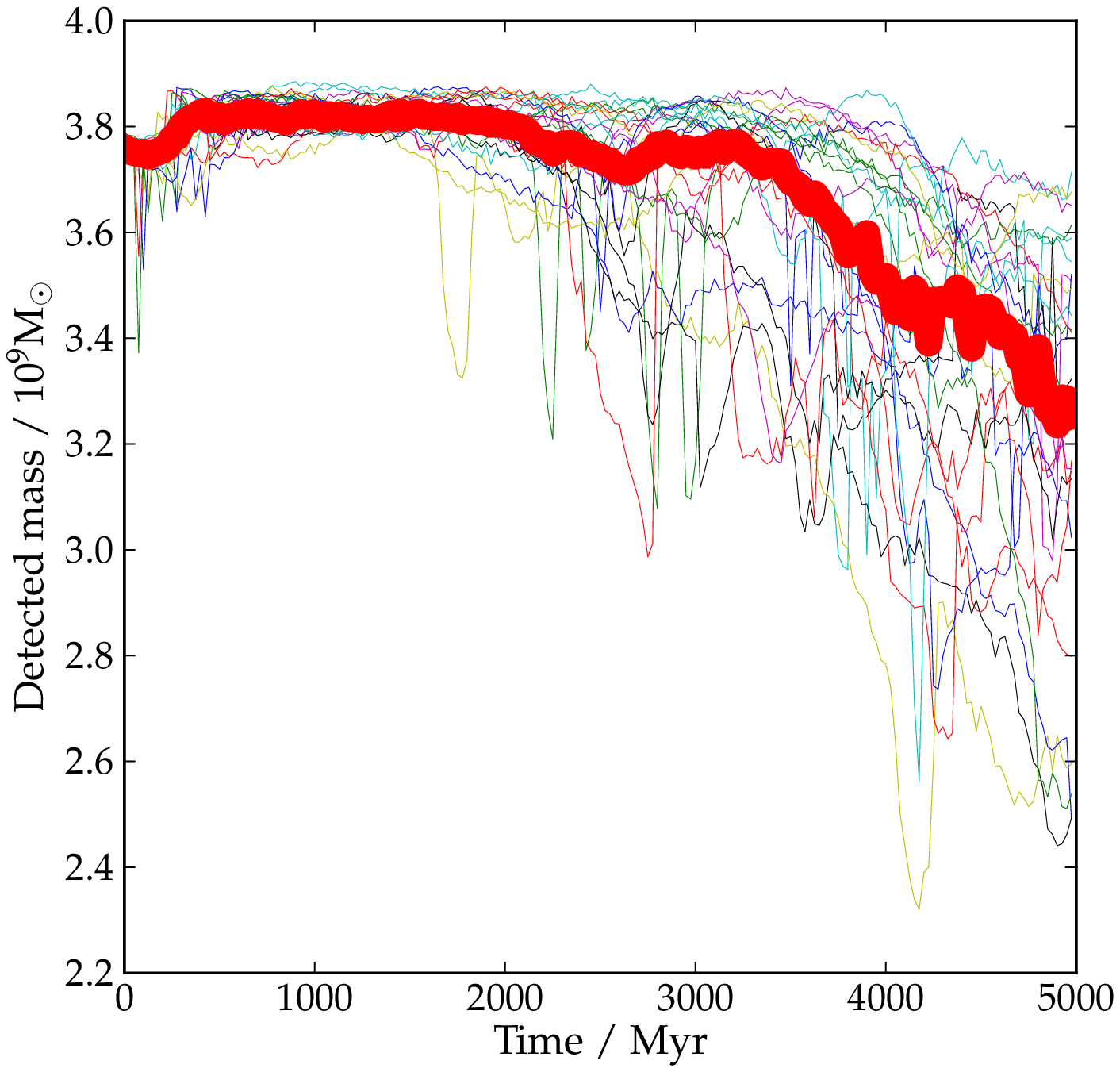}}
  \subfloat[]{\includegraphics[height=55mm]{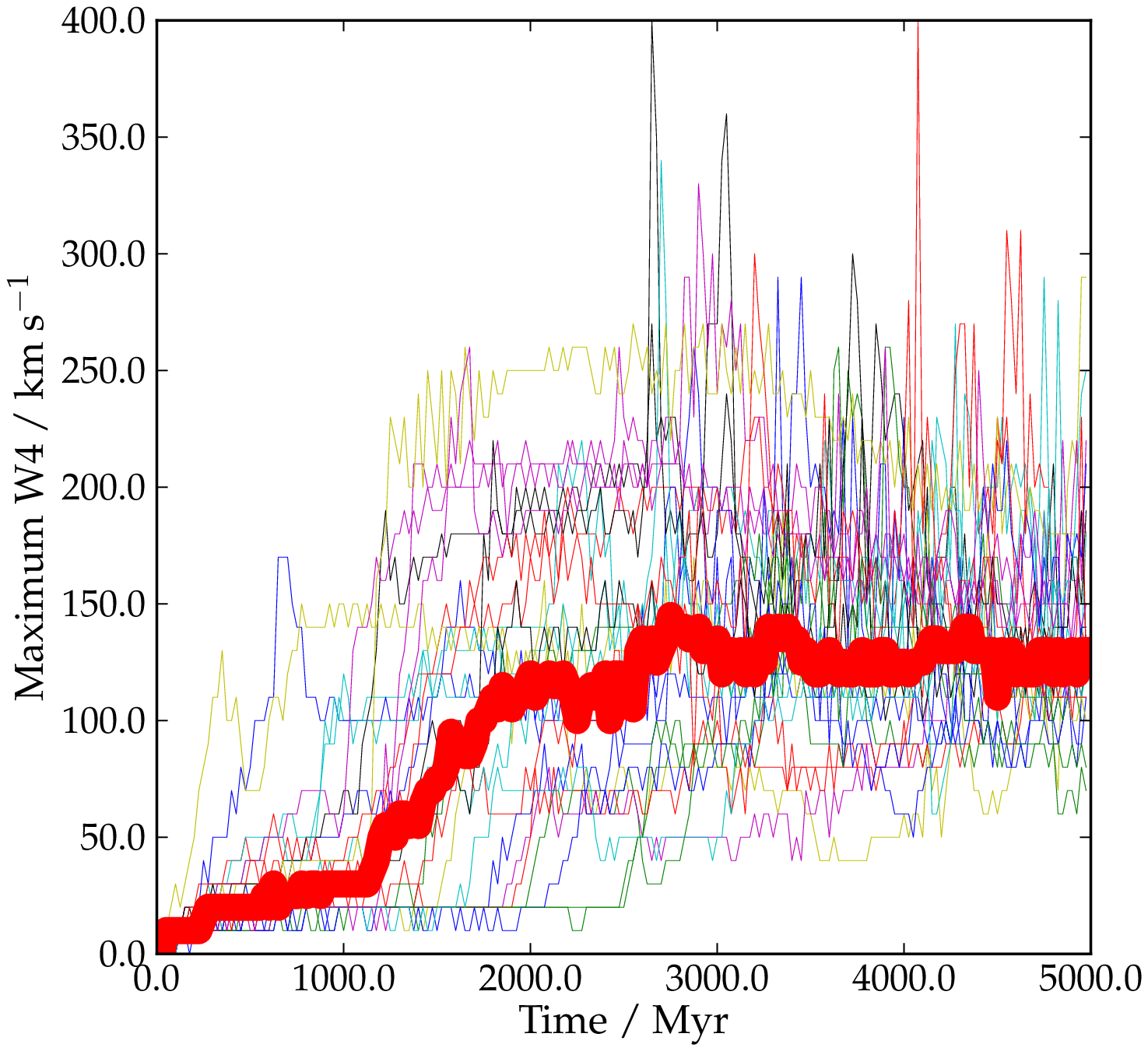}} 
  \subfloat[]{\includegraphics[height=55mm]{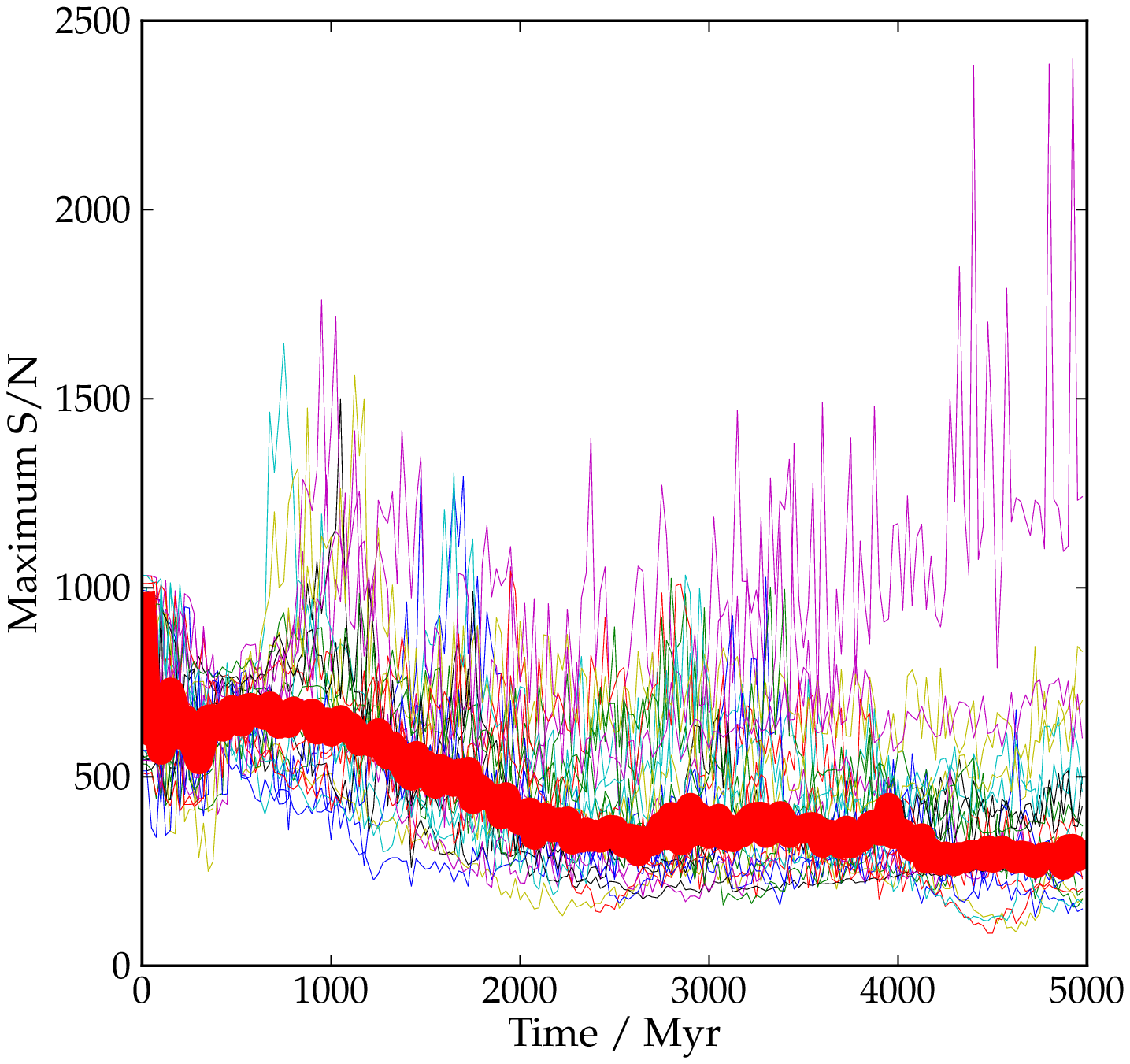}}
\caption[]{Evolution of the properties of 4$\times$10$^{9}$ \Msolar{} streams initially at 0.5 Mpc from the cluster centre. The top panel shows the measurements using an ALFALFA sensitivity level and beam size while the bottom panel shows the equivalent sensitivity of AGES. From left to right : detected mass, maximum $W4$ of any part of the stream, and peak SNR. Each simulation is shown using a different colour; the thick red line shows the median value of all 26 simulations.}
\label{fig:massivestream0.5Mpc}
\end{figure*}

\begin{figure*}
\centering 
  \subfloat[]{\includegraphics[height=55mm]{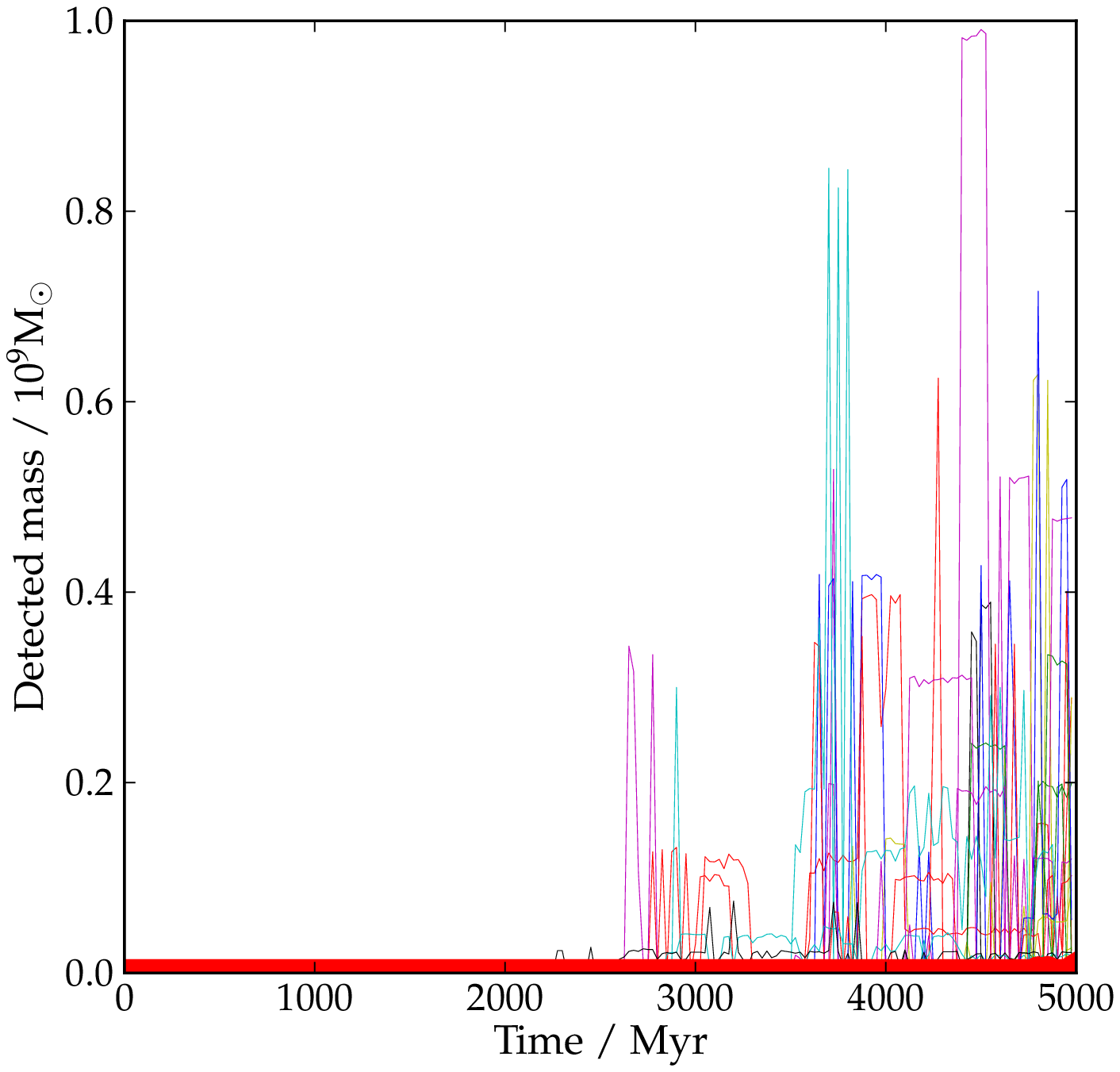}}
  \subfloat[]{\includegraphics[height=55mm]{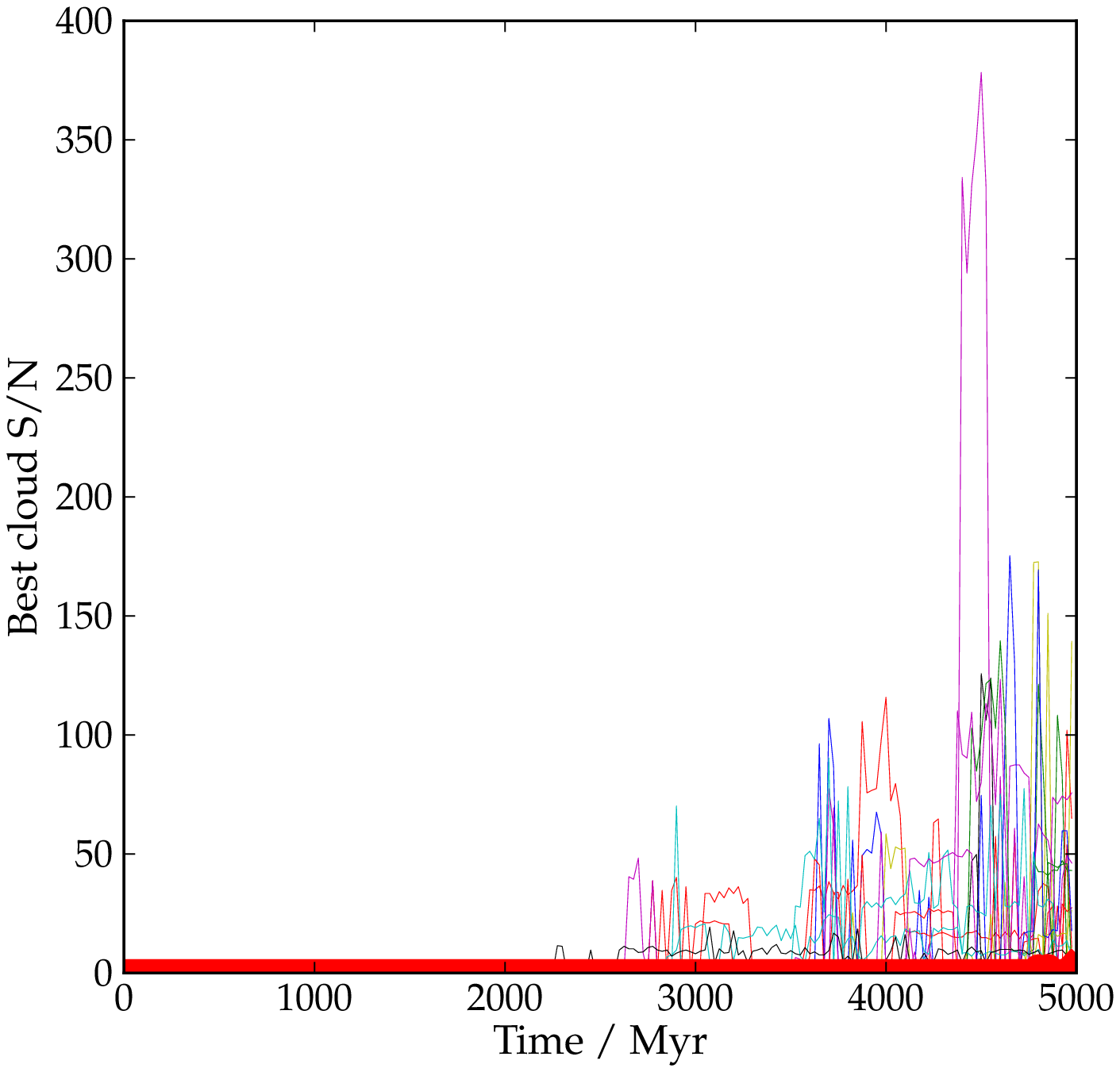}} 
  \subfloat[]{\includegraphics[height=55mm]{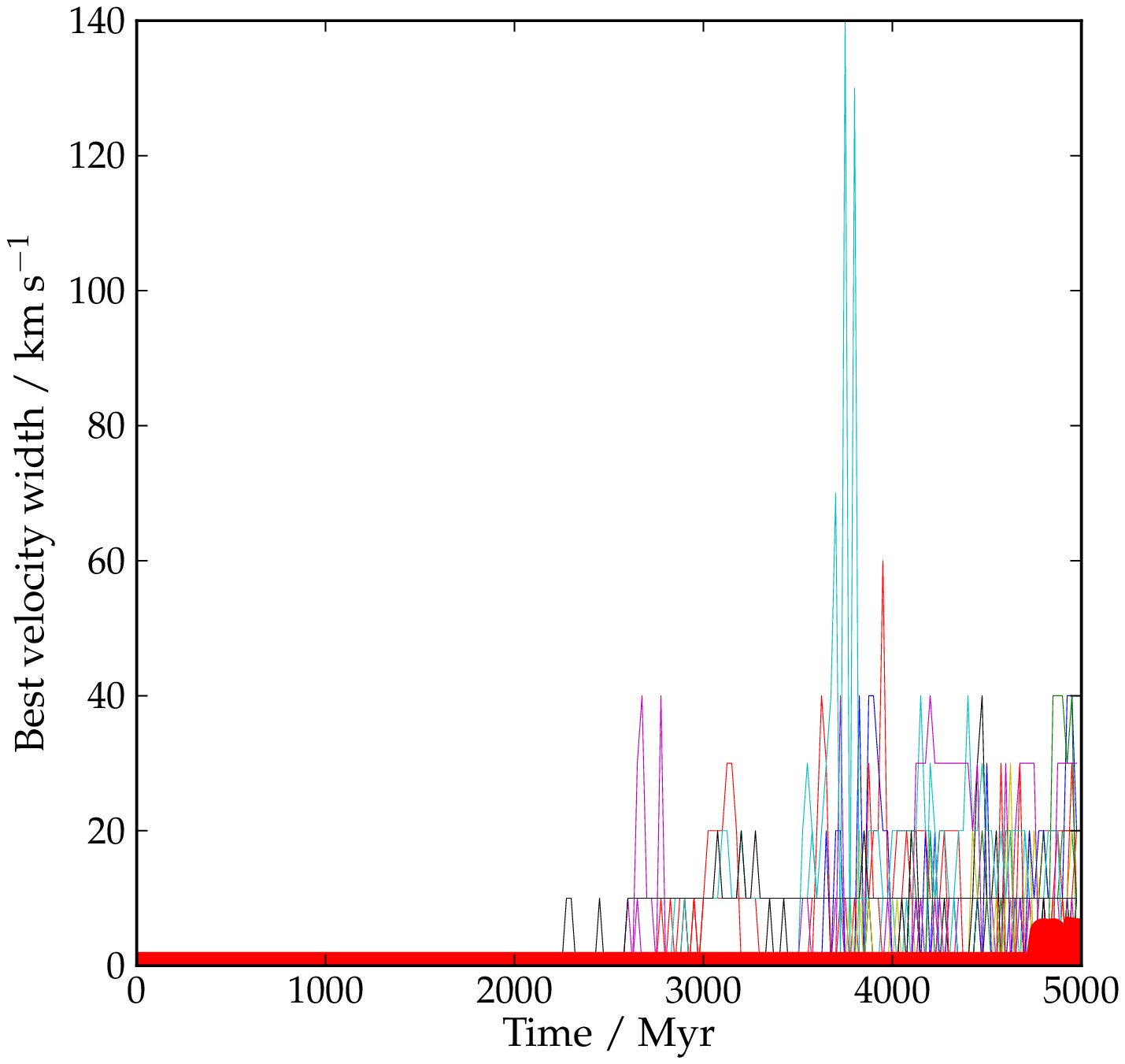}}\\
  \subfloat[]{\includegraphics[height=55mm]{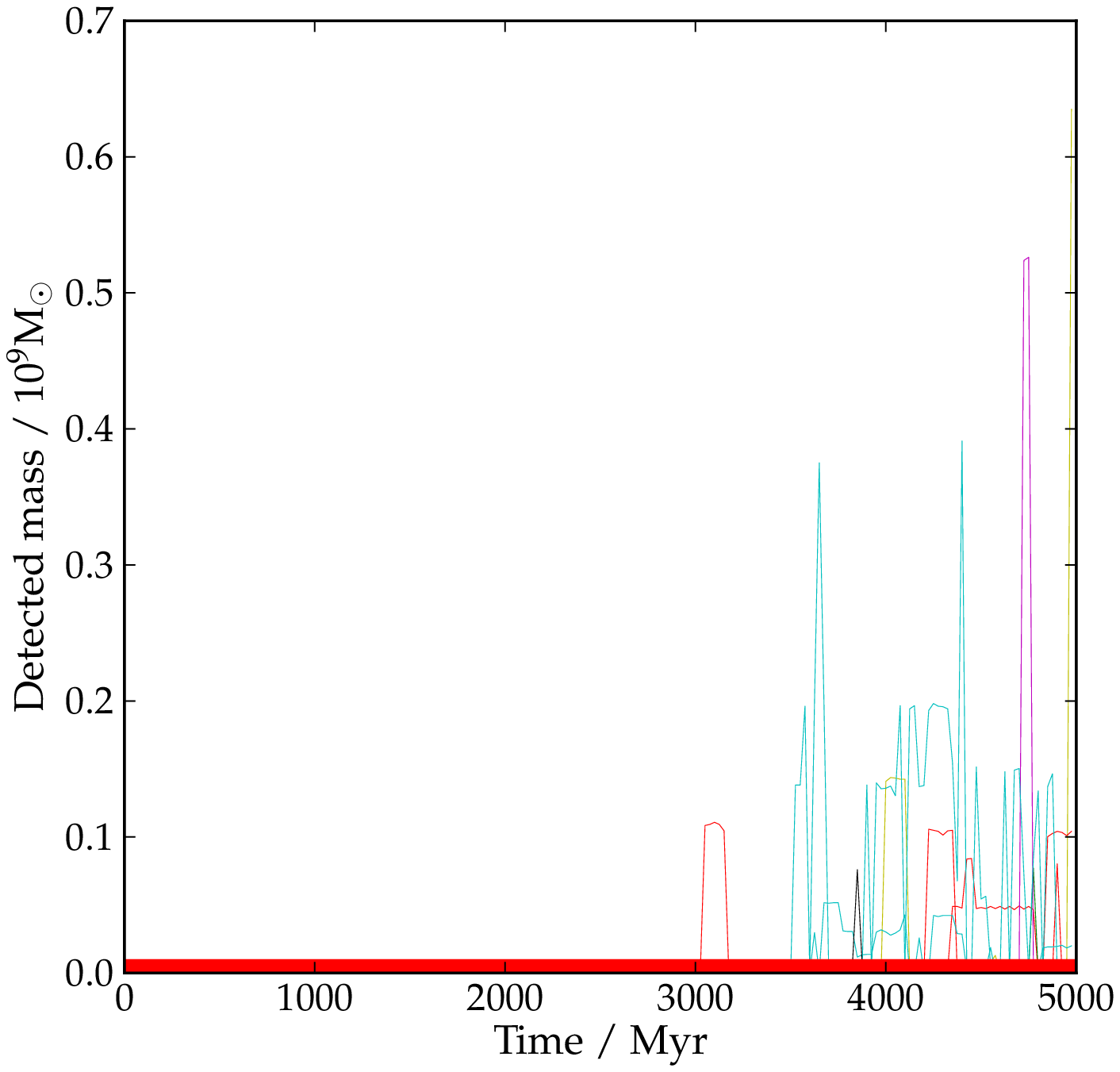}}
  \subfloat[]{\includegraphics[height=55mm]{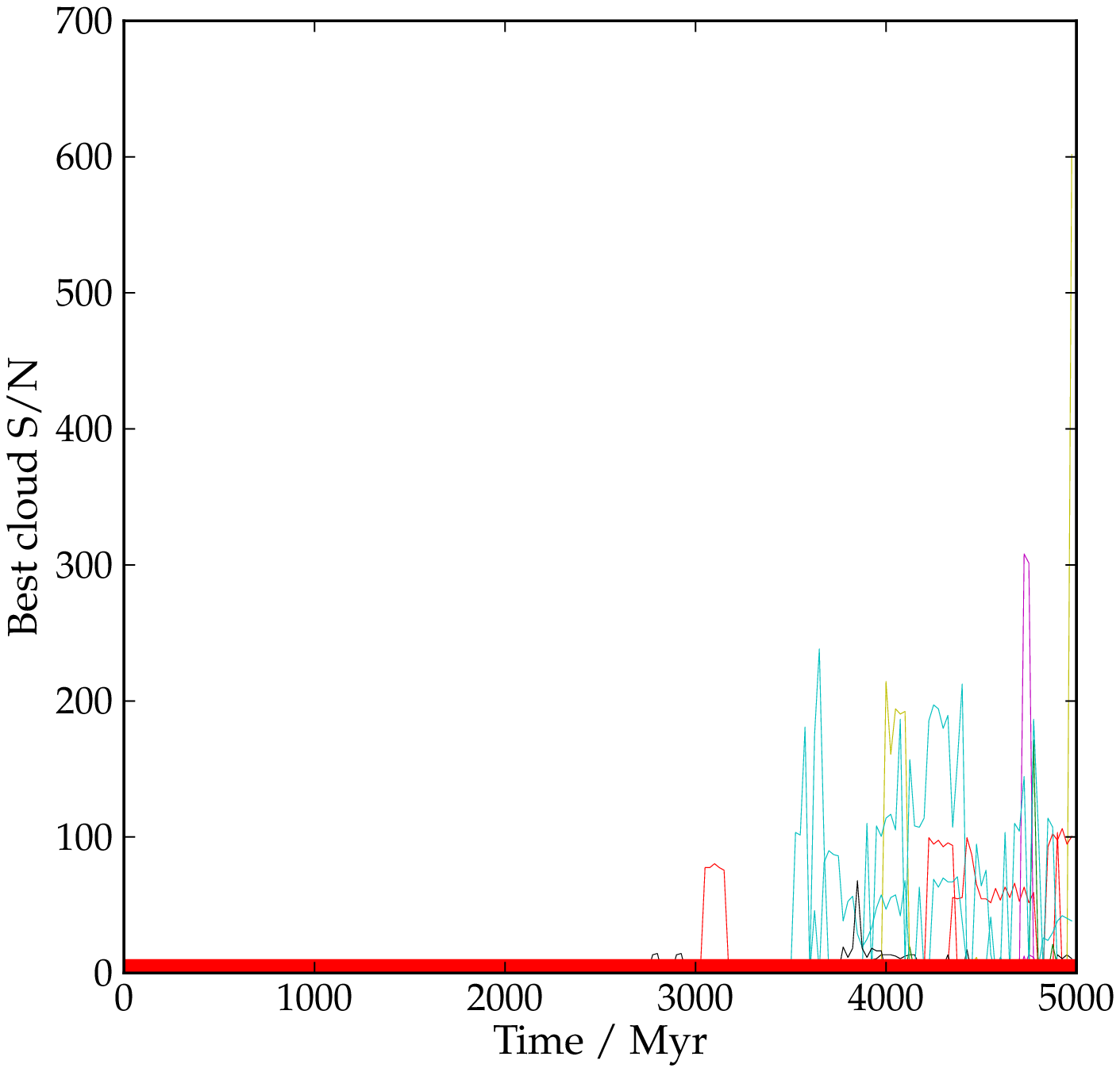}} 
  \subfloat[]{\includegraphics[height=55mm]{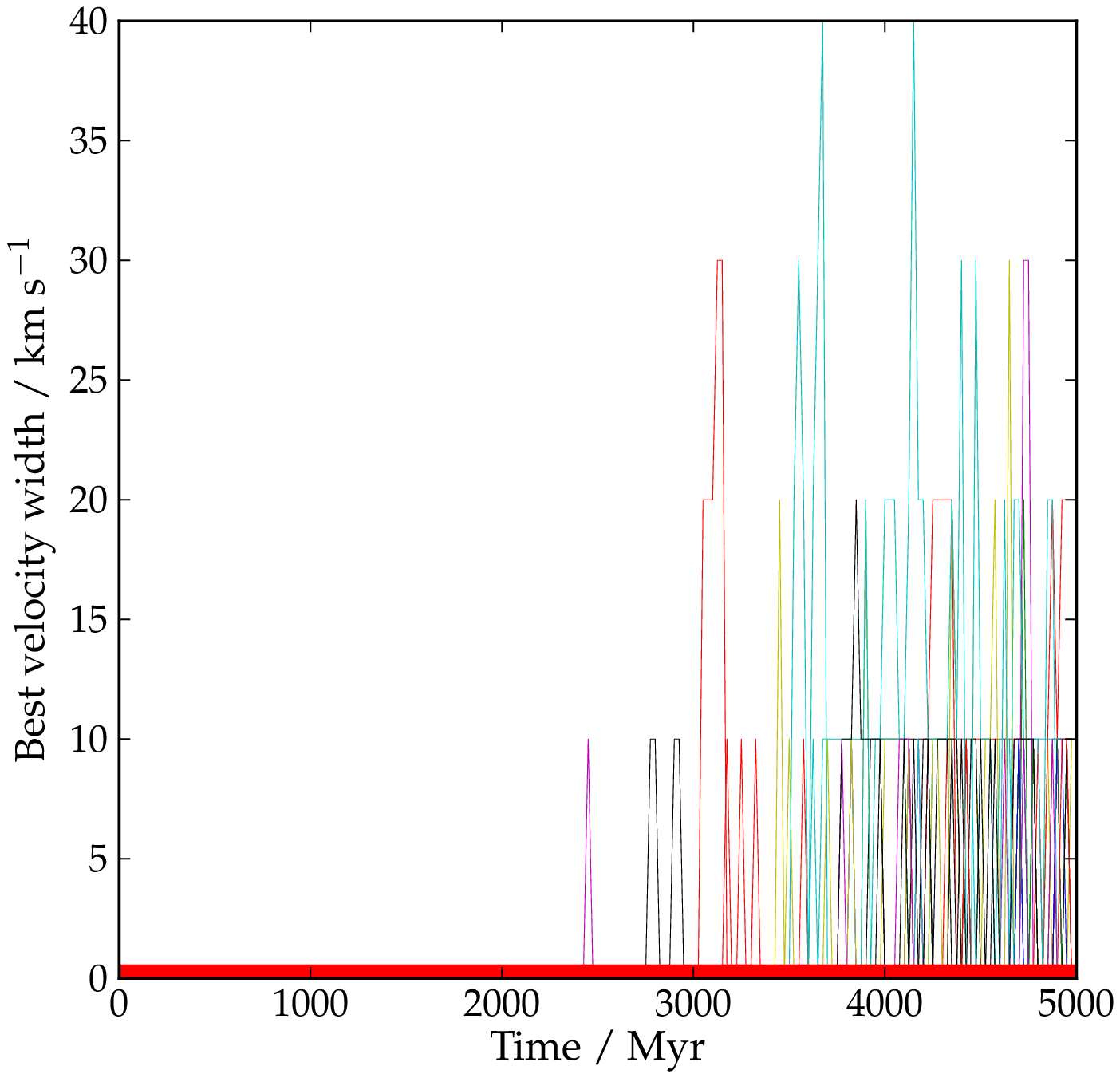}}
\caption[]{Evolution of the properties of the isolated cloud with the highest velocity width, produced from the 4$\times$10$^{9}$ \Msolar{} streams initially at 0.5 Mpc from the cluster centre. The top panel shows the measurements using an ALFALFA sensitivity level and beam size while the bottom panel shows the equivalent sensitivity of AGES. From left to right : detected mass, peak SNR, and $W50$. Each simulation is shown using a different colour; the thick red line shows the median value of all 26 simulations.}
\label{fig:massivestreamclouds0.5Mpc}
\end{figure*}

\begin{figure*}
\centering
\includegraphics[width=160mm]{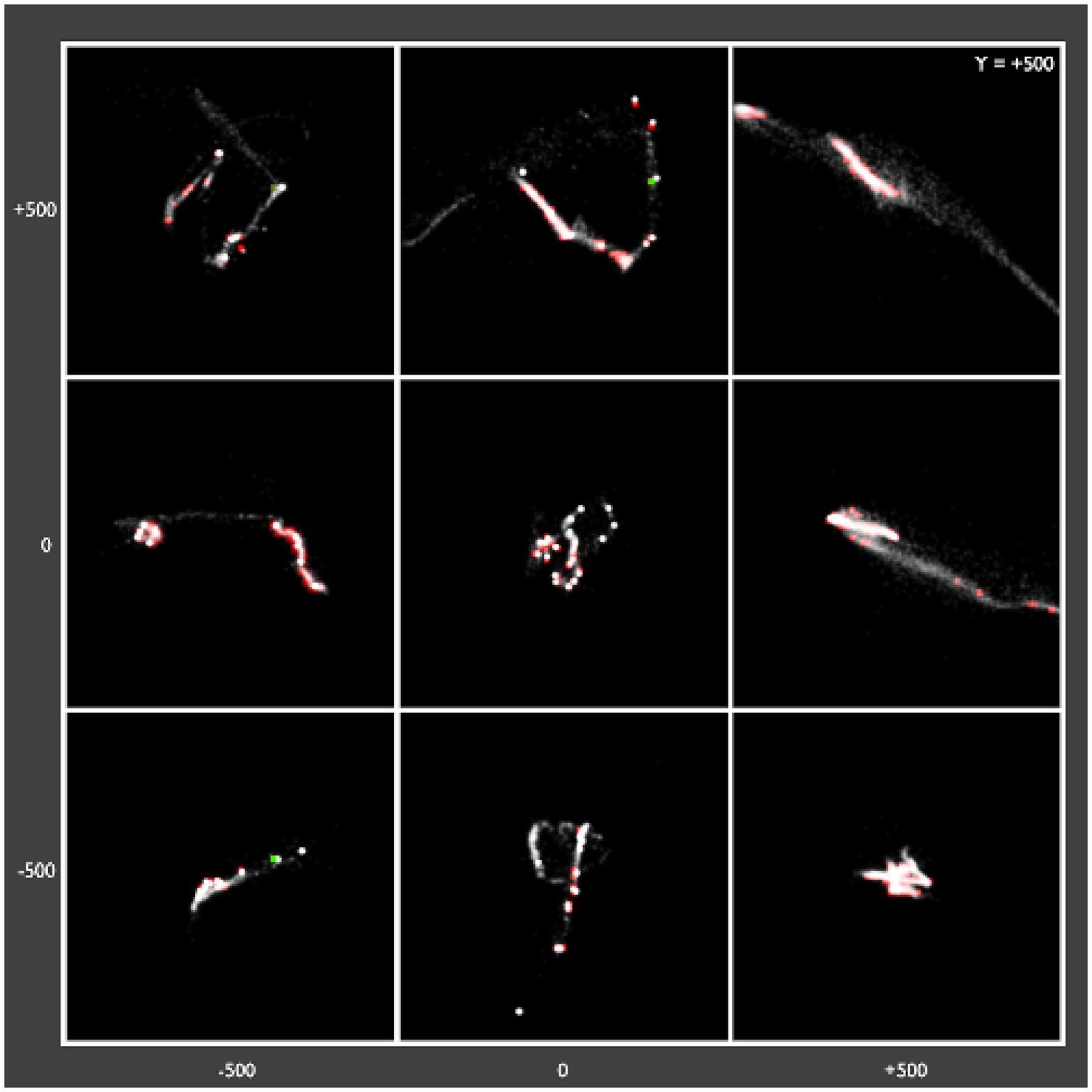}
\caption[hienv]{Final timestep (5 Gyr) of the simulation of a sample of 4$\times$10$^{9}$ \Msolar{} streams entering the cluster from an initial distance of 0.5 Mpc. Each box spans 1 Mpc and is centred on the mean particle position. White shows the raw particle data. Red shows all gridded data in which the emission would exceed a SNR of 4.0 with an ALFALFA sensitivity level; green indicates detectable clouds at least 100 kpc from the nearest other detection. Movies of the simulations can be see at \href{http://tinyurl.com/gonowbj}{this url} : http://tinyurl.com/gonowbj.}
\label{fig:mmovien}
\end{figure*}

\begin{figure*}
\centering 
  \subfloat[]{\includegraphics[height=55mm]{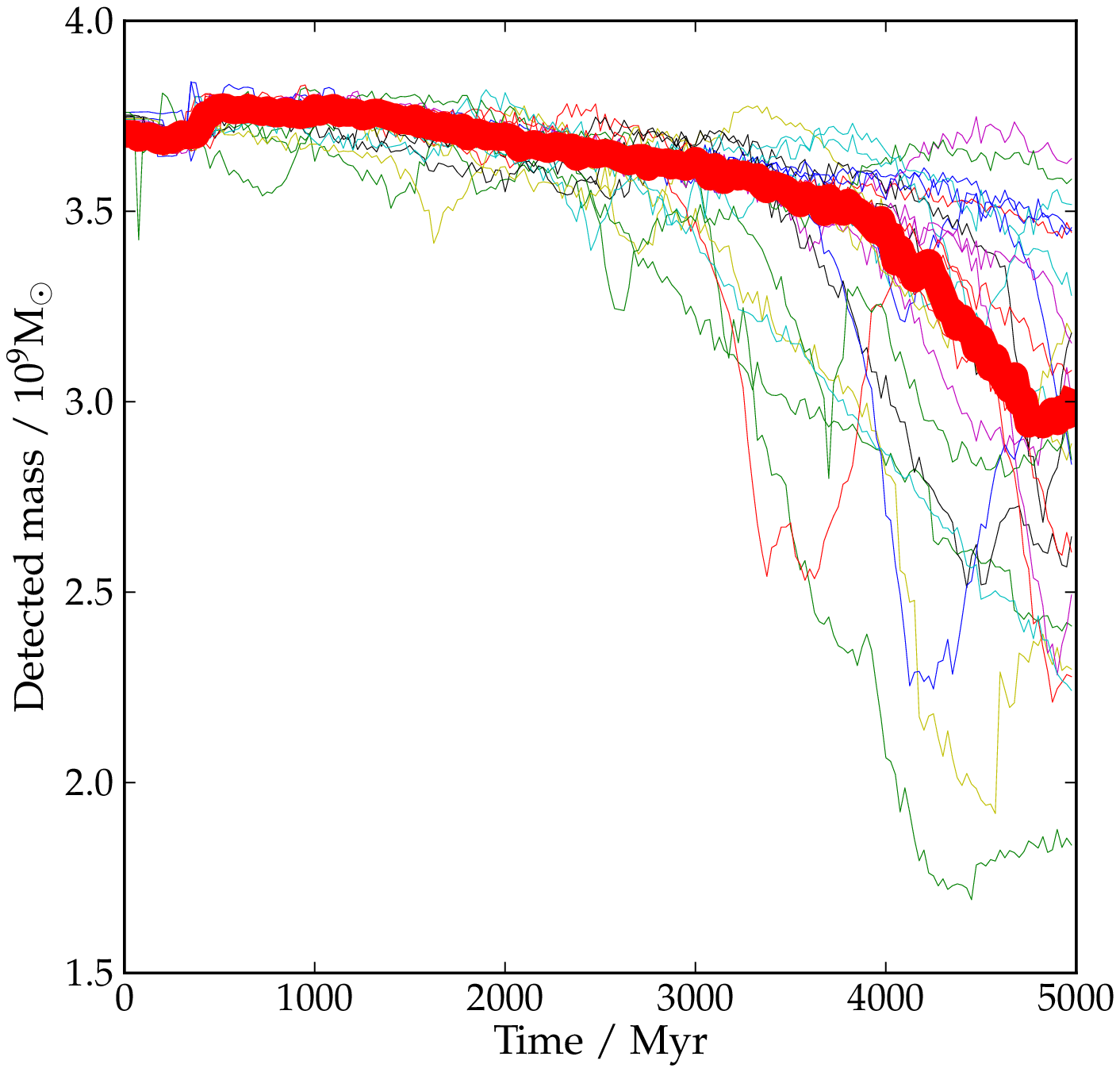}}
  \subfloat[]{\includegraphics[height=55mm]{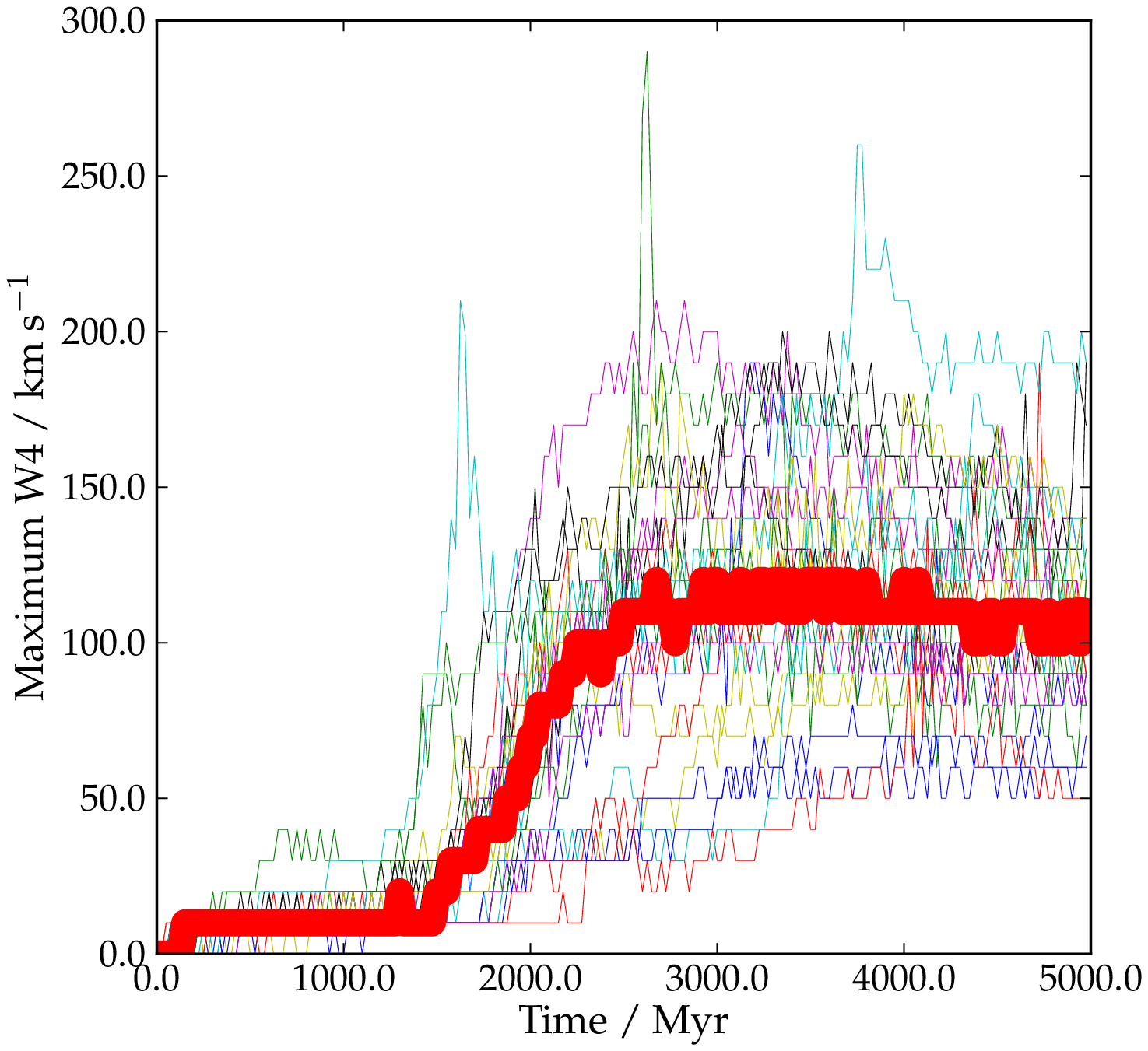}} 
  \subfloat[]{\includegraphics[height=55mm]{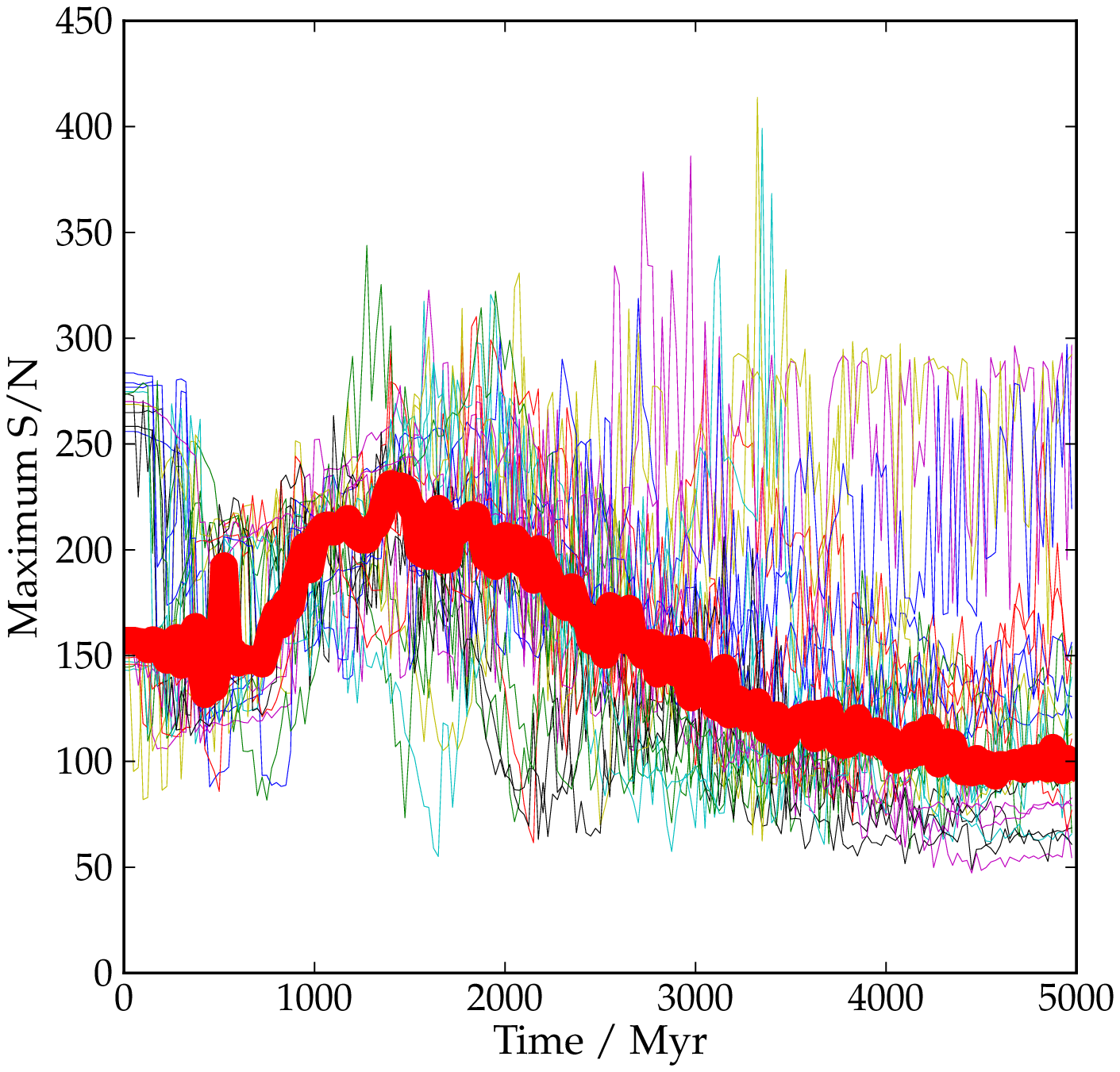}}\\
  \subfloat[]{\includegraphics[height=55mm]{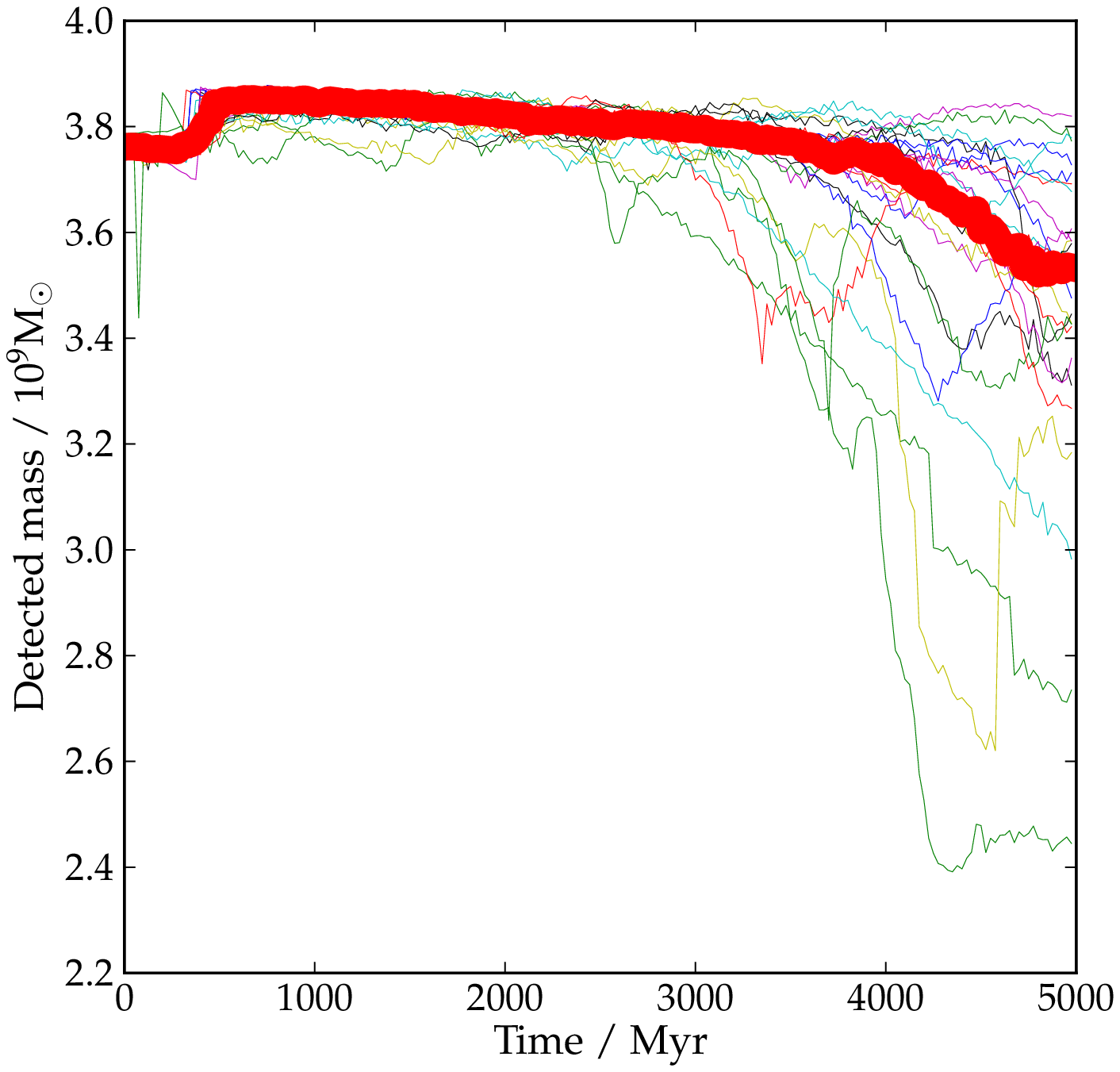}}
  \subfloat[]{\includegraphics[height=55mm]{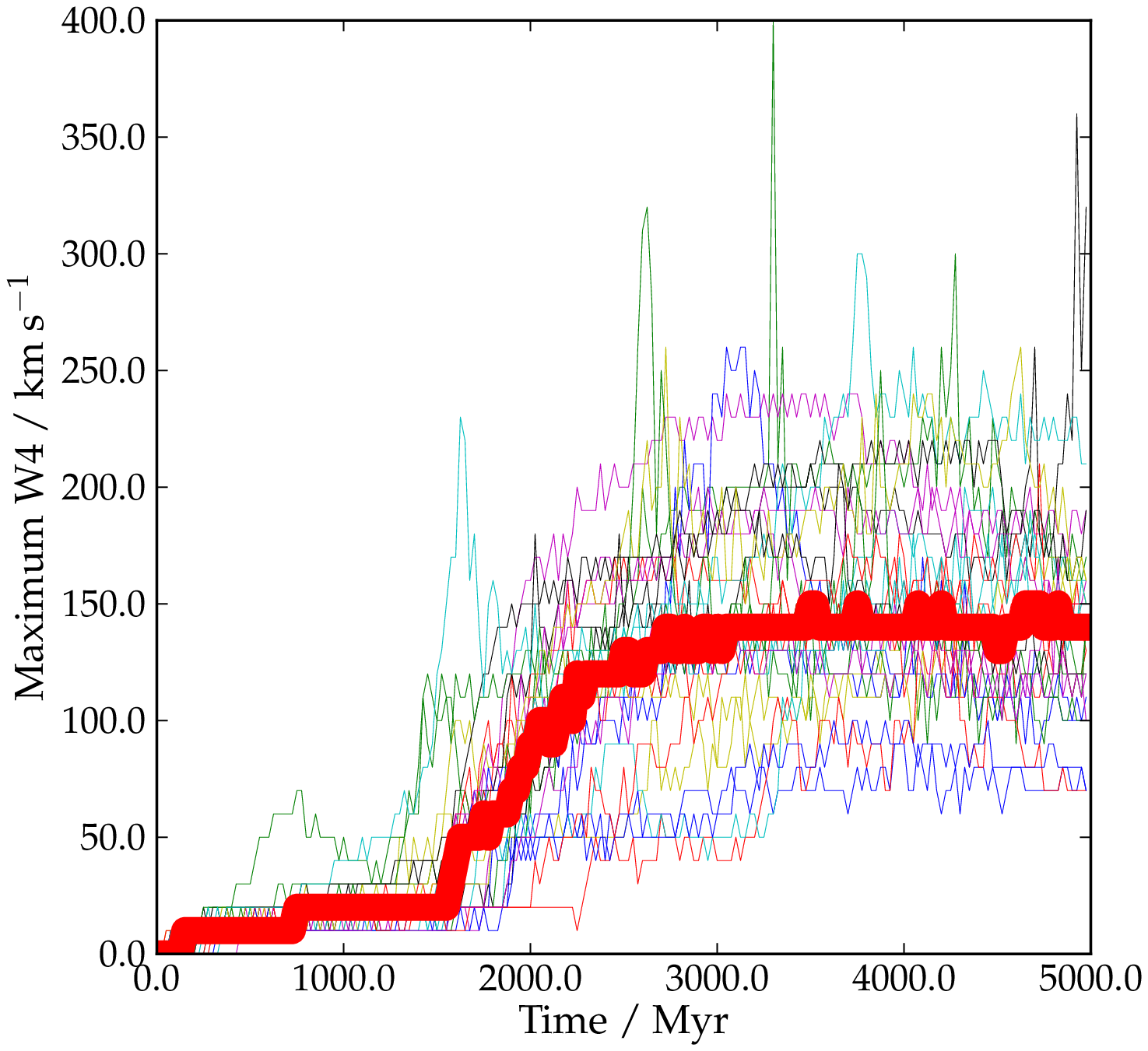}} 
  \subfloat[]{\includegraphics[height=55mm]{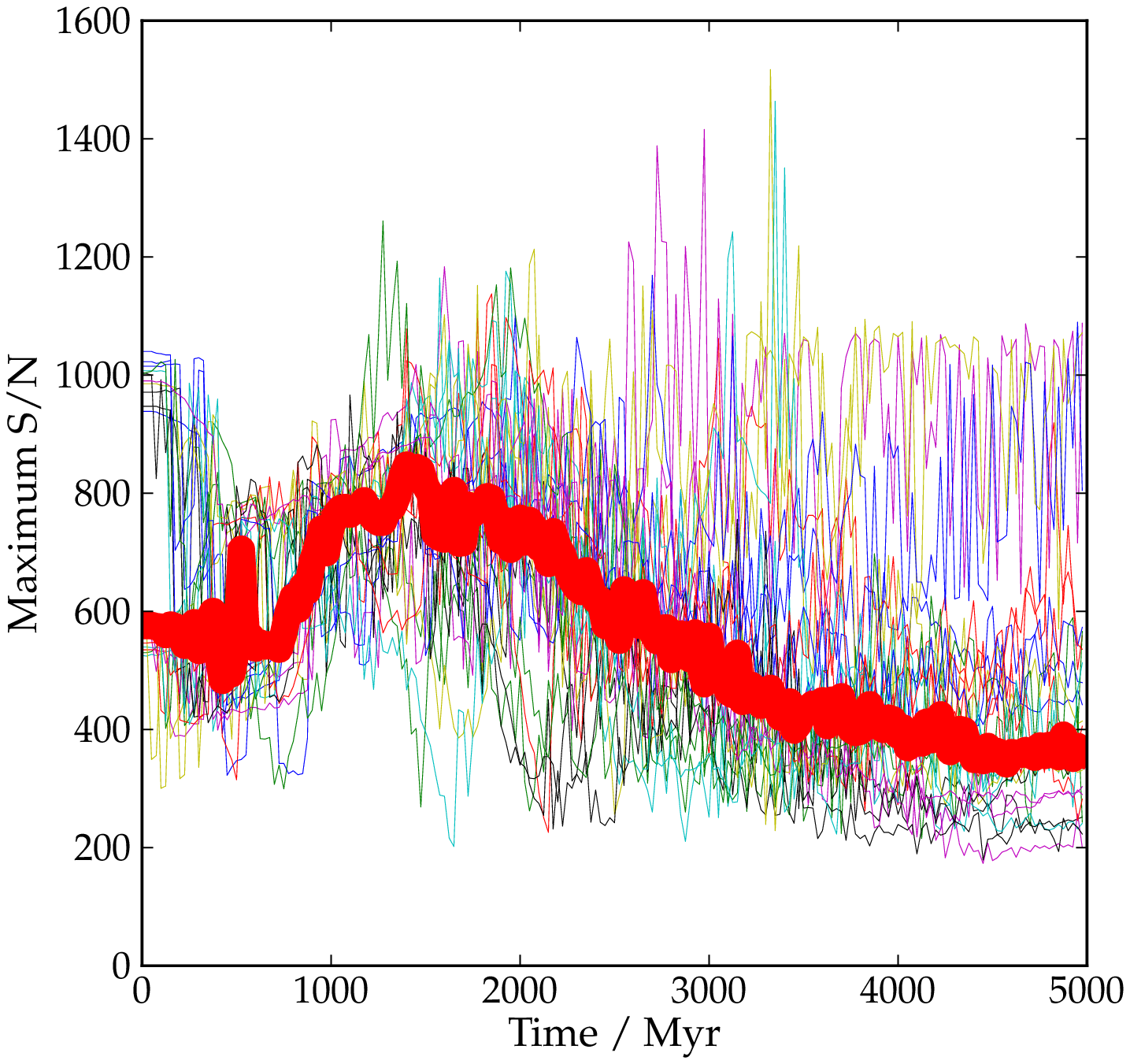}}
\caption[]{Evolution of the properties of 4$\times$10$^{9}$ \Msolar{} streams initially at 1.0 Mpc from the cluster centre. The top panel shows the measurements using an ALFALFA sensitivity level and beam size while the bottom panel shows the equivalent sensitivity of AGES. From left to right : detected mass, maximum $W4$ of any part of the stream, and peak SNR. Each simulation is shown using a different colour; the thick red line shows the median value of all 26 simulations.}
\label{fig:massivestream1.0Mpc}
\end{figure*}

\begin{figure*}
\centering  
  \subfloat[]{\includegraphics[height=55mm]{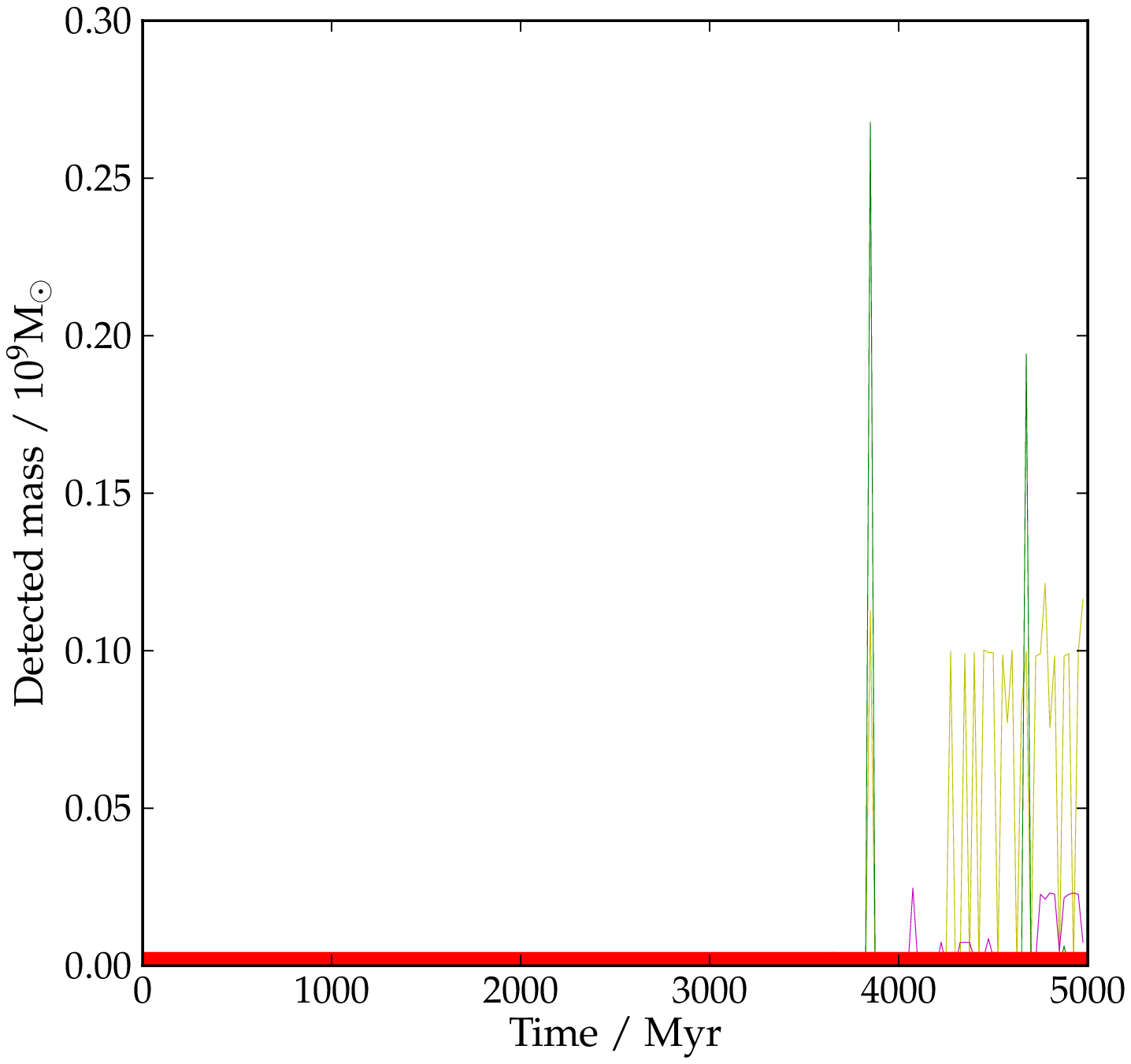}}     
  \subfloat[]{\includegraphics[height=55mm]{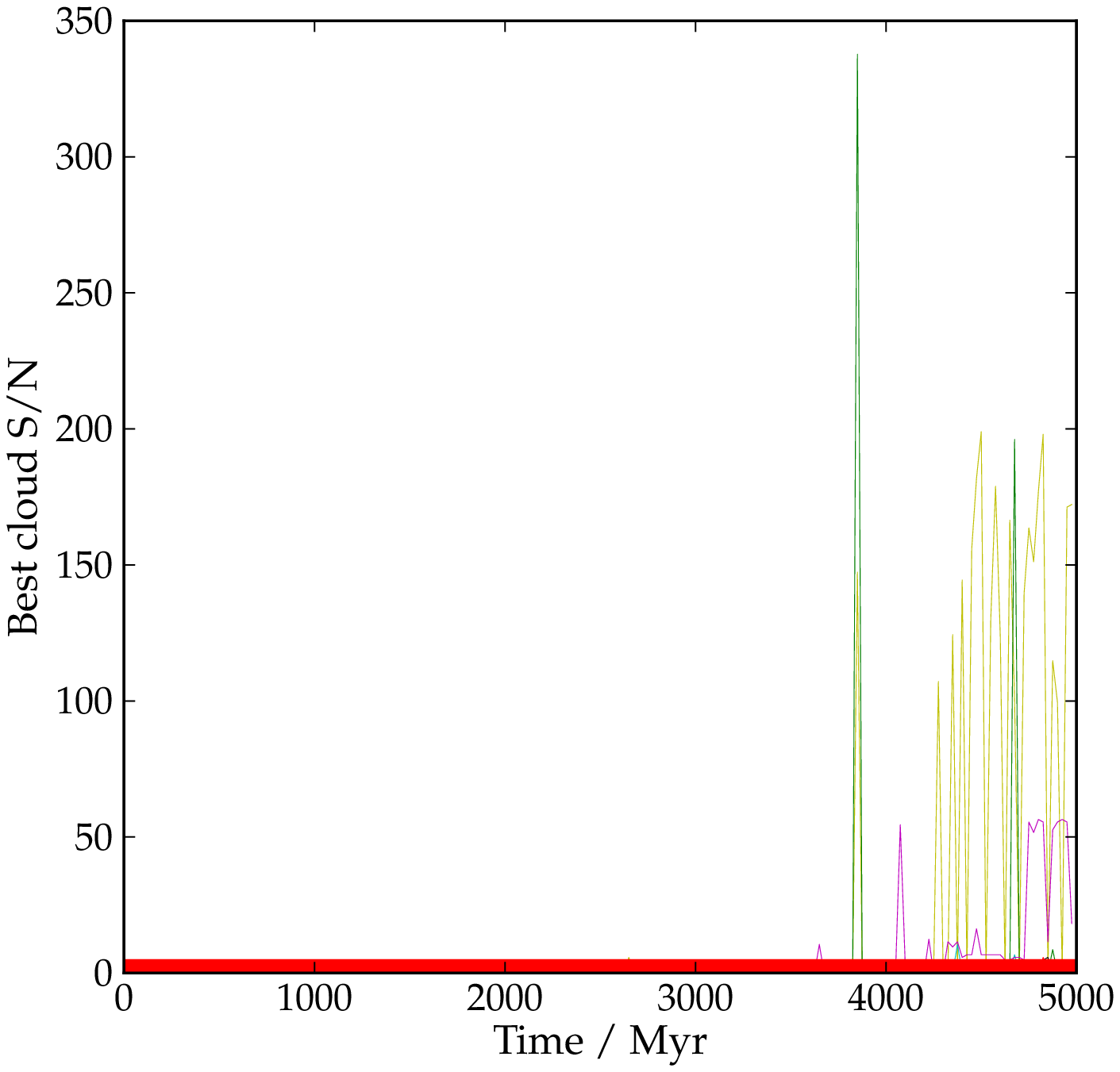}} 
  \subfloat[]{\includegraphics[height=55mm]{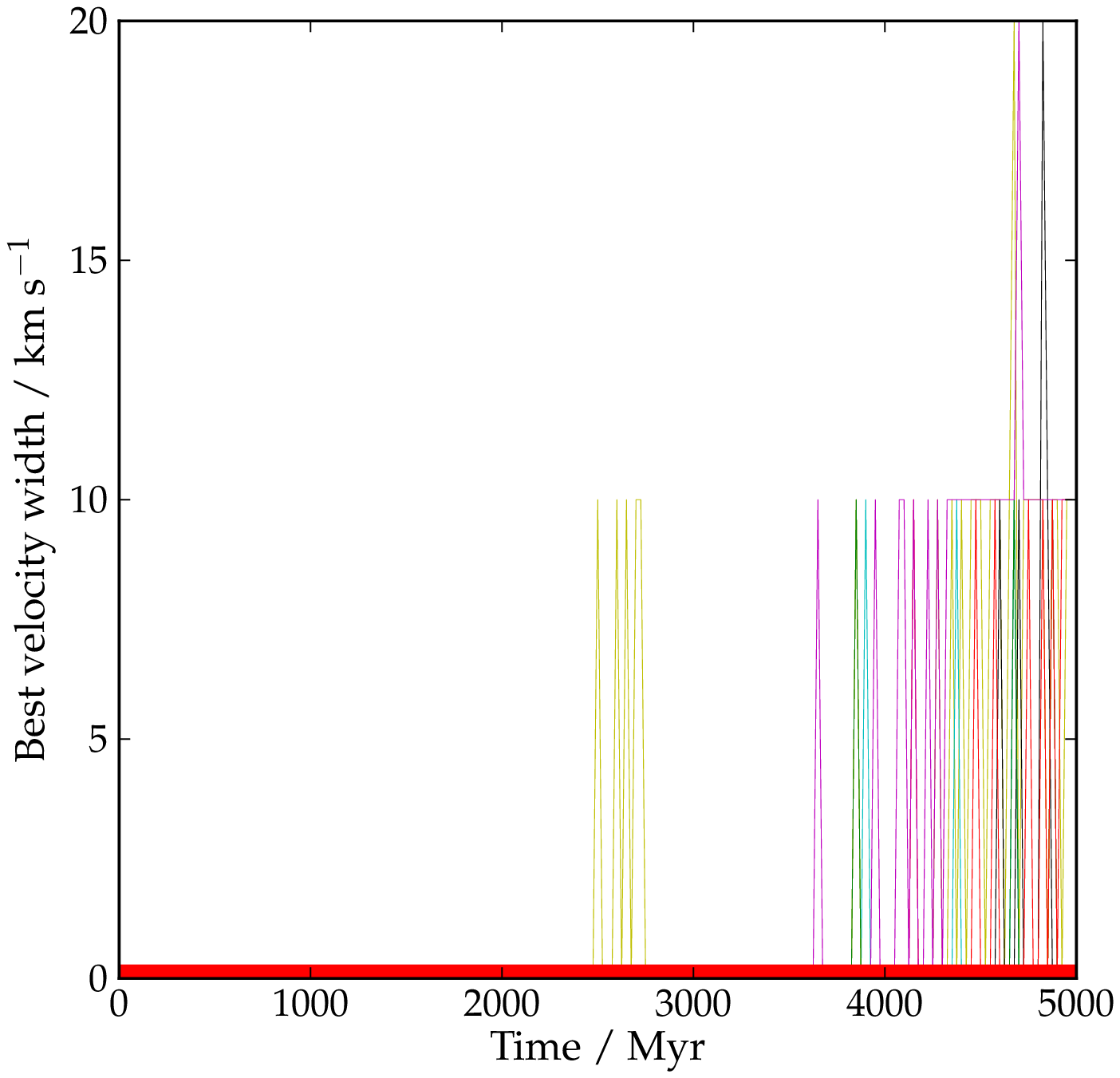}}
\caption[]{Evolution of the properties of the isolated cloud with the highest velocity width, produced from the 4$\times$10$^{9}$ \Msolar{} streams initially at 1.0 Mpc from the cluster centre, using an AGES sensitivity level. No isolated clouds were detected using an ALFALFA sensitivity level. From left to right : detected mass, peak SNR, and $W50$. Each simulation is shown using a different colour; the thick red line shows the median value of all 26 simulations.}
\label{fig:massivestreamclouds1.0Mpc}
\end{figure*}

\begin{figure*}
\centering
\includegraphics[width=160mm]{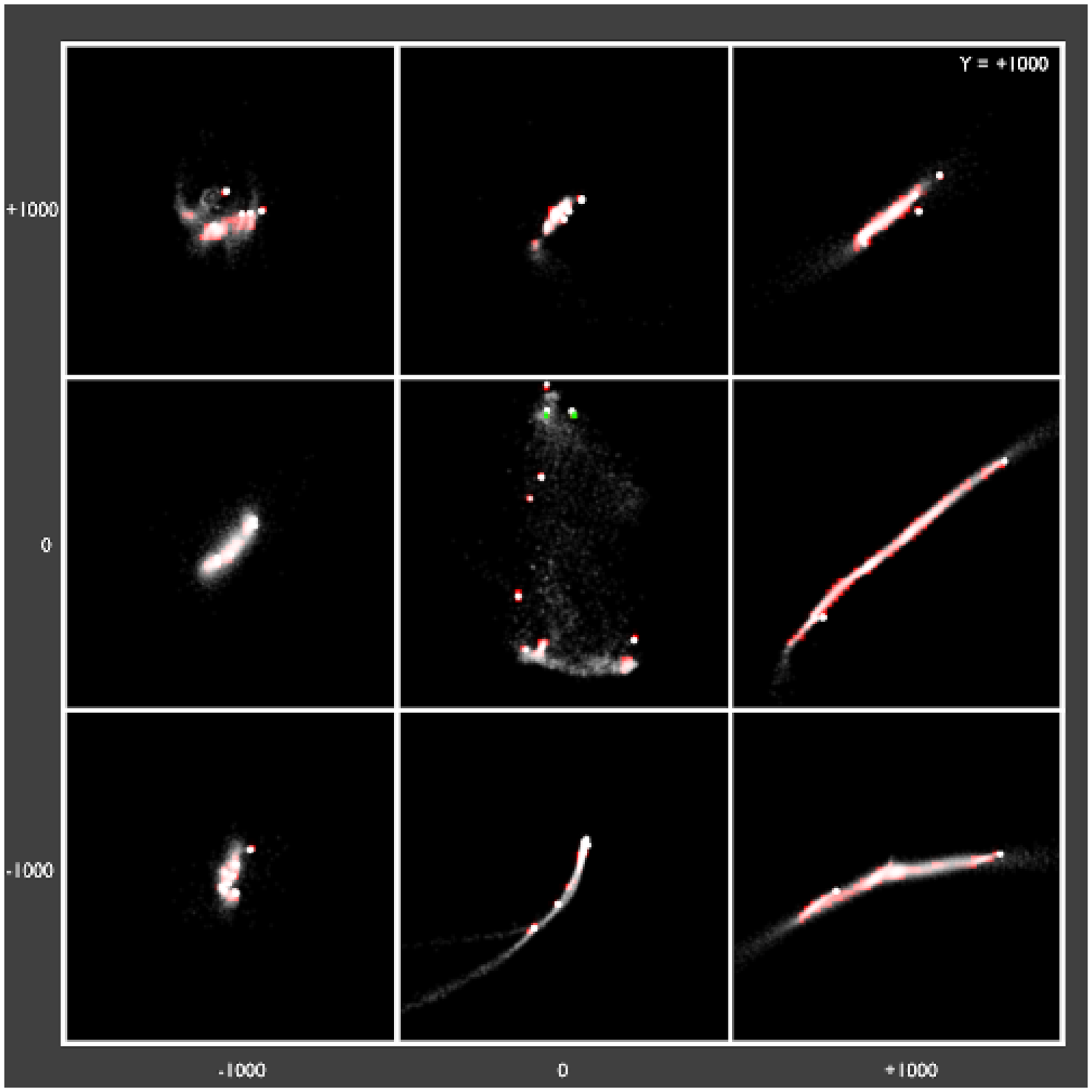}
\caption[hienv]{Final timestep (5 Gyr) of the simulation of a sample of 4$\times$10$^{9}$ \Msolar{} streams entering the cluster from an initial distance of 1.0 Mpc. Each box spans 1 Mpc and is centred on the mean particle position. White shows the raw particle data. Red shows all gridded data in which the emission would exceed a SNR of 4.0 with an ALFALFA sensitivity level; green indicates detectable clouds at least 100 kpc from the nearest other detection. Movies of the simulations can be see at \href{http://tinyurl.com/zhs7yyd}{this url} : http://tinyurl.com/zhs7yyd.}
\label{fig:mmovien}
\end{figure*}

\begin{figure*}
\centering  
  \subfloat[]{\includegraphics[height=55mm]{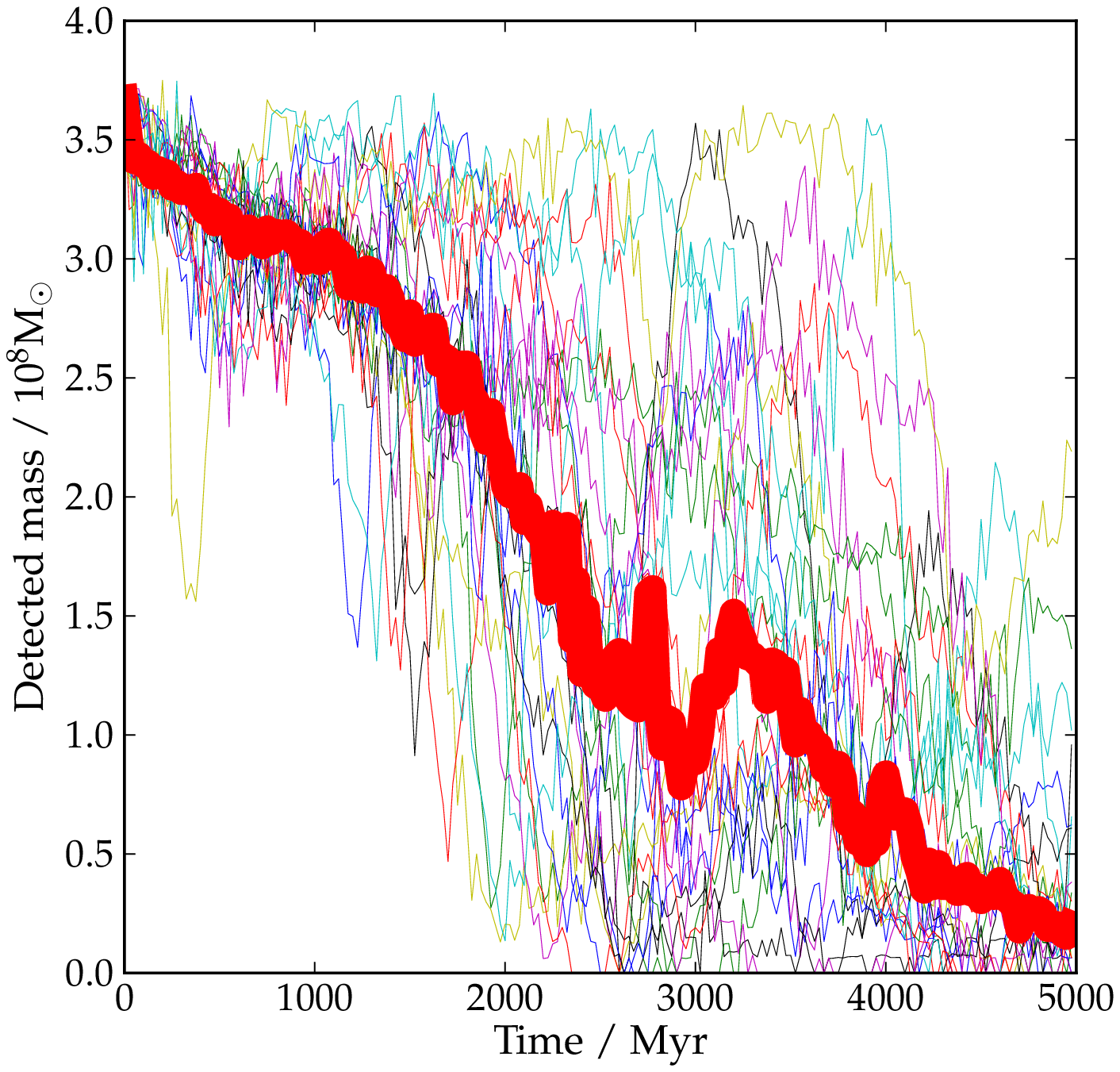}}
  \subfloat[]{\includegraphics[height=55mm]{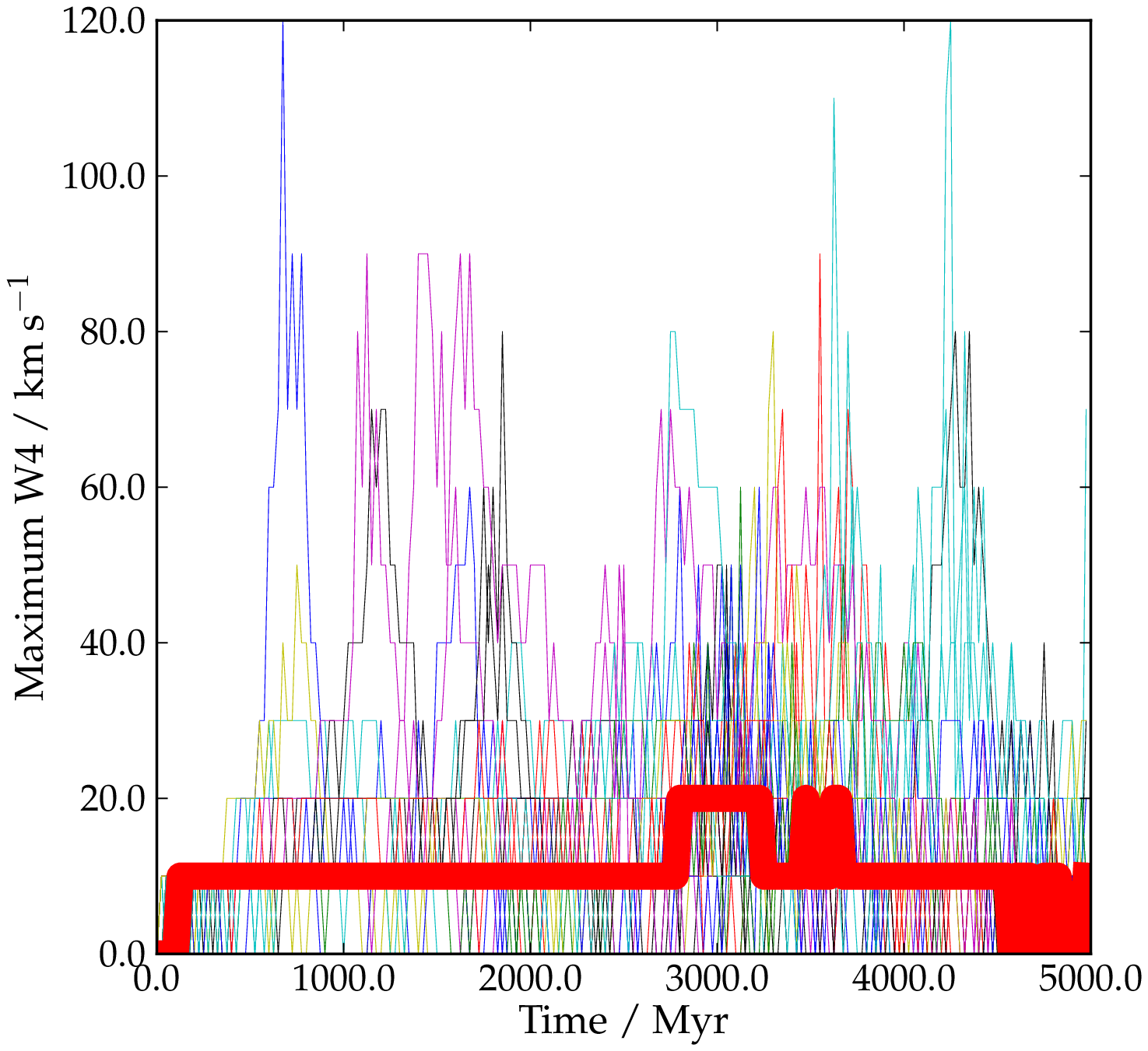}} 
  \subfloat[]{\includegraphics[height=55mm]{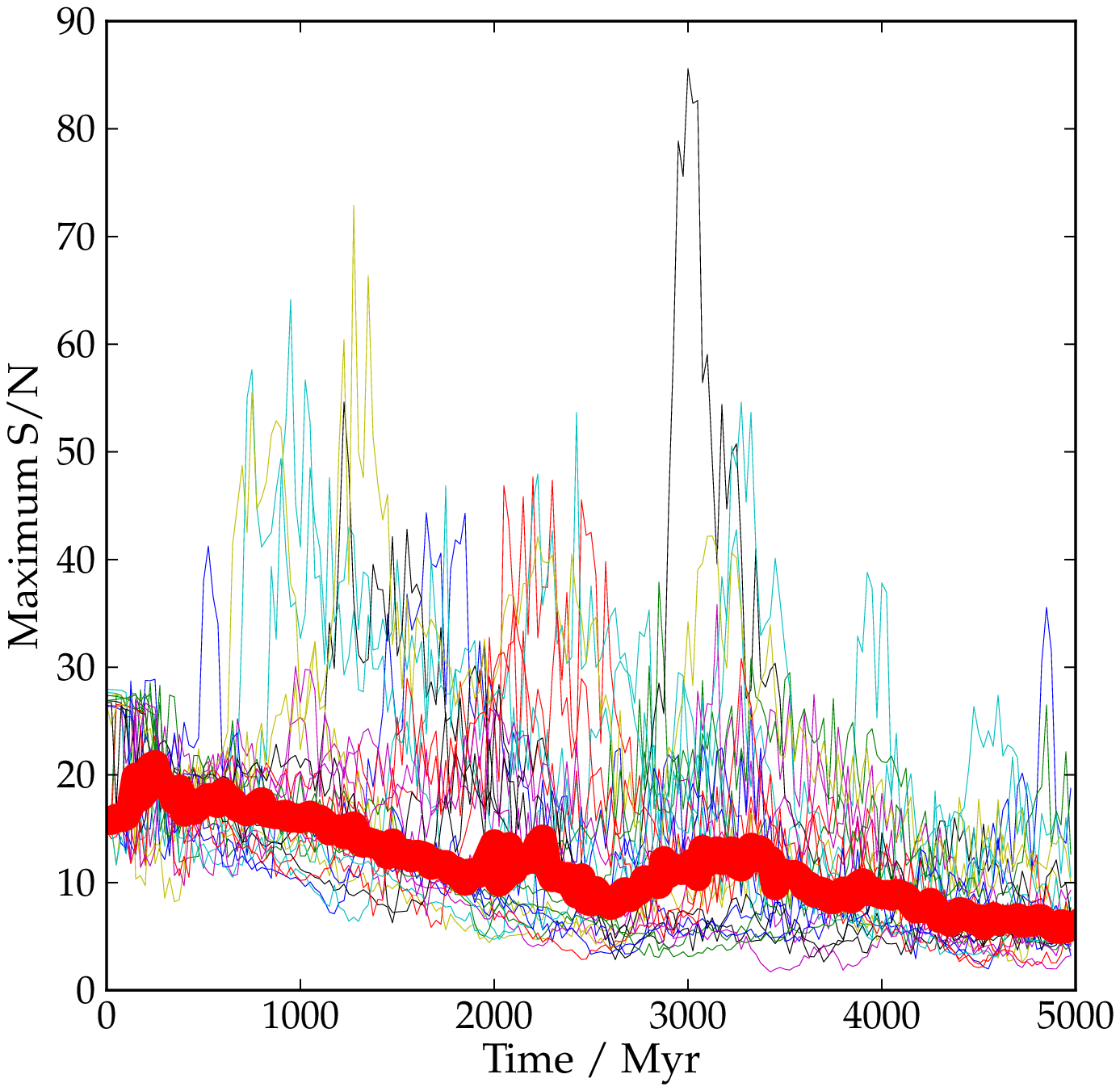}}\\
  \subfloat[]{\includegraphics[height=55mm]{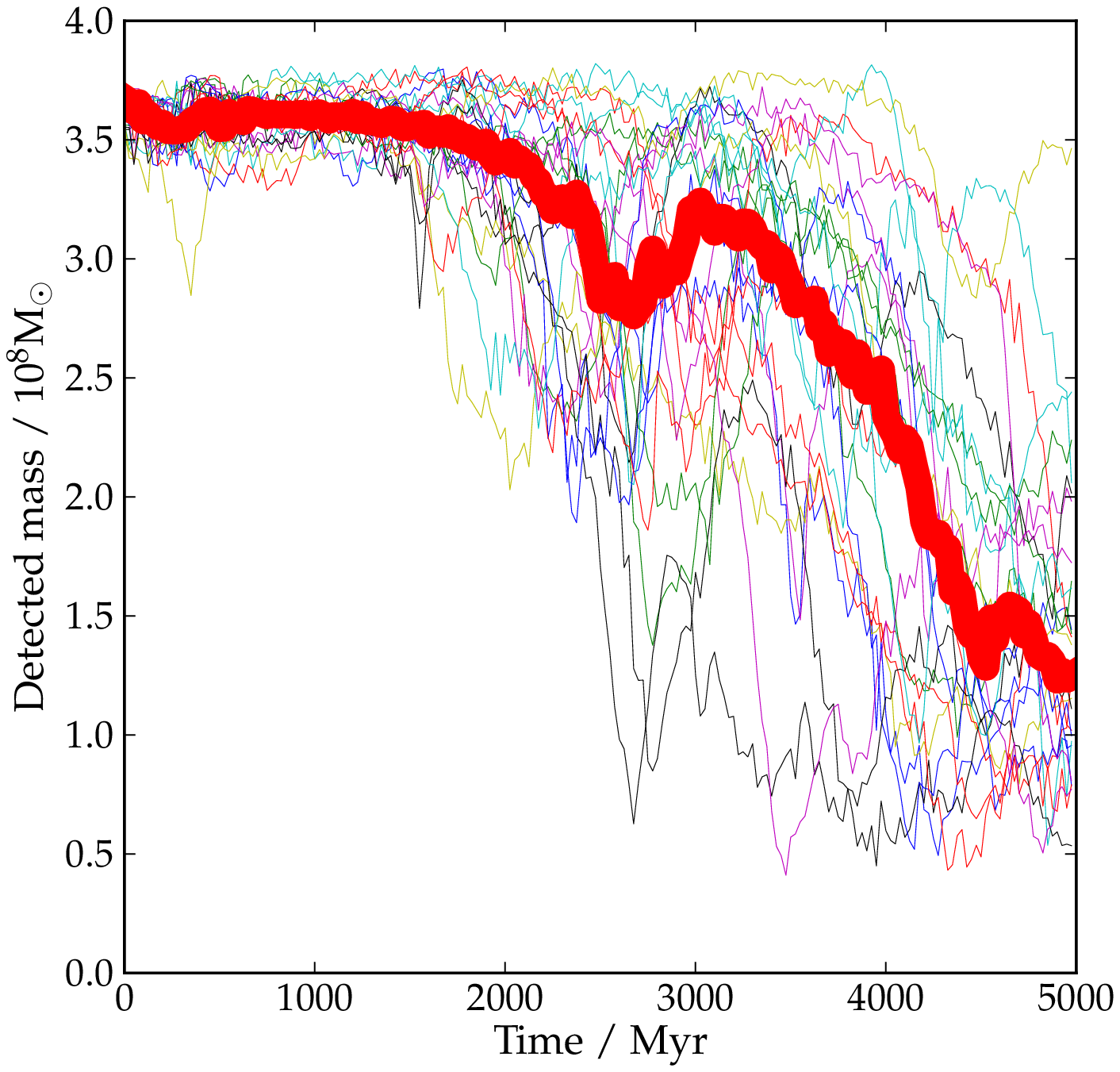}}
  \subfloat[]{\includegraphics[height=55mm]{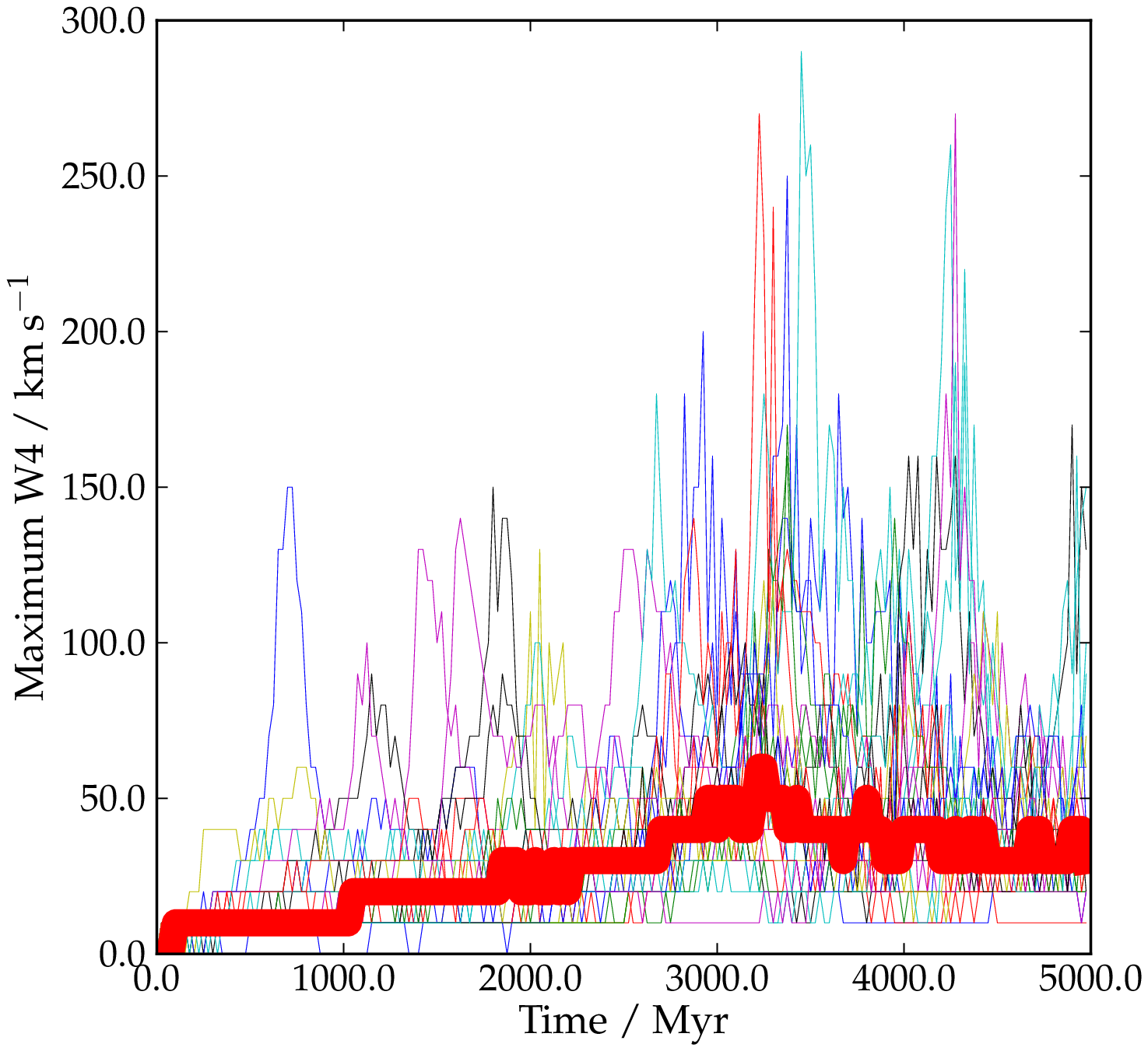}} 
  \subfloat[]{\includegraphics[height=55mm]{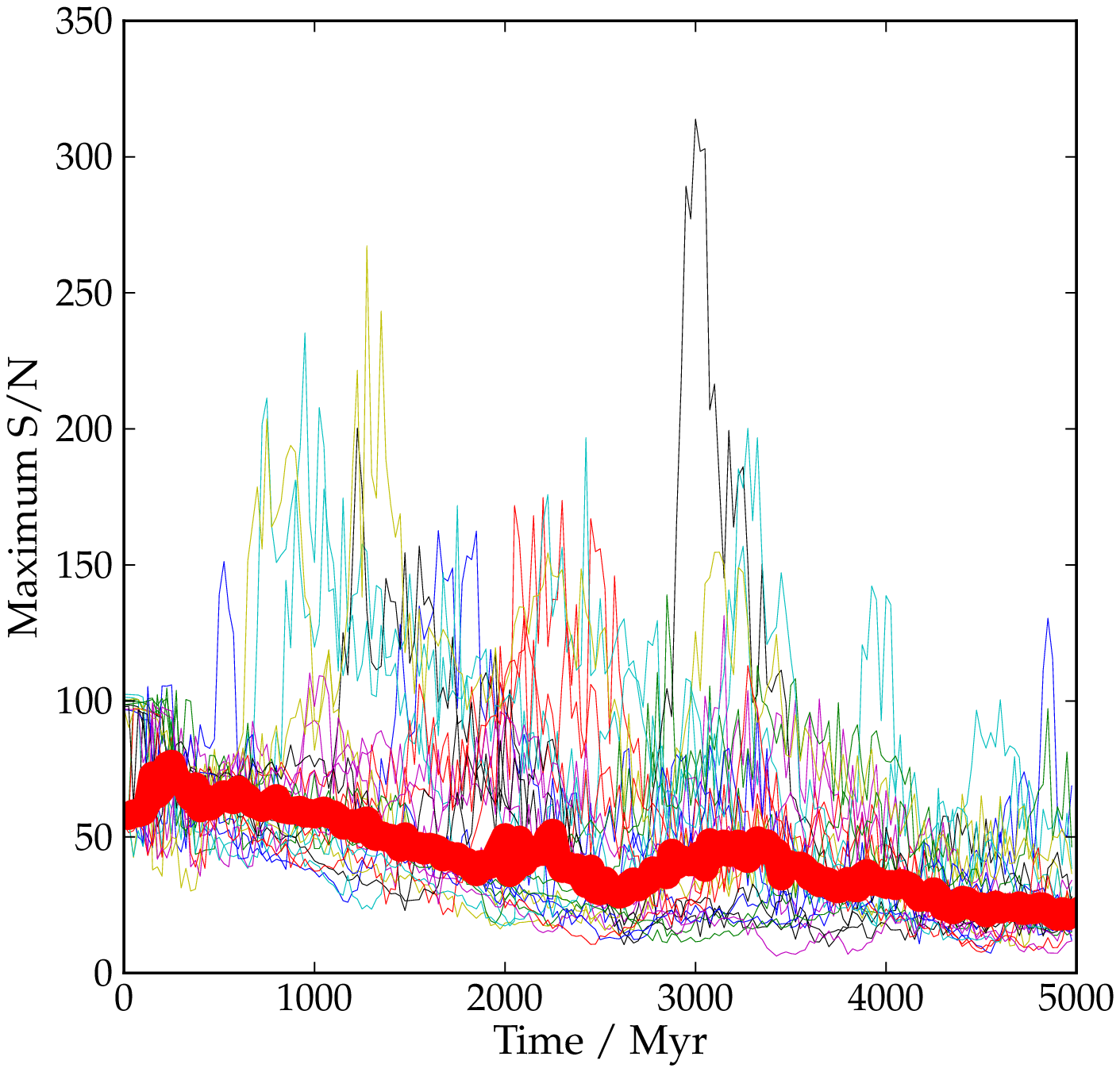}}
\caption[]{Evolution of the properties of 4$\times$10$^{8}$ \Msolar{}, 1500 K streams initially at 0.5 Mpc from the cluster centre. The top panel shows the measurements using an ALFALFA sensitivity level and beam size while the bottom panel shows the equivalent sensitivity of AGES. From left to right : detected mass, maximum $W4$ of any part of the stream, and peak SNR. Each simulation is shown using a different colour; the thick red line shows the median value of all 26 simulations.}
\label{fig:smallcoolstream0.5Mpc}
\end{figure*}

\begin{figure*}
\centering 
  \subfloat[]{\includegraphics[height=55mm]{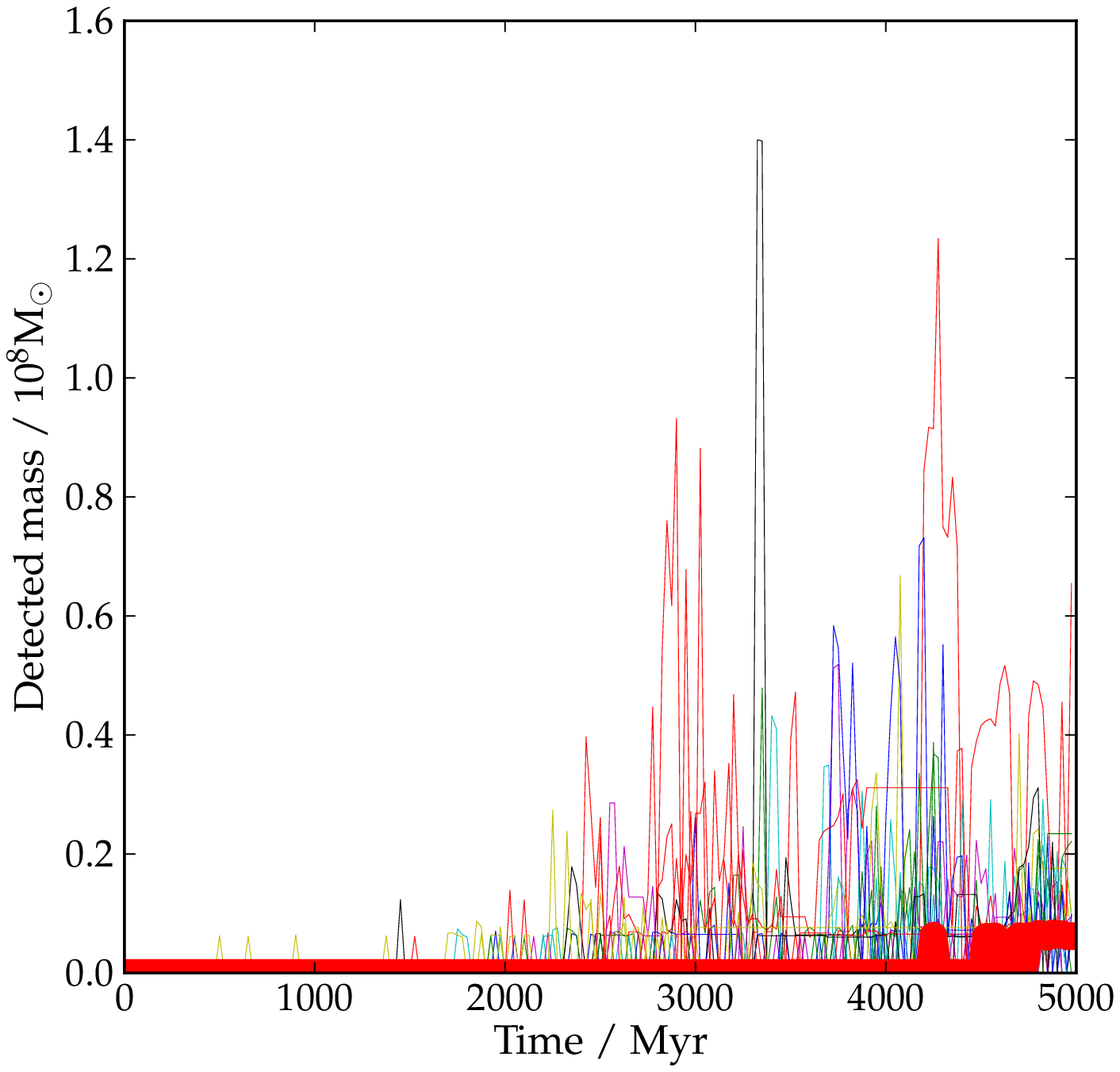}}
  \subfloat[]{\includegraphics[height=55mm]{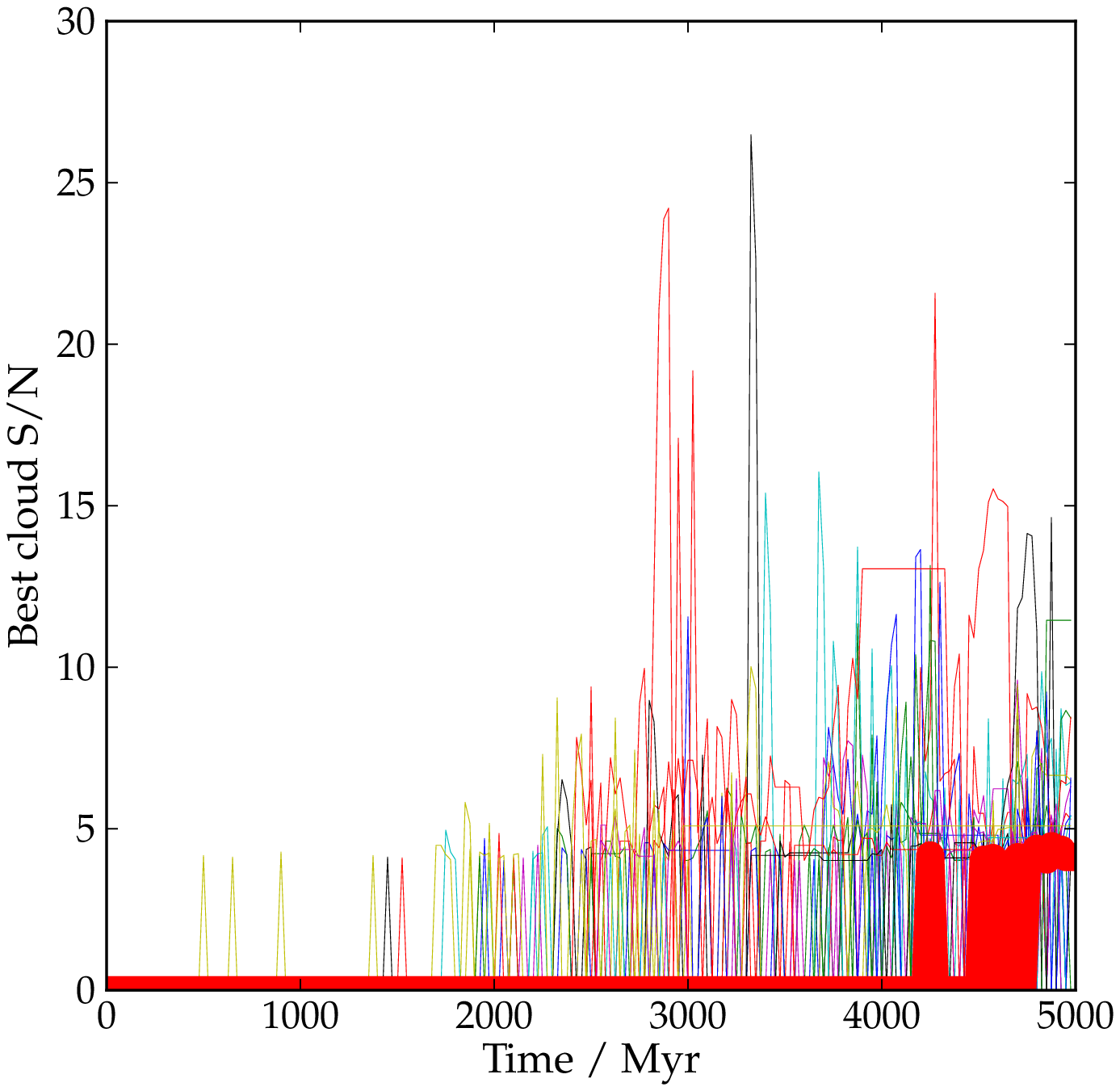}} 
  \subfloat[]{\includegraphics[height=55mm]{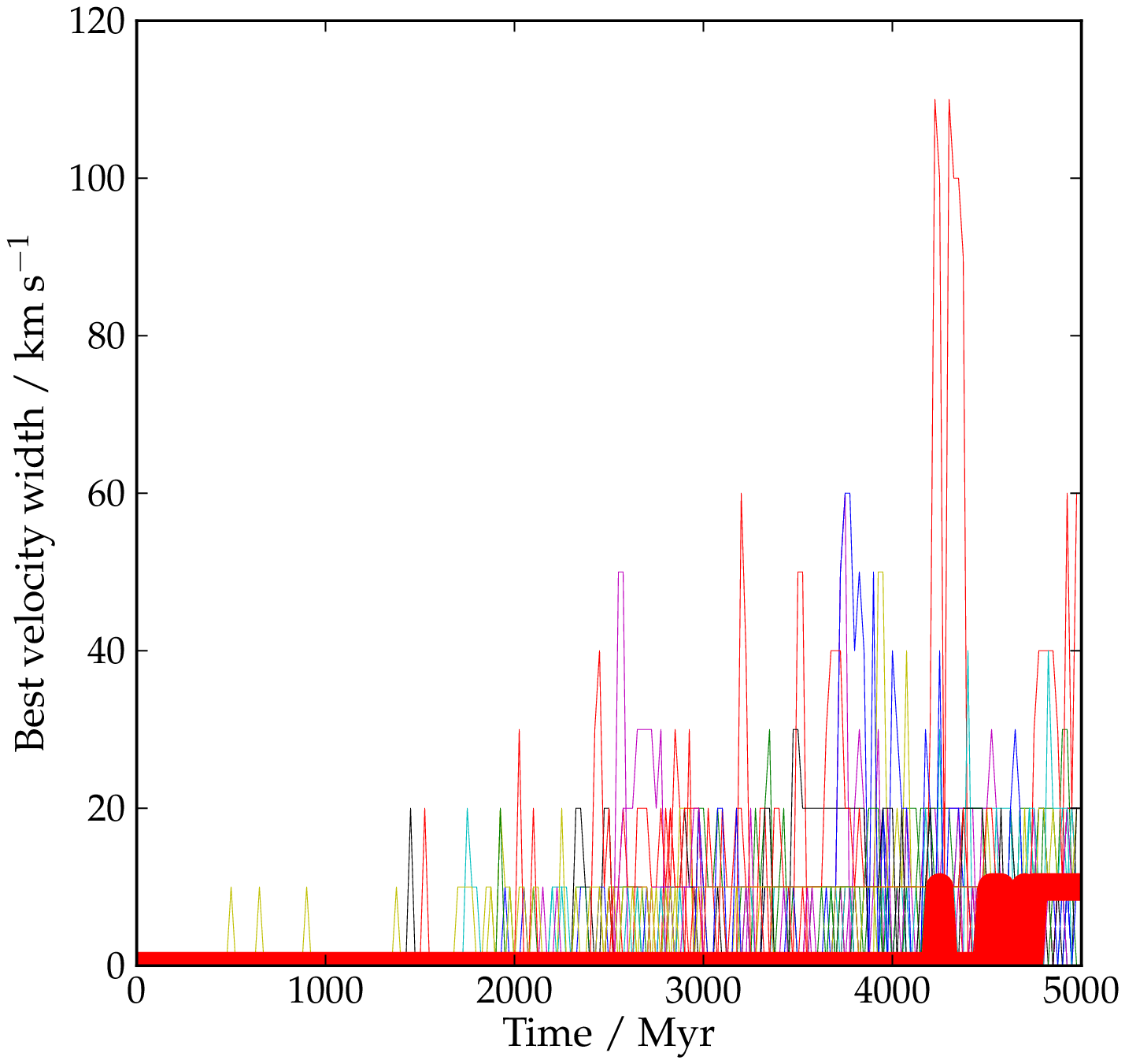}}\\
  \subfloat[]{\includegraphics[height=55mm]{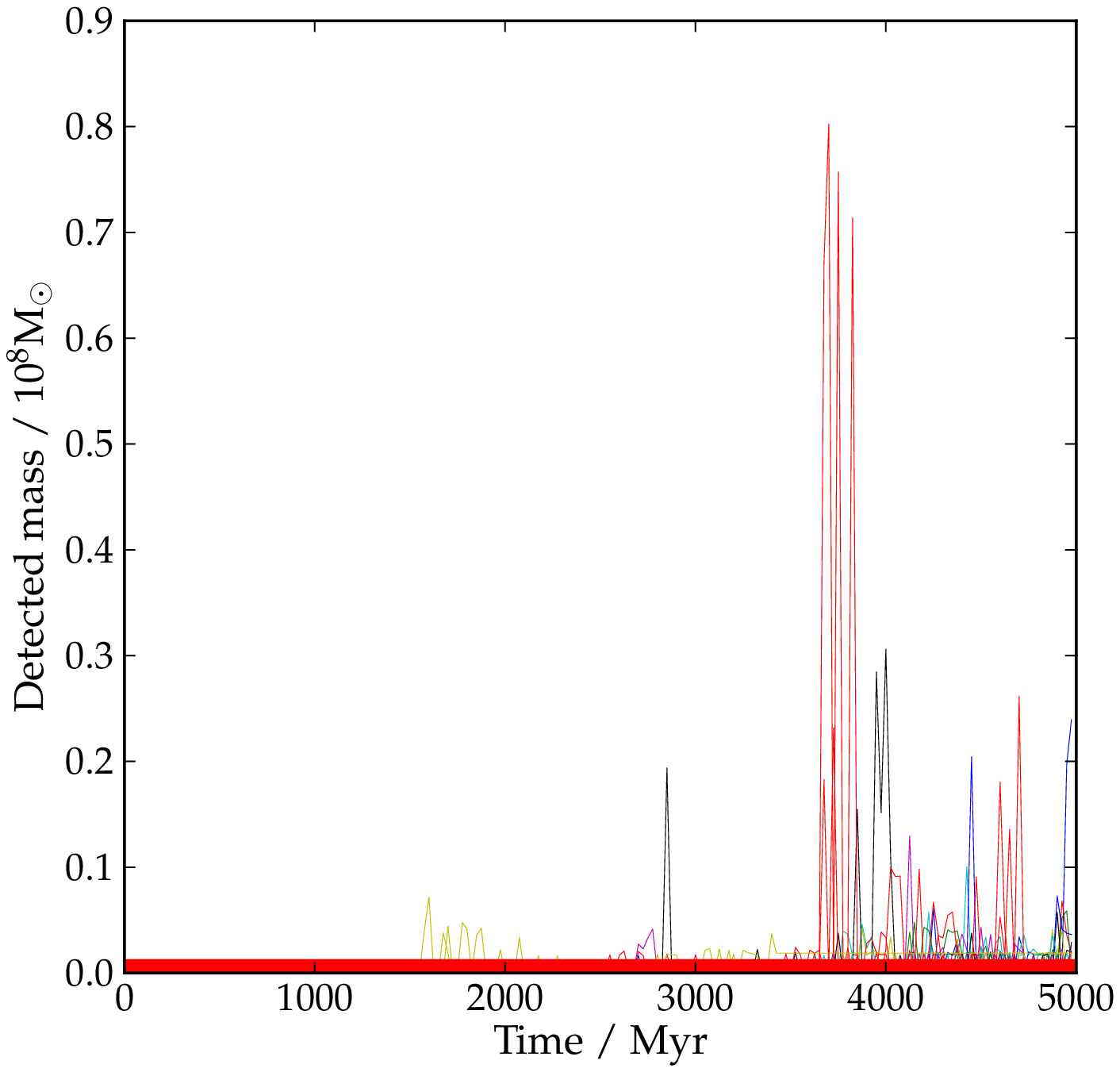}}
  \subfloat[]{\includegraphics[height=55mm]{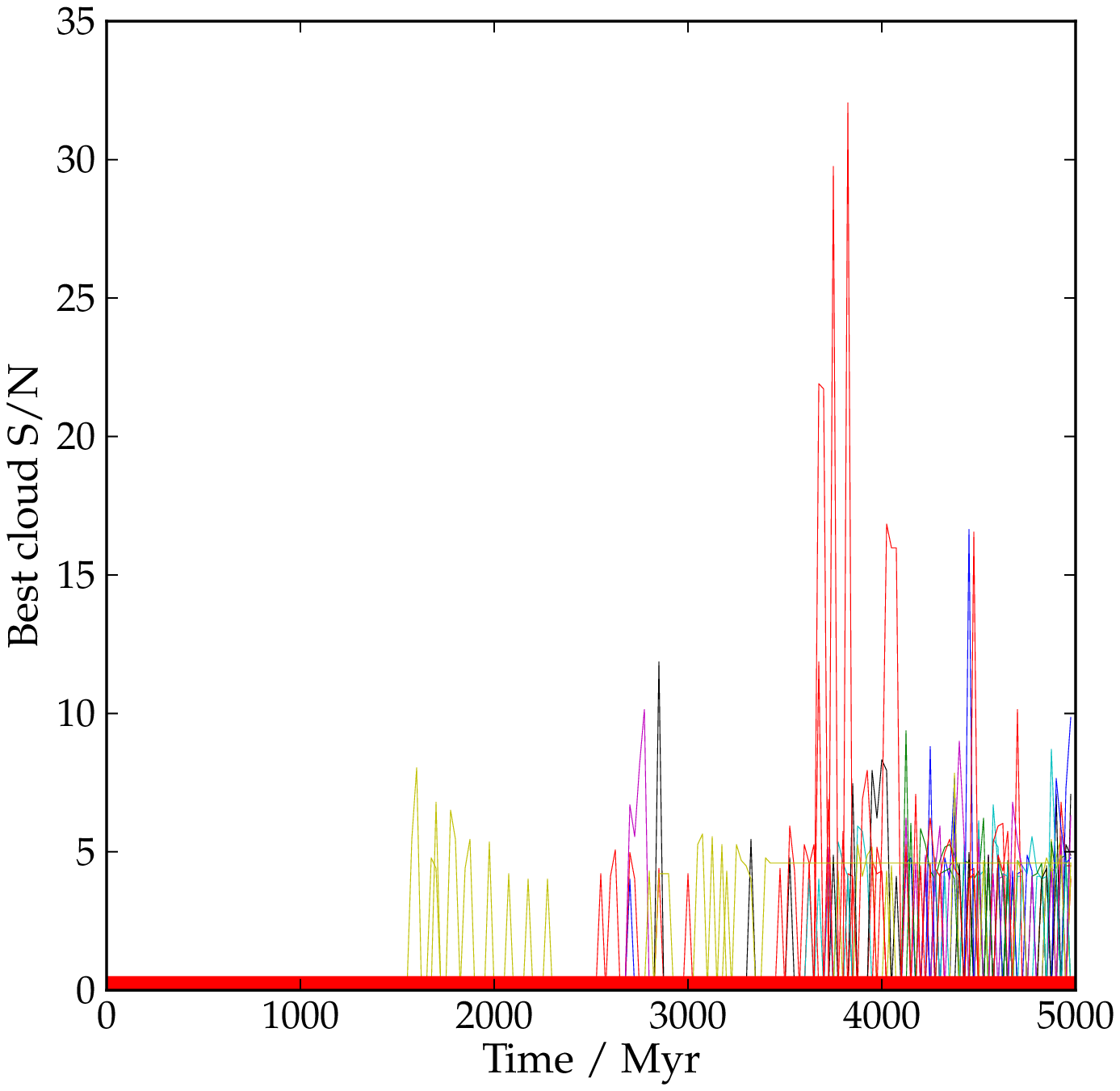}} 
  \subfloat[]{\includegraphics[height=55mm]{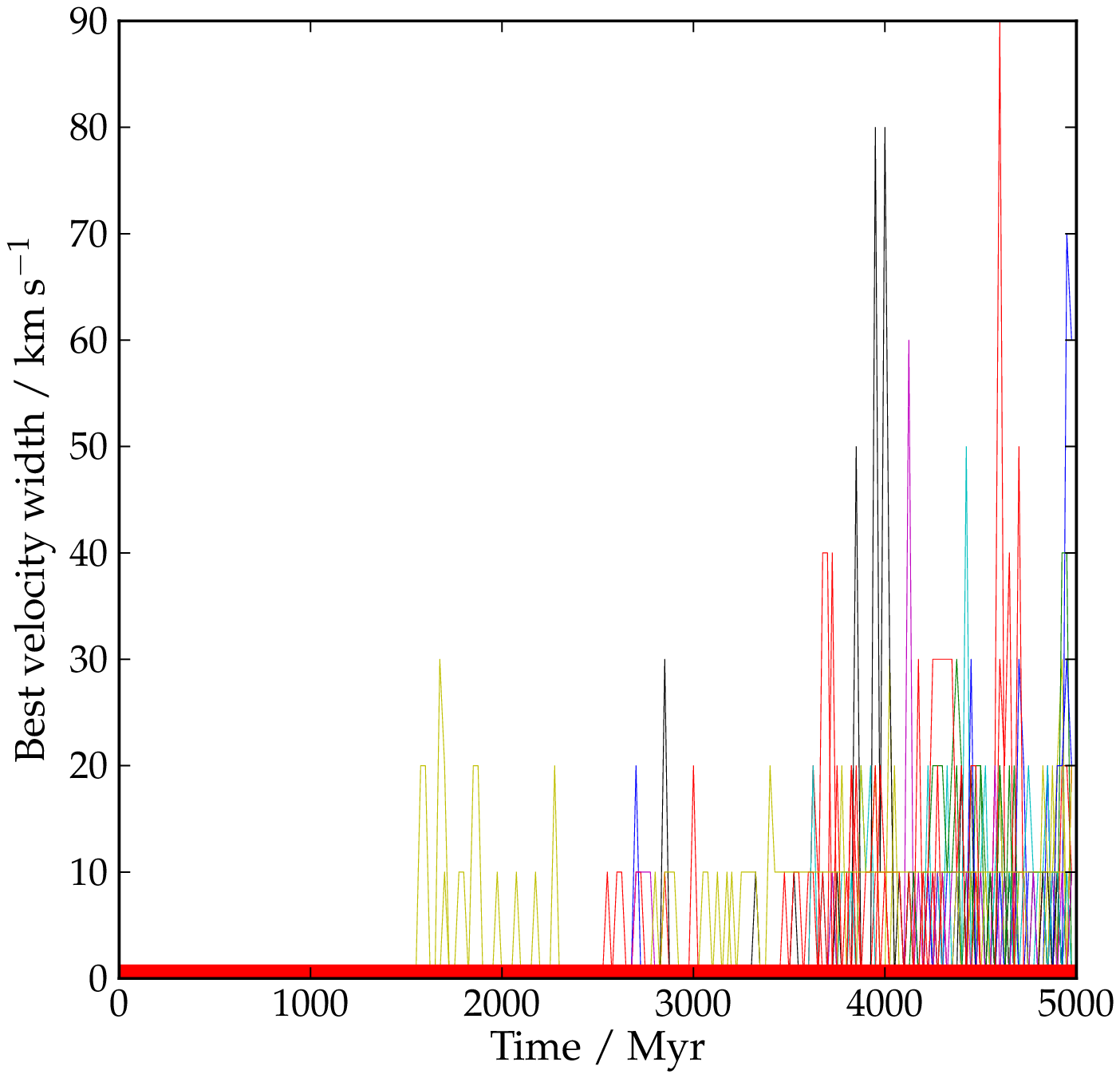}}
\caption[]{Evolution of the properties of the isolated cloud with the highest velocity width, produced from the 4$\times$10$^{8}$ \Msolar{}, 1500 K streams initially at 0.5 Mpc from the cluster centre. The top panel shows the measurements using an ALFALFA sensitivity level and beam size while the bottom panel shows the equivalent sensitivity of AGES. From left to right : detected mass, peak SNR, and $W50$. Each simulation is shown using a different colour; the thick red line shows the median value of all 26 simulations.}
\label{fig:smallcoolstreamclouds0.5Mpc}
\end{figure*}

\begin{figure*}
\centering
\includegraphics[width=160mm]{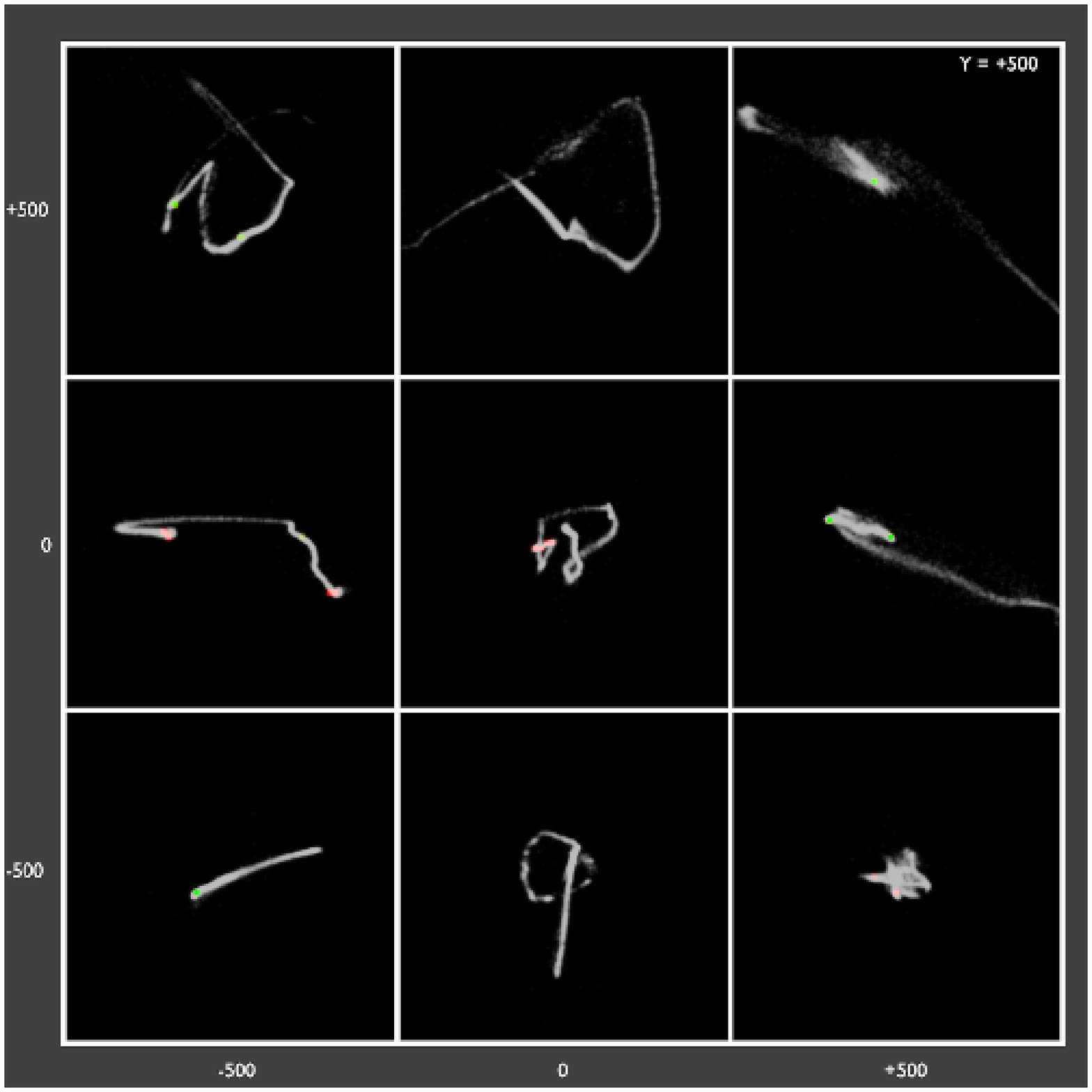}
\caption[hienv]{Final timestep (5 Gyr) of the simulation of a sample of 4$\times$10$^{8}$ \Msolar{}, 1500 K streams entering the cluster from an initial distance of 0.5 Mpc. Each box spans 1 Mpc and is centred on the mean particle position. White shows the raw particle data. Red shows all gridded data in which the emission would exceed a SNR of 4.0 with an ALFALFA sensitivity level; green indicates detectable clouds at least 100 kpc from the nearest other detection. Movies of the simulations can be see at \href{http://tinyurl.com/zu673ky}{this url} : http://tinyurl.com/zu673ky.}
\label{fig:lmcmovien}
\end{figure*}

\begin{figure*}
\centering
  \subfloat[]{\includegraphics[height=55mm]{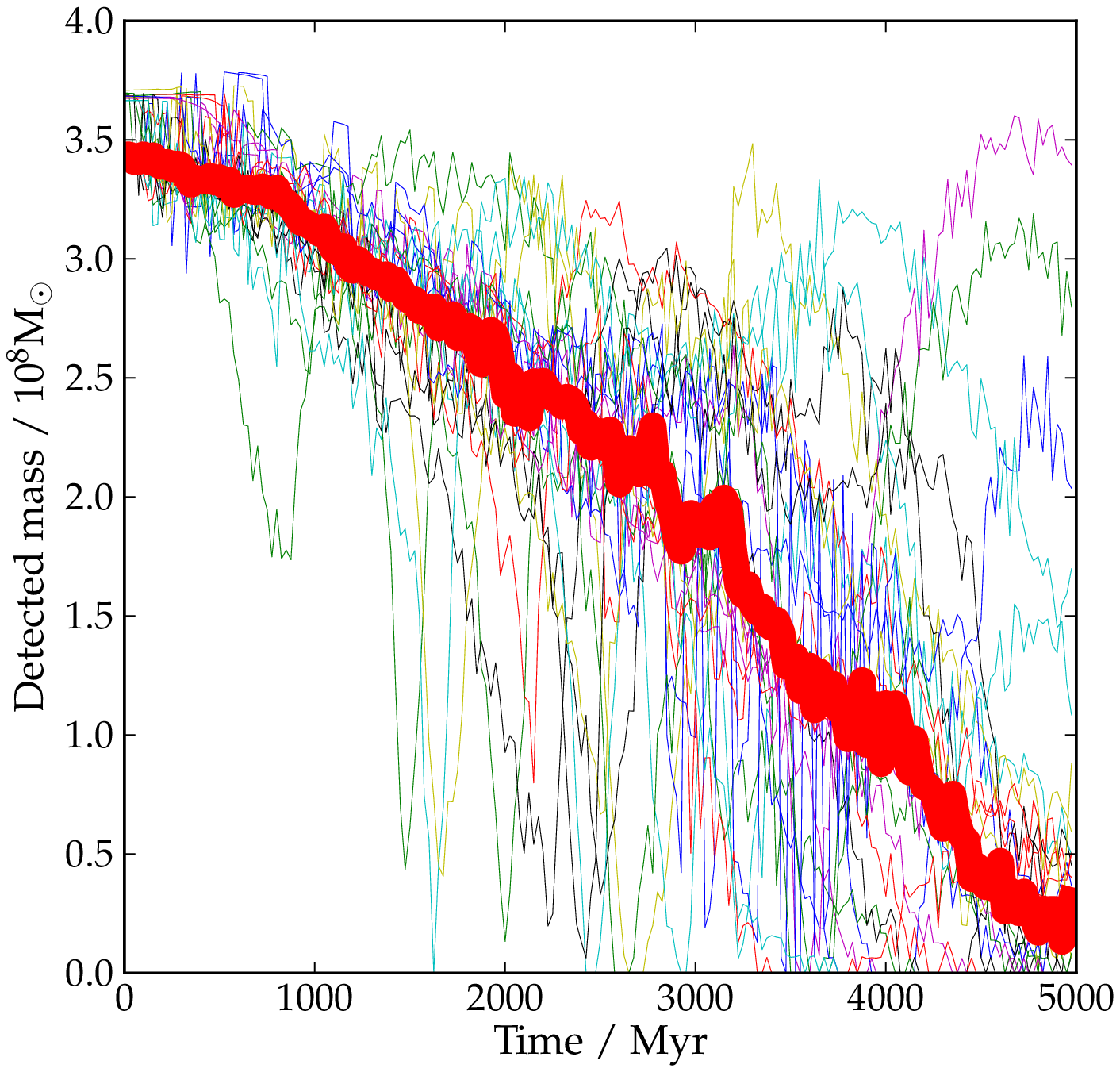}}
  \subfloat[]{\includegraphics[height=55mm]{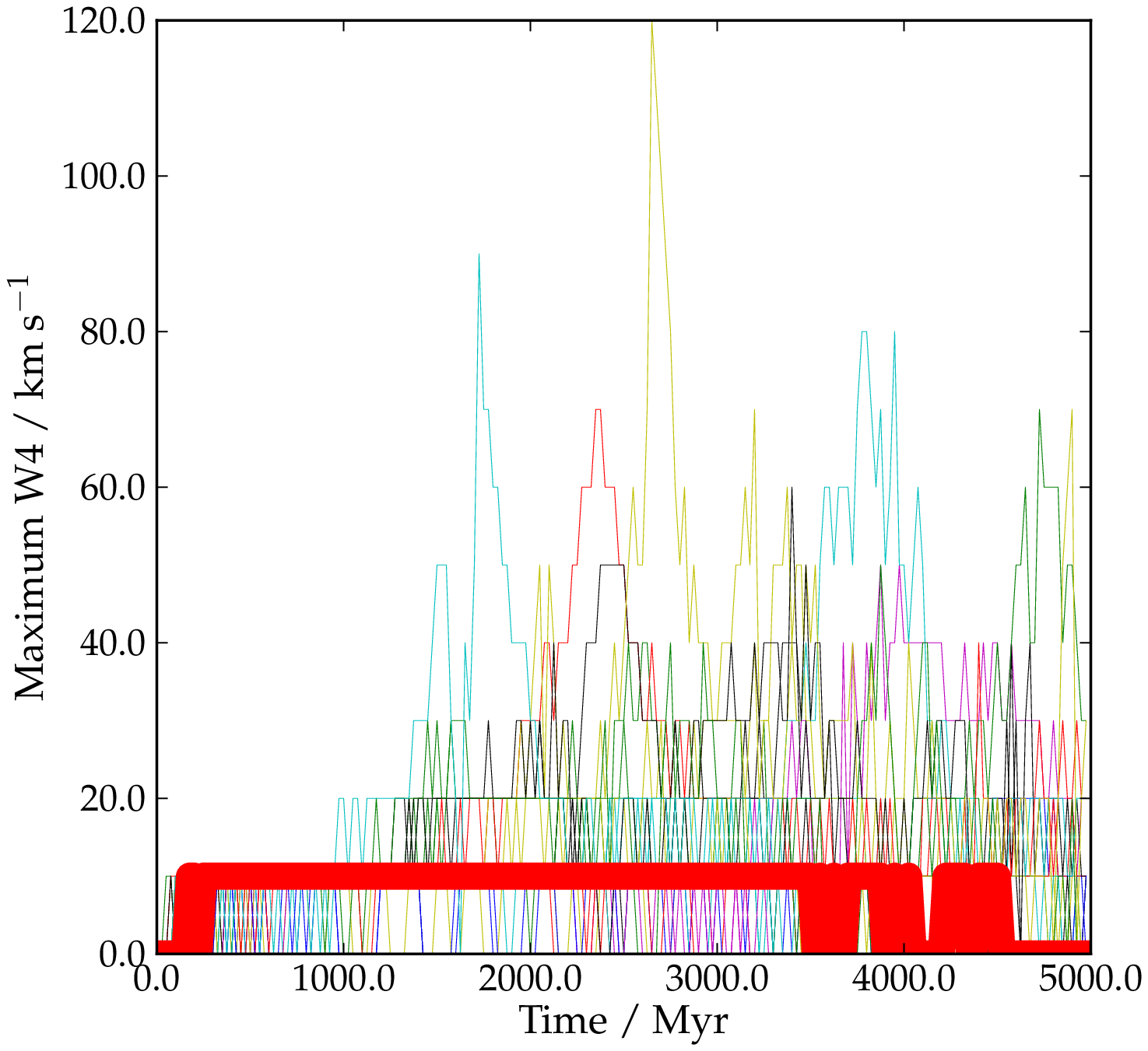}} 
  \subfloat[]{\includegraphics[height=55mm]{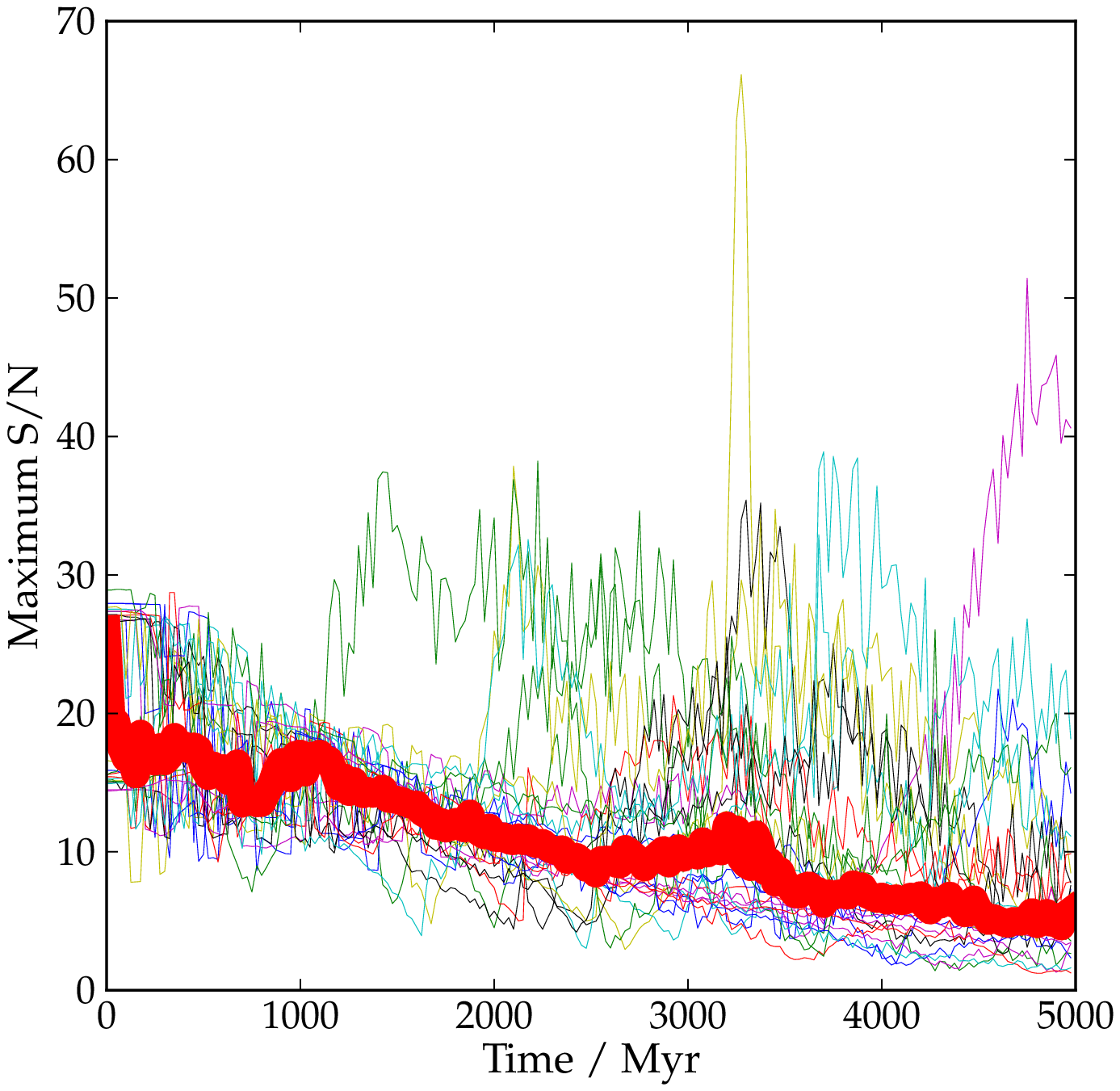}}\\
  \subfloat[]{\includegraphics[height=55mm]{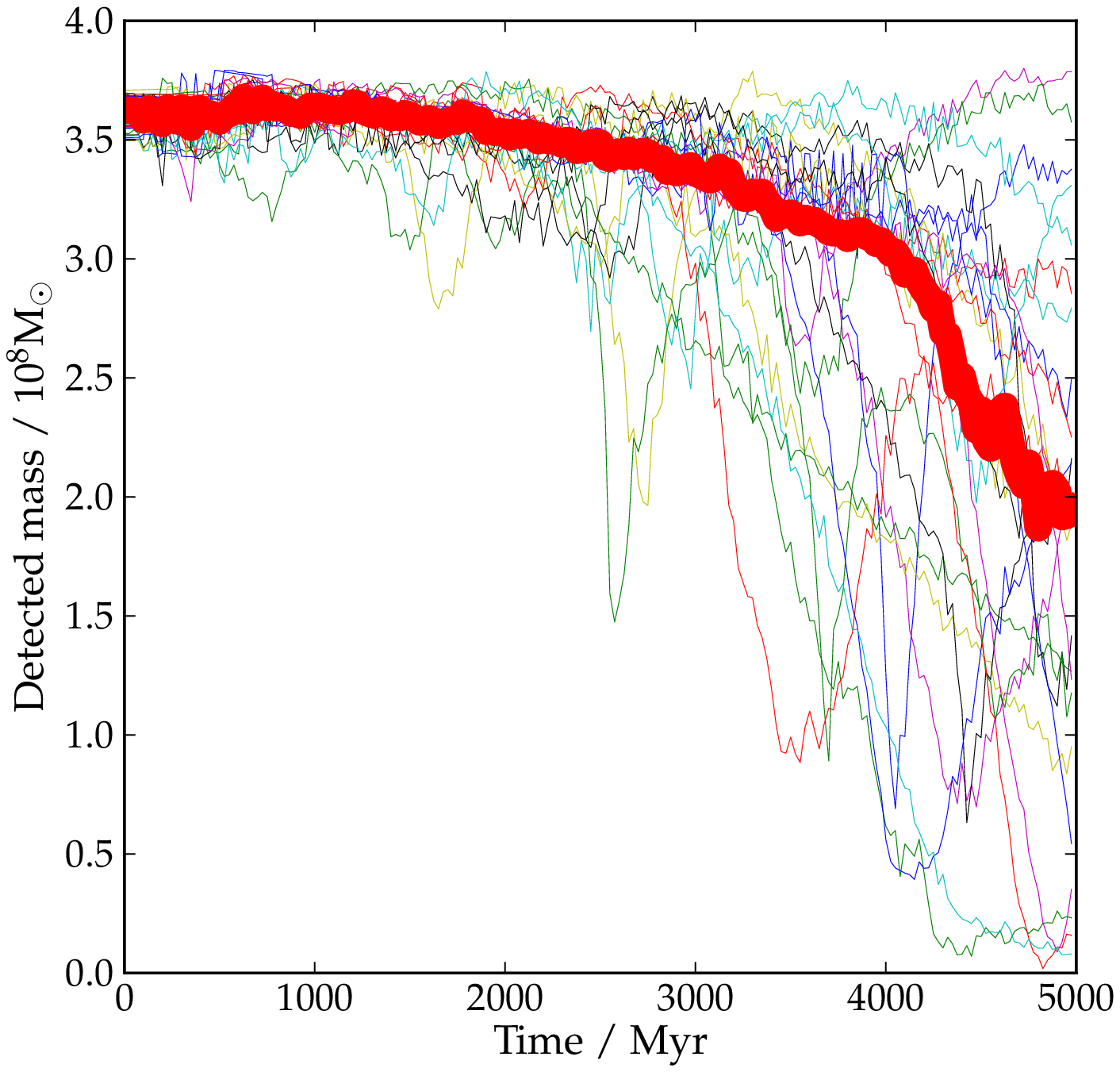}}
  \subfloat[]{\includegraphics[height=55mm]{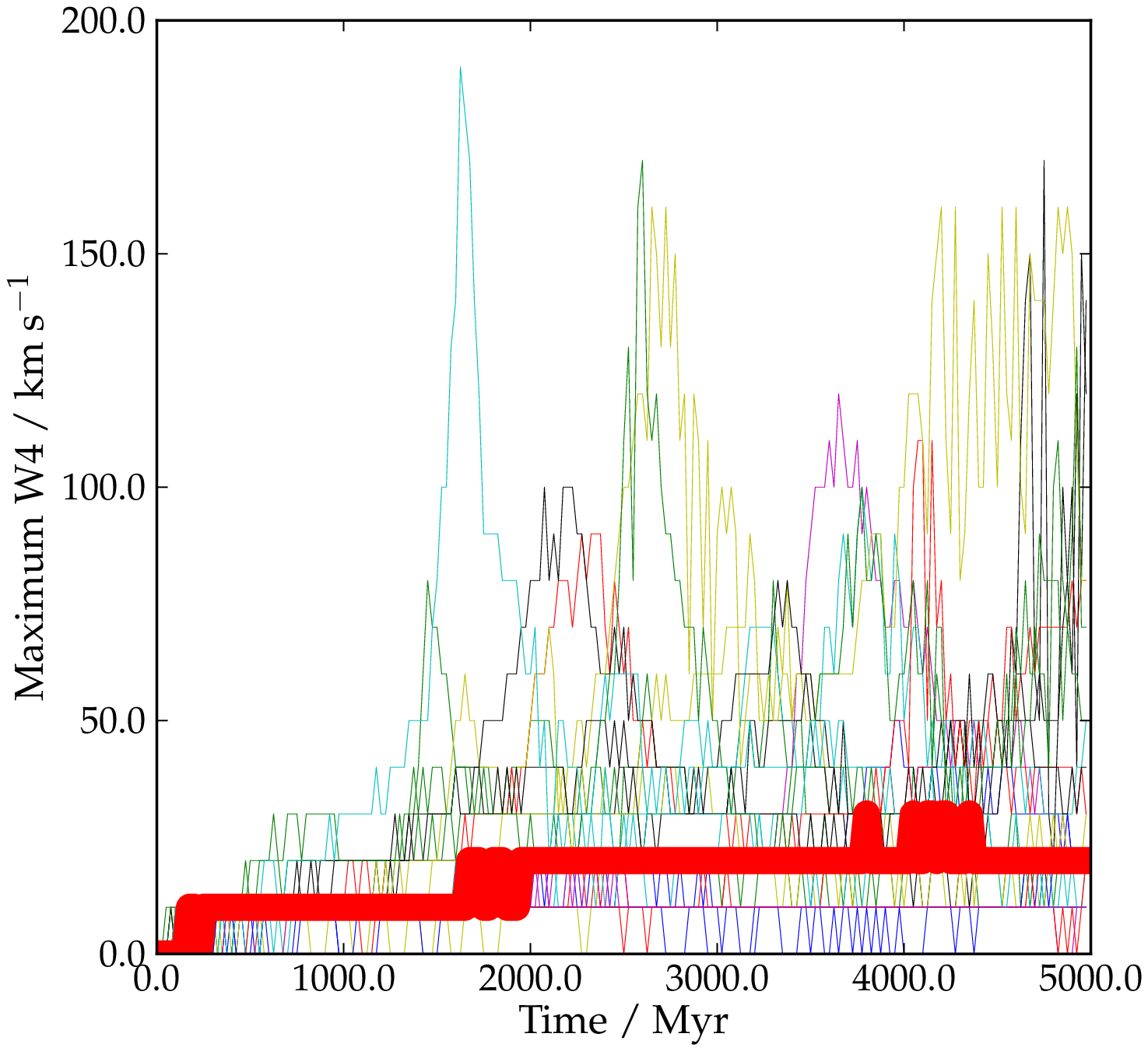}} 
  \subfloat[]{\includegraphics[height=55mm]{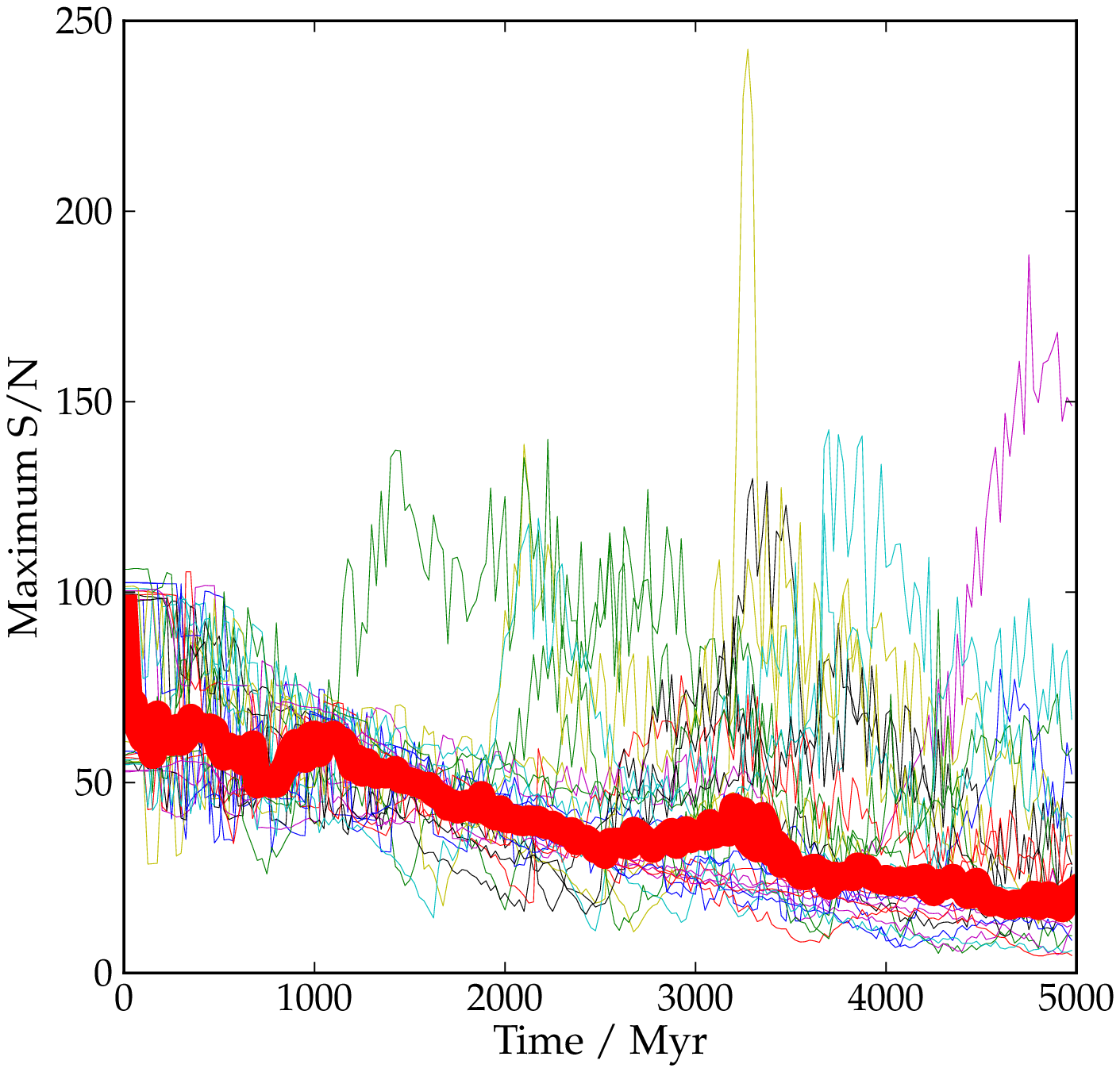}}
\caption[]{Evolution of the properties of 4$\times$10$^{8}$ \Msolar{}, 1500 K streams initially at 1.0 Mpc from the cluster centre. The top panel shows the measurements using an ALFALFA sensitivity level and beam size while the bottom panel shows the equivalent sensitivity of AGES. From left to right : detected mass, maximum $W4$ of any part of the stream, and peak SNR. Each simulation is shown using a different colour; the thick red line shows the median value of all 26 simulations.}
\label{fig:smallcoolstream1.0Mpc}
\end{figure*}

\begin{figure*}
\centering  
  \subfloat[]{\includegraphics[height=55mm]{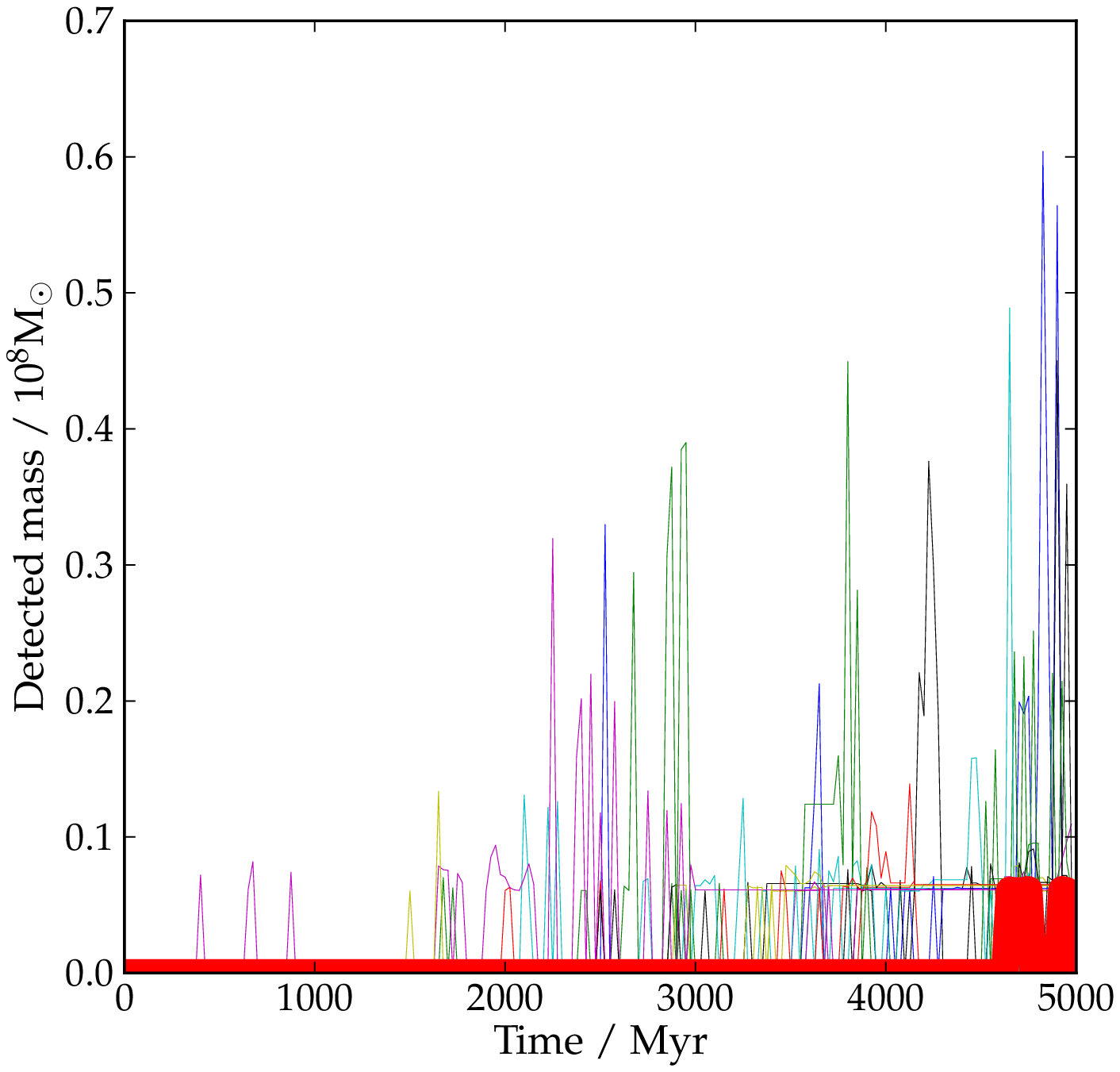}}
  \subfloat[]{\includegraphics[height=55mm]{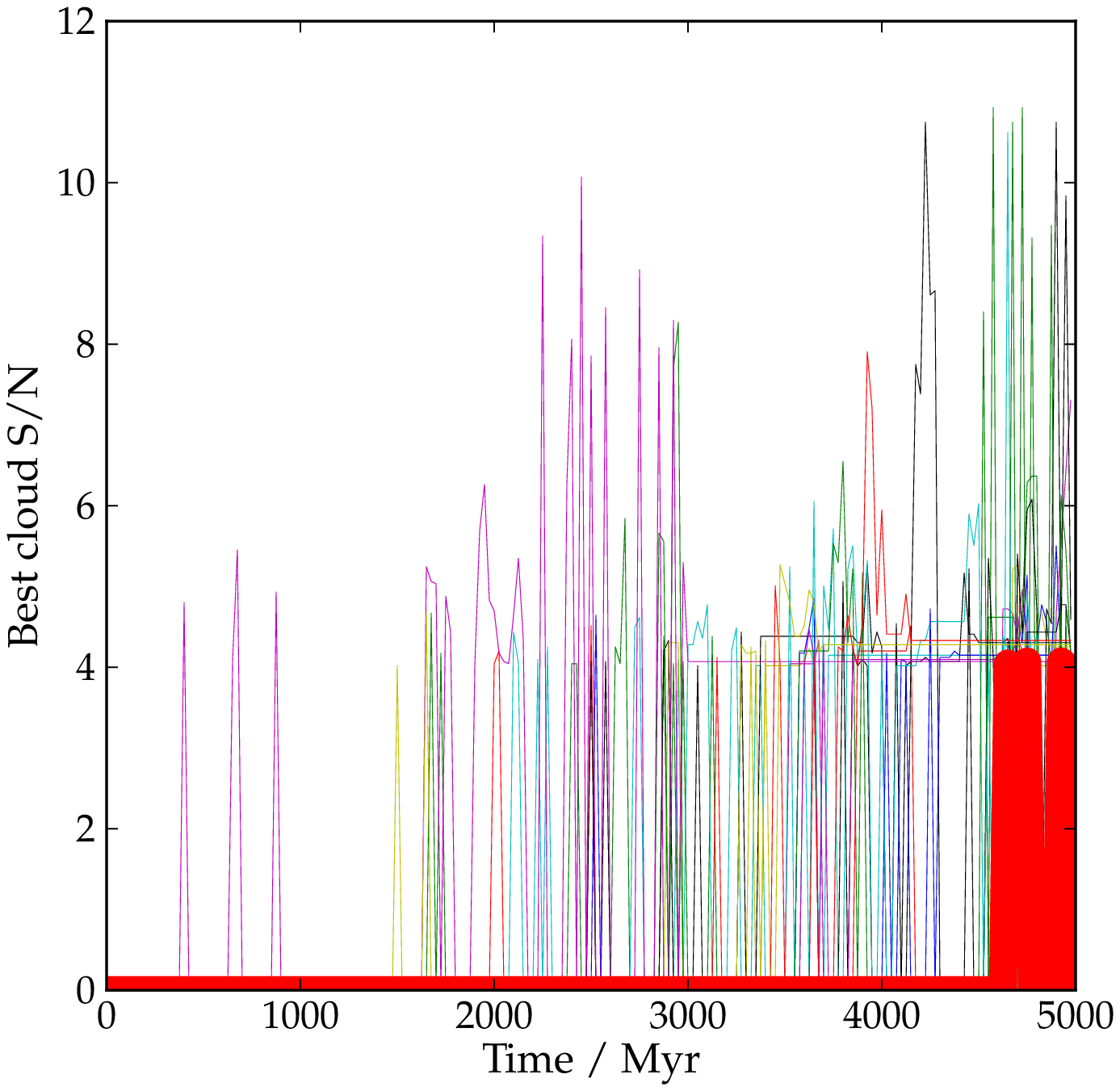}} 
  \subfloat[]{\includegraphics[height=55mm]{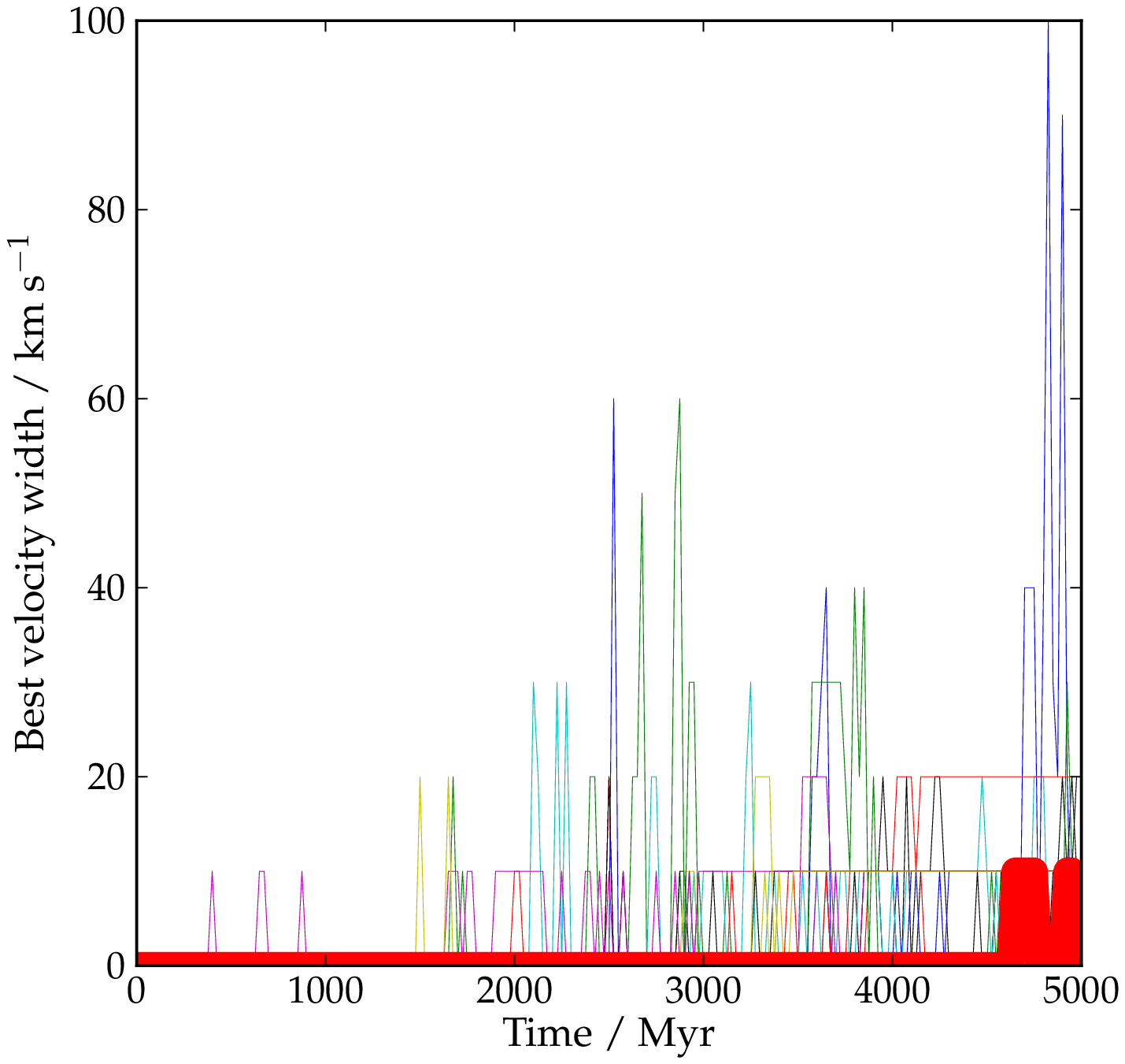}}\\
  \subfloat[]{\includegraphics[height=55mm]{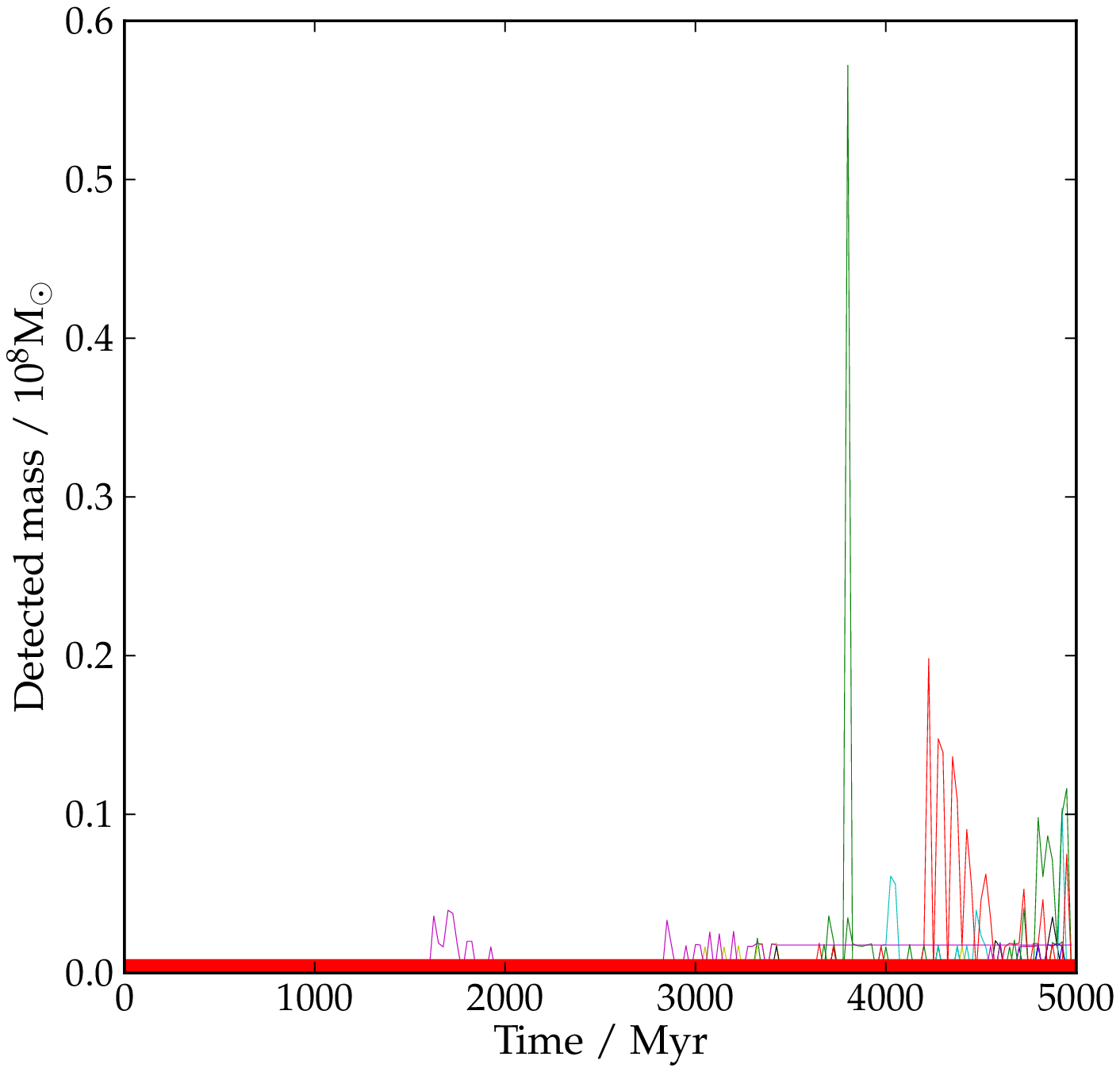}}
  \subfloat[]{\includegraphics[height=55mm]{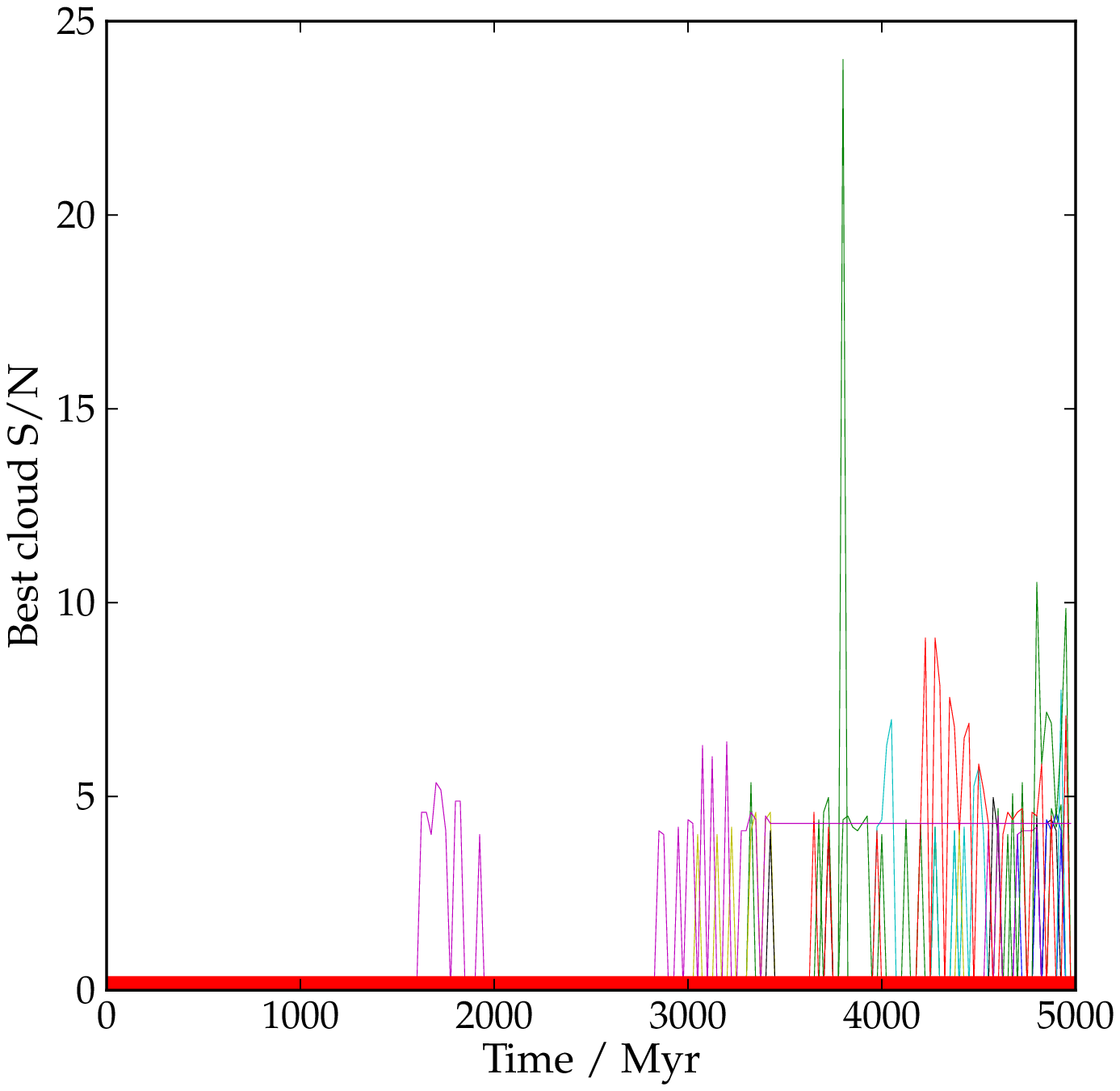}} 
  \subfloat[]{\includegraphics[height=55mm]{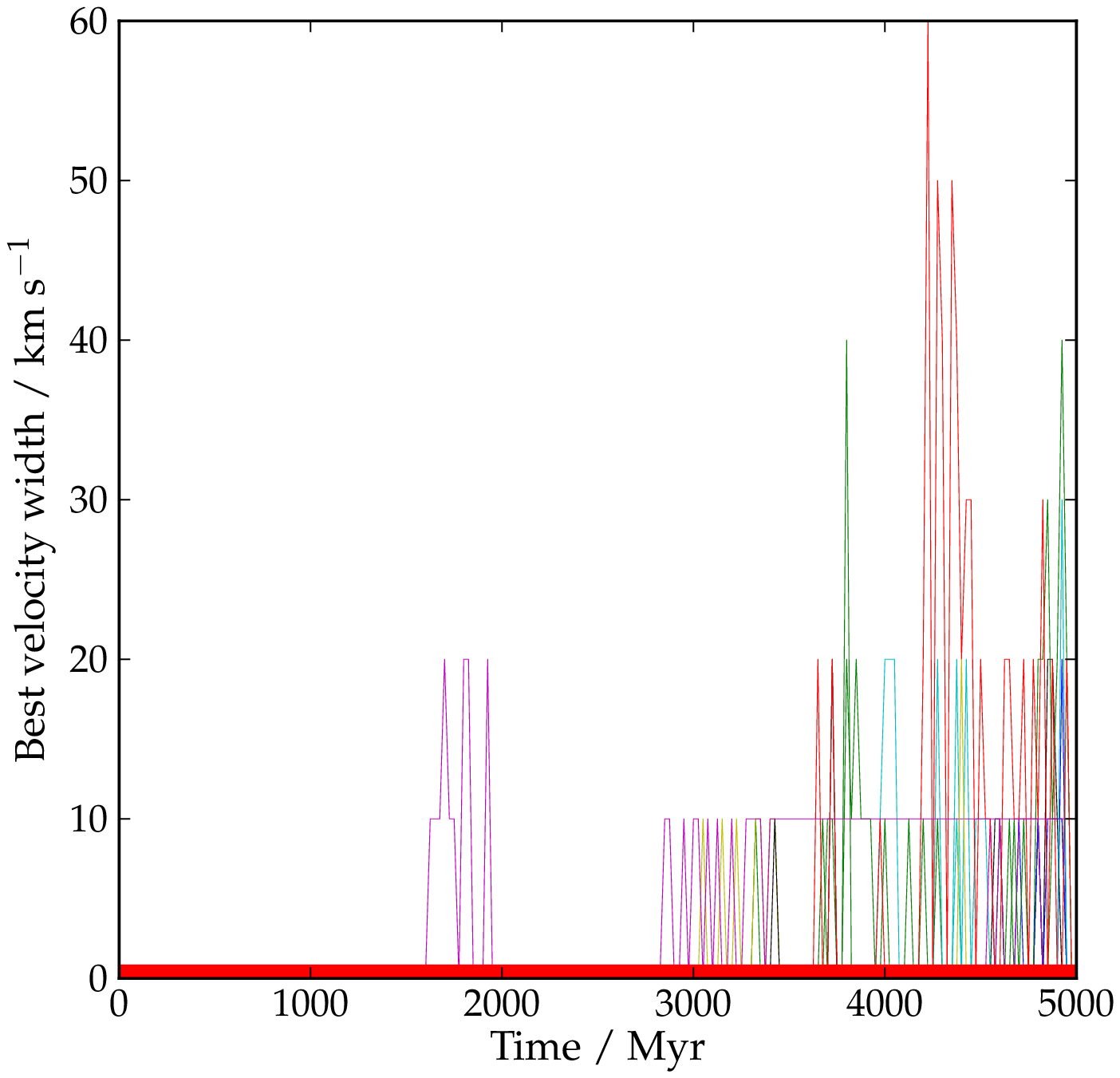}}
\caption[]{Evolution of the properties of the isolated cloud with the highest velocity width, produced from the 4$\times$10$^{8}$ \Msolar{}, 1500 K streams initially at 1.0 Mpc from the cluster centre. The top panel shows the measurements using an ALFALFA sensitivity level and beam size while the bottom panel shows the equivalent sensitivity of AGES. From left to right : detected mass, peak SNR, and $W50$. Each simulation is shown using a different colour; the thick red line shows the median value of all 26 simulations.}
\label{fig:smallcoolstreamclouds1.0Mpc}
\end{figure*}

\begin{figure*}
\centering
\includegraphics[width=160mm]{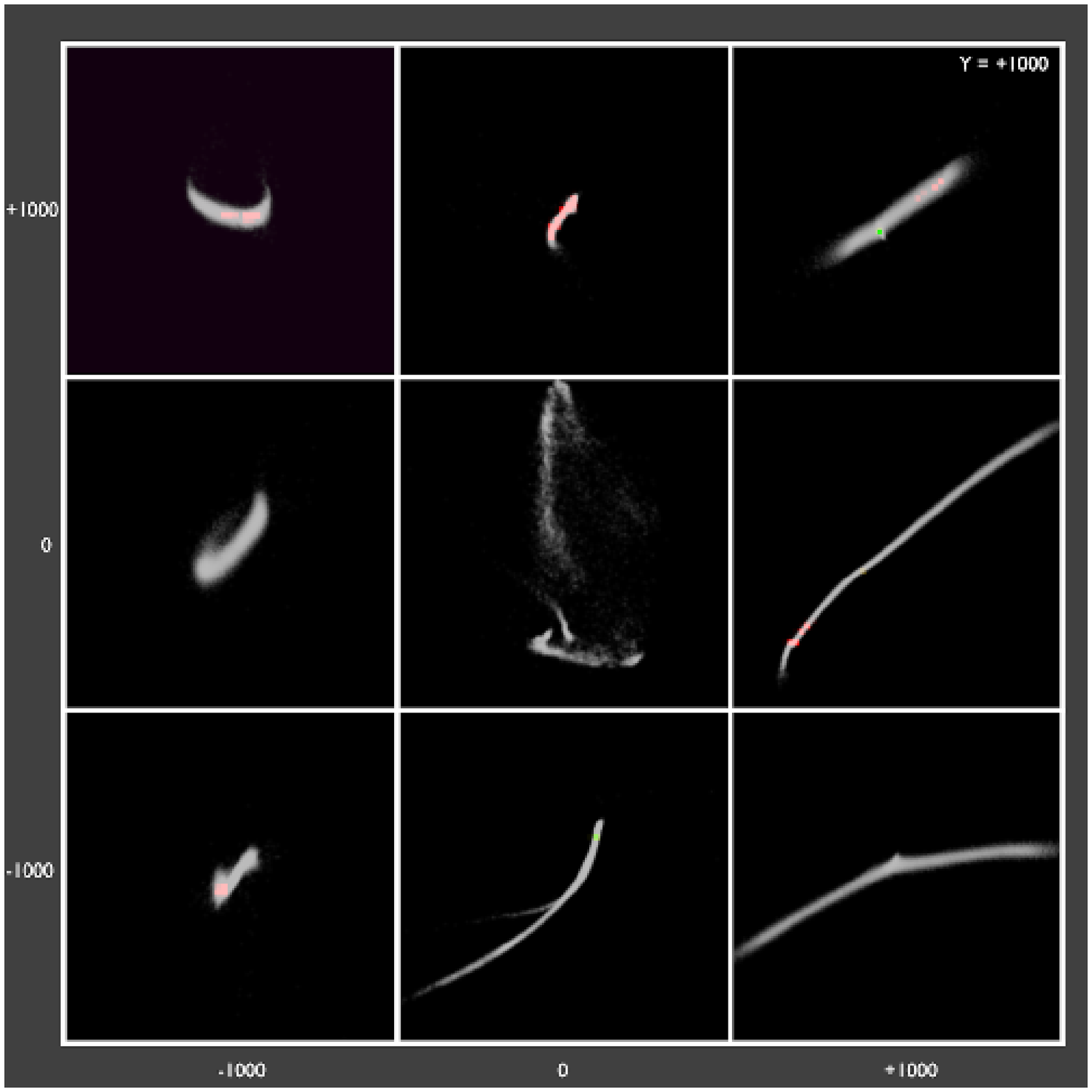}
\caption[hienv]{Final timestep (5 Gyr) of the simulation of a sample of 4$\times$10$^{8}$ \Msolar{}, 1500 K streams entering the cluster from an initial distance of 1.0 Mpc. Each box spans 1 Mpc and is centred on the mean particle position. White shows the raw particle data. Red shows all gridded data in which the emission would exceed a SNR of 4.0 with an ALFALFA sensitivity level; green indicates detectable clouds at least 100 kpc from the nearest other detection. Movies of the simulations can be see at \href{http://tinyurl.com/jbjtvtb}{this url} : http://tinyurl.com/jbjtvtb.}
\label{fig:lmcmovied}
\end{figure*}

\begin{figure*}
\centering 
  \subfloat[]{\includegraphics[height=55mm]{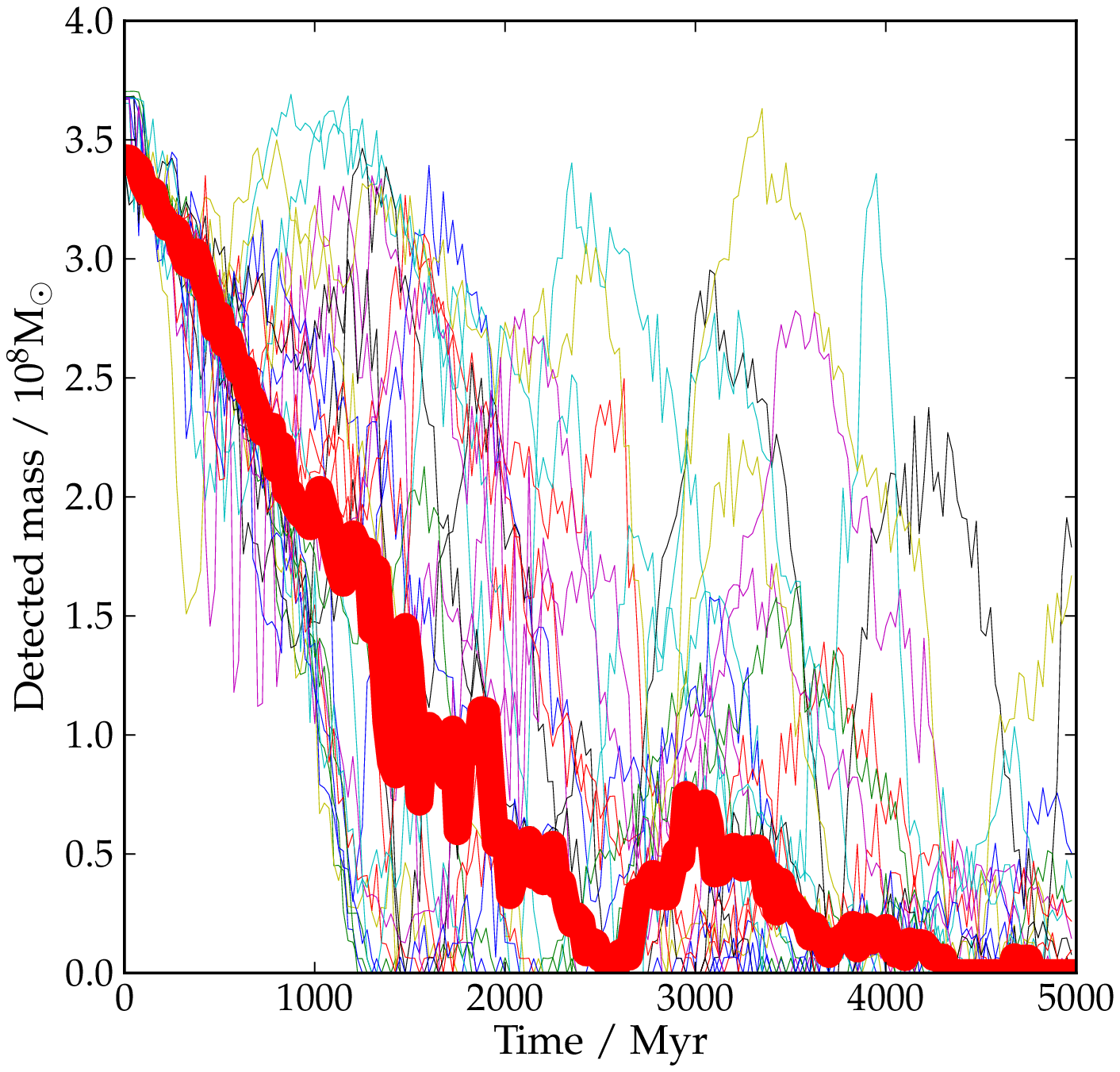}}
  \subfloat[]{\includegraphics[height=55mm]{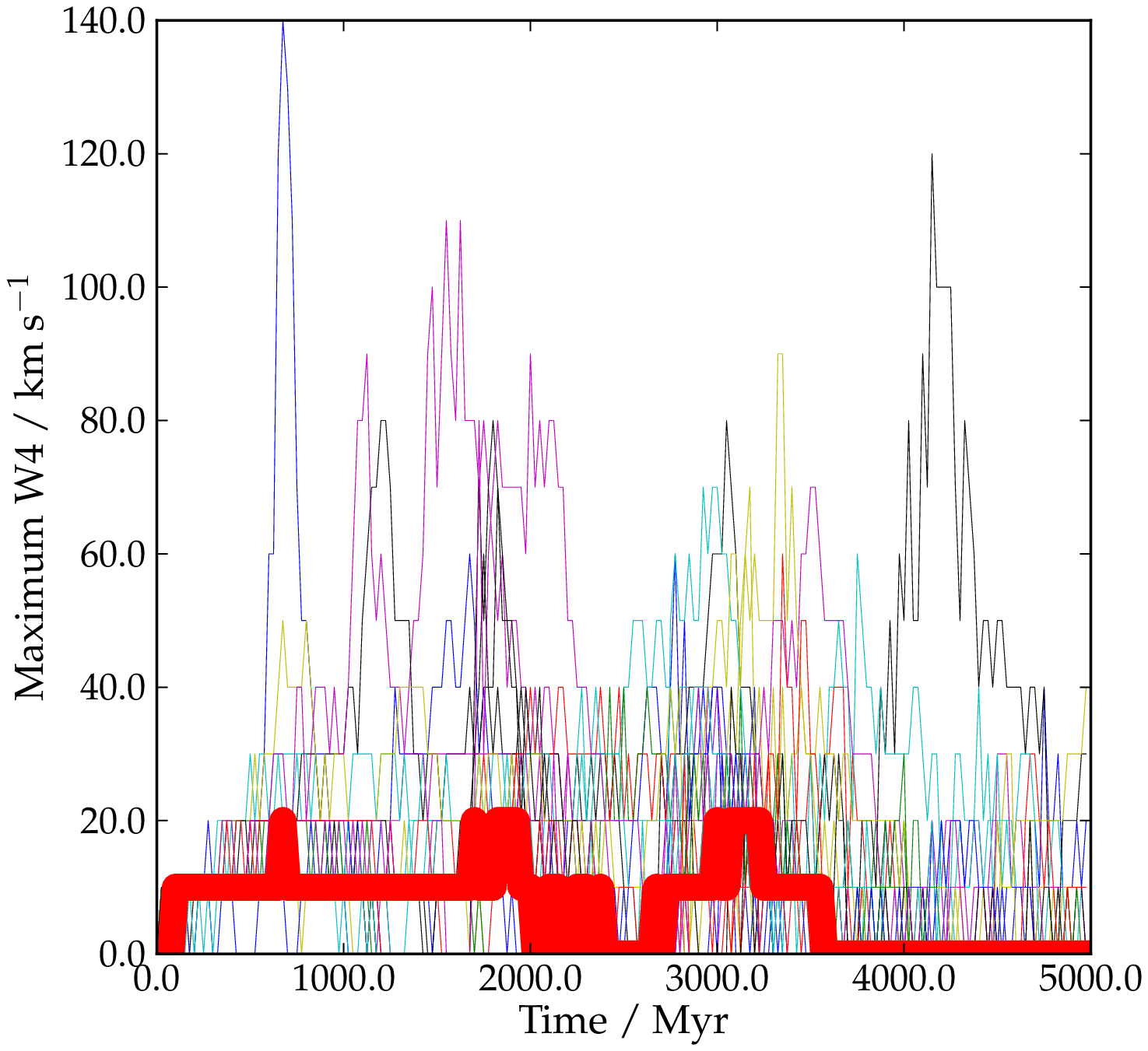}} 
  \subfloat[]{\includegraphics[height=55mm]{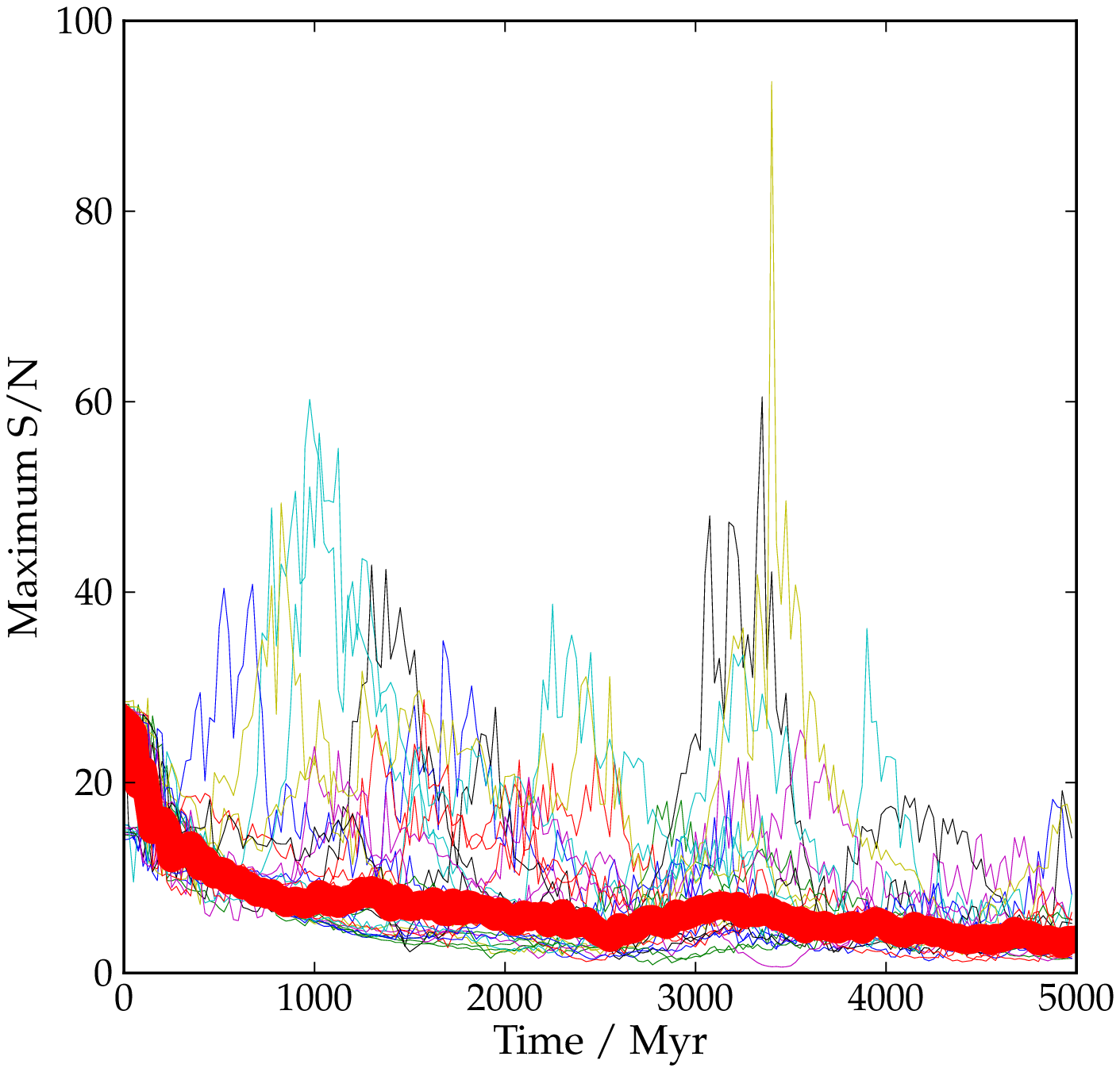}}\\
  \subfloat[]{\includegraphics[height=55mm]{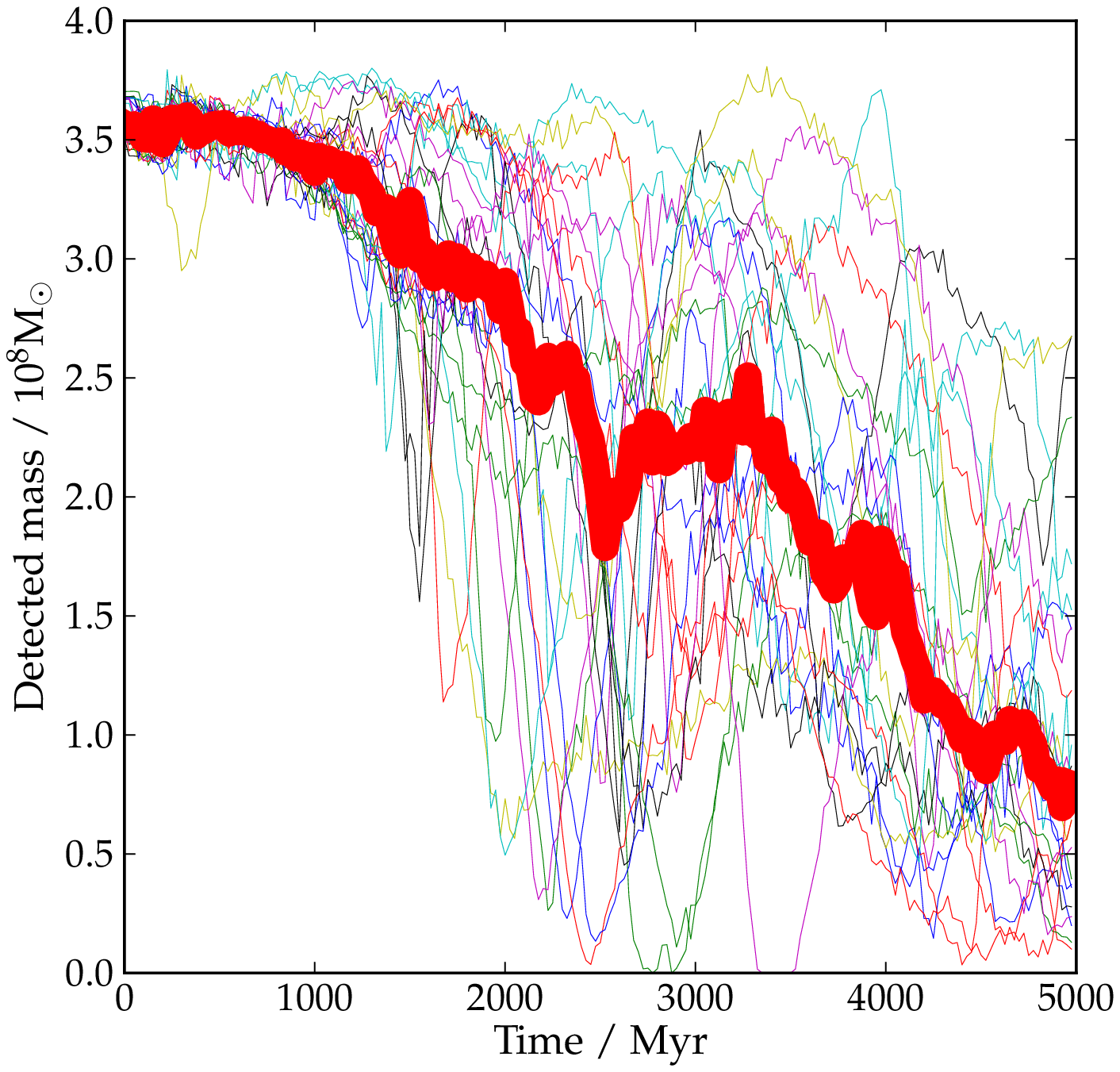}}
  \subfloat[]{\includegraphics[height=55mm]{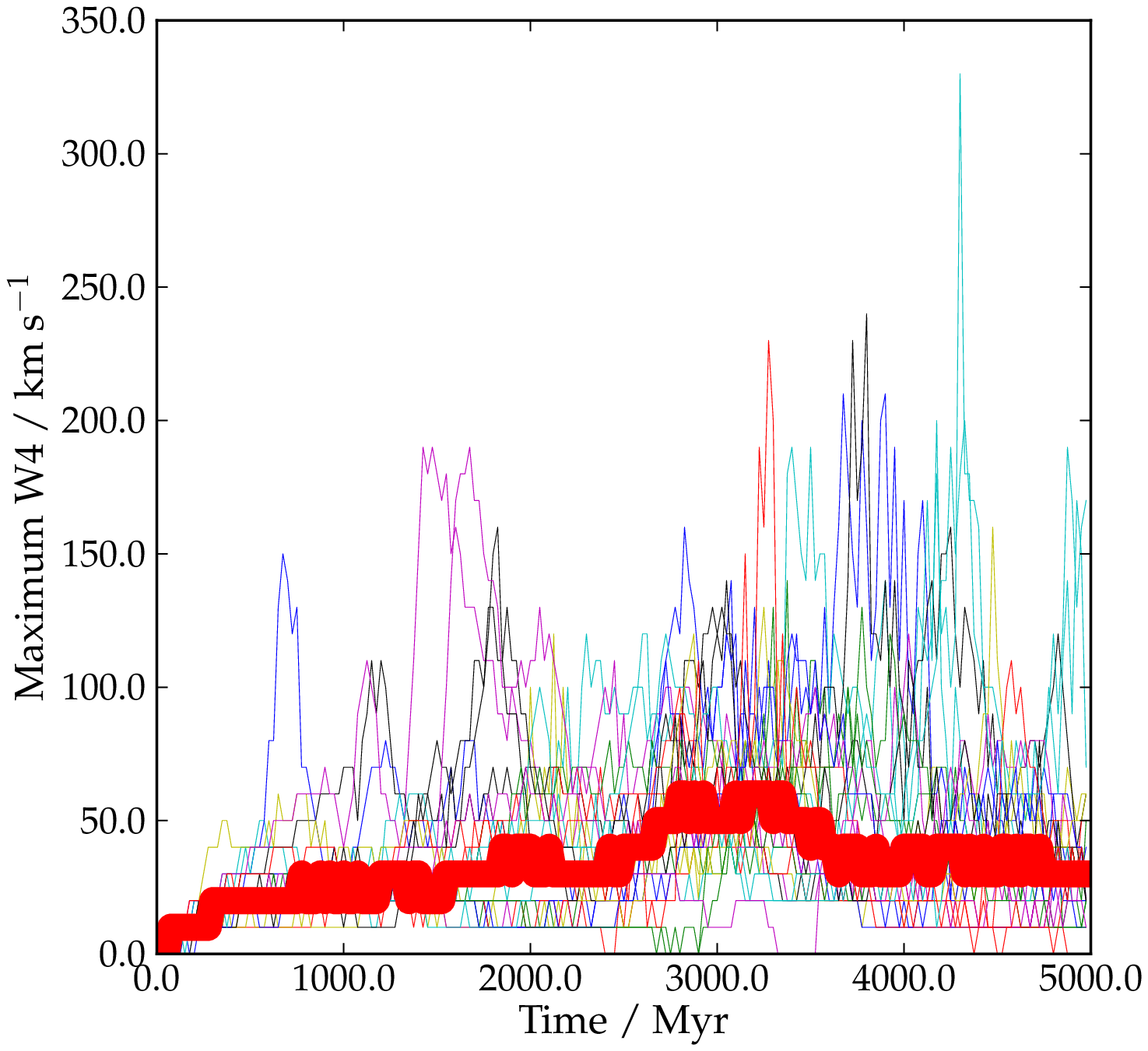}} 
  \subfloat[]{\includegraphics[height=55mm]{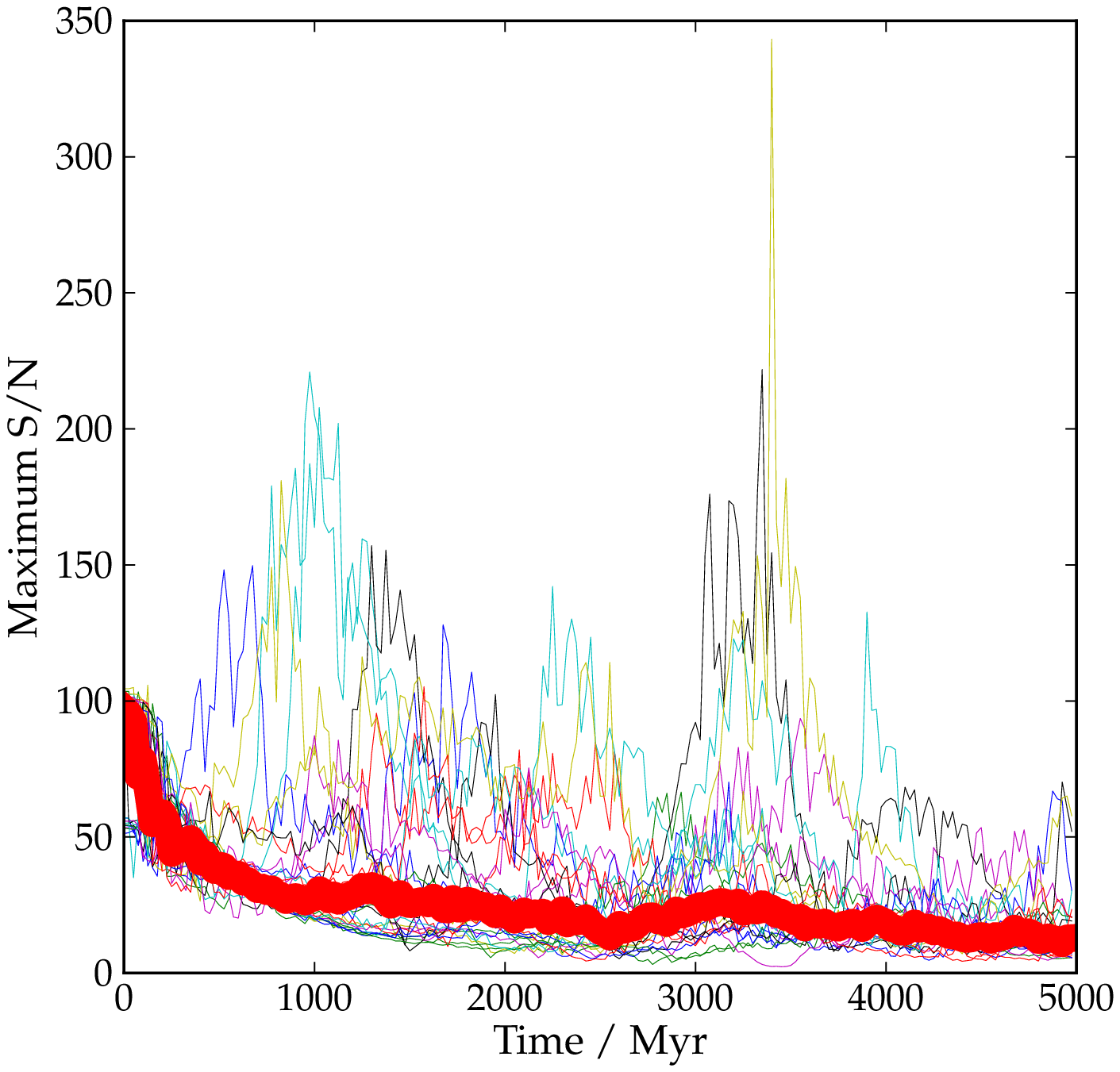}}
\caption[]{Evolution of the properties of 4$\times$10$^{8}$ \Msolar{}, 5100 K streams initially at 0.5 Mpc from the cluster centre. The top panel shows the measurements using an ALFALFA sensitivity level and beam size while the bottom panel shows the equivalent sensitivity of AGES. From left to right : detected mass, maximum $W4$ of any part of the stream, and peak SNR. Each simulation is shown using a different colour; the thick red line shows the median value of all 26 simulations.}
\label{fig:smallhotstream0.5Mpc}
\end{figure*}

\begin{figure*}
\centering  
  \subfloat[]{\includegraphics[height=55mm]{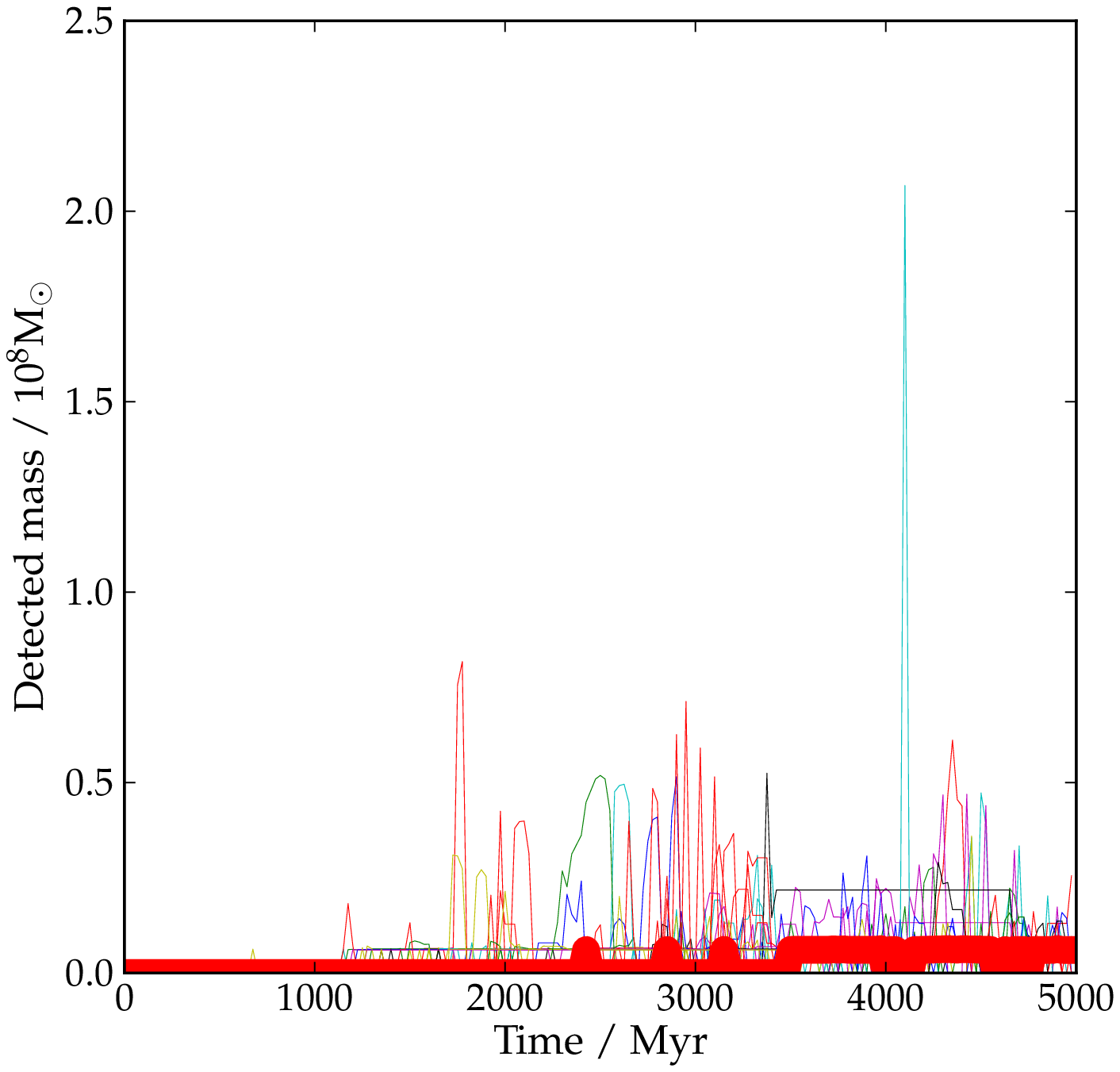}}
  \subfloat[]{\includegraphics[height=55mm]{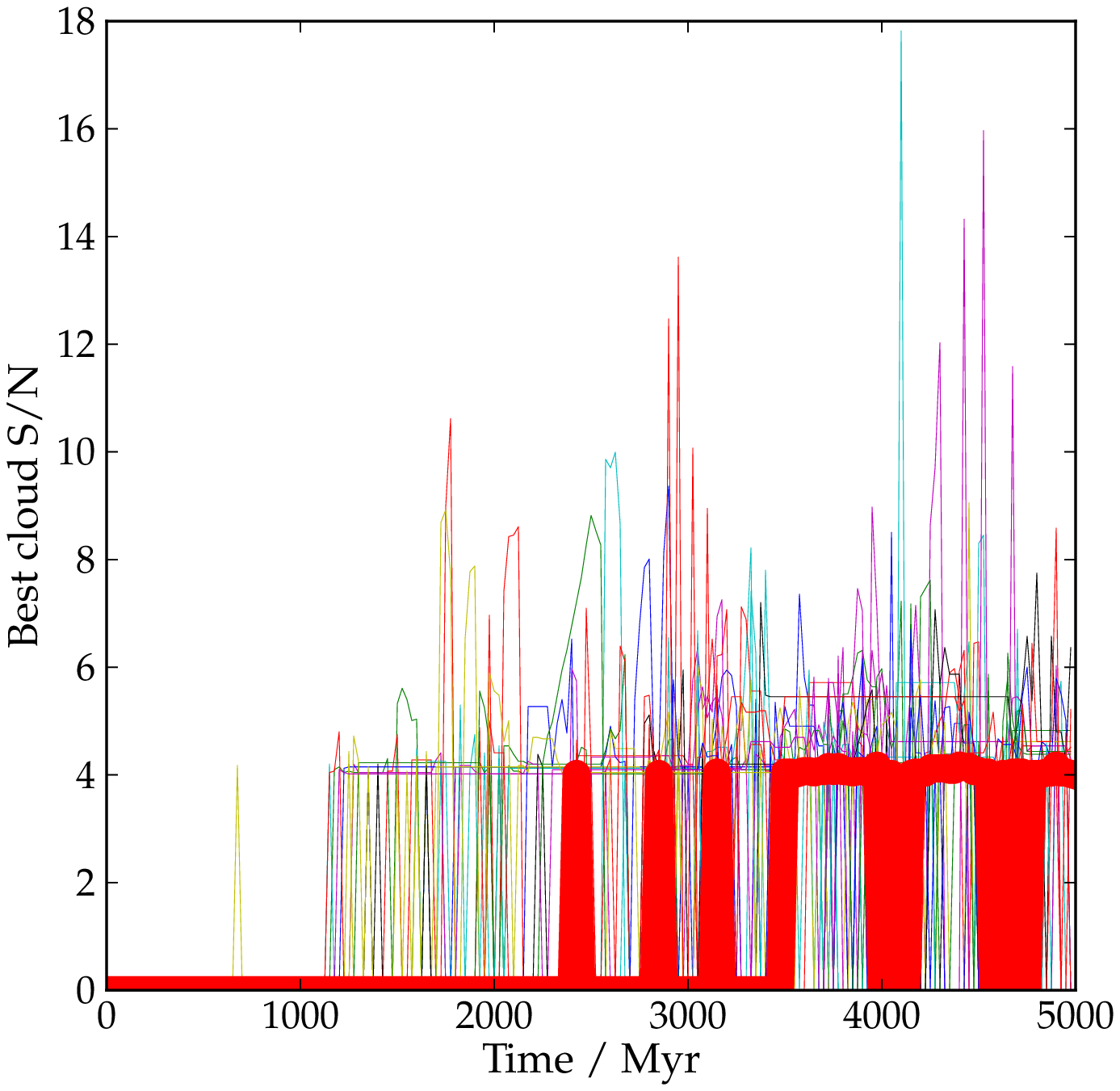}} 
  \subfloat[]{\includegraphics[height=55mm]{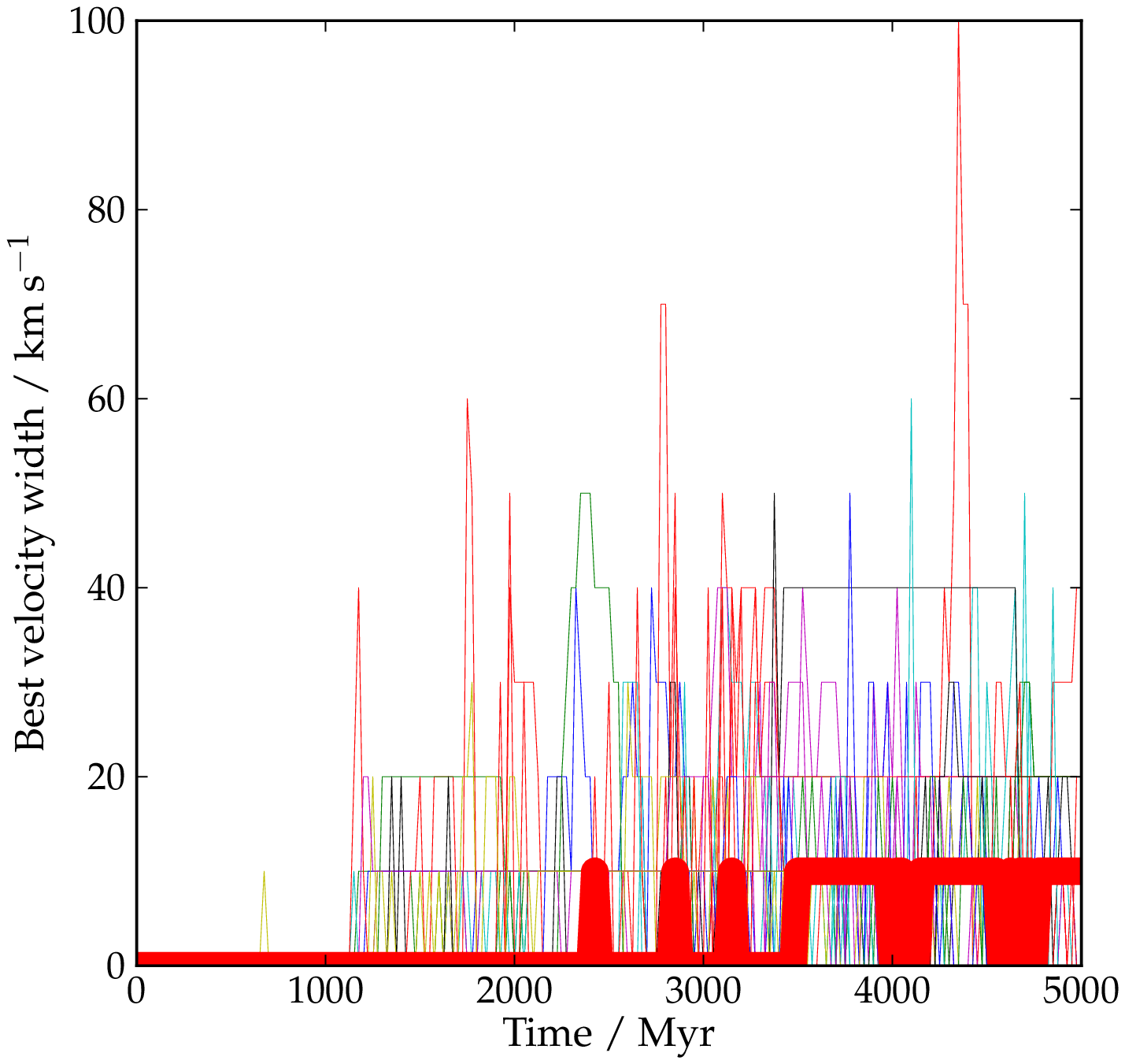}}\\
  \subfloat[]{\includegraphics[height=55mm]{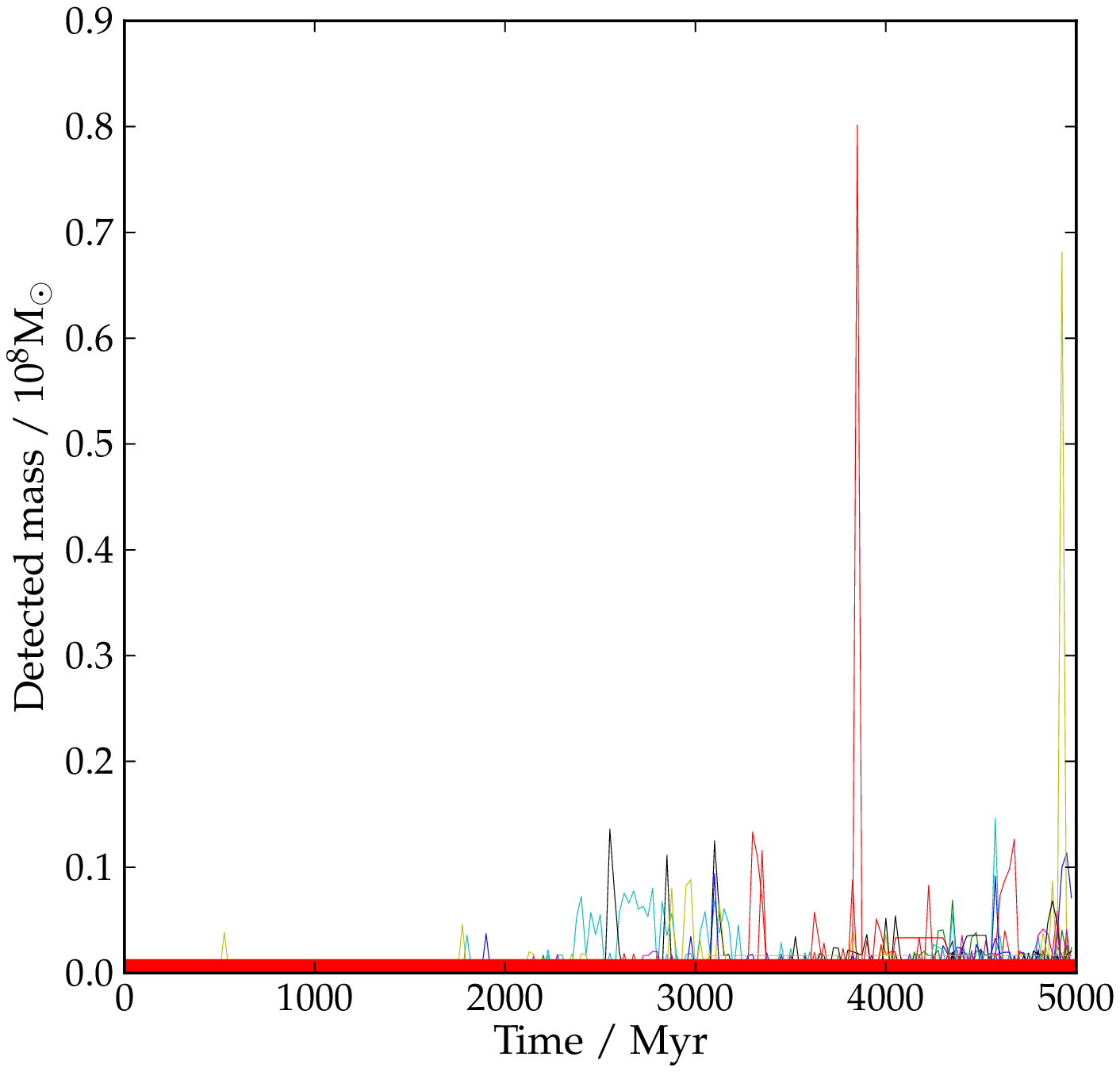}}
  \subfloat[]{\includegraphics[height=55mm]{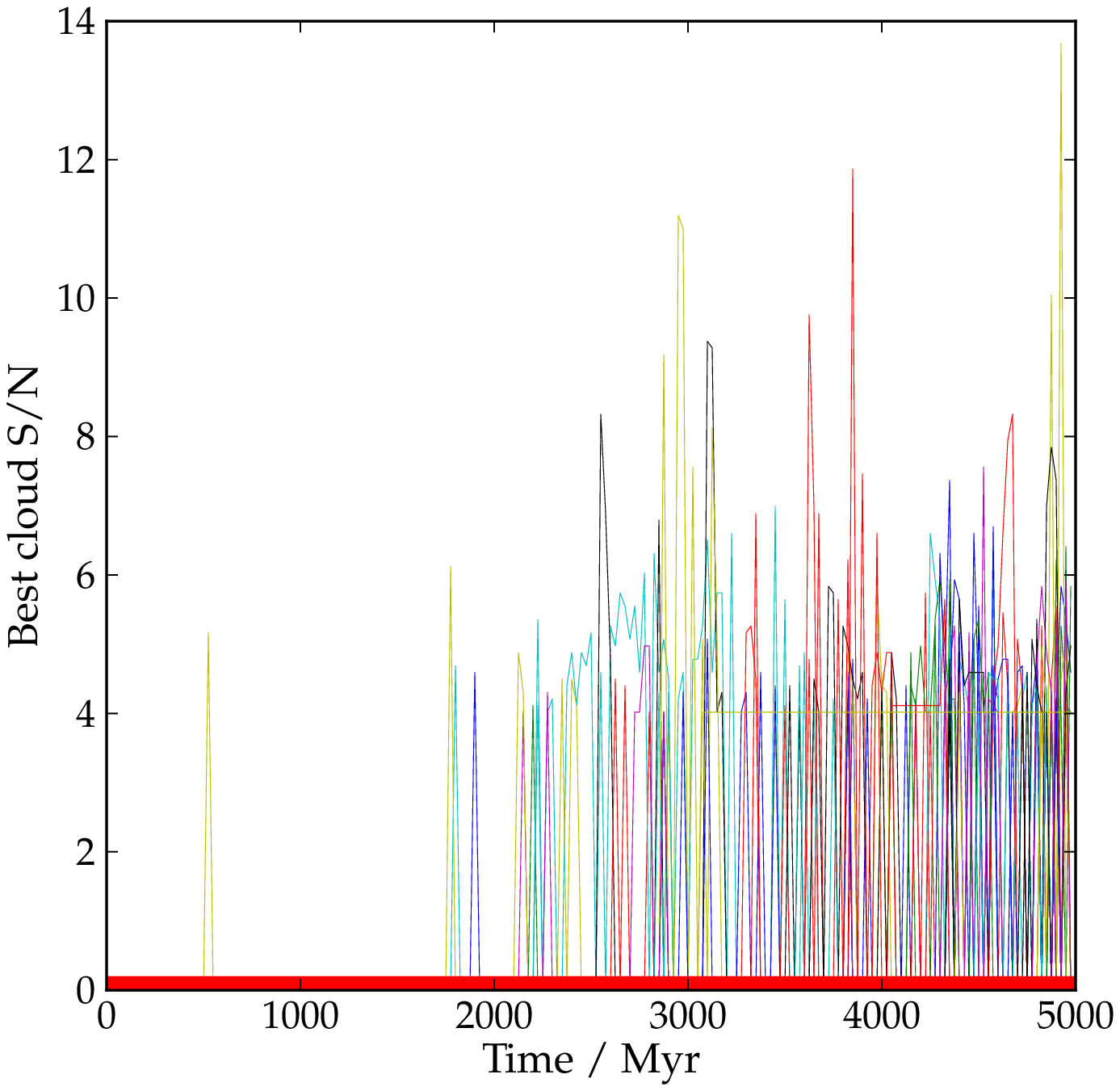}} 
  \subfloat[]{\includegraphics[height=55mm]{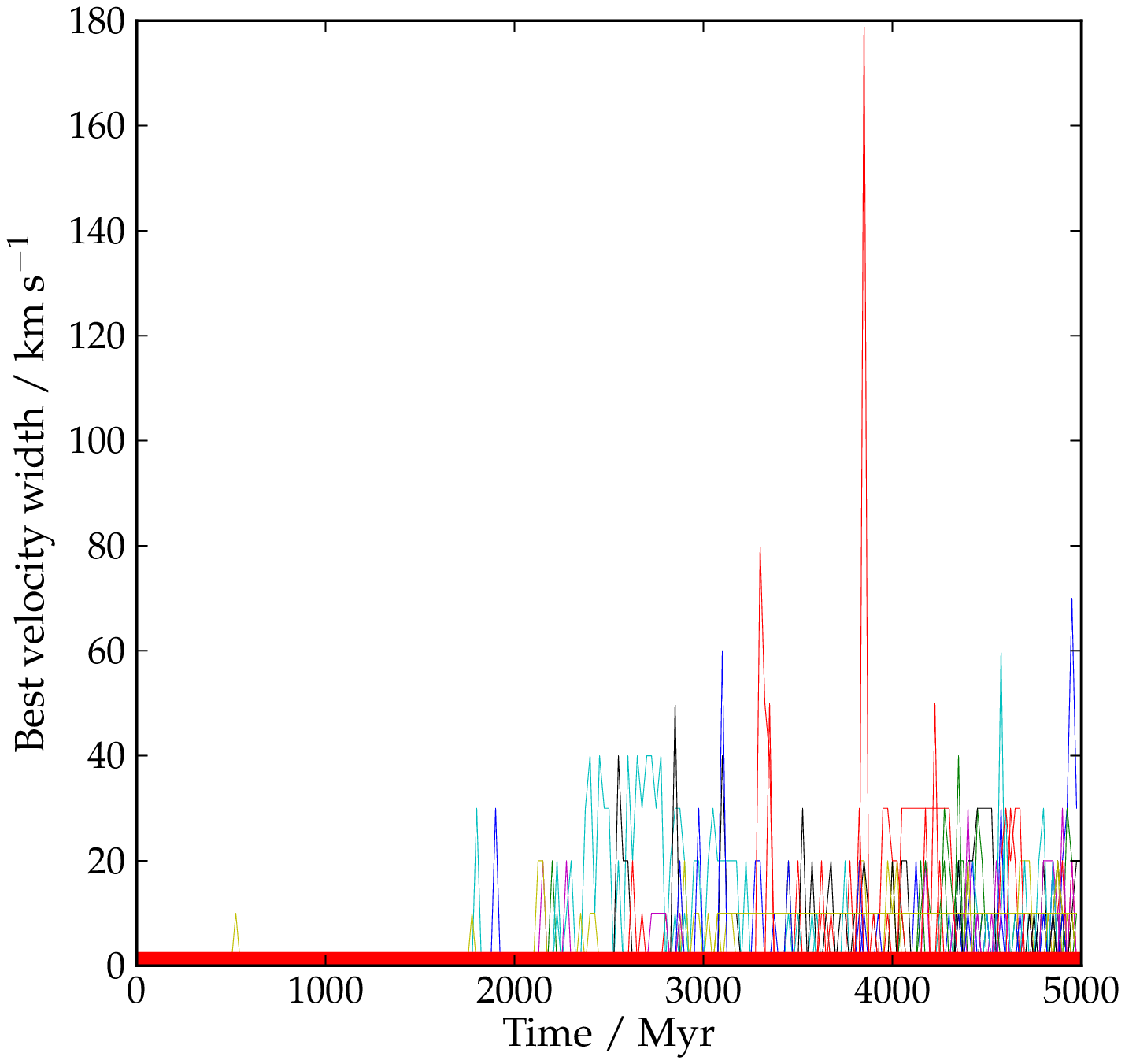}}
\caption[]{Evolution of the properties of the isolated cloud with the highest velocity width, produced from the 4$\times$10$^{8}$ \Msolar{}, 5100 K streams initially at 0.5 Mpc from the cluster centre. The top panel shows the measurements using an ALFALFA sensitivity level and beam size while the bottom panel shows the equivalent sensitivity of AGES. From left to right : detected mass, peak SNR, and $W50$. Each simulation is shown using a different colour; the thick red line shows the median value of all 26 simulations.}
\label{fig:smallhotstreamclouds0.5Mpc}
\end{figure*}

\begin{figure*}
\centering
\includegraphics[width=160mm]{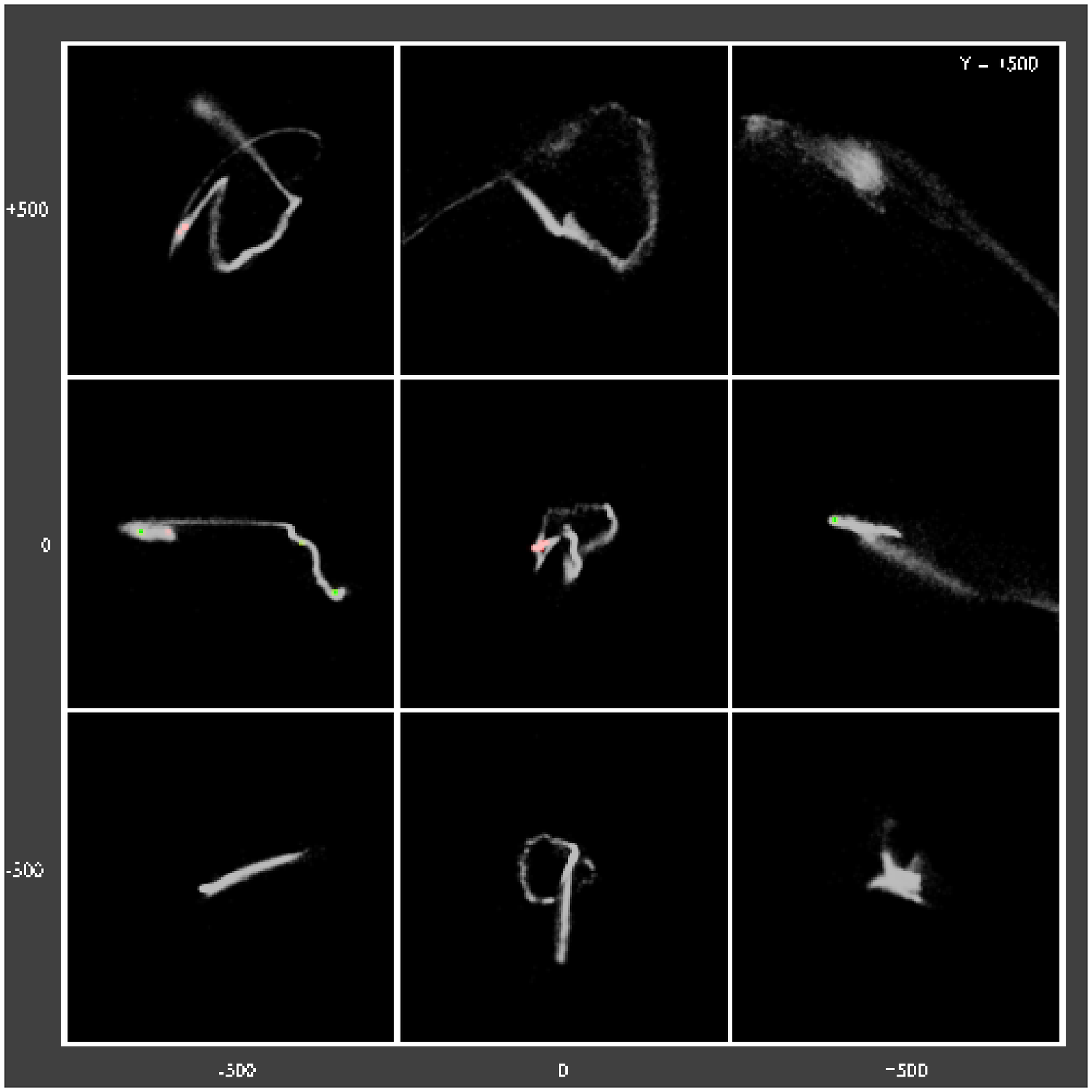}
\caption[hienv]{Final timestep (5 Gyr) of the simulation of a sample of 4$\times$10$^{8}$ \Msolar{}, 5100 K streams entering the cluster from an initial distance of 0.5 Mpc. Each box spans 1 Mpc and is centred on the mean particle position. White shows the raw particle data. Red shows all gridded data in which the emission would exceed a SNR of 4.0 with an ALFALFA sensitivity level; green indicates detectable clouds at least 100 kpc from the nearest other detection. Movies of the simulations can be see at \href{http://tinyurl.com/gos27cs}{this url} : http://tinyurl.com/gos27cs.}
\label{fig:lmcmovied}
\end{figure*}

\begin{figure*}
\centering  
  \subfloat[]{\includegraphics[height=55mm]{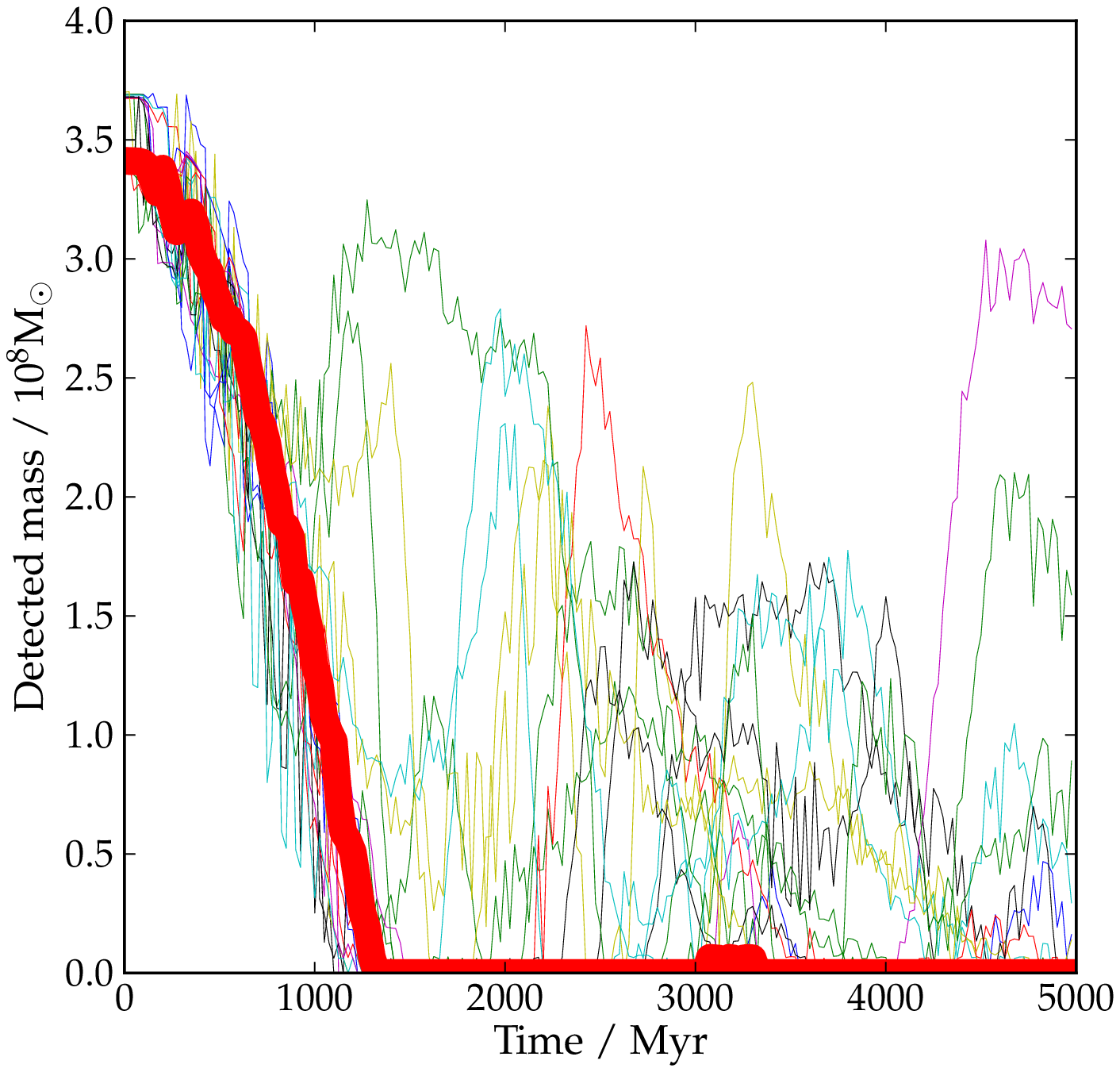}}
  \subfloat[]{\includegraphics[height=55mm]{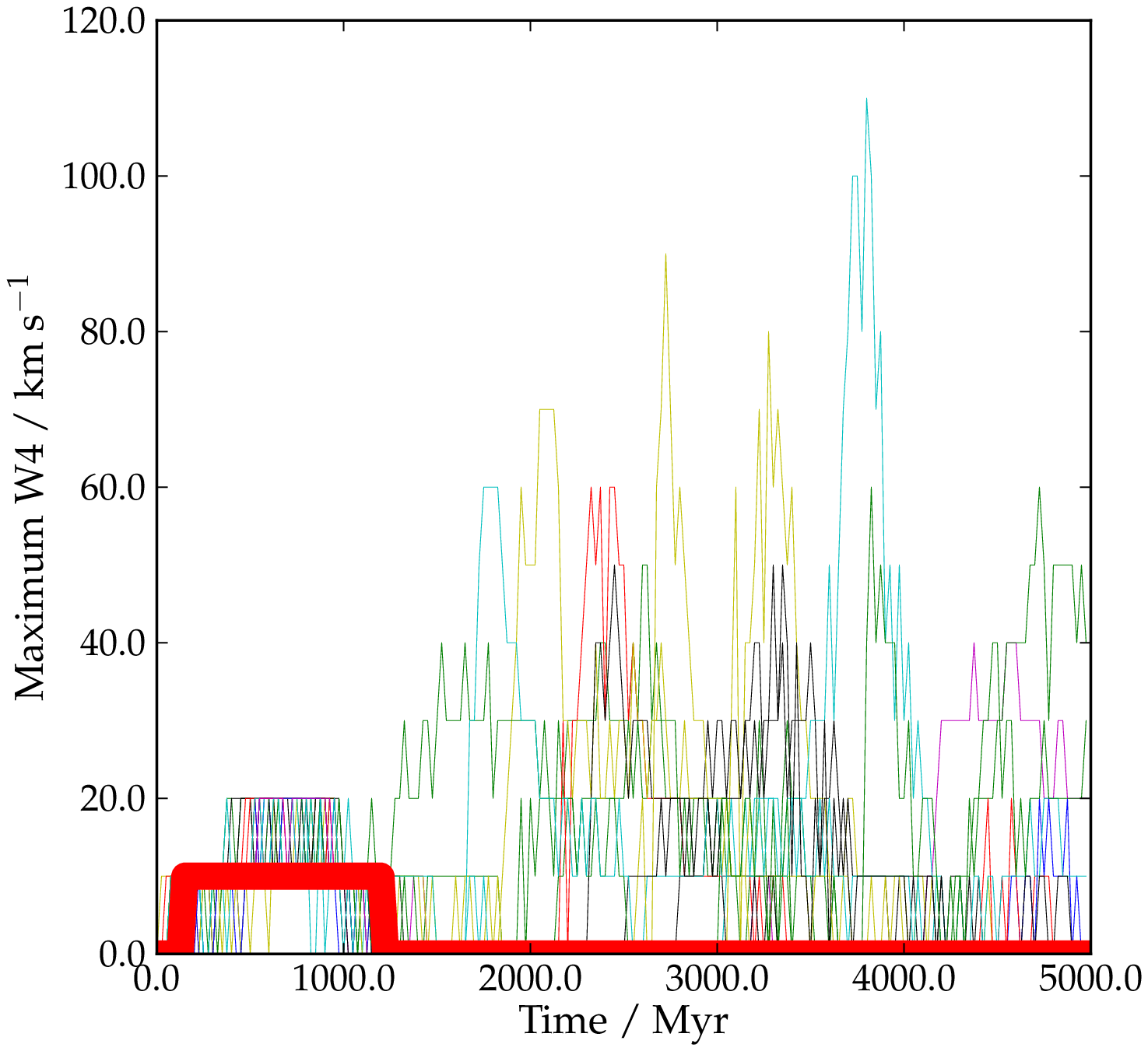}} 
  \subfloat[]{\includegraphics[height=55mm]{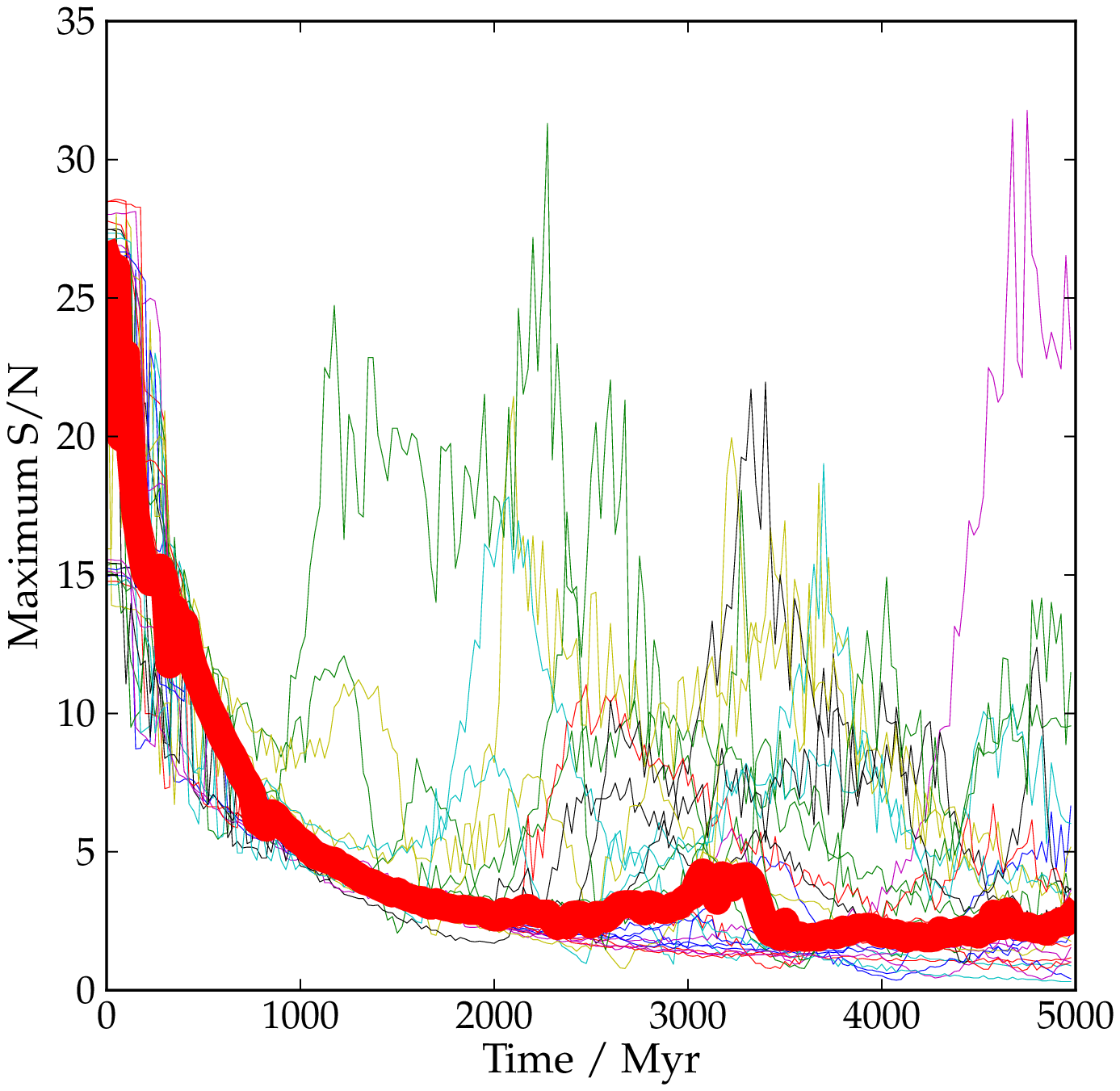}}\\
  \subfloat[]{\includegraphics[height=55mm]{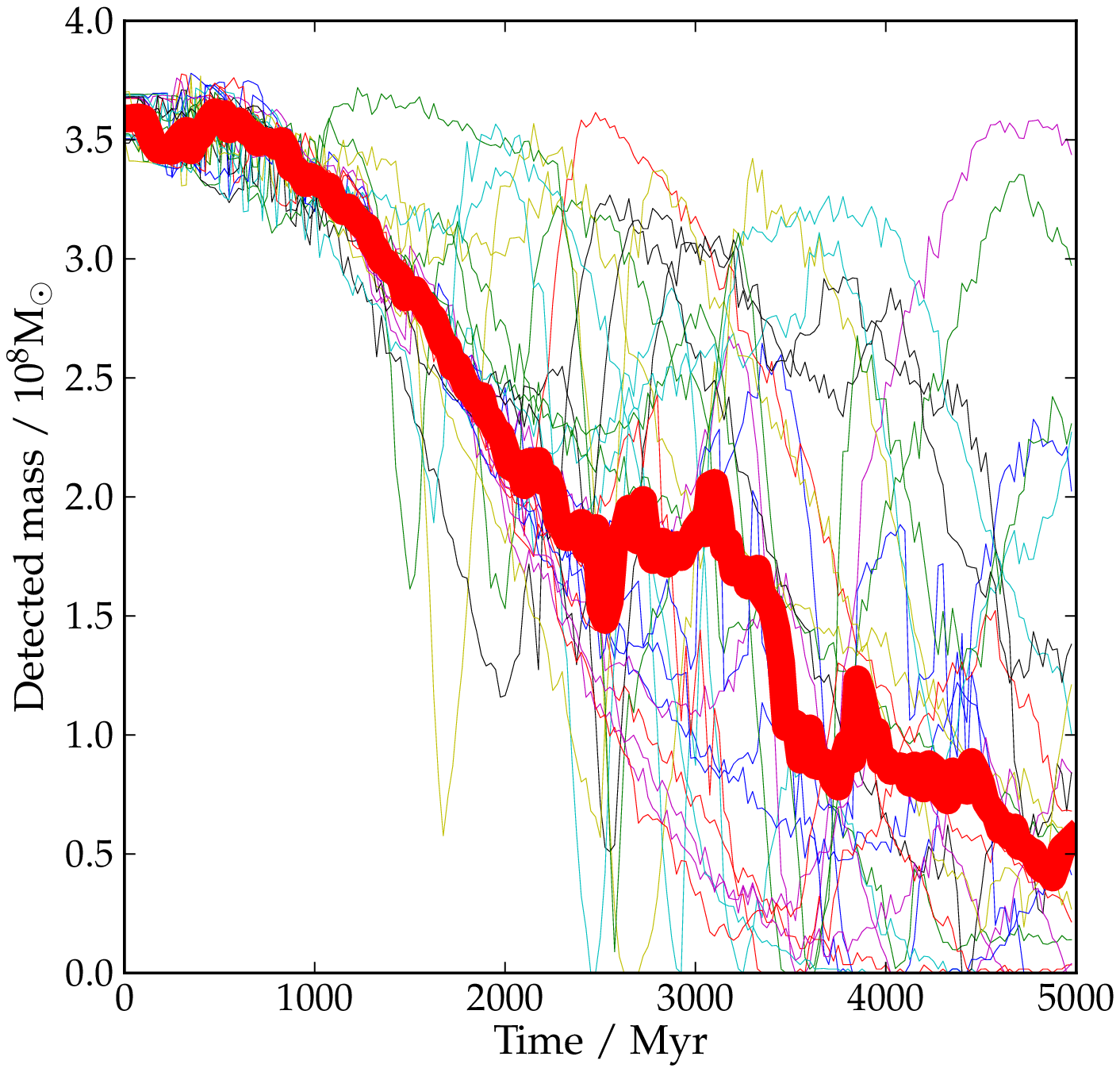}}
  \subfloat[]{\includegraphics[height=55mm]{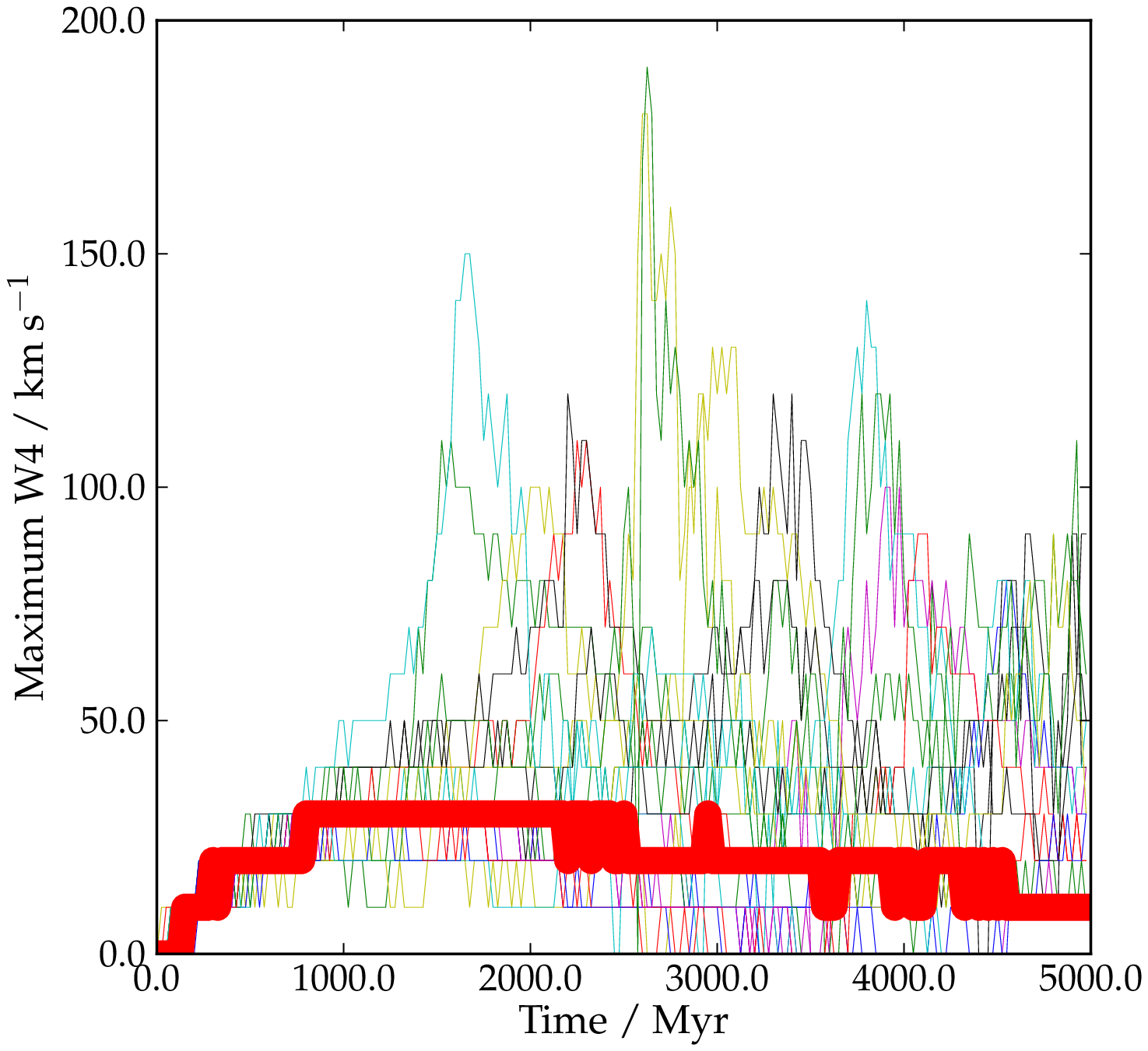}} 
  \subfloat[]{\includegraphics[height=55mm]{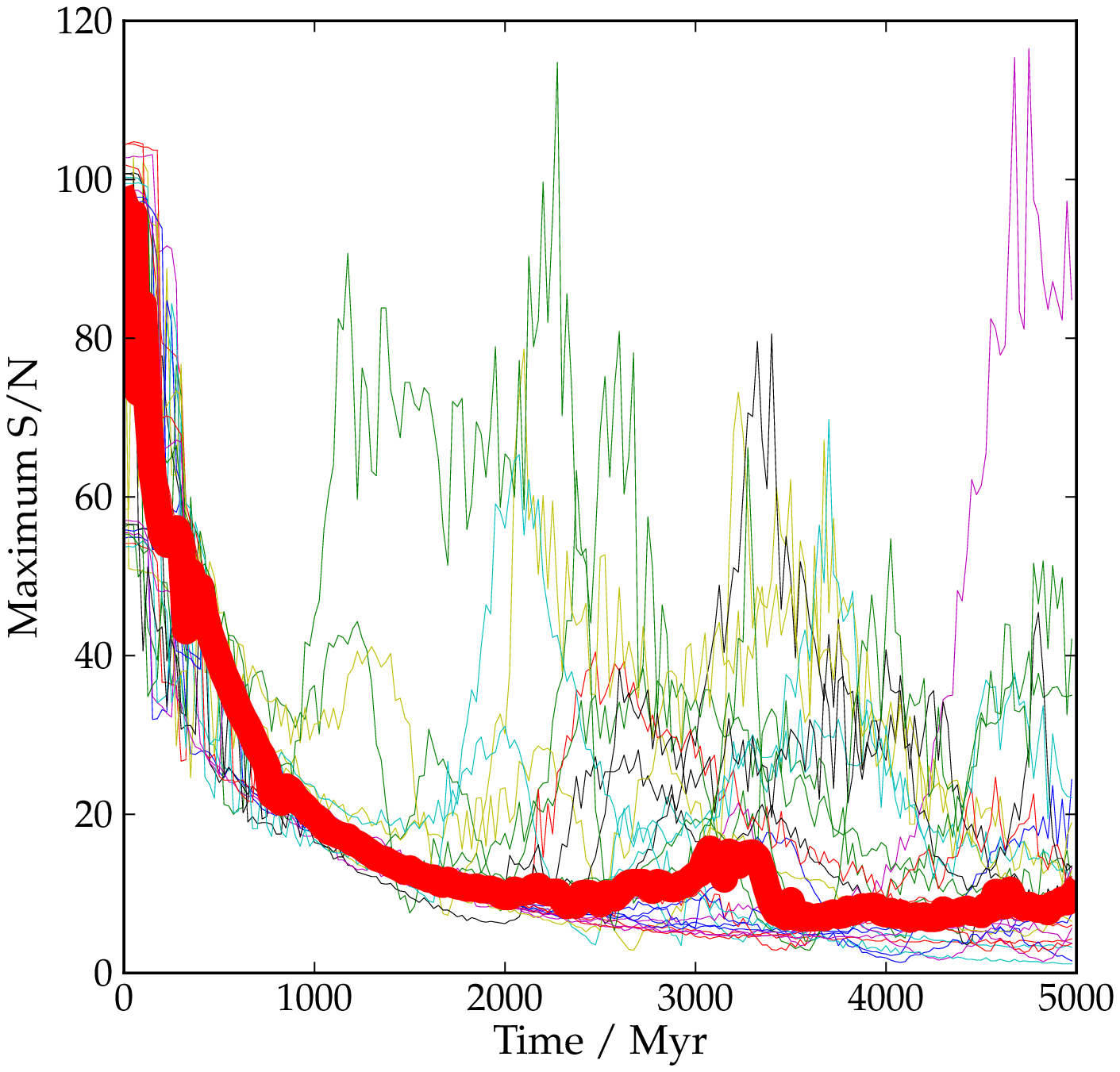}}
\caption[]{Evolution of the properties of 4$\times$10$^{8}$ \Msolar{}, 5100 K streams initially at 1.0 Mpc from the cluster centre. The top panel shows the measurements using an ALFALFA sensitivity level and beam size while the bottom panel shows the equivalent sensitivity of AGES. From left to right : detected mass, maximum $W4$ of any part of the stream, and peak SNR. Each simulation is shown using a different colour; the thick red line shows the median value of all 26 simulations.}
\label{fig:smallhotstream1.0Mpc}
\end{figure*}

\begin{figure*}
\centering 
  \subfloat[]{\includegraphics[height=55mm]{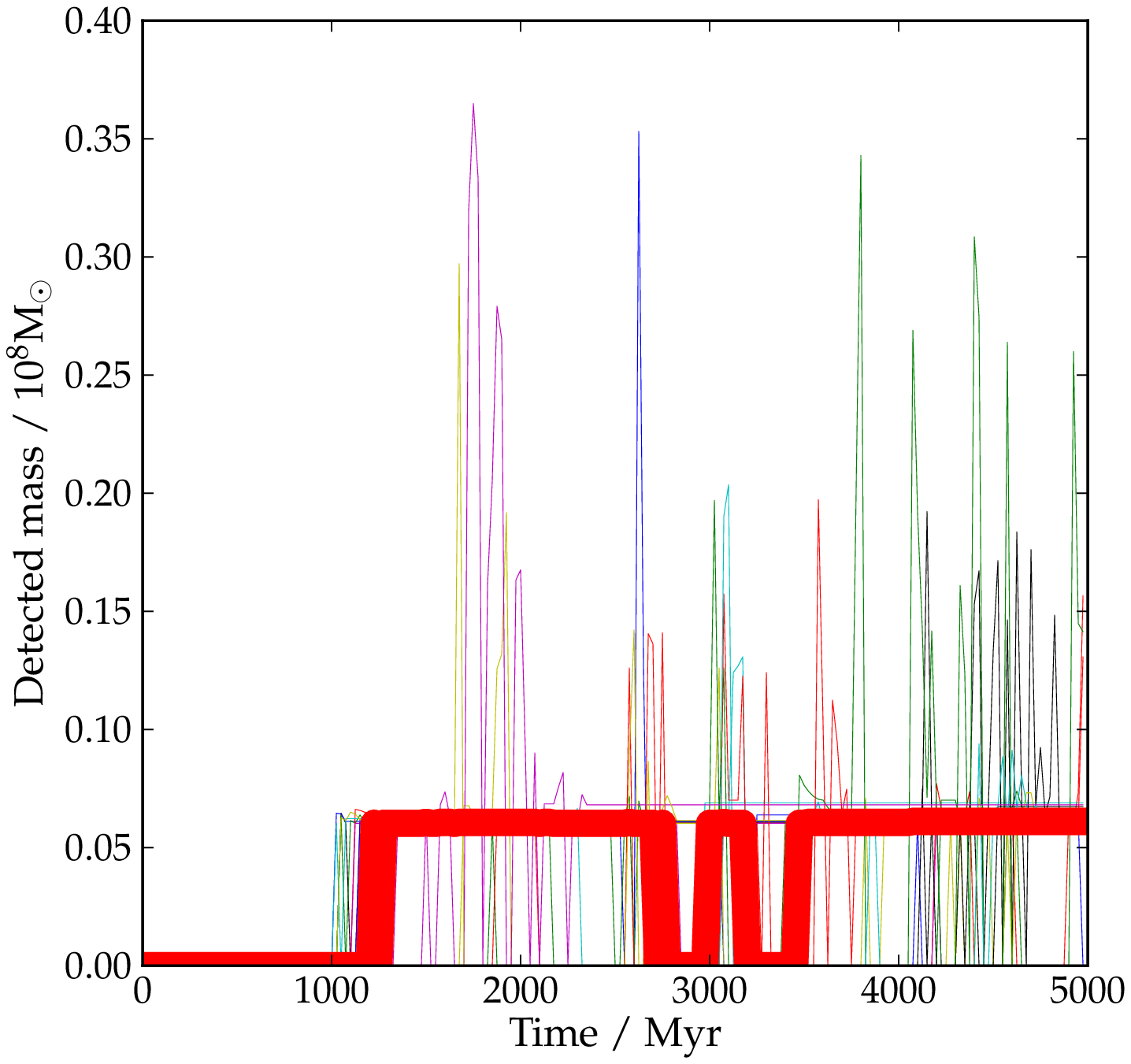}}
  \subfloat[]{\includegraphics[height=55mm]{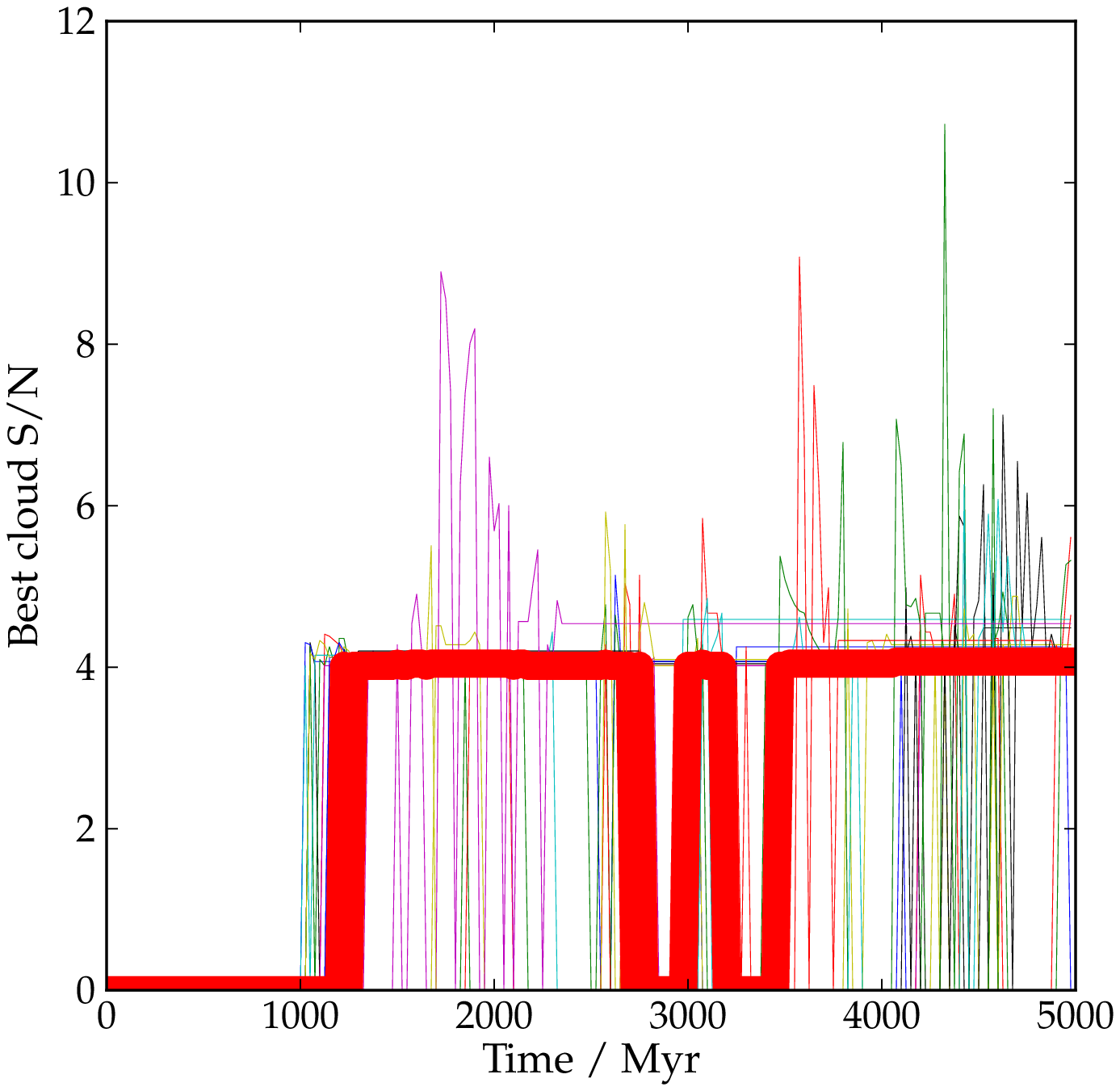}} 
  \subfloat[]{\includegraphics[height=55mm]{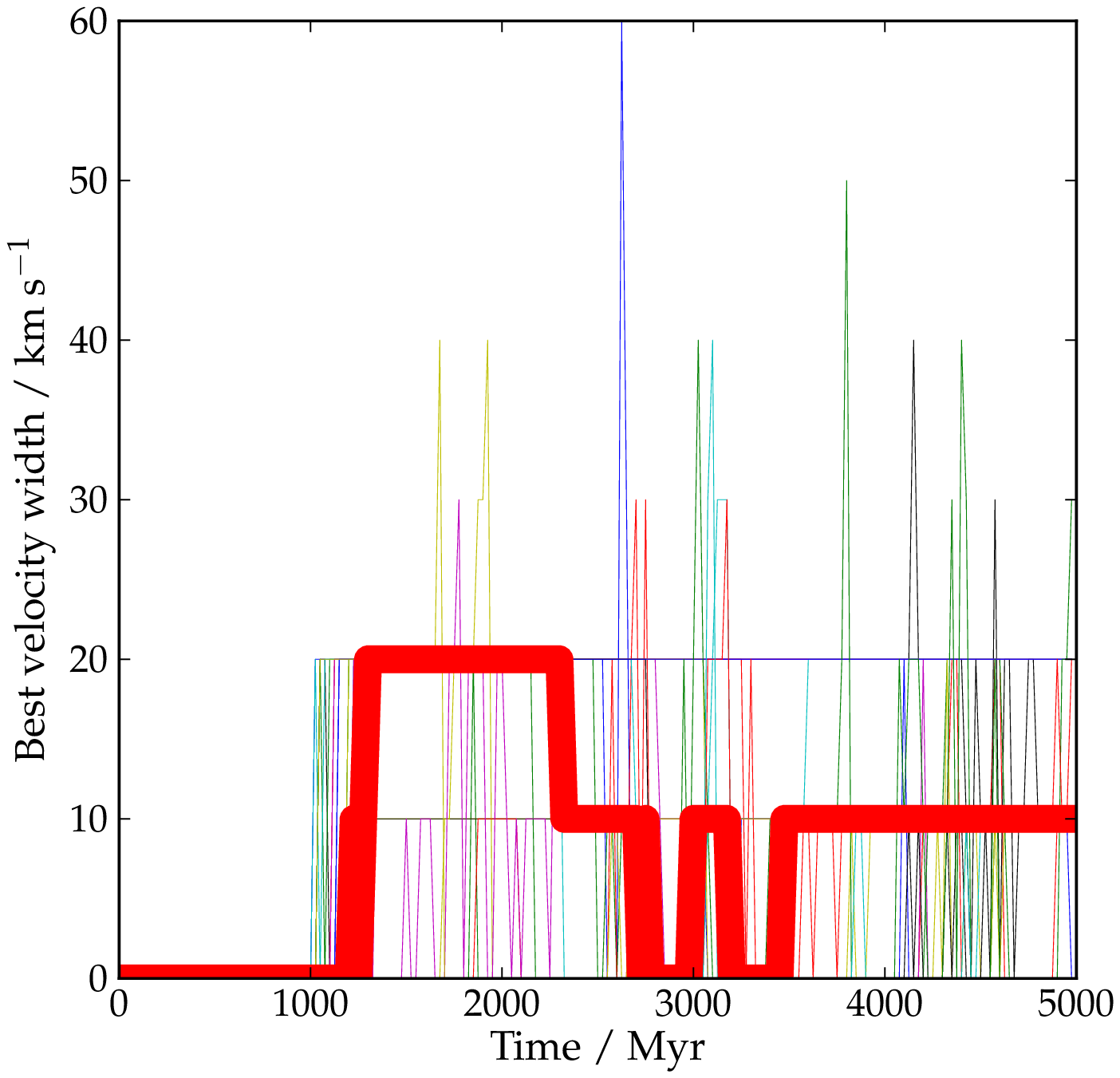}}\\
  \subfloat[]{\includegraphics[height=55mm]{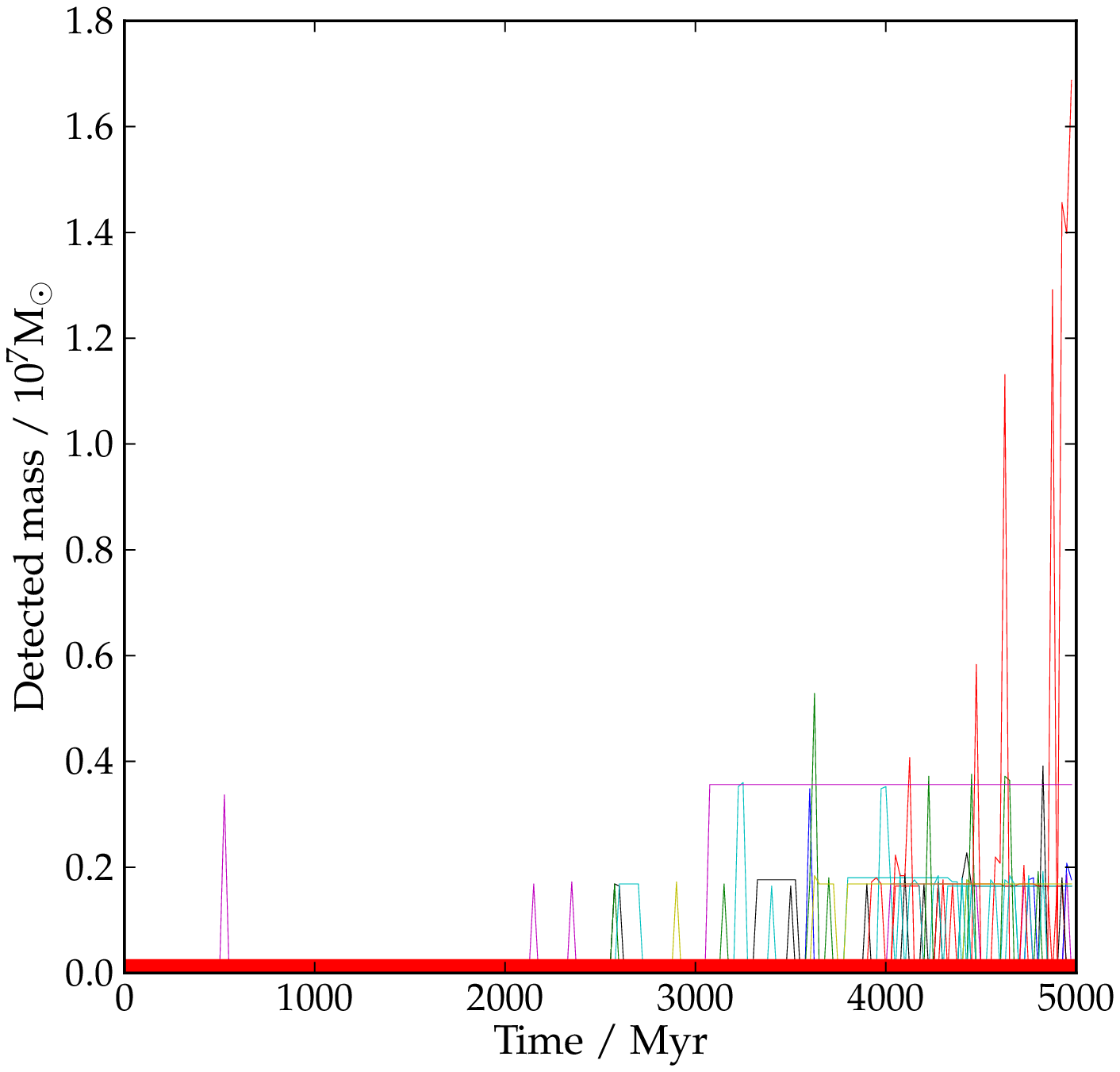}}
  \subfloat[]{\includegraphics[height=55mm]{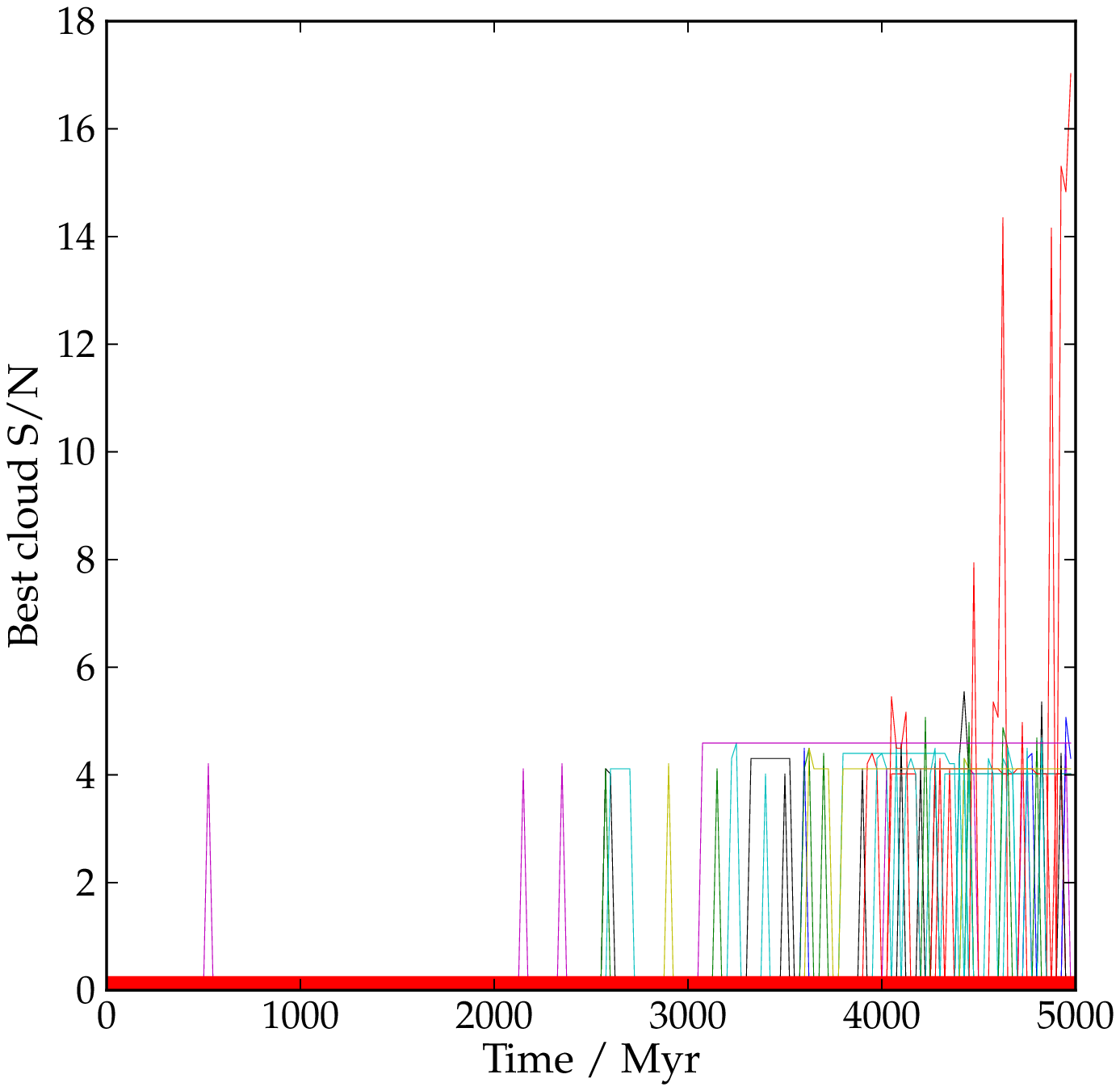}} 
  \subfloat[]{\includegraphics[height=55mm]{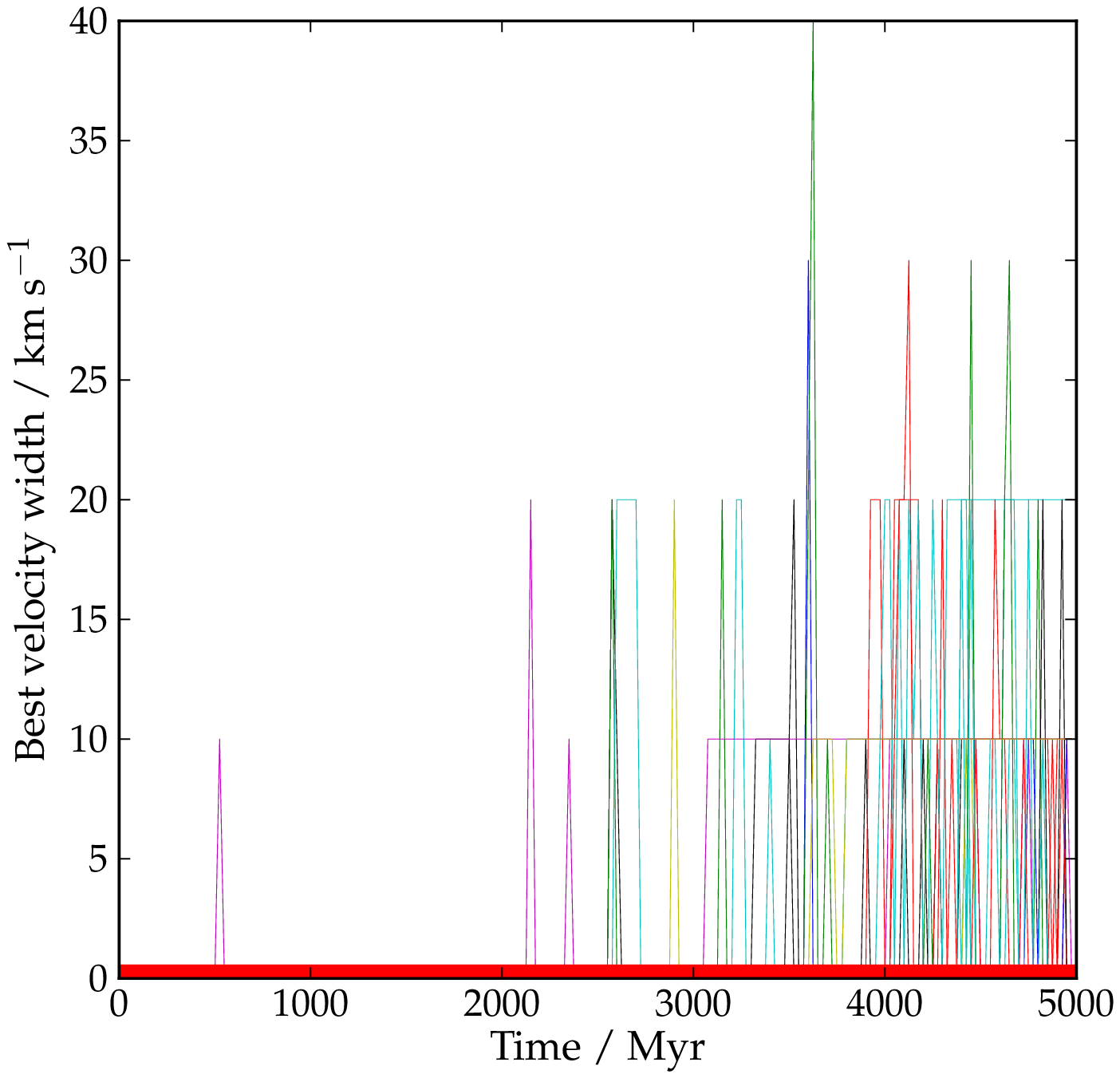}}
\caption[]{Evolution of the properties of the isolated cloud with the highest velocity width, produced from the 4$\times$10$^{8}$ \Msolar{}, 5100 K streams initially at 1.0 Mpc from the cluster centre. The top panel shows the measurements using an ALFALFA sensitivity level and beam size while the bottom panel shows the equivalent sensitivity of AGES. From left to right : detected mass, peak SNR, and $W50$. Each simulation is shown using a different colour; the thick red line shows the median value of all 26 simulations.}
\label{fig:smallhotstreamclouds1.0Mpc}
\end{figure*}

\begin{figure*}
\centering
\includegraphics[width=160mm]{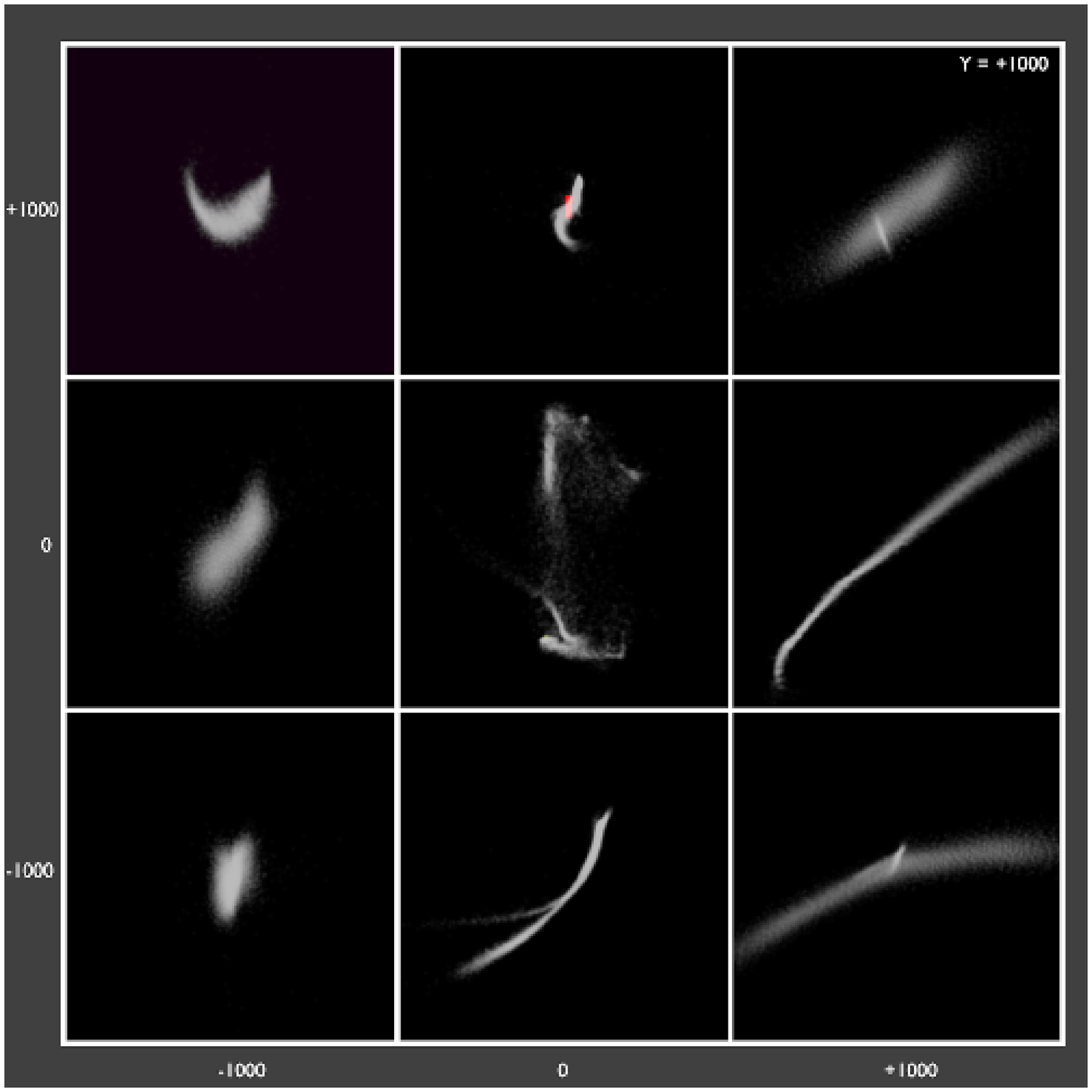}
\caption[hienv]{Final timestep (5 Gyr) of the simulation of a sample of 4$\times$10$^{8}$ \Msolar{}, 5100 K streams entering the cluster from an initial distance of 1.0 Mpc. Each box spans 1 Mpc and is centred on the mean particle position. White shows the raw particle data. Red shows all gridded data in which the emission would exceed a SNR of 4.0 with an ALFALFA sensitivity level; green indicates detectable clouds at least 100 kpc from the nearest other detection. Movies of the simulations can be see at \href{http://tinyurl.com/j4tdrh8}{this url} : http://tinyurl.com/j4tdrh8.}
\label{fig:lmcmovied}
\end{figure*}

\label{lastpage}

\end{document}